\documentclass[structabstract]{aa}  

\usepackage{color}
\usepackage{graphicx}
\usepackage{txfonts}
\usepackage{longtable}
\usepackage{natbib}

\begin{document}

   \title{Gravitational potential and X-ray luminosities of 
   early-type galaxies observed with XMM-Newton and Chandra}

   \subtitle{}

   \author{R. Nagino
          \inst{1}
          \and
          K. Matsushita\inst{2}
          }

   \institute{Tokyo University of Science, 1-3 Kagurazaka,
   Shinjyuku-ku, Tokyo, Japan, 162-8601\\
              \email{j1207705@ed.kagu.tus.ac.jp}
         \and
             Tokyo University of Science, 1-3 Kagurazaka,
   Shinjyuku-ku, Tokyo, Japan, 162-8601\\
             \email{matusita@rs.kagu.tus.ac.jp}
             }

   \titlerunning{Gravitational potential and X-ray luminosities of early-type galaxies}

   \date{Received ; accepted }

  \abstract
   {}
   {We study dark matter content in early-type galaxies and
   investigate whether X-ray luminosities of early-type galaxies are
   determined by the surrounding gravitational potential.}
   {We derived gravitational mass profiles of 22 early-type galaxies
   observed with XMM-Newton and Chandra.}
   {Sixteen galaxies show constant or
   decreasing radial temperature profiles, and their X-ray luminosities
   are consistent with kinematical energy input from stellar mass
   loss. The temperature profiles of the other 6 galaxies increase with
   radius, and their X-ray luminosities are significantly higher.
   The integrated mass-to-light ratio of each galaxy is constant at
   that of stars within $0.5\sim 1 r_e$, and increases with radius, where
   $r_e$ is the effective radius of a galaxy. The scatter of the central
   mass-to-light ratio of galaxies was less in K-band light. At 3$r_e$, the
   integrated mass-to-light ratios of galaxies with flat or decreasing
   temperature profiles are twice the value at 0.5$r_e$, where the stellar
   mass dominates, and at 6$r_e$, these increase to three times the value
   at 0.5$r_e$.}
   {This feature should reflect common dark and stellar mass distributions
   in early-type galaxies: Within 3$r_e$, the mass of dark matter is similar
   to the stellar mass, while within 6$r_e$, the former is larger than the
   latter by a factor of two. By contrast, X-ray luminous galaxies have
   higher gravitational mass in the outer regions than X-ray faint
   galaxies. We describe these X-ray luminous galaxies as the central
   objects of large potential structures; the presence or absence of this
   potential is the main source of the large scatter in the X-ray
   luminosity.}

   \keywords{galaxies: elliptical and lenticular, cD --- galaxies:
   ISM --- X-rays: galaxies --- X-rays: ISM}

   \maketitle

\section{Introduction}

The bottom-up hierarchical theory of galaxy formation
predicts that galaxies should be embedded in massive dark matter halos 
(e.g. \citealt{Nava1997}). The presence of dark matter in spiral
galaxies has been revealed through observations of stellar rotation
curves \citep{Rubin1978,vanA1985}. However, the study
of halos in early-type galaxies is limited due to a lack of suitable
and easy tracers such as rotation curves. Recently, observations of
stellar velocity dispersion of early-type galaxies have reached to
$1\sim2~ r_e$, and a correlation between mass-to-light ratio and optical
luminosity was found (e.g., \citealt{Kron2000,Gerh2001}). Here, $r_e$ is
the effective radius of a galaxy. For a small
number of galaxies, mass profiles up to several $r_e$ have been obtained
using test particles such as globular clusters or planetary nebulae
(e.g., \citealt{Roma2003,Chak2008}). 

X-ray observations provide a powerful tool to study the shape of the
gravitational potential, and hence dark matter distributions, of
early-type galaxies. These galaxies have a hot, X-ray-emitting
interstellar medium (ISM), which is considered to be gravitationally
confined (e.g., \citealt{Form1985,Matsu2001,Fuka2006}). The ISM
luminosities of early-type galaxies vary by two
orders of magnitude for the same optical B-band luminosity
($L_{\rm{B}}$) (e.g., \citealt{Cani1987,Beui1999,Matsu2000,Matsu2001}),
whereas optical observations indicate that these
galaxies are dynamically uniform systems
(\citealt{Djor1987,Bend1993,Kron2000}). A key to solving this
discrepancy is the extended X-ray emissions that have been detected
around X-ray luminous early-type galaxies (\citealt{Matsu1998,Matsu2001}).
 On the basis of ROSAT data, \citet{Matsu2001}
discovered that, for most early-type galaxies, ISM luminosities within
the optical radius agree with kinematical energy input from the
stellar mass loss $L_{\sigma}$. These galaxies have flat or
decreasing temperature profiles against radius. By contrast, galaxies
with ISM luminosities much larger than $L_{\sigma}$ show largely extended
emission with a radius of a few tens of $r_e$ with positive temperature
gradients. 
XMM-Newton RGS observations provided evidence of a
weak positive temperature gradient in the inner region of the ISM
in NGC 4636, which has a much
higher ISM luminosity than $L_{\sigma}$ \citep{Xu2002}.
The correlation between the temperature gradient and
spatial distribution was confirmed with Chandra observations
(\citealt{Fuka2006}). These features suggest that X-ray luminous early-type
galaxies commonly sit in the center of a large-scale (a few hundred
kpc) potential well, which leads to their high luminosities. Other
galaxies may lack such a large-scale potential. On the basis of the
extent of the ISM brightness, \citet{Matsu2001} denoted galaxies as
either X-ray extended galaxies or X-ray compact galaxies. The
gravitational mass profile of cD galaxies also shows two distinct
contributions that can be assigned to the gravitational potential of
the cD galaxy and that of the cluster (\citealt{Matsu2002}). Thus,
the only way to measure the gravitational mass profile of pure
early-type galaxies is to observe the X-ray compact galaxies. 

With ROSAT PSPC observations, \citet{O'Sul2003} found that
 the relation between the  central stellar velocity dispersion 
and the temperature obtained from X-ray emission 
is similar to that for clusters and 
the relation between the X-ray luminosity and the temperature has a
steep slope comparable with that found for galaxy groups.

Chandra and XMM-Newton have already observed several tens of
early-type galaxies. Most of the analysis was done for X-ray luminous
and extended objects, and the number of X-ray compact galaxies with
accurately derived gravitational mass profiles is still limited. For
X-ray luminous galaxies, mass profiles are easily obtained over 10$r_e$
(\citealt{Fuka2006}; \citealt{Hump2006}). Using XMM-Newton, even
for several X-ray compact galaxies, mass profiles can be derived up to
several $r_e$ (\citealt{Fuka2006}), and observed gravitational mass
profiles of X-ray extended and X-ray compact galaxies are similar when
plotted against radius in units of $r_{200}$. The dark matter profiles are
well described by the NFW model (\citealt{Nava1996,Nava1997}), which is
based on numerical simulations assuming cold dark
matter (CDM) as well as galaxy clusters (\citealt{Fuka2006}; \citealt{Zapp2006}). 
Chandra observations suggested that the
shape of the X-ray isophotes is unrelated to the shape of the
gravitational potential \citep{Dieh2007,Dieh2008}.

In this study, we obtained gravitational mass profiles of 22
early-type galaxies observed with XMM-Newton and Chandra to
investigate whether X-ray luminosities of early-type galaxies are
determined by the surrounding gravitational potential and to study
dark matter content in early-type galaxies. Throughout this paper, we
adopt the solar abundances of \citet{Ande1989}. Unless
otherwise specified, errors are quoted at $1\sigma$ confidence.

\section{Targets and Observations}

We analyzed archival data of 22 early-type galaxies with distances
less than 40 Mpc and B-band luminosities $L_B>10^{10.1}L_{\odot}$
observed with XMM-Newton. The values of $L_B$ and the
distances to the galaxies are taken from \citet{Tully1988}. The
characteristics and observational log of the sample galaxies are
summarized in Tables \ref{tab:tar_list} and \ref{tab:obs_data},
respectively. The sample includes 15 elliptical and 7 S0
galaxies. Eight are located in the Virgo Cluster, 2 are in the Fornax
Cluster, and the others are either in the field or in small
groups. All observations were carried out with MOS1, MOS2, and PN
together. 

We also used Chandra data for 19 of the sample galaxies with good
signal-to-noise ratios to derive the mass profile at their central
regions. As summarized in Table \ref{tab:obs_data}, 15 galaxies were
observed with ACIS-S, and 4 galaxies were observed with ACIS-I. 

\begin{table*}
\caption{Galaxy sample in the XMM-Newton archive data}
\label{tab:tar_list}
\centering
\begin{tabular}{lccccccl}
\hline \hline
Galaxy & Type$^a$ & $D^b$ & $r_e$$^c$ & $\log L_B$$^d$ & $\sigma$$^e$ & $N_H$$^f$ & Note\\
 & & (Mpc) & (arcmin) & ($L_{\odot}$) & (km/s) & ($10^{20}{\rm cm}^{-2}$) & \\ \hline
IC1459 & -5.0 & 20.0 & 0.58 & 10.42 & 311 & 1.18 & $X_C$\\
NGC720 & -5.0 & 20.3 & 0.60 & 10.34 & 240 & 1.54 & $X_C$\\
NGC1316 & -2.0 & 16.9 & 1.35 & 10.78 & 250 & 1.89 & $X_C$,Fornax\\
NGC1332 & -2.0 & 17.7 & 0.47 & 10.22 & 319 & 2.23 & $X_C$\\
NGC1395 & -5.0 & 20.0 & 0.81 & 10.33 & 254 & 1.99 & $X_E$\\
NGC1399 & -5.0 & 16.9 & 0.68 & 10.31 & 362 & 1.34 & $X_E$,Fornax\\
NGC1549 & -5.0 & 13.4 & 0.78 & 10.10 & 213 & 1.46 & $X_C$\\
NGC3585 & -5.0 & 21.6 & 0.60 & 10.56 & 227 & 5.58 & $X_C$\\
NGC3607 & -2.0 & 19.9 & 0.73 & 10.41 & 223 & 1.48 & $X_C$\\
NGC3665 & -2.0 & 32.4 & 0.48 & 10.54 & 186 & 2.06 & $X_C$\\
NGC3923 & -5.0 & 25.8 & 0.83 & 10.68 & 241 & 6.21 & $X_C$\\
NGC4365 & -5.0 & 16.8 & 0.83 & 10.40 & 266 & 1.62 & $X_C$,Virgo\\
NGC4382 & -2.0 & 16.8 & 0.91 & 10.64 & 187 & 2.52 & $X_C$,Virgo\\
NGC4472 & -5.0 & 16.8 & 1.74 & 10.92 & 302 & 1.66 & $X_E$,Virgo\\
NGC4477 & -2.0 & 16.8 & 0.63 & 10.14 & 175 & 2.64 & $X_C$,Virgo\\
NGC4526 & -2.0 & 16.8 & 0.74 & 10.41 & 260 & 1.65 & $X_C$,Virgo\\
NGC4552 & -5.0 & 16.8 & 0.49 & 10.35 & 262 & 2.57 & $X_C$,Virgo\\
NGC4636 & -5.0 & 17.0 & 1.48 & 10.46 & 208 & 1.81 & $X_E$,Virgo\\
NGC4649 & -5.0 & 16.8 & 1.15 & 10.74 & 343 & 2.20 & $X_C$,Virgo\\
NGC5044 & -5.0 & 38.9 & 0.89 & 10.60 & 237 & 4.93 & $X_E$\\
NGC5322 & -5.0 & 31.6 & 0.56 & 10.86 & 233 & 1.81 & $X_C$\\
NGC5846 & -5.0 & 28.5 & 1.05 & 10.66 & 251 & 4.26 & $X_E$\\
\hline
\multicolumn{8}{l}{$^a$Morphological type code from \citet{Tully1988}.}\\
\multicolumn{8}{l}{$^b$Distance to the galaxy from \citet{Tully1988}.}\\
\multicolumn{8}{l}{$^c$Effective radius from RC3 Catalog 
 \citep{deVau1991}.}\\
\multicolumn{8}{l}{$^d$Total B-band luminosity from \citet{Tully1988}.}\\
\multicolumn{8}{l}{$^e$Central stellar velocity dispersion from
 \citet{Prug1996}.}\\
\multicolumn{8}{l}{$^f$Column density of the Galactic absorption from
 \citet{Dick1990}.}\\
\end{tabular}
\end{table*}

\begin{table*}
\caption{Observational log of the sample galaxies}
\label{tab:obs_data}
\centering
\begin{tabular}{lccccccc}
\hline \hline
 & \multicolumn{3}{c}{XMM-Newton} & & \multicolumn{3}{c}{Chandra} \\ 
\cline{2-4} \cline{6-8}
Galaxy & ObsID$^a$ & exposure$^b$ & total counts$^c$ & & ObsID$^a$ & ACIS & exposure \\
 & & (ksec) & ($\times10^3$) & & & & (ksec) \\ \hline
IC1459  & 0135980201 & 25,26,21  & 8.9,14  & & 2196 & ACIS-S & 54  \\
NGC720  & 0112300101 & 17,18,11  & 5.6,7.1 & & 492  & ACIS-S & 38  \\
NGC1316 & 0302780101 & 51,62,29  & 32,38   & & 2022 & ACIS-S & 26  \\
NGC1332 & 0304190101 & 54,54,41  & 10,16   & & 4372 & ACIS-S & 54  \\
NGC1395 & 0305930101 & 43,46,31  & 15,21   & & 799  & ACIS-I & 15  \\
NGC1399 & 0012830101 & 3,3,3     & 7.1,10  & & 240  & ACIS-S & 43  \\
NGC1549 & 0205090201 & 8,8,6     & 0.4,1.0 & & ---  & ---    & --- \\
NGC3585 & 0071340201 & 11,11,7   & 0.6,0.9 & & 2078 & ACIS-S & 35  \\
NGC3607 & 0099030101 & 14,14,6   & 2.5,1.9 & & 2073 & ACIS-I & 38  \\
NGC3665 & 0052140201 & 23,24,19  & 2.3,4.2 & & 3222 & ACIS-I & 18  \\
NGC3923 & 0027340101 & 32,32,24  & 11,18   & & 1563 & ACIS-S & 19  \\
NGC4365 & 0205090101 & 25,25,21  & 3.8,7.1 & & 5923 & ACIS-S & 37  \\
NGC4382 & 0201670101 & 16,16,14  & 3.0,5.6 & & 2016 & ACIS-S & 39  \\
NGC4472 & 0200130101 & 79,80,--- & 300,--- & & 321  & ACIS-S & 37  \\
NGC4477 & 0112552101 & 13,13,7   & 1.7,1.9 & & ---  & ---    & --- \\
NGC4526 & 0205010201 & 20,20,16  & 1.7,3.0 & & 3925 & ACIS-S & 41  \\
NGC4552 & 0141570101 & 21,22,15  & 10,15   & & 2072 & ACIS-S & 53  \\
NGC4636 & 0111190701 & 58,58,50  & 210,320 & & 4415 & ACIS-I & 73  \\
NGC4649 & 0021540201 & 46,46,36  & 84,120  & & 785  & ACIS-S & 31  \\
NGC5044 & 0037950101 & 17,17,8   & 130,92  & & 3225 & ACIS-S & 82  \\
NGC5322 & 0071340501 & 16,15,12  & 1.0,1.5 & & ---  & ---    & --- \\
NGC5846 & 0021540501 & 13,13,9   & 25,28   & & 788  & ACIS-S & 24  \\
\hline
\multicolumn{8}{l}{$^a$Observation number of the XMM-Newton and Chandra
 data.}\\
\multicolumn{8}{l}{$^b$Exposure time of the EPIC-MOS1, MOS2, and PN,
 respectively.}\\
\multicolumn{8}{l}{$^c$Total counts within 4$r_e$ at 0.3-2.0 keV of the
 MOS (MOS1 + MOS2) and PM, respectively.}\\
\end{tabular}
\end{table*}

\section{Data Reduction}
\label{sec:dp}

\subsection{XMM-Newton}

We analyzed MOS1, MOS2, and PN data of 21 galaxies. For NGC 4472, only
MOS1 and MOS2 data were used, since PN data for this galaxy did not
exist in the archive. We used XMMSAS version 7.0.0 for the data
reduction. 

We selected events with $FLAG = 0$ and pattern smaller than 4 and 12 for
the PN and MOS, respectively. A significant fraction of XMM-Newton
observations is contaminated by soft proton flares. To filter the flares, for each
observation, we made a count rate histogram of each detector, fitted
the histogram with a Gaussian, and selected times within 2.5$\sigma$ of the
mean of the histogram. The total exposure times after screening the
flare events are summarized in Table \ref{tab:obs_data}.

The spectra were accumulated within rings centered on the center of
each galaxy. Hereafter, we denote $r$ as the projected radius from the
galaxy center. We excluded point sources with MOS1 and MOS2 count
rates larger than 0.01 ${\rm count/sec}$. The edetect\_chain command was used to
detect point sources. The response matrix file and the auxiliary
response file corresponding to each spectrum were calculated using SAS
version 7.0.0. 

\begin{figure}
  \centering
  \includegraphics[width=7.0cm]{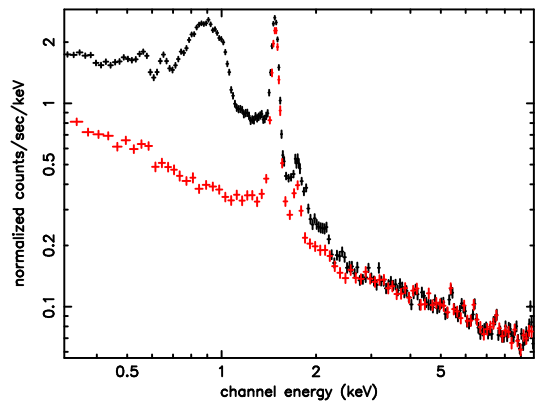}
  \caption{Raw MOS (MOS1 + MOS2) spectrum at $r$=6--14' of NGC 4636
  (black), and the adopted background spectrum (red).}
  \label{fig:bgd_n4636}
 \end{figure}

The background spectrum was calculated for each spectrum by
integrating blank sky data in the same detector regions. Among deep
sky observations with the XMM, we selected data with the most similar
background to that of each galaxy, after screening background flare
events in the same way. Each background agrees well with the data at
higher energies, as shown in Figure \ref{fig:bgd_n4636} for NGC 4636. 

Table \ref{tab:obs_data} also summarizes 
 total counts of MOS and PN within 4$r_e$ centered on each galaxy.
Here,  an annular region, 10--14', from the center of each galaxy was
used as a background after  subtracting a  blank sky data.
The total counts of sum of those of MOS and PN have a wide range from 1000 to 600000.
We analyzed projected annular spectra of all of the sample galaxies, while
fittings of deprojected spectra were performed for thirteen
galaxies with the total counts $>$ 12000.

\subsection{Chandra}

Chandra data analysis was performed with the CIAO software package,
version 3.3. We excluded time regions with a high background rate. We
also eliminated point sources identified with the tool wavedetect. The
outer region of each data set was subtracted as background. 

\section{Spectral Analysis and Results}

\subsection{Spectral fit}
\label{sec:anspec}

\subsubsection{Projected annular spectra}
\label{sec:allfit}

To derive the gravitational mass profiles of individual
galaxies, we need temperature and density profiles of the ISM.
First, we fitted projected annular spectra
 centered on each galaxy from MOS
(MOS1 + MOS2) and PN simultaneously, except NGC 4472. To exclude
possible emissions from our Galaxy and surrounding clusters, we also
subtracted the spectrum in an annular region, 10'--14',
from the annular spectra of
each galaxy. The fitting model is a sum of a
vAPEC model \citep{Smith2001} and a power-law model. The vAPEC
model represents thin thermal emission from the ISM, and the
power-law model represents
the contribution from unresolved low-mass
X-ray binaries (LMXBs),
where we fixed the power-law index at 1.6.
Since Chandra observations found that total spectra of discrete sources
in early-type galaxies are well
described with this power-law model \citep{Blan2001,Rand2004}. The two components were subjected to a
common absorption with fixed column density, $N_H$, at the Galactic value
from \citet{Dick1990}. We organized heavy element abundances
into three groups: the $\alpha$-element O group (O, Ne, and Mg), Si group (Si
and S), and Fe group (Fe and Ni). 
The abundances of the three elemental
groups were allowed to vary. 
For the innermost region  of IC 1459, we added a power-law model 
from the central nuclei  found by Chandra 
 \citep{Fabb2003}.

For brightest galaxies, NGC 4472, NGC 4636, 
NGC 4649, and NGC 5044,  whose
total counts  within 4$r_e$ larger than 200000, we performed spectral
fitting on each annular region. In order to derive accurate
temperature profiles of the ISM in the other galaxies with lower
signal-to-noise ratio, the spectra of all annular regions
were fitted simultaneously, where the ISM abundances 
 were assumed to have common values.

\begin{figure*}
    \centering
    \includegraphics[width=13.5cm]{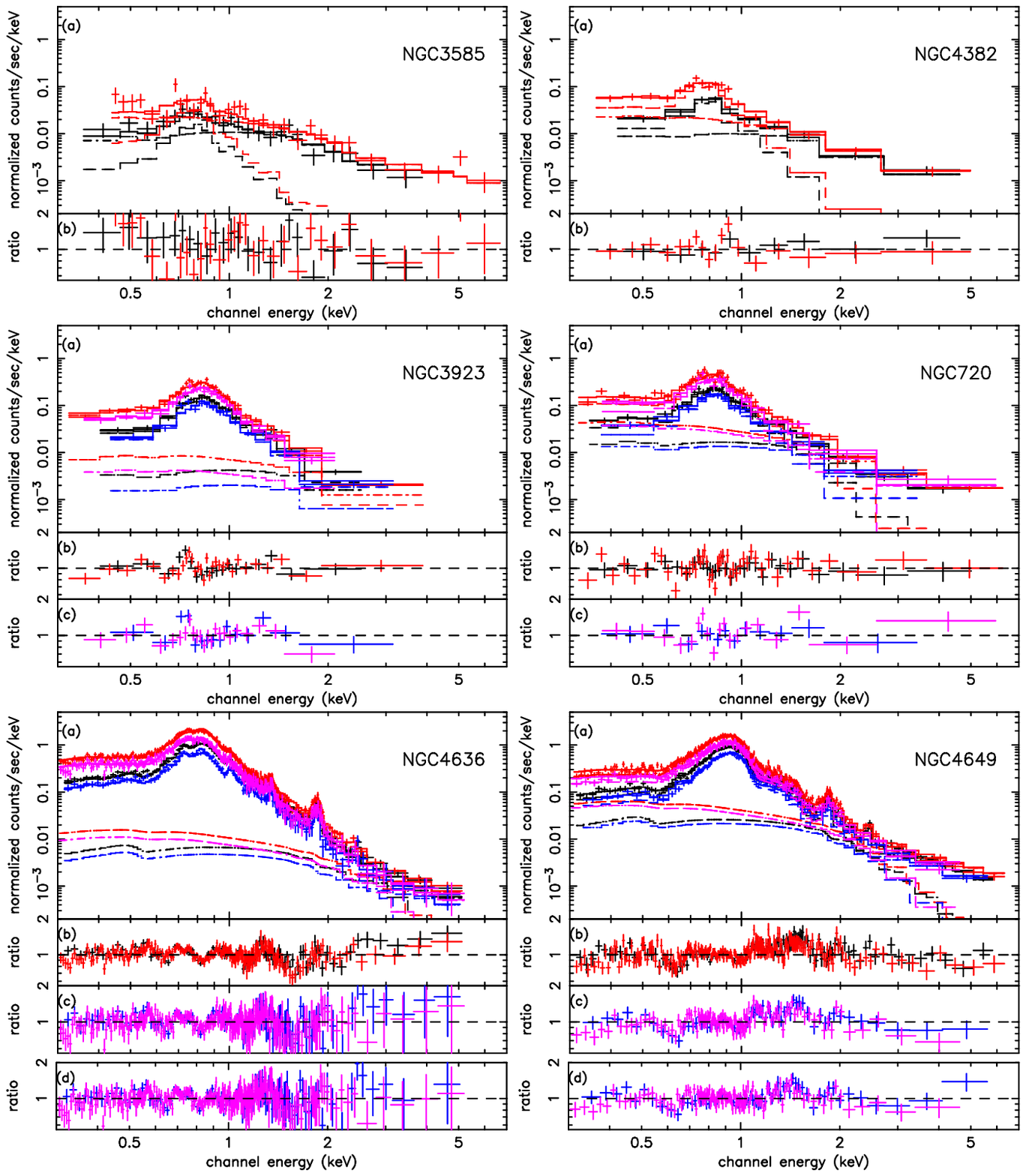}
    \caption{
    (a) Innermost projected spectra observed with MOS (black) and
    PN (red), and innermost deprojected spectra with MOS (blue) and PN
    (magenta). These spectra are fitted with a vAPEC model plus
    power-law multiplied by the Galactic absorption (solid line). Dashed lines
    correspond to the contribution from each component. (b)(c) The actual
    data to model ratio from the fit in panel a. The meanings of colors
    correspond to that in panel a. (d) The same as panel b and c,
    but when using vAPEC + vAPEC model plus power-law multiplied by the
    Galactic absorption as fitting model.}\label{fig:spct_fit}
   \end{figure*}

Table \ref{tab:fit_pj} and Figure A.1
summarize the results of the spectral fitting, ISM temperature,
abundance, and ISM luminosity. Figure \ref{fig:spct_fit} shows MOS and PN
spectra of the innermost regions several representative
galaxies. The spectra of
X-ray faint galaxies, whose

total counts within $4r_e$ are smaller than
12000 counts, 
provide gradually smaller values of reduced-$\chi^2$
with this single-temperature model (hereafter
1T model) for the ISM.
However, 
a representative spectrum of NGC 4382 whose ISM temperature is $\sim$
0.4 keV,  there are residual structures around 0.9 keV.
The X-ray brighter galaxies show
larger reduced-$\chi^2$. The spectra of galaxies with
kT$\sim 0.6$ keV,  NGC 4636, NGC 720 and NGC 3923, show common residuals
at 0.7-0.9 keV (Figure \ref{fig:spct_fit}). While, NGC 4649, which the
galaxy with kT$\sim 0.8$ keV, has different residual structures
from these three  galaxies.

\addtocounter{table}{1}

\subsubsection{Deprojected Spectra}
\label{sec:depjspec}

To consider projection effects, we performed spectral fittings of
deprojected spectra.
For the data with
high statistics,
 total counts
of sum of those of MOS and PN within 4$r_e$ larger
than 12000, deprojected spectra were calculated using ``onion
peeling'' methods by subtracting the contribution from the outer
shell regions for all spectral components, assuming the ISM is
spherically symmetric as described by \citet{Taka2004}.
We then fitted
the deprojected spectra with the 1T model in the same way as in Section
\ref{sec:allfit}.

 Table \ref{tab:fit_dpj} and Figure A.1
summarize the results.

The reduced $\chi^2$ reduced to $\sim 1$ except the innermost regions
of three brightest galaxies, NGC 4472, NGC 4636, and NGC 4649.
However,  residual structures around 0.7--0.9 keV in the projected
spectra still remain in the deprojected spectra fitted with the 1T model
(Figure \ref{fig:spct_fit}).
We also fitted the spectra of innermost regions
of the three brightest galaxies with a two-temperature vAPEC model
for the ISM (hereafter 2T model). 
The results are summarized in Table  \ref{tab:fit_dpj}.
Then, reduced $\chi^2$  reduced to 1.2--1.5, and 
even the 2T model gives similar residual structures (Figure \ref{fig:spct_fit}).
These discrepancies in the Fe-L energy range are also seen in the RGS
spectrum of the X-ray luminous elliptical galaxy, NGC 4636 \citep{Xu2002}.
Suzaku observations of NGC 720 and NGC 1404 whose ISM temperatures
are also $\sim$ 0.6 keV also give similar residual structures
\citep{Matsu2007, Tawa2008}.
Therefore, these residual structures are likely to be related to poorly
modeled Fe-L lines.

The spectral fittings of the deprojected spectra give mostly same
temperatures with those of the projected ones (Figure A.1).
Therefore,  for the fainter galaxies
 which were not performed the deprojected analysis,
we used  the temperature profiles derived  from the projected spectra
 to derive gravitational mass profiles.

\addtocounter{table}{1}

\subsection{Results}

\subsubsection{ISM luminosities within 4$r_e$}

We derived the absorption-corrected ISM luminosities ($L_{{\rm ISM}}$) and 
the  luminosities  of the power-law component ($L_{hard}$)
in the energy band of 0.3-2.0 keV within $4r_e$ (Table \ref{tab:lx}).
The values of  $L_{{\rm ISM}}$ derived from the deprojected 
spectra are close to those from projected annular spectra.
Hereafter, we use  $L_{{\rm ISM}}$ from the deprojected analysis.
For the X-ray fainter galaxies without deprojected analysis,
we use the values derived from the projected analysis.

\begin{figure*}
    \centering
    \includegraphics[width=13.0cm]{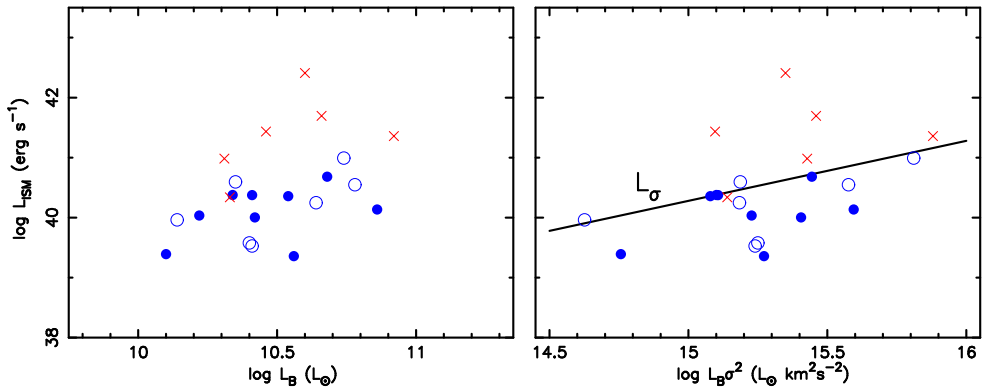}
    \caption{$L_{{\rm ISM}}(<4r_e)$ of galaxies plotted against
    $L_B$ (left panel) and $L_B\sigma^2$ (right panel). 
    Symbols indicate galaxy categories defined in Section
    \ref{sec:temp} for $X_E$ galaxies (red crosses), field $X_C$
    galaxies (filled blue circles), and $X_C$ galaxies in the clusters
    (open blue circles). The solid line represents the kinetic
    heating rate by stellar mass loss ($L_\sigma$).}\label{fig:lxlbs2}
   \end{figure*}

The relationship of  $L_{{\rm ISM}}$  to $L_B$ is shown in the
left panel of Figure \ref{fig:lxlbs2}. The sample galaxies have $L_B$
from $10^{10.1}$ to $10^{10.9}$ ${\rm L_\odot}$.
On the other hand, $L_{{\rm ISM}}$ scatters from $10^{39.4}$ erg/s to $10^{42.4}$
erg/s. 

The right panel of Figure \ref{fig:lxlbs2} shows the correlation between
$L_{{\rm ISM}}$ within 4$r_e$ and $L_B\sigma^2$, with $\sigma$ denoting the
central stellar velocity dispersion
in each galaxy. If stellar motion is the main heat source for the hot
ISM, its X-ray luminosity should be approximated by the input rate of
the kinetic energy of the gas from stellar mass loss. This is
proportional to $L_B\sigma^2$, because the mass-loss rate is thought to be
proportional to $L_B$ (e.g., \citealt{Ciot1991}). Figure
\ref{fig:lxlbs2} also shows the
expected energy input from stellar mass loss, $L_\sigma$, assuming a mass-loss
rate of $1.5 \times 10^{-11} (L_B/L_{\odot})~
t_{15}^{-1.3} M_{\odot}~{\rm yr}^{-1}$ \citep{Ciot1991}. Here,
$t_{15}$ is the stellar age in units of 15 Gyr, and we assumed a stellar
age of 12 Gyr. Several galaxies have larger $L_{{\rm ISM}}$ than $L_\sigma$, while $L_{{\rm ISM}}$ of
the other galaxies are similar to or smaller than $L_\sigma$. 

\begin{table}
\caption{X-ray luminosities within $4r_e$ of each sample galaxies}
\label{tab:lx}
\centering
\begin{tabular}{lccc}
\hline \hline
Galaxy & $\log L_{{\rm ISM}}$$^a$ & $\log L_{hard}$$^b$ & $\log L_{{\rm ISM}}$$^a$  \\
 & projection & projection & deprojection \\
 & (erg/s) & (erg/s) & (erg/s) \\ \hline
IC1459  & 40.08 & 40.28 & 40.00 \\
NGC720  & 40.46 & 39.91 & 40.38 \\
NGC1316 & 40.63 & 40.20 & 40.55 \\
NGC1332 & 40.12 & 39.74 & 40.04 \\
NGC1395 & 40.43 & 40.16 & 40.34 \\
NGC1399 & 41.13 & 40.32 & 40.98 \\
NGC1549 & 39.39 & 39.09 & ---   \\
NGC3585 & 39.36 & 39.39 & ---   \\
NGC3607 & 40.37 & 39.86 & ---   \\
NGC3665 & 40.36 & 39.96 & ---   \\
NGC3923 & 40.74 & 40.27 & 40.68 \\
NGC4365 & 39.58 & 39.86 & ---   \\
NGC4382 & 40.25 & 39.83 & ---   \\
NGC4472 & 41.40 & 40.51 & 41.36 \\
NGC4477 & 39.96 & 39.54 & ---   \\
NGC4526 & 39.53 & 39.64 & ---   \\
NGC4552 & 40.62 & 40.27 & 40.60 \\
NGC4636 & 41.46 & 40.27 & 41.44 \\
NGC4649 & 41.00 & 40.41 & 40.99 \\
NGC5044 & 42.49 & 41.01 & 42.41 \\
NGC5322 & 40.14 & 39.82 & ---   \\
NGC5846 & 41.72 & 40.56 & 41.70 \\
\hline
\multicolumn{4}{l}{$^a$The X-ray luminosity of the thermal emission in the}\\
\multicolumn{4}{l}{~~range of 0.3-2.0 keV derived from the projected}\\
\multicolumn{4}{l}{~~and deprojected spectra.}\\
\multicolumn{4}{l}{$^b$The X-ray luminosity of the non thermal emission in}\\
\multicolumn{4}{l}{~~the range of 0.3-2.0 keV derived from the projected}\\
\multicolumn{4}{l}{~~spectra.}\\
\end{tabular}
\end{table}

\subsubsection{Temperature Profiles of the ISM and Classification with
   X-Ray Extended and X-Ray Compact galaxies}
\label{sec:temp}

\begin{figure*}
    \centering
    \includegraphics[width=6.5cm]{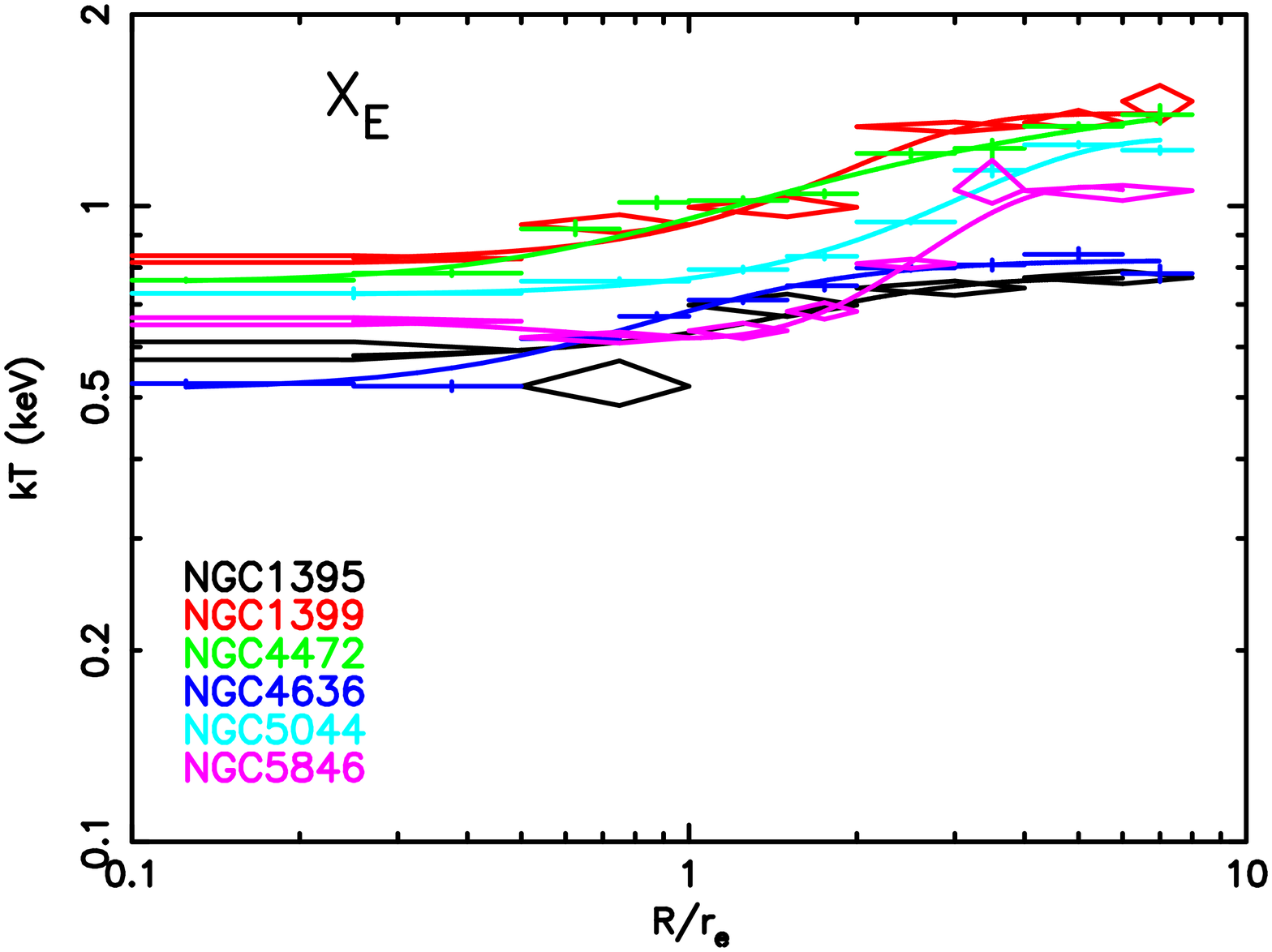}
    \includegraphics[width=6.5cm]{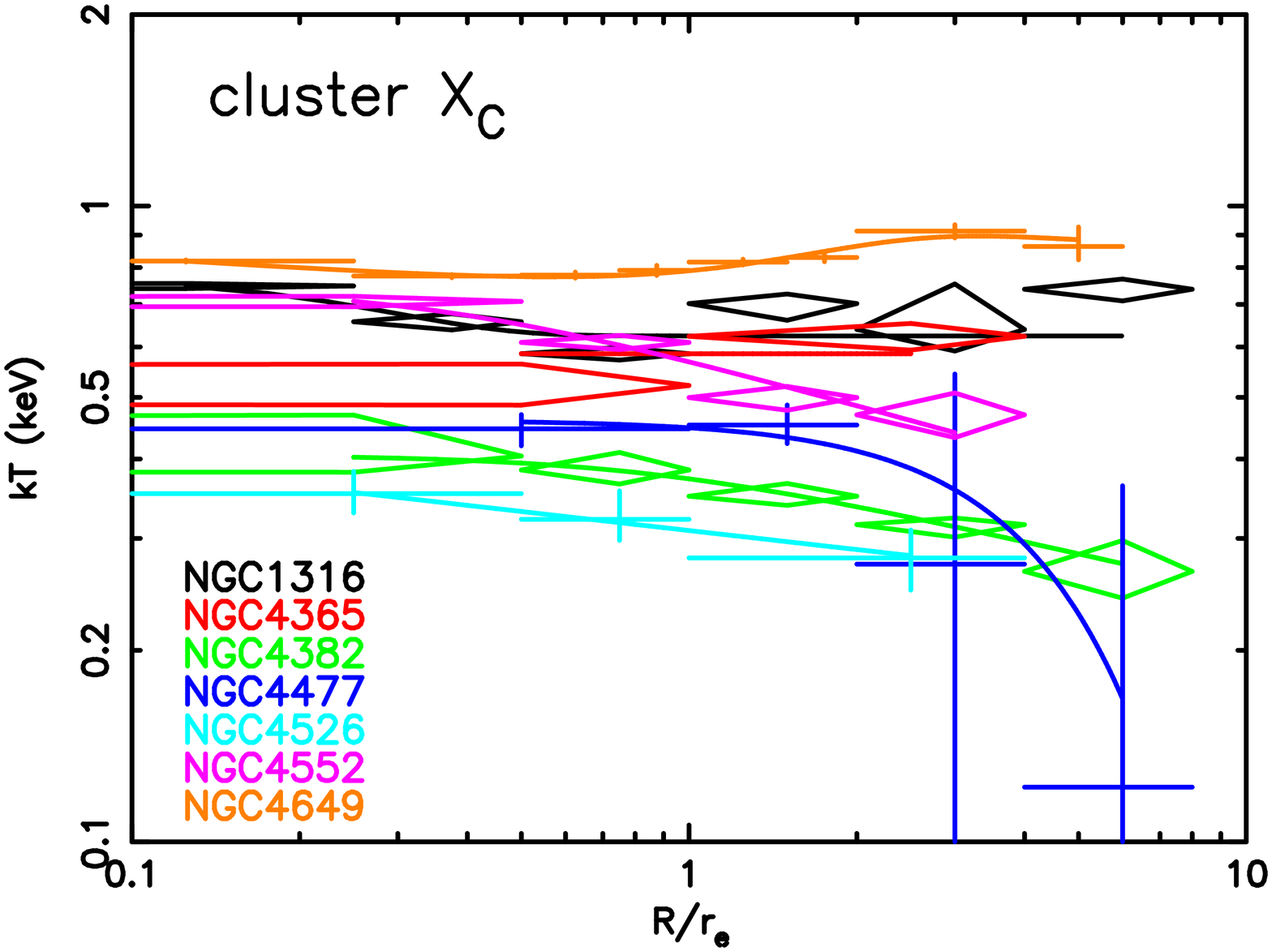}
    \includegraphics[width=6.5cm]{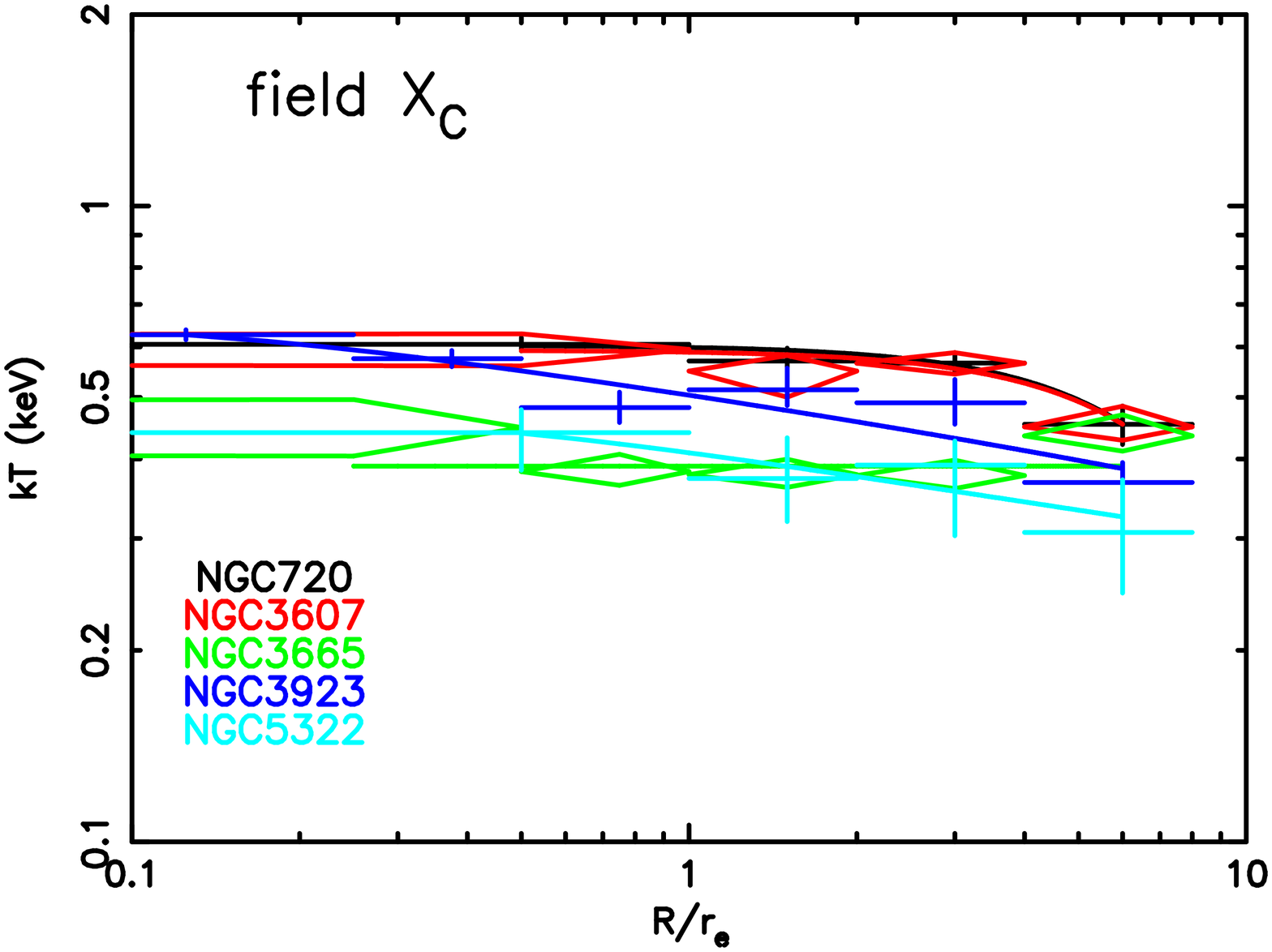}
    \includegraphics[width=6.5cm]{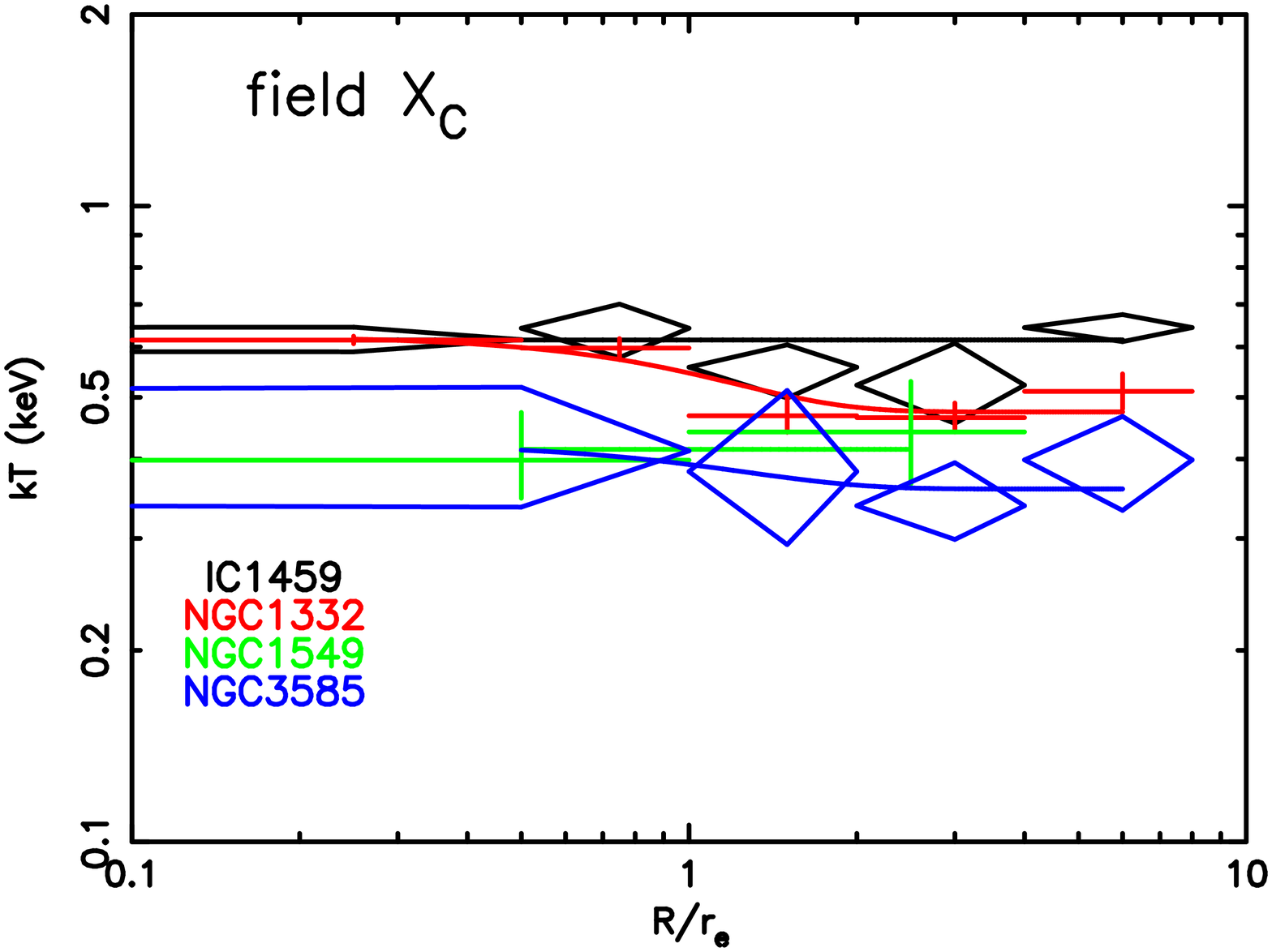}
  \caption{The derived temperature profiles of the ISM. The $X_E$
 galaxies are plotted in the top left panel, the $X_C$ galaxies in
 clusters are in the top right, and the field $X_C$ galaxies are in the
 bottom panels. Colors indicate individual galaxies. The solid lines
 represent the best-fit function.}\label{fig:kt_all}
   \end{figure*}

Figure \ref{fig:kt_all} shows the derived radial temperature profiles of the
ISM. Some galaxies have gradually increasing temperature profiles
toward the outer radius. By contrast, other galaxies have flat or
decreasing radial temperature profiles. The galaxies with positive
temperature gradients have high ISM temperatures of 0.8-1.5 keV
at a radius of several times $r_e$, which are comparable to those of
galaxy groups. On the other hand, the temperatures of the other
galaxies are systematically lower at 0.2-0.6 keV. 
The derived temperature profiles were fitted with the sum of a constant
and single- or double-$\beta$ functions.

The derived temperature profiles 
are mostly consistent with previous results
from  ROSAT  \citep{Matsu2001}  and
Chandra  \citep{Fuka2006, Athe2007}, except
central regions of several brightest galaxies with
positive temperature gradients, due to higher angular 
resolution of Chandra.

\begin{figure*}
    \centering
    \includegraphics[width=6.5cm]{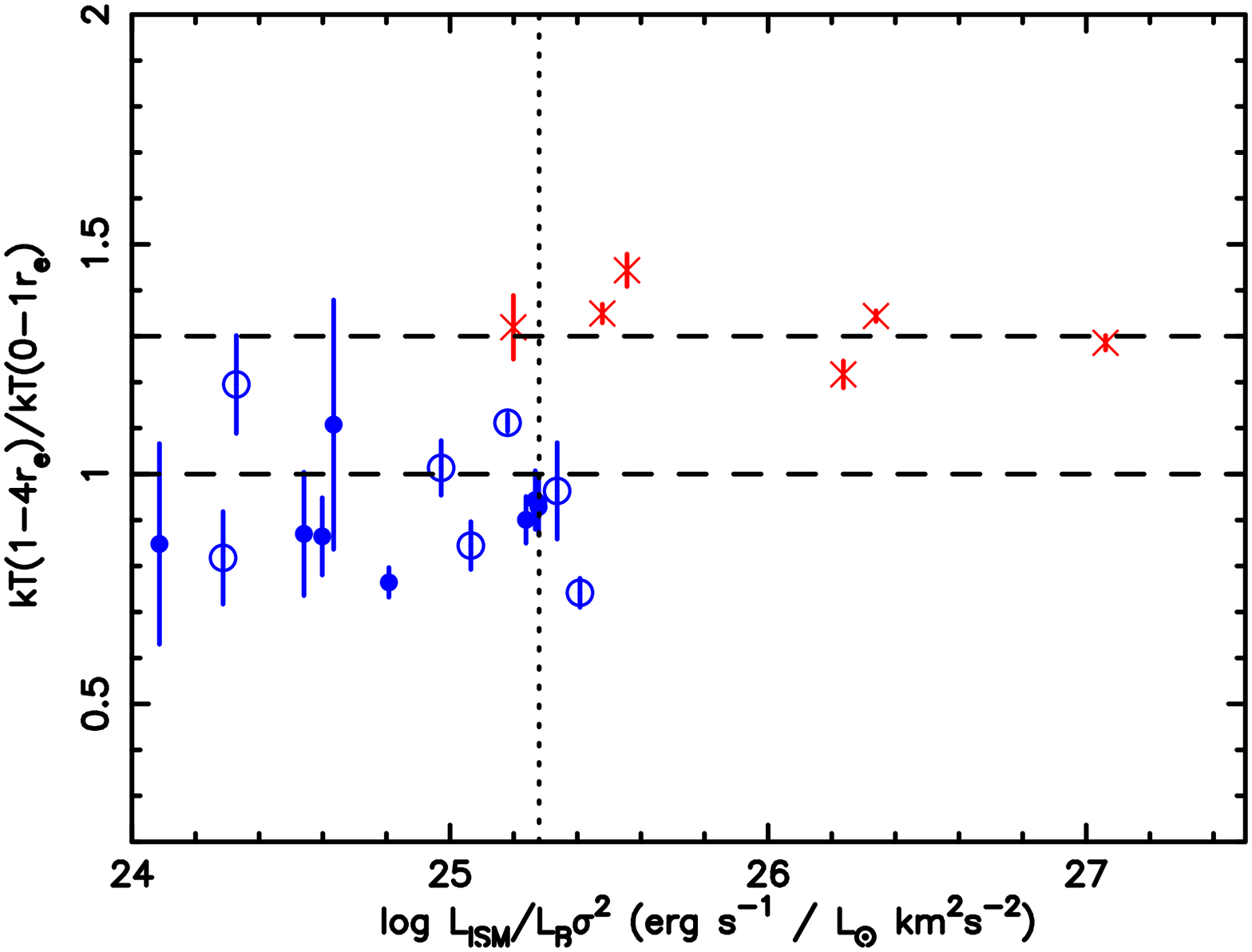}
    \includegraphics[width=6.5cm]{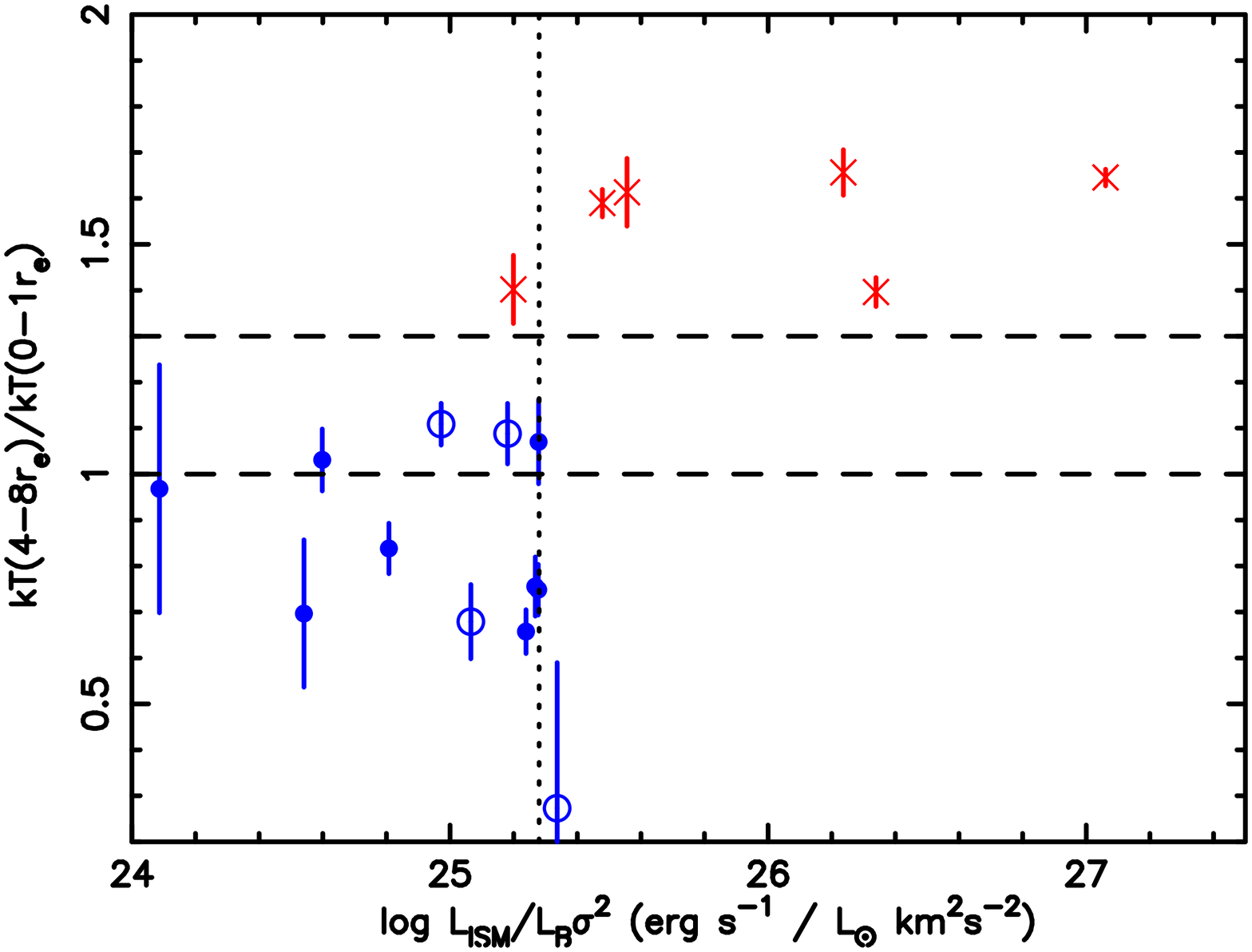}
 \caption{The ISM temperature gradients parametrized by $kT(1$-$4r_e)/kT(<1r_e)$
 (left panel) and $kT(4$-$8r_e)/kT(<1r_e)$ (right panel) plotted
 against $L_{{\rm ISM}}/L_B\sigma^2$. Meanings of the symbols are the same as
 those in Figure \ref{fig:lxlbs2}. The dotted line represents the rate
 of kinetic heating by stellar mass loss.
 The dashed lines indicate that $kT(1$-$4r_e)$ (or
 $kT(4$-$8r_e)$) is 1 or 1.3 times higher than $kT(<1r_e)$. }
 \label{fig:kt_grad}
   \end{figure*}

In Figure \ref{fig:kt_grad}, we plotted $kT(1$--$4r_e)/kT(<1r_e)$ and
$kT(4$--$8r_e)/kT(<1r_e)$ against $L_{{\rm ISM}}/L_B\sigma^2$. Here, $kT(<1r_e)$,
$kT(1$--$4r_e)$, and $kT(4$--$8r_e)$ correspond to the emission-weighted
ISM temperatures of regions in $r<1r_e$,  $1r_e<r<4r_e$ and  $4r_e<r<8r_e$,
respectively. In general, the temperature profiles of galaxies
with $L_{{\rm ISM}}\lesssim L_\sigma$ have smaller $kT(1$-$4r_e)$ and $kT(4$-$8r_e)$
than $kT(<1r_e)$. By contrast, $L_{{\rm ISM}}$ of galaxies with
$kT(4$-$8r_e)/kT(<1r_e)>1.3$ are systematically larger than $L_\sigma$. 

Our sample galaxies are divided into two types: X-ray faint galaxies
with flat or negative temperature gradients and X-ray luminous
galaxies with positive temperature gradients. The ISM luminosities in
the former type are consistent with heating by stellar motion, while
galaxies of the latter type need additional sources of
heating. Hereafter, we denote galaxies with
$kT(4$--$8r_e)/kT(<1r_e)>1.3$ as X-ray extended ($X_E$) galaxies and others
as X-ray compact ($X_C$) galaxies. The classification of each galaxy is
summarized in Table \ref{tab:tar_list}. 

\section{Spatial Analysis and Results}

\subsection{X-ray surface brightness and gas density profiles}

\begin{figure*}
    \centering
    \includegraphics[width=7.0cm]{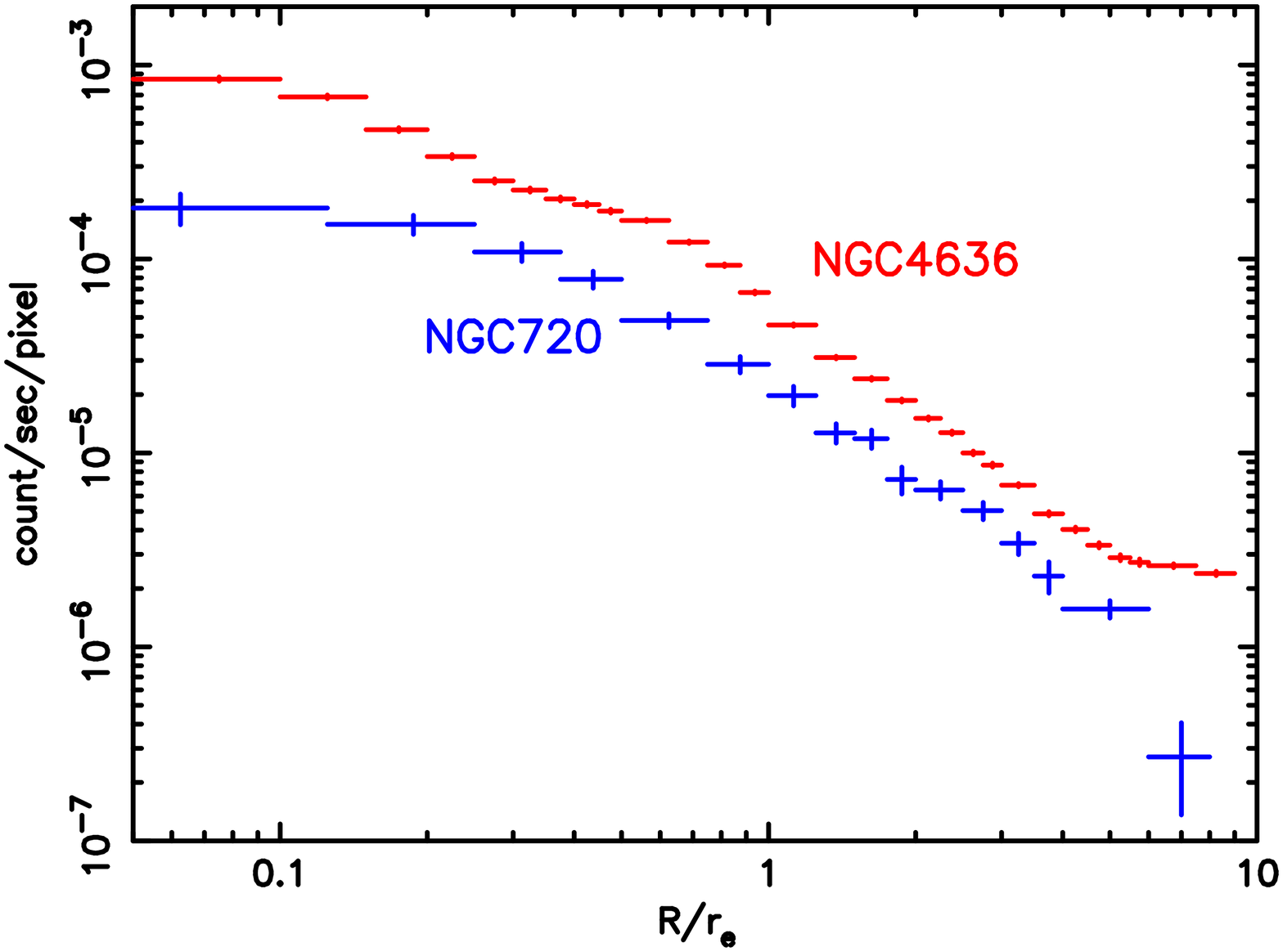}
  \caption{X-ray surface brightness profiles derived from MOS images of
NGC 720 (blue) and NGC 4636 (red).}
\label{fig:radp_n720_n4636}
   \end{figure*}

We derived radial profiles of X-ray surface brightness from
background-subtracted and vignetting-corrected X-ray images from MOS1
and MOS2. We considered only photons in the energy band 0.8--2.0
keV, where the ISM emission dominates. PN data were not used for this
analysis because of gaps between CCD chips. Figure
\ref{fig:radp_n720_n4636} shows two
representative X-ray surface brightness profiles, those of an $X_E$
galaxy, NGC 4636, and an $X_C$ galaxy, NGC 720.

\begin{figure*}
    \centering
    \includegraphics[width=6.5cm]{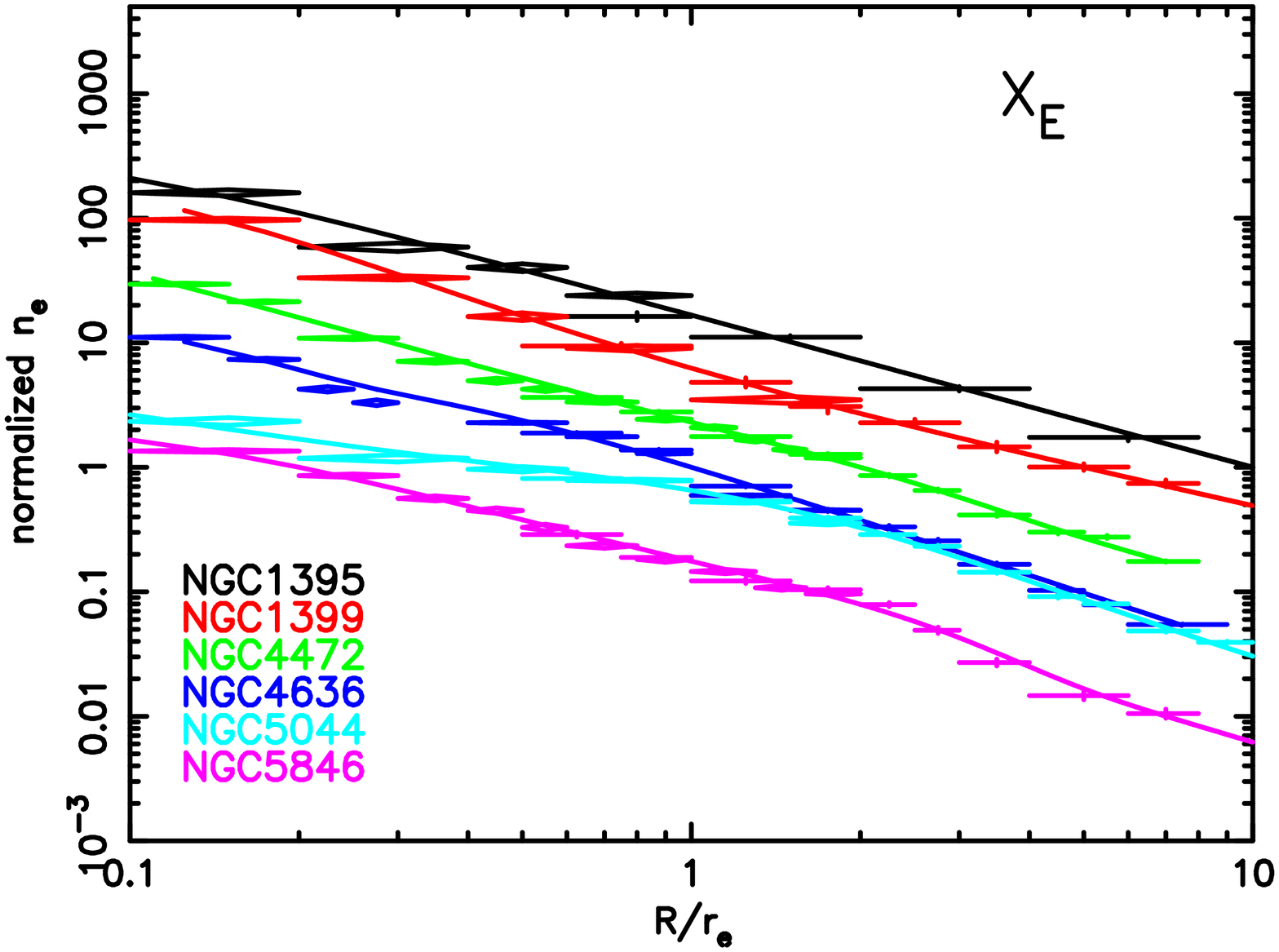}
    \includegraphics[width=6.5cm]{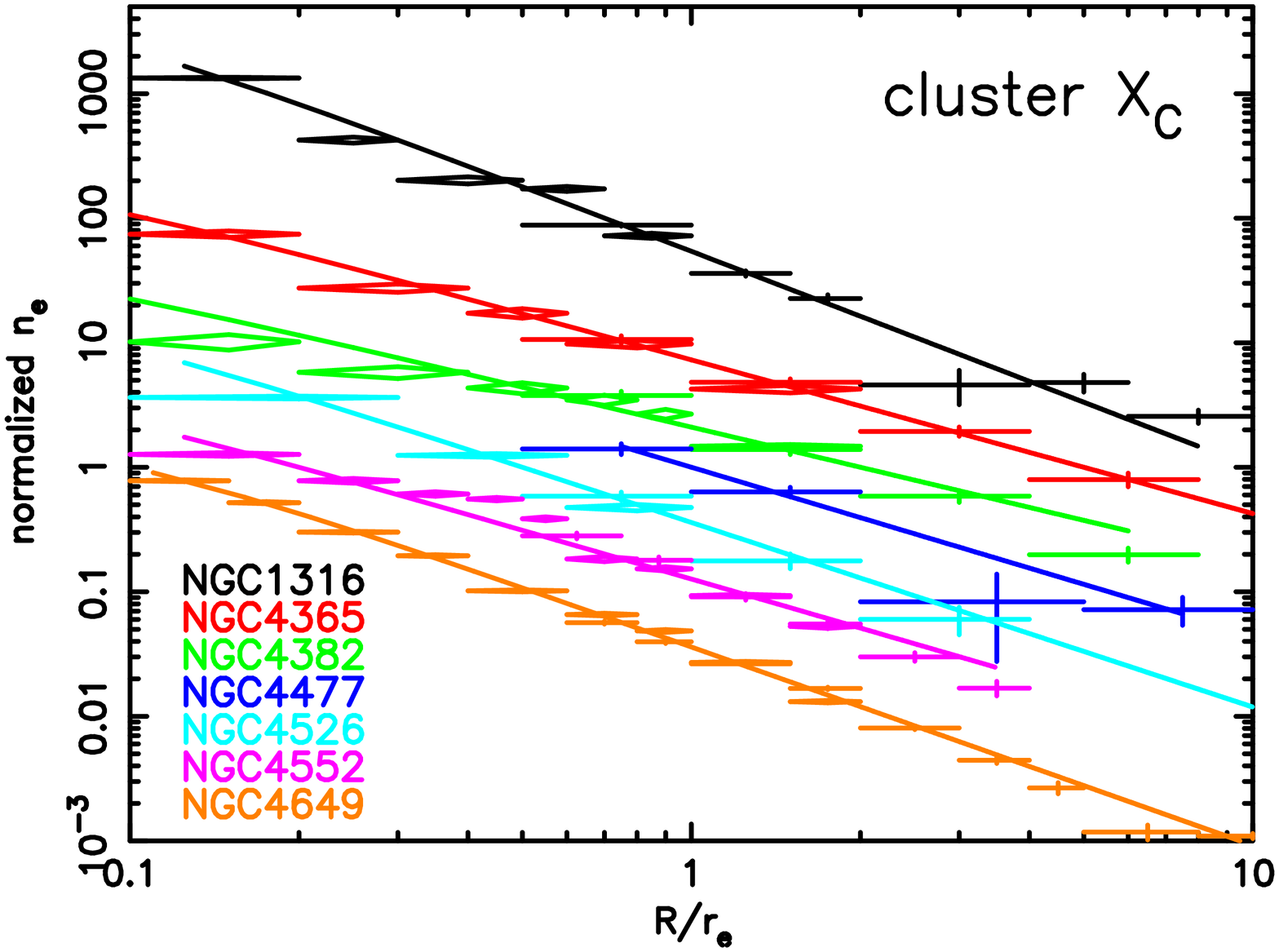}
    \includegraphics[width=6.5cm]{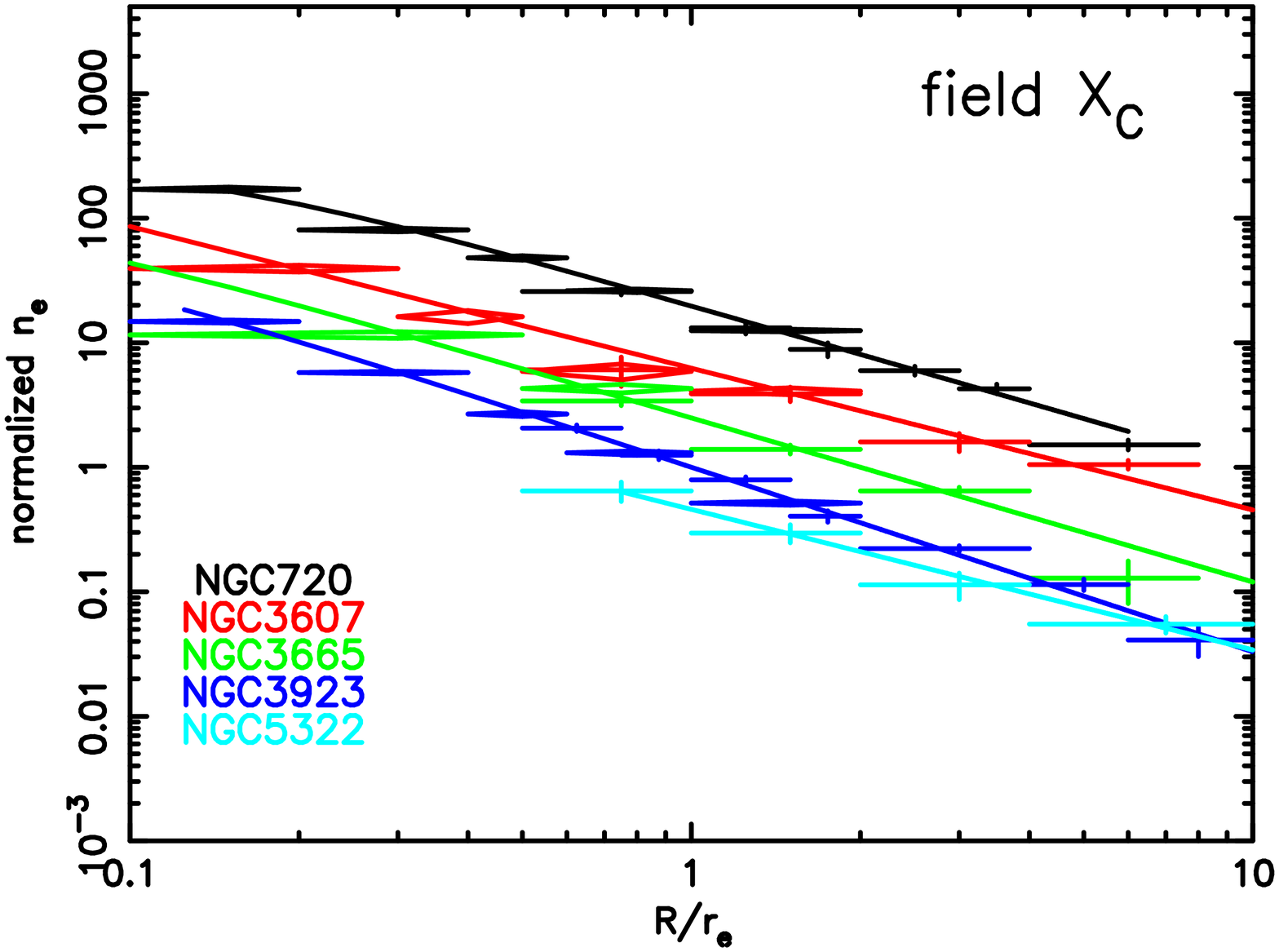}
    \includegraphics[width=6.5cm]{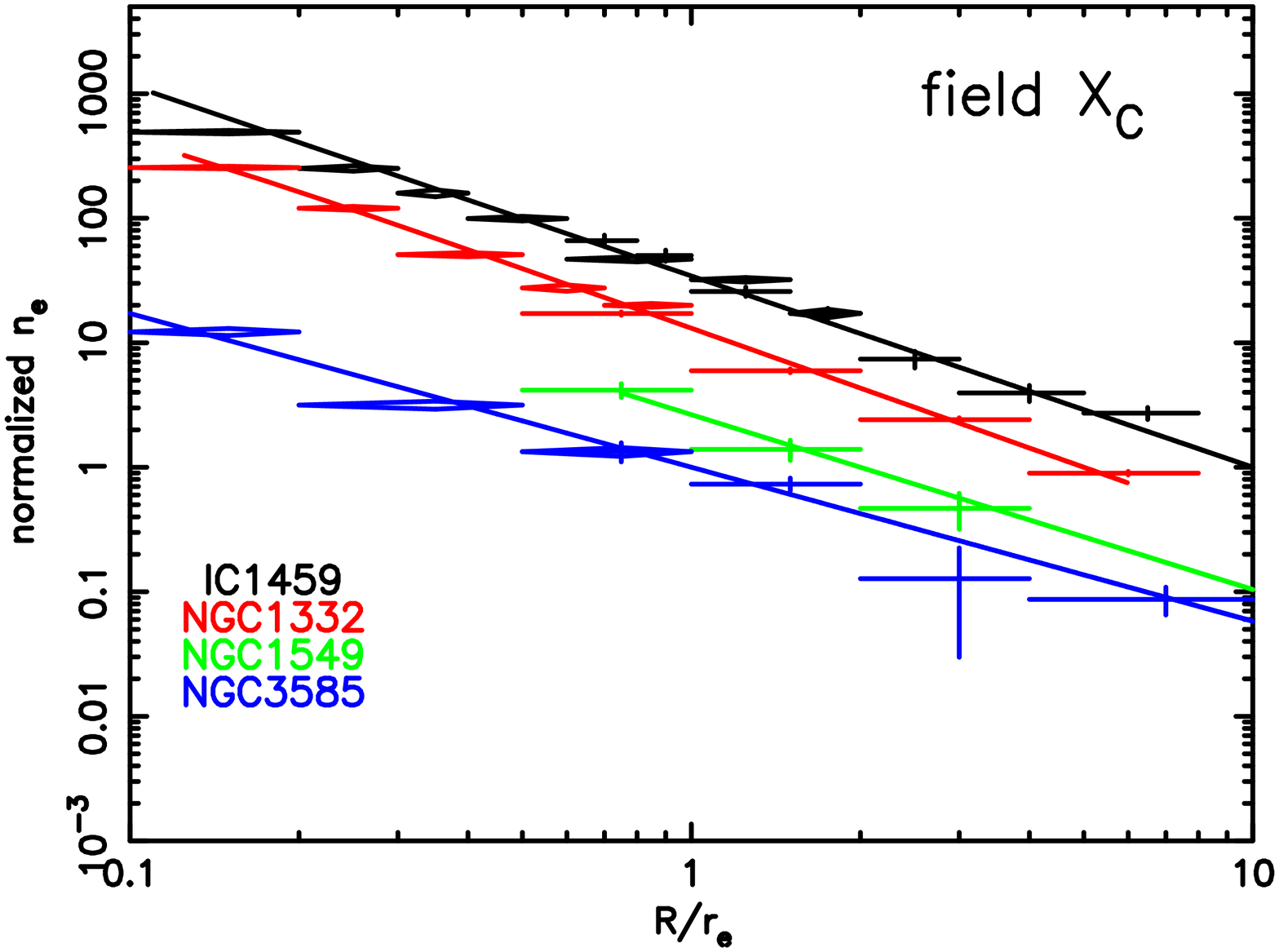}
  \caption{ISM density profiles. The $X_E$ galaxies are plotted in the top
left panel, the $X_C$ galaxies in clusters are in the top right, and the
field $X_C$ galaxies are in the bottom panels. These profiles are
arbitrarily normalized. Colors indicate individual galaxies. Solid
lines represent the best-fit function. Crosses and diamonds correspond
to XMM-Newton and Chandra data, respectively.}\label{fig:n_all}
   \end{figure*}

Assuming circular symmetry, we deprojected the X-ray surface
brightness profiles to derive gas density profiles. In order to
subtract emission from outside the field of view, we
first fitted the radial surface brightness profile within 8$r_e$ of each
galaxy and assumed that the profile extends outside the field of
view. The radial profiles of most of the galaxies were fitted with a
$\beta$-model. The brightness of several galaxies in clusters are clearly
constant at the outer regions because of the surrounding intra-cluster
medium (ICM). Therefore, we fitted these profiles with the sum of a $\beta$
model and a constant. Several X-ray luminous galaxies need a double-$\beta$
model to fit the surface brightness profiles. Because at $>5r_e$, one
of the $\beta$ models dominates, the profiles at $r>5r_e$ are fitted with a
single-$\beta$ model. The derived density profiles are summarized in Figure
\ref{fig:n_all}. Since we need only the gradient of a density profile to derive a
gravitational mass profile, the plotted density profiles were
arbitrarily normalized.

We also used Chandra data for 19 galaxies with sufficiently high
signal-to-noise ratios to derive accurate X-ray surface brightness
profiles  within 1--2$r_e$. Radial profiles of X-ray surface
brightness were derived from ACIS X-ray images in the energy band
0.3-2.0 keV. Then, we deprojected X-ray surface brightness
profiles and derived gas density profiles in the same way as for the
XMM-Newton data. As summarized in Figure \ref{fig:n_all}, normalized density
profiles of the ISM derived from XMM-Newton and Chandra are mostly
consistent with each other from 0.5$r_e$ to  2$r_e$. 

We then fitted the derived gas density profiles of XMM-Newton at
$r>0.5r_e$ and that of Chandra within $r<2r_e$ of each galaxy
simultaneously with a $\beta$ model, as $f(R)=S_0(1+(R/R_c)^{2})^{-IN}$.
Most of the density profiles were well
fitted with this single-$\beta$ model (Figure \ref{fig:n_all}).
The derived $R_c$ values of galaxies with Chandra data
are almost about 0.05$r_e$.
Therefore, we fixed $R_c$ value  to 0.05$r_e$ to fit the  gas density
profiles  of galaxies without Chandra data.
For NGC 4552 we also fixed $R_c$ value to 0.05$r_e$, because of
edge like structure at 0.4--0.6$r_e$ 
observed with Chandra \citep{Mach2006}.
 Several
X-ray luminous galaxies need a double-$\beta$ model to fit the density
profiles.

The ISM and the hard component may have different
surface brightness profiles. Therefore,
we also derived gas density profiles of galaxies whose $L_{ISM}$ is
smaller than 80\% of the total luminosity within 4$r_e$,
directly from the normalization of the ISM component derived from
spectral fittings. The best-fit $\beta$-model of the density profile of
each galaxy is plotted in Figure A.1. 
The best-fit values of $\beta$ derived  in this way
are mostly consistent with those derived from surface brightness profiles,
although several galaxies show discrepancies of a few tens of \%.

\section{Mass profiles}

We then calculated the total mass profile $M(R)$ within a
three-dimensional radius $R$ from the obtained best-fit functions of
ISM temperature $T(R)$ and gas density $n(R)$ profiles
from the surface brightness, assuming
hydrostatic equilibrium and circular symmetry, by the equation 
\begin{eqnarray}
M(R)=-\frac{kT(R)\cdot R}{G\mu m_p} \left(\frac{d\ln n(R)}{d\ln
				     R}+\frac{d\ln T(R)}{d\ln R}\right),
\label{eq:mass}
\end{eqnarray}
where $m_p$ is the proton mass, $k$ is the Boltzmann constant, $G$ is the
constant of gravity, and $\mu \sim$ 0.62 is the mean particle mass in units of
$m_p$. Figure \ref{fig:mass_all} summarizes the derived mass
profiles. For NGC 1549, NGC
4477, and NGC 5322, the mass profiles within 0.5$r_e$ were not plotted in
the figure, since we used only XMM-Newton data for those galaxies. 

In addition, the upper and lower limits of the mass profiles were
calculated considering the errors in the temperature and the
temperature and density gradients of each data bin. The upper and
lower limits of the temperature and density gradients of the $i$-th
shell were obtained from the ratio of the value within $i+1$-th shell
to that within $i-1$-th shell. Here, we used the temperature
profiles derived from XMM-Newton. At $r<1 r_e$ and $r>1 r_e$, density
profiles derived from Chandra and XMM-Newton, respectively, were
used. For NGC 4636, the density at 0.3--0.4$r_e$ is significantly
smaller than the best-fit function, due to the existence of
complicated structure discovered by Chandra (\citealt{Jones2002}). 
Therefore, we ignored this shell when deriving the mass
profiles. As summarized in Figure \ref{fig:mass_all}, the total masses
derived in
this way are mostly consistent with those using the best-fit
functions. 

We also derived stellar mass profiles, using the deprojected de
Vaucouleurs profile of \citet{mell1987}, assuming stellar $M/L_B$
is in the range from 3 to 8 in solar units. We plotted these profiles
in Figure \ref{fig:mass_all}. Within 1$r_e$, the gravitational mass is
consistent with the
stellar mass. Further, the gradients of the gravitational mass are
similar to those of the stellar mass. By contrast, outside the radius
of a few $r_e$, the derived gravitational mass becomes much larger than
the stellar mass. These results indicate the existence of dark matter
in the outer regions of early-type galaxies.

When $L_{ISM}$ is smaller than 80\% of the total luminosity within 4$r_e$,
the total mass profile is calculated using the best-fit
$\beta$-model of the density profile from the spectral fitting.
The results are compared with those from the
surface brightness in Figure A.1.
The two methods give similar total mass profiles within 10--20\%.
For NGC 3585 and NGC 5322, these discrepancies are $\sim$30\%,
but within the large errors of the mass profiles of the two galaxies.
Thus, the mass profiles of these galaxies obtained from two methods are
consistent within the error.
Hereafter, we use the total mass profiles derived from the surface
brightness profiles.

   \begin{figure*}
    \centering
    \includegraphics[width=4.5cm]{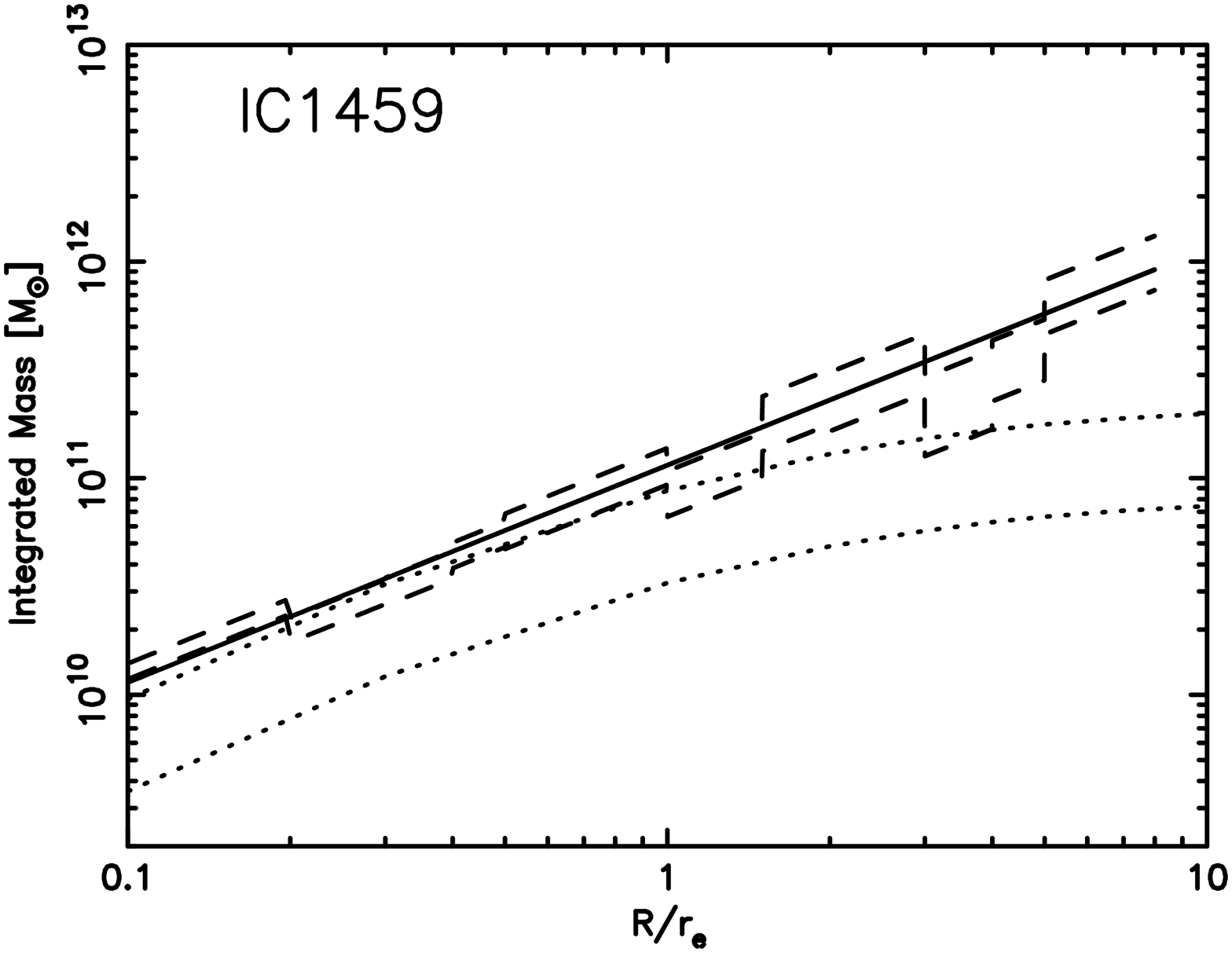}
    \includegraphics[width=4.5cm]{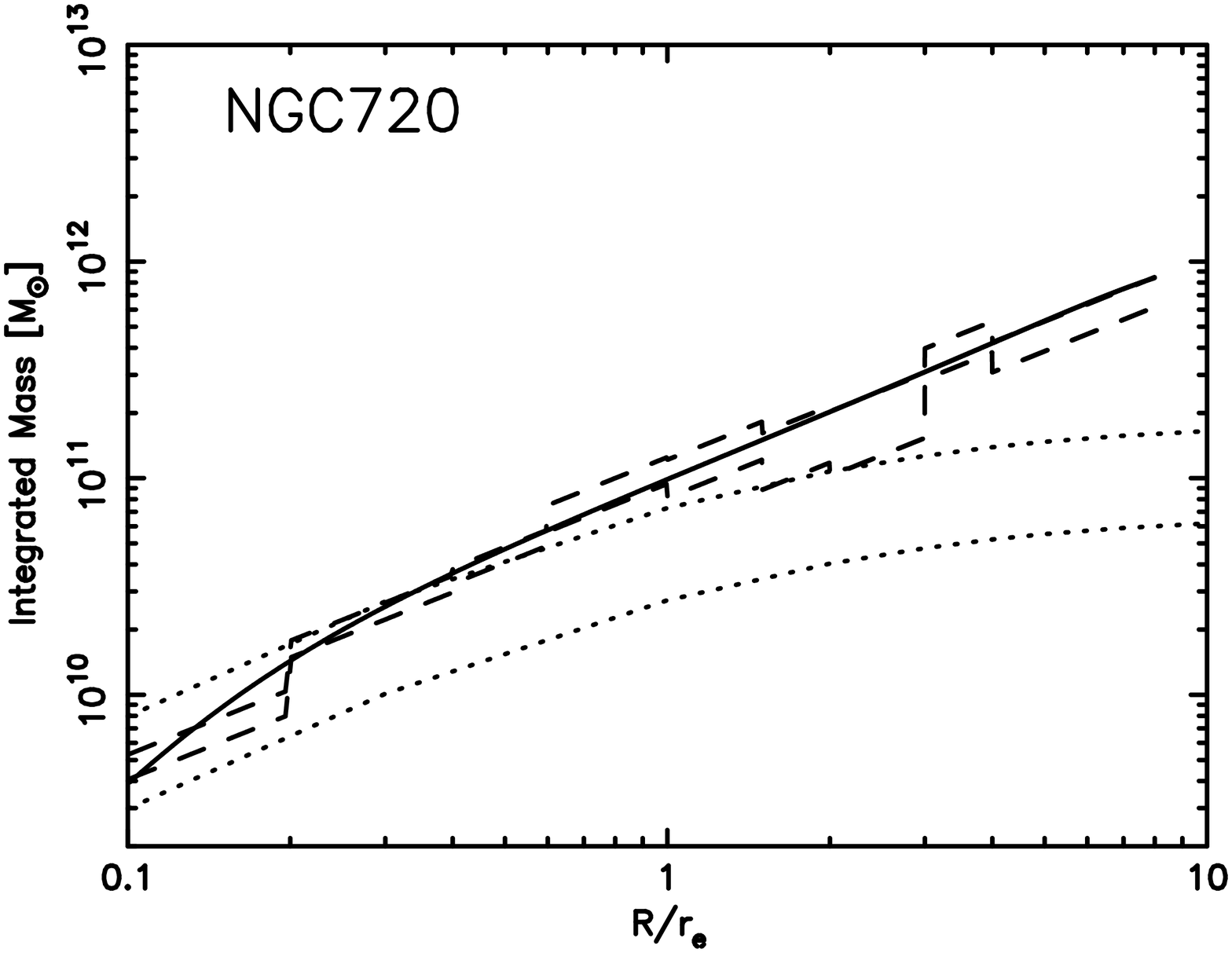}
    \includegraphics[width=4.5cm]{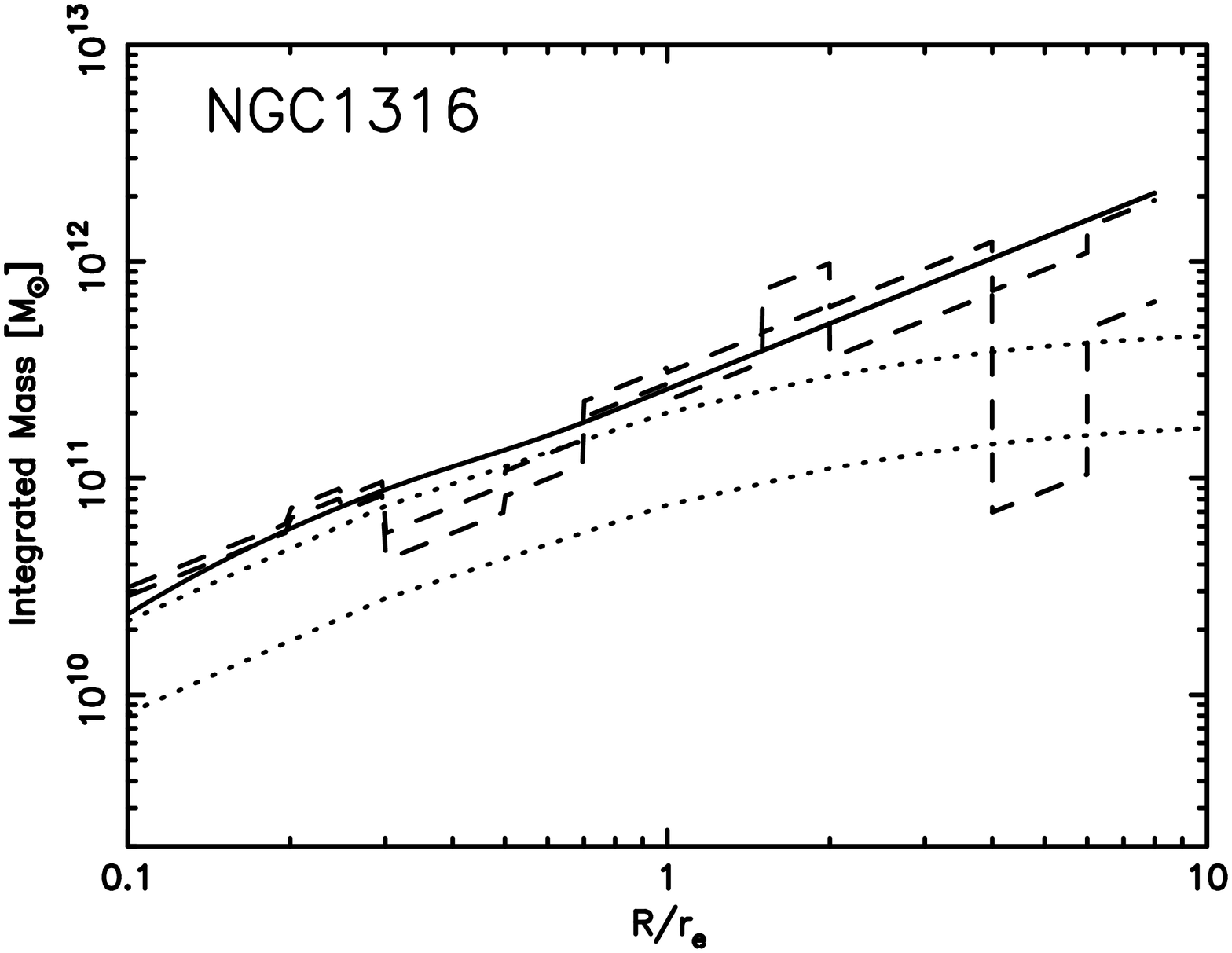}
    \includegraphics[width=4.5cm]{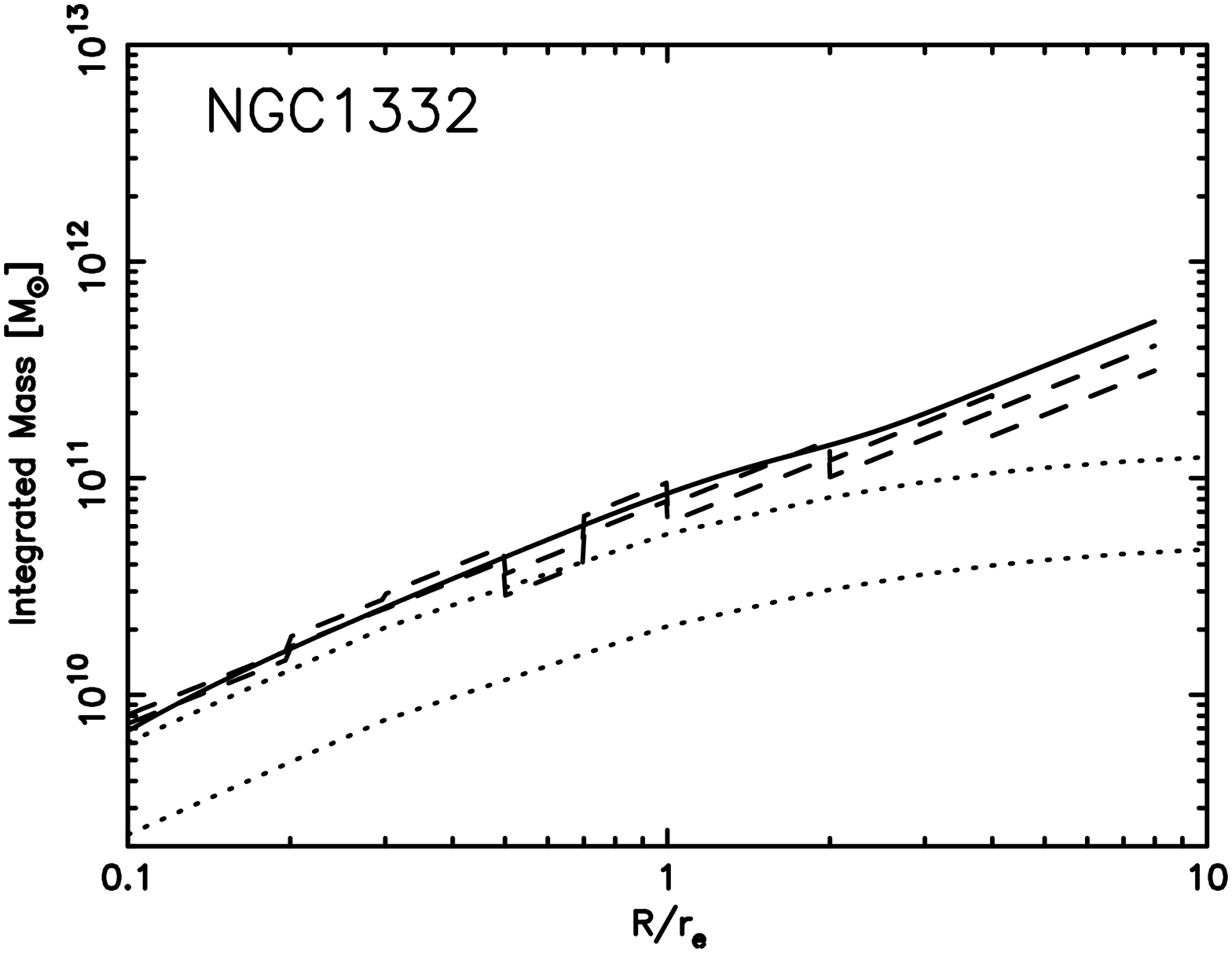}
    \includegraphics[width=4.5cm]{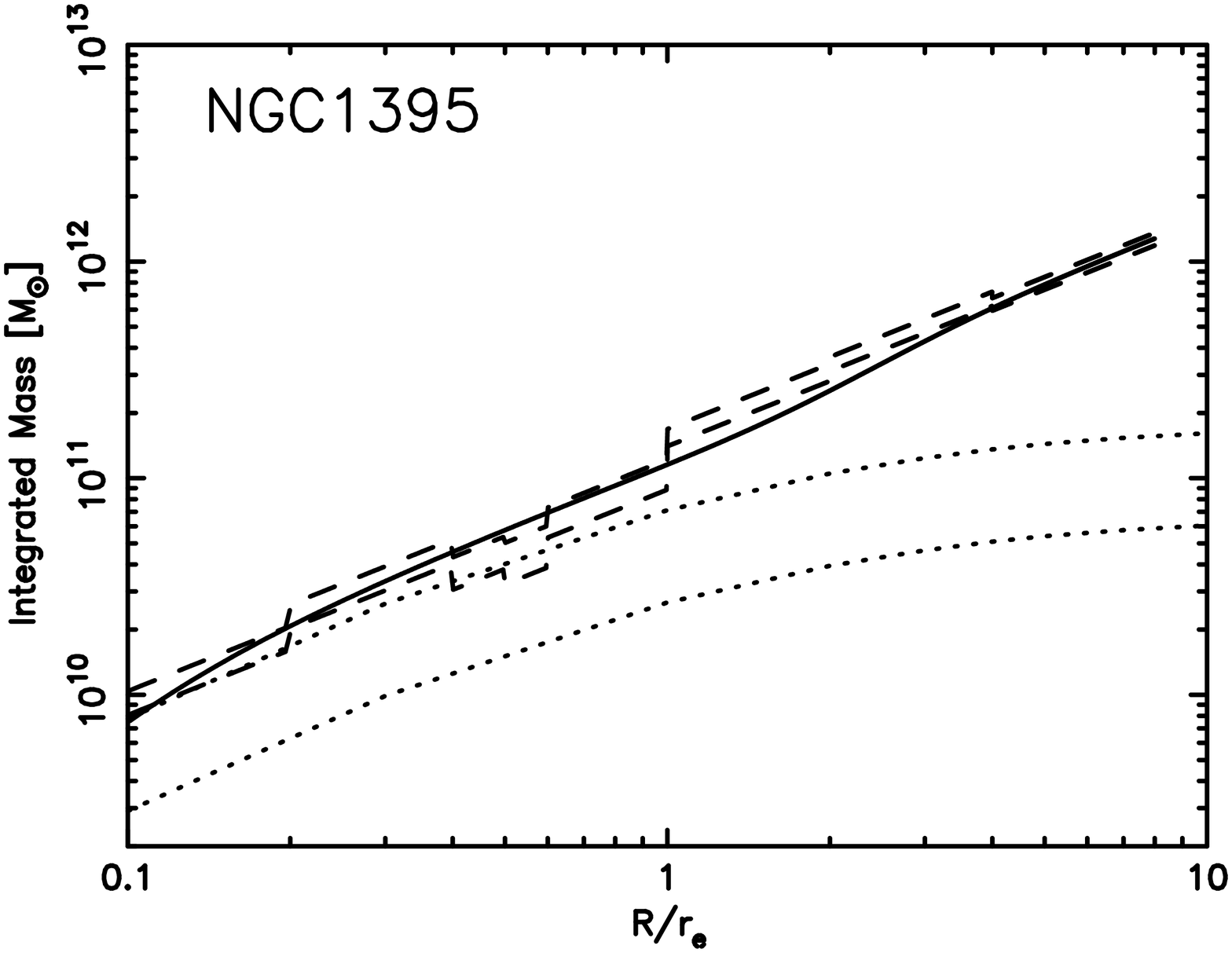}
    \includegraphics[width=4.5cm]{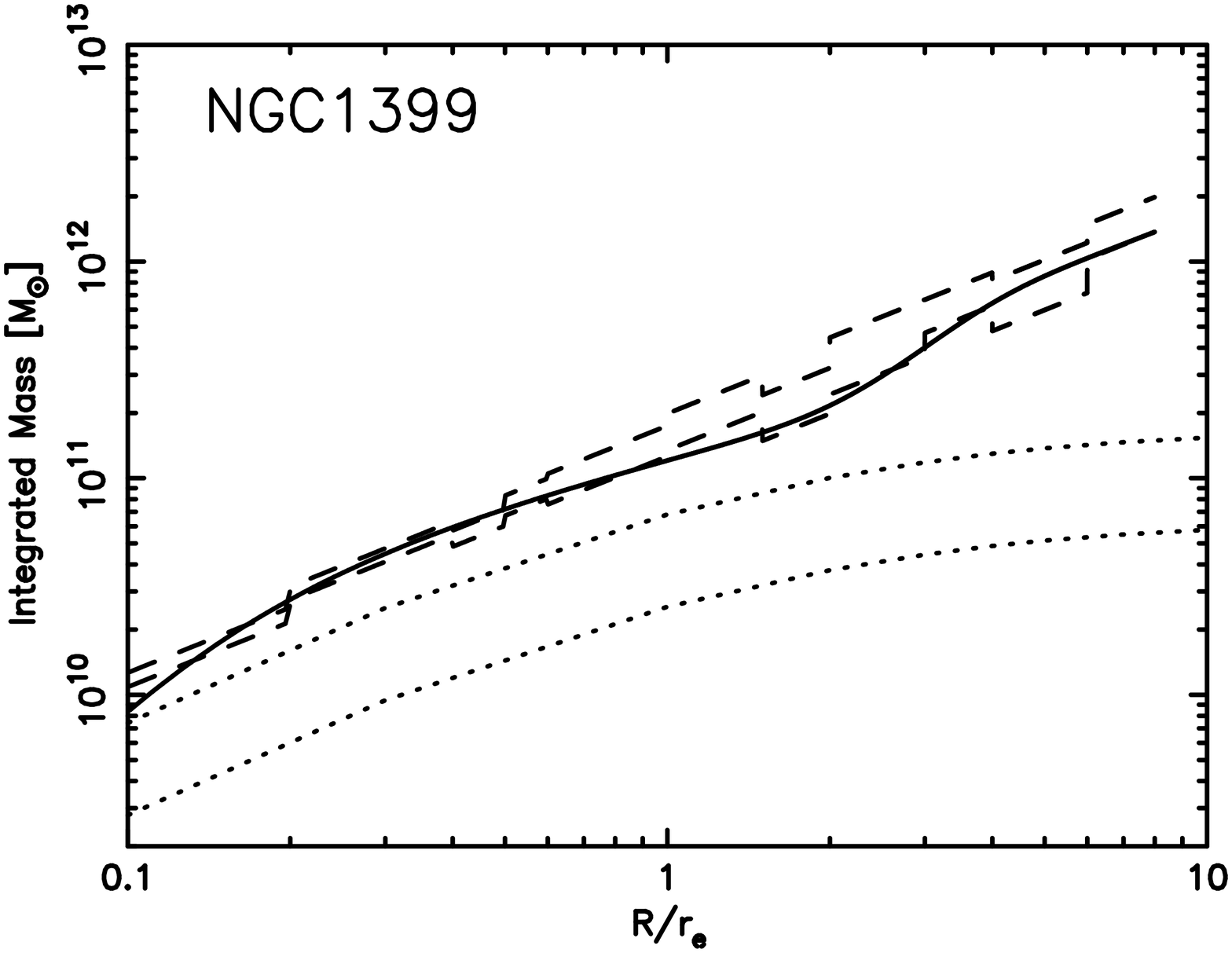}
    \includegraphics[width=4.5cm]{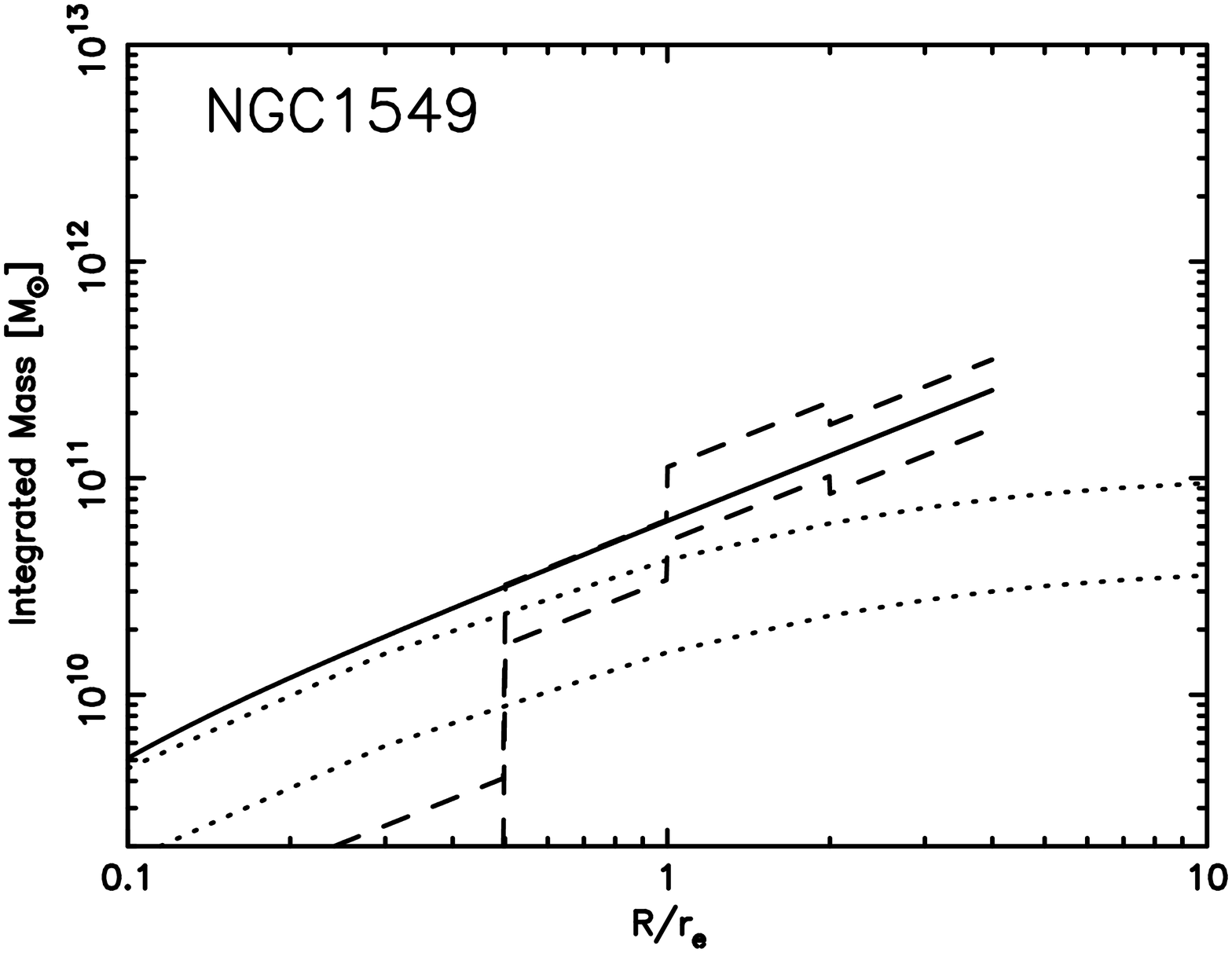}
    \includegraphics[width=4.5cm]{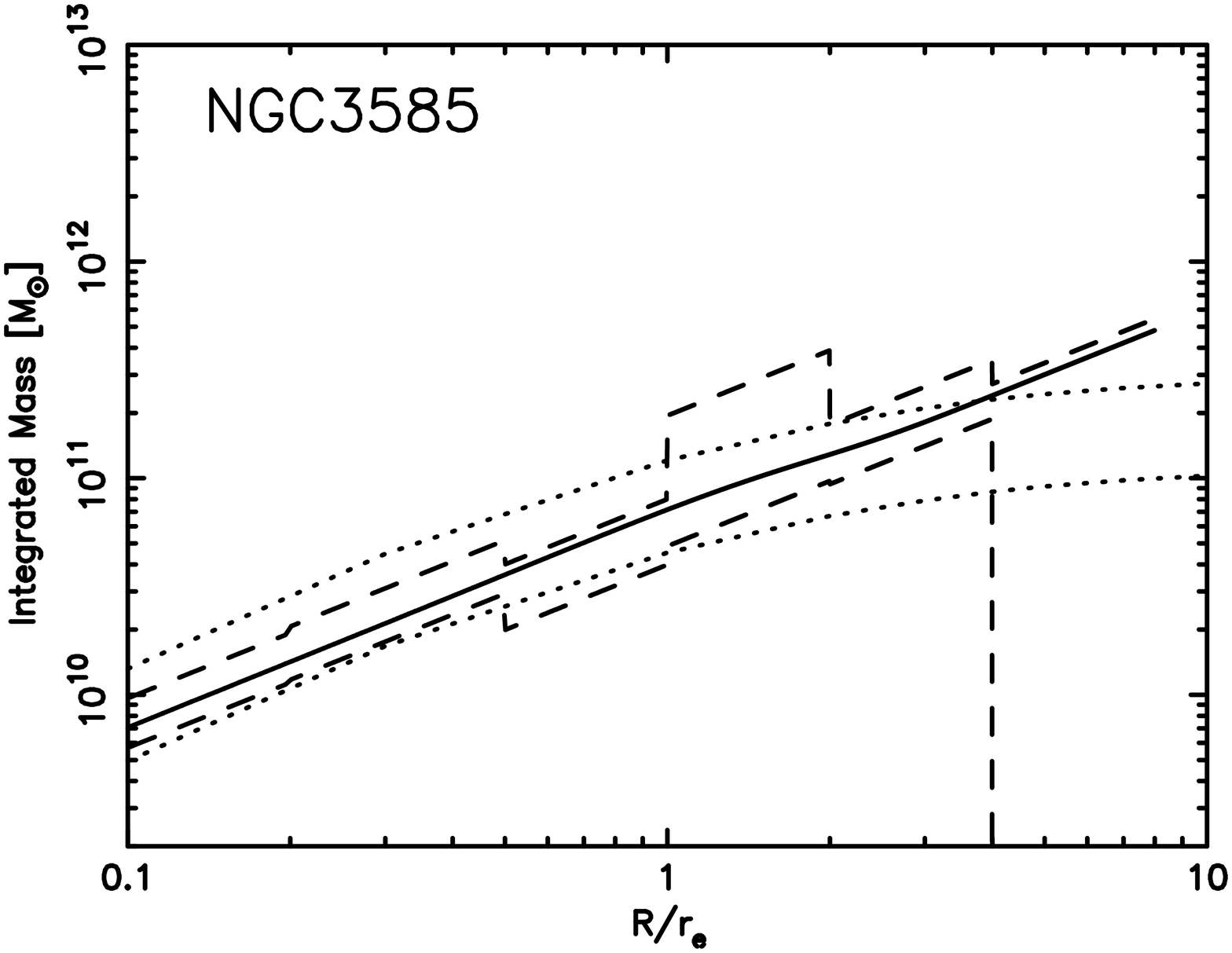}
    \includegraphics[width=4.5cm]{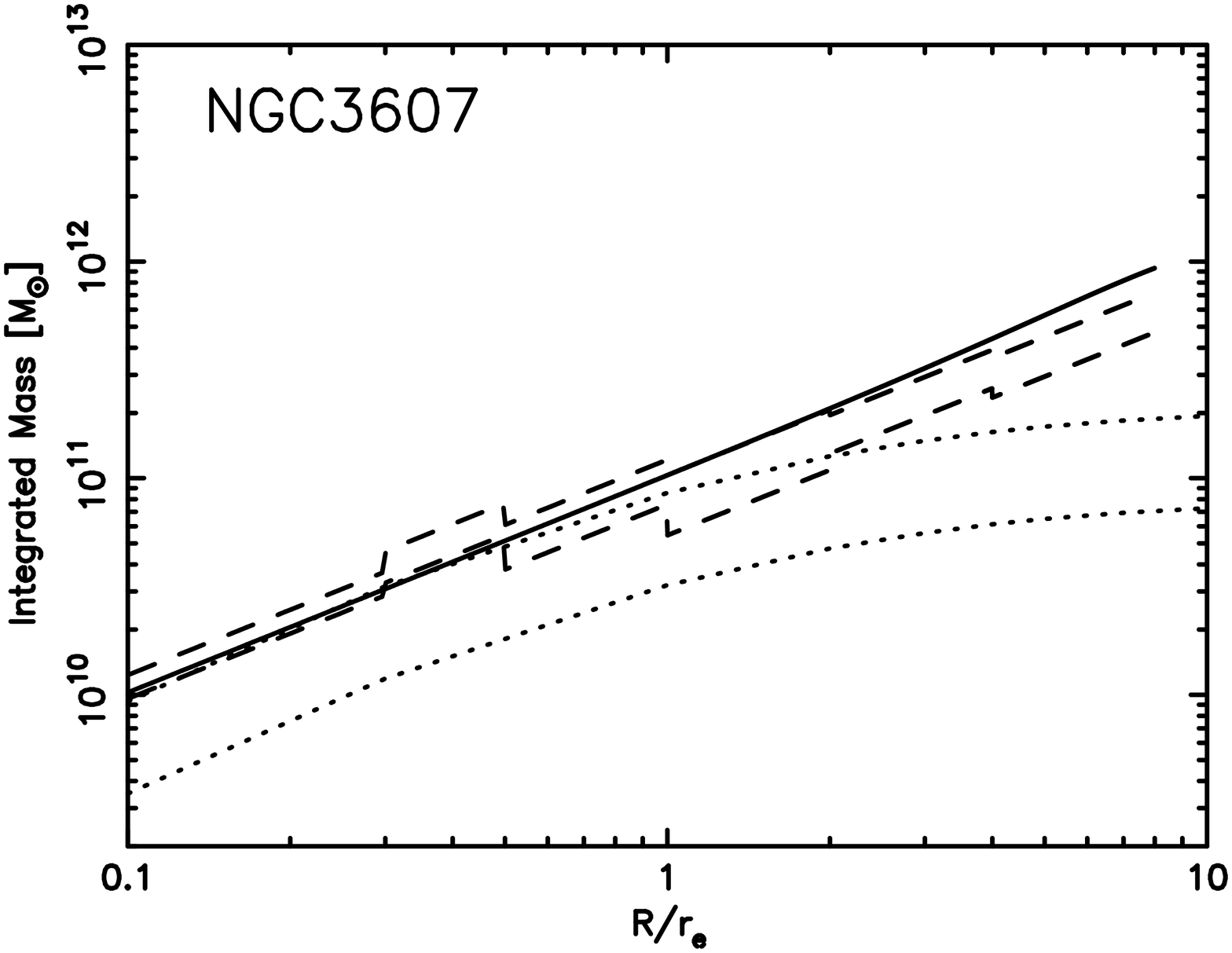}
    \includegraphics[width=4.5cm]{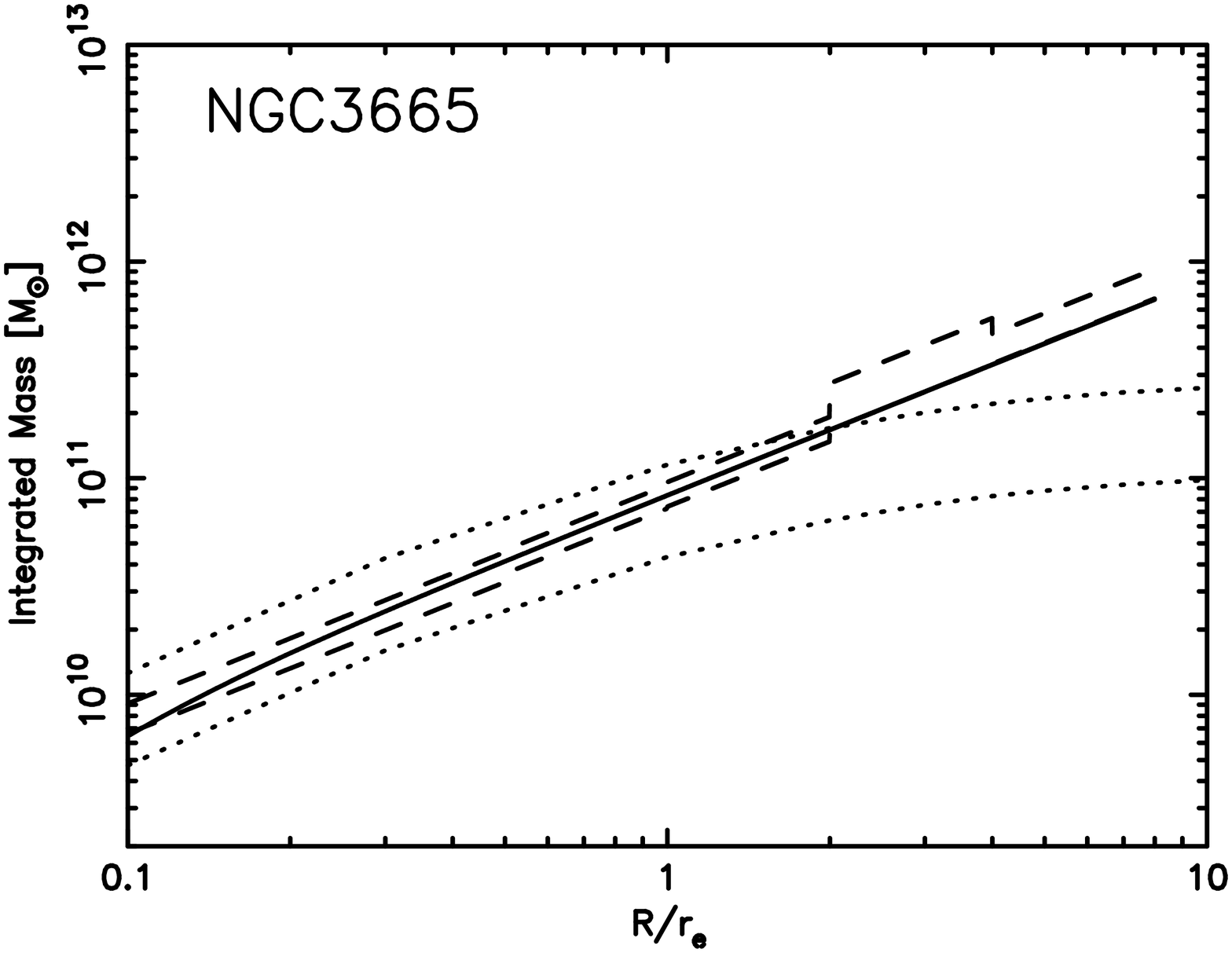}
    \includegraphics[width=4.5cm]{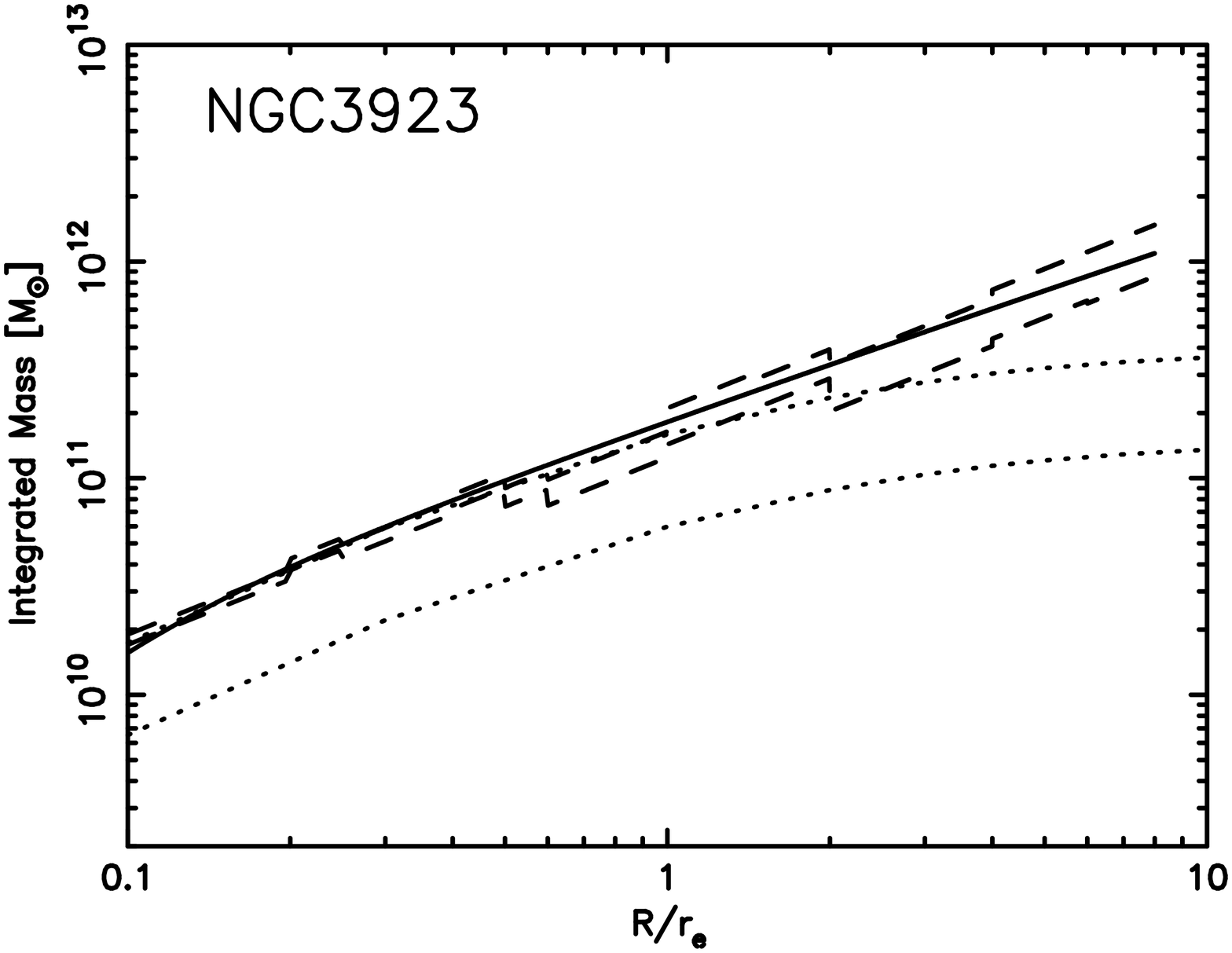}
  \caption{Integrated mass profiles of galaxies. The solid lines are
total gravitational mass obtained from the best-fit functions of
temperature and ISM density profiles. The dashed lines correspond to
the upper and lower limits derived from the local gradients of
temperature and density. We also plotted the stellar mass profiles
assuming stellar $M/L_B$ to be 3 and 8 (dotted lines).}\label{fig:mass_all}
   \end{figure*}

\addtocounter{figure}{-1}

   \begin{figure*}
    \centering
    \includegraphics[width=4.5cm]{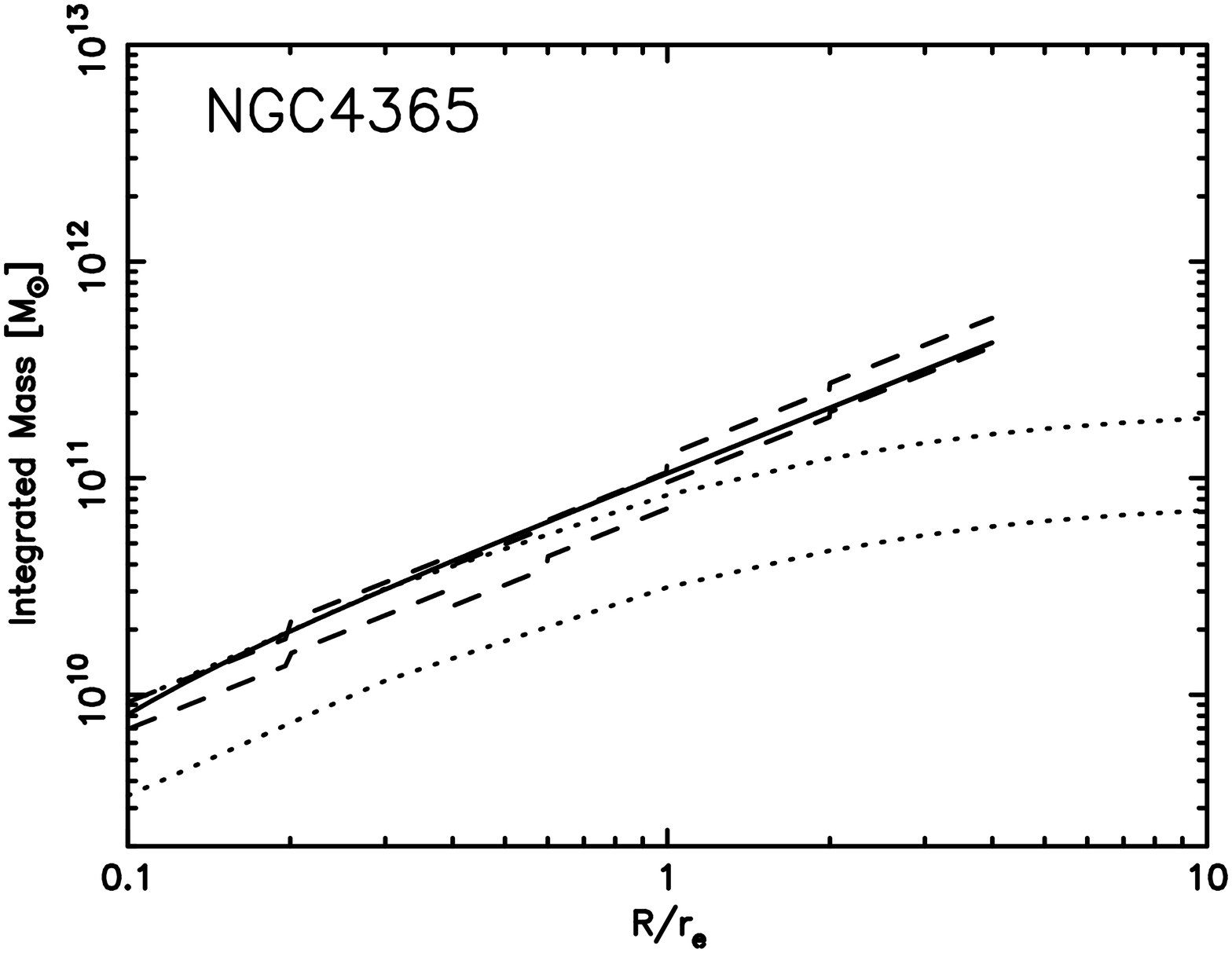}
    \includegraphics[width=4.5cm]{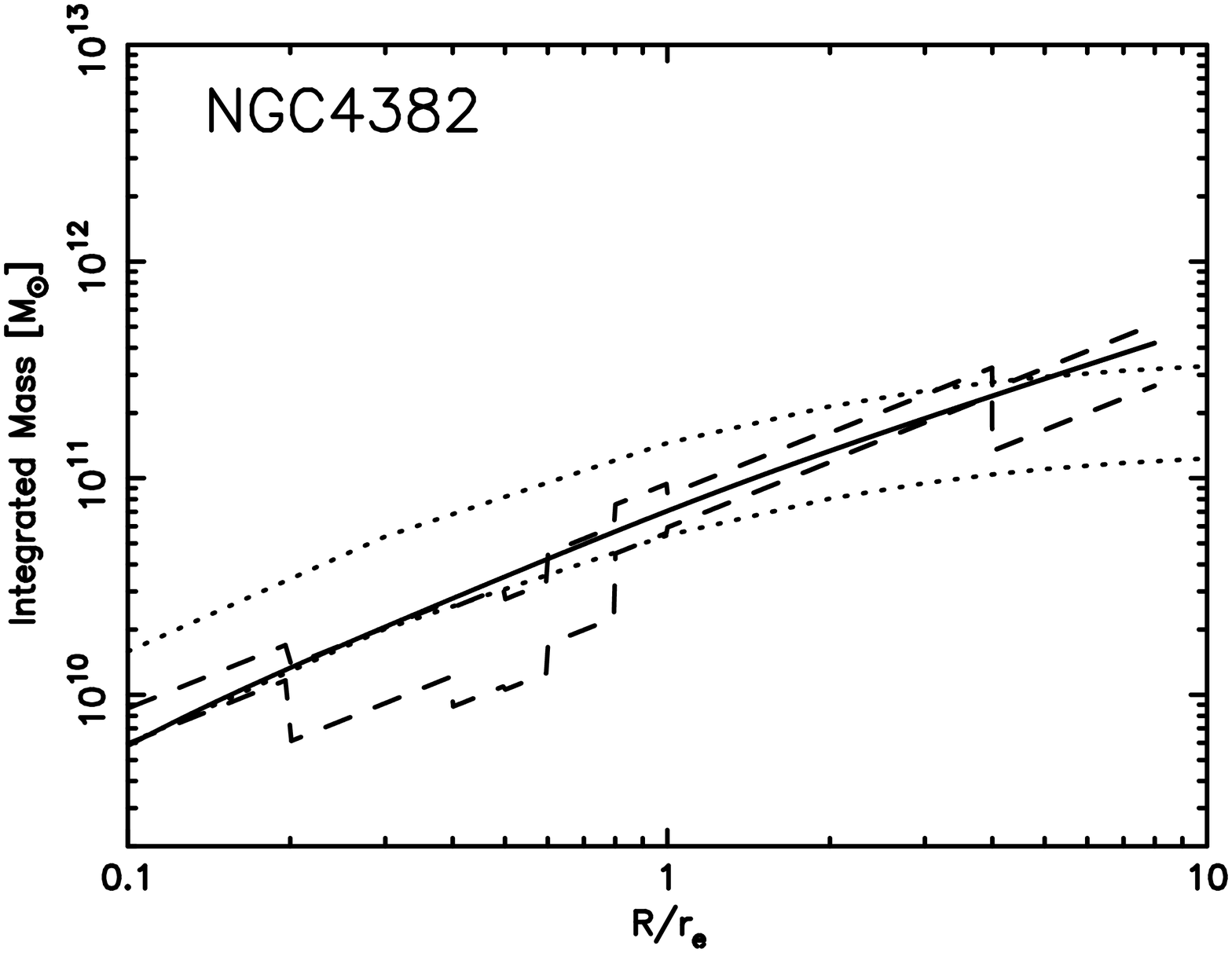}
    \includegraphics[width=4.5cm]{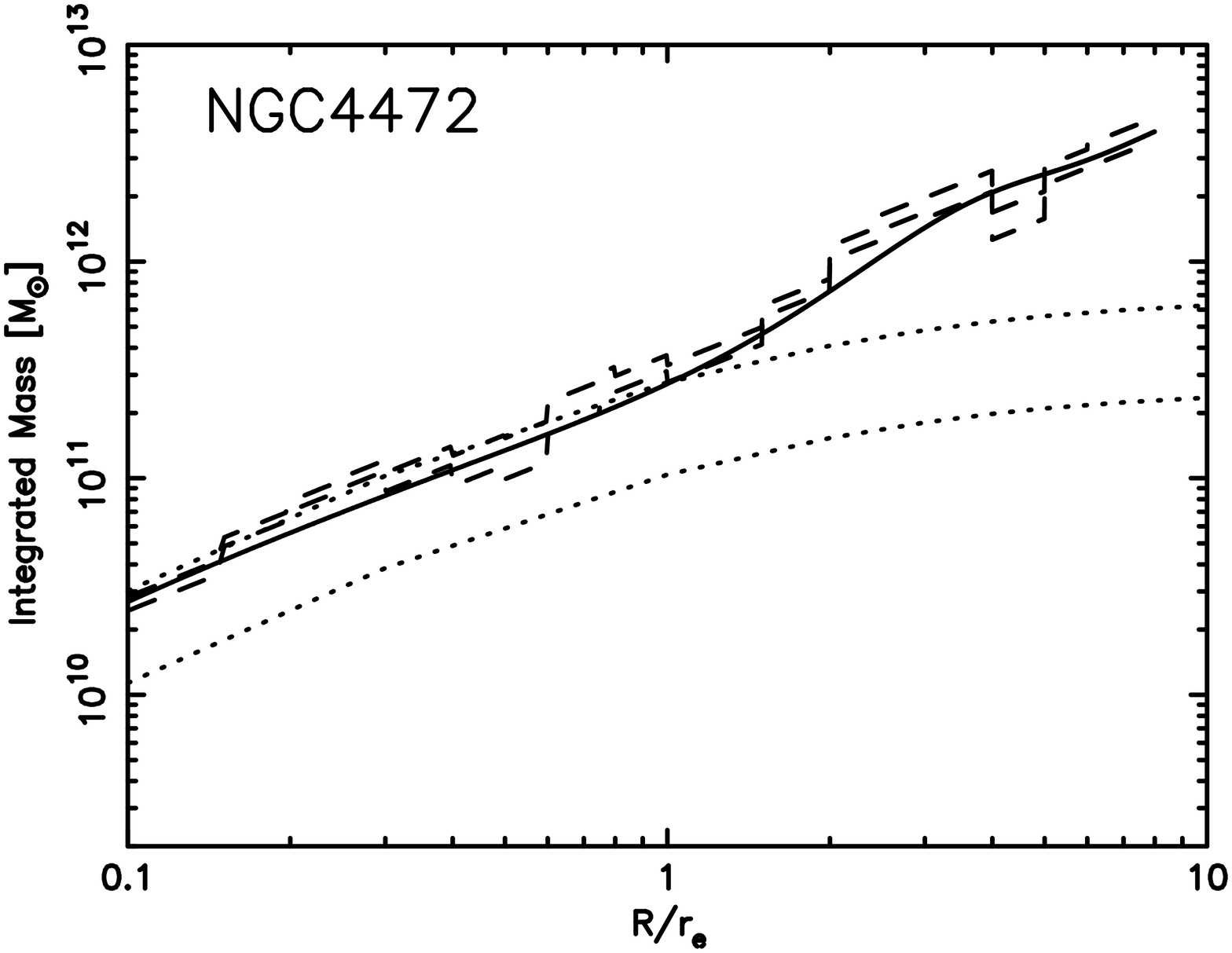}
    \includegraphics[width=4.5cm]{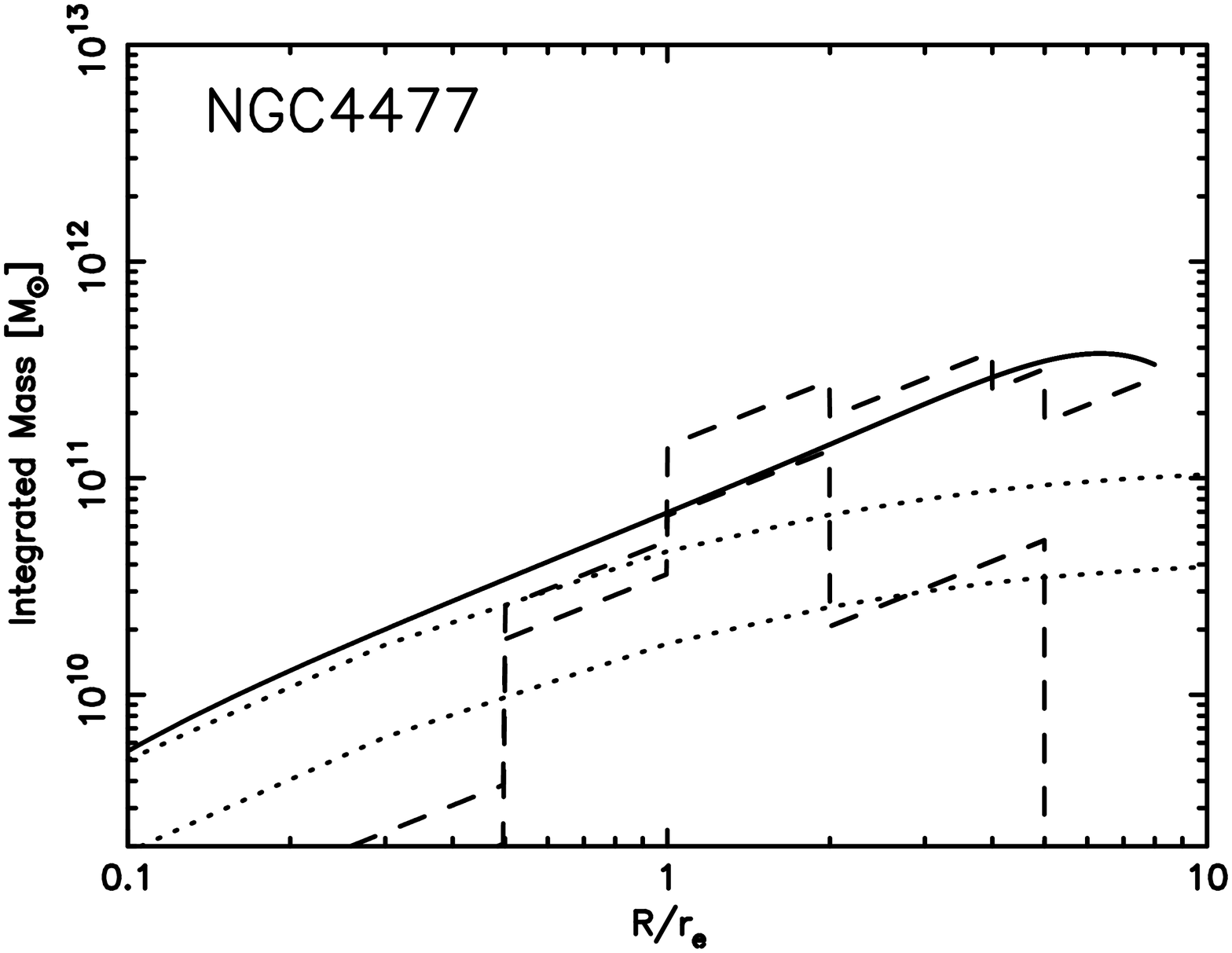}
    \includegraphics[width=4.5cm]{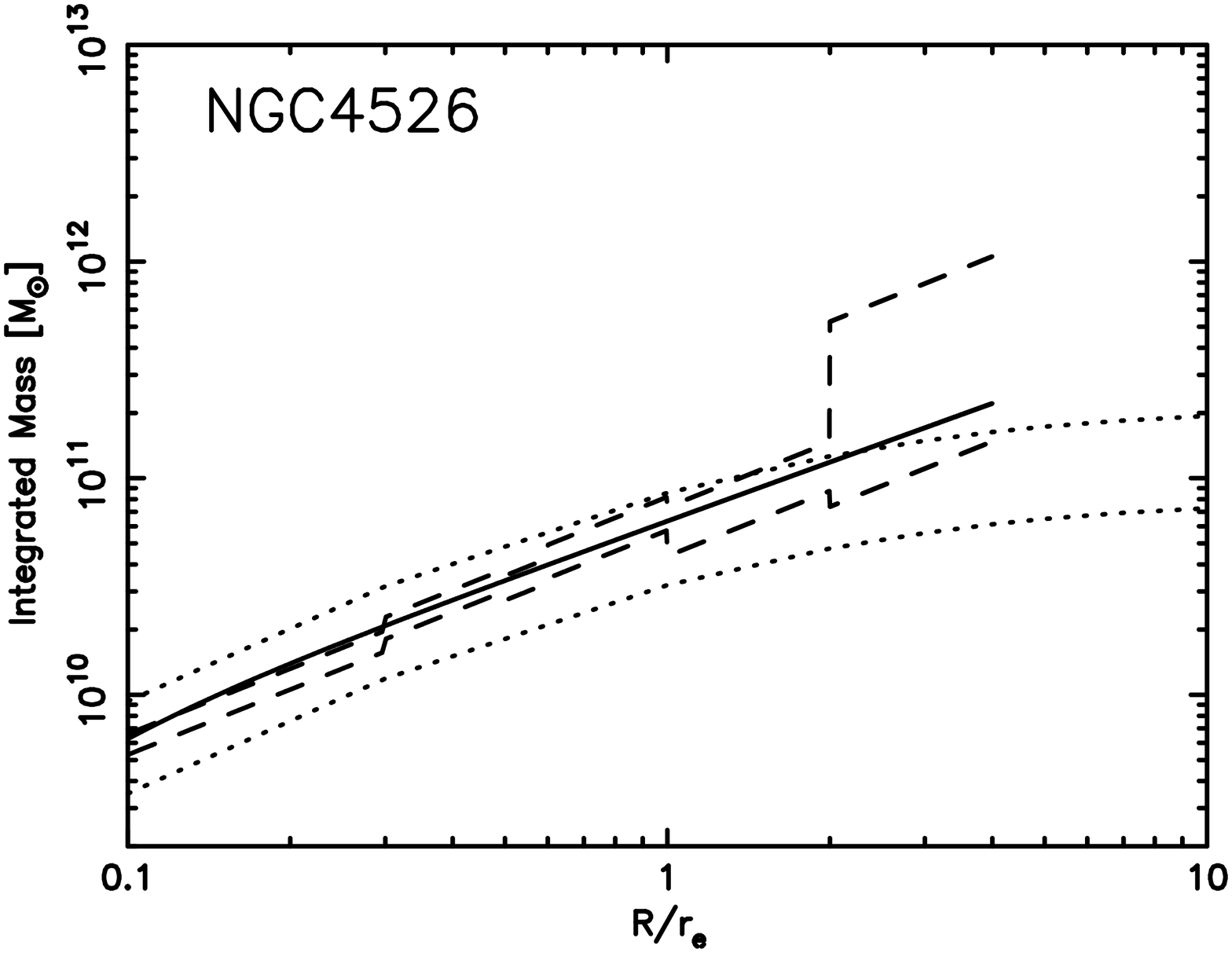}
    \includegraphics[width=4.5cm]{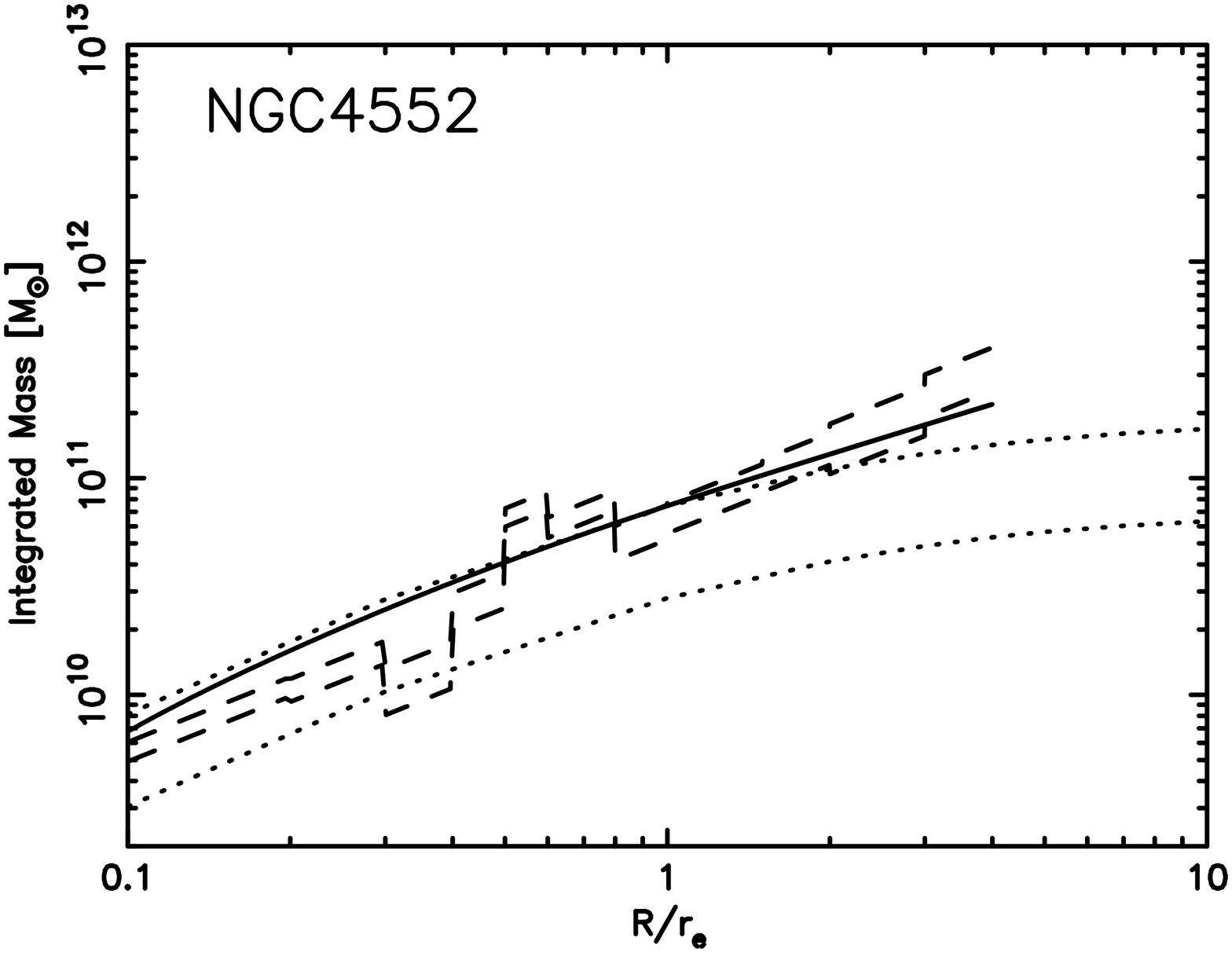}
    \includegraphics[width=4.5cm]{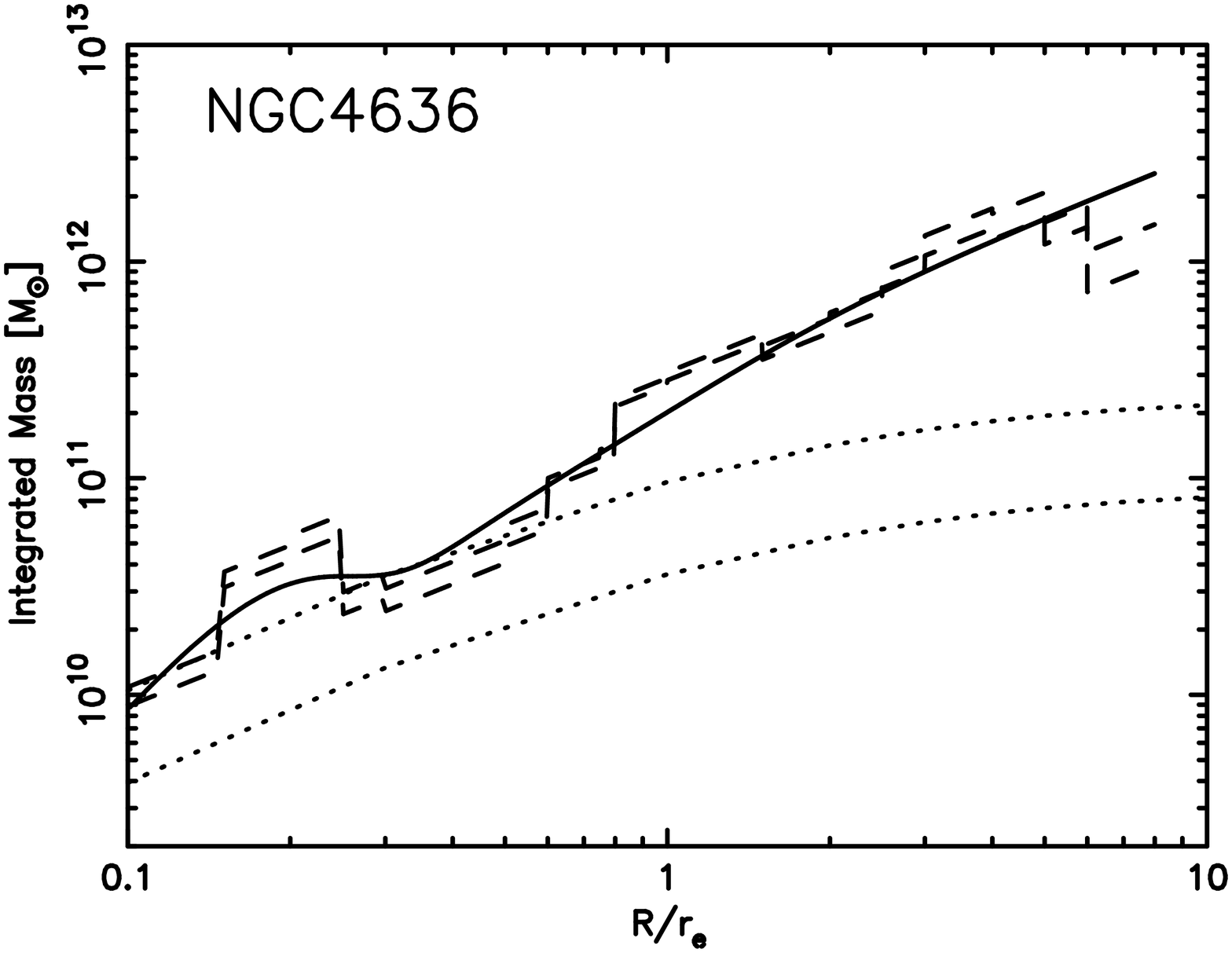}
    \includegraphics[width=4.5cm]{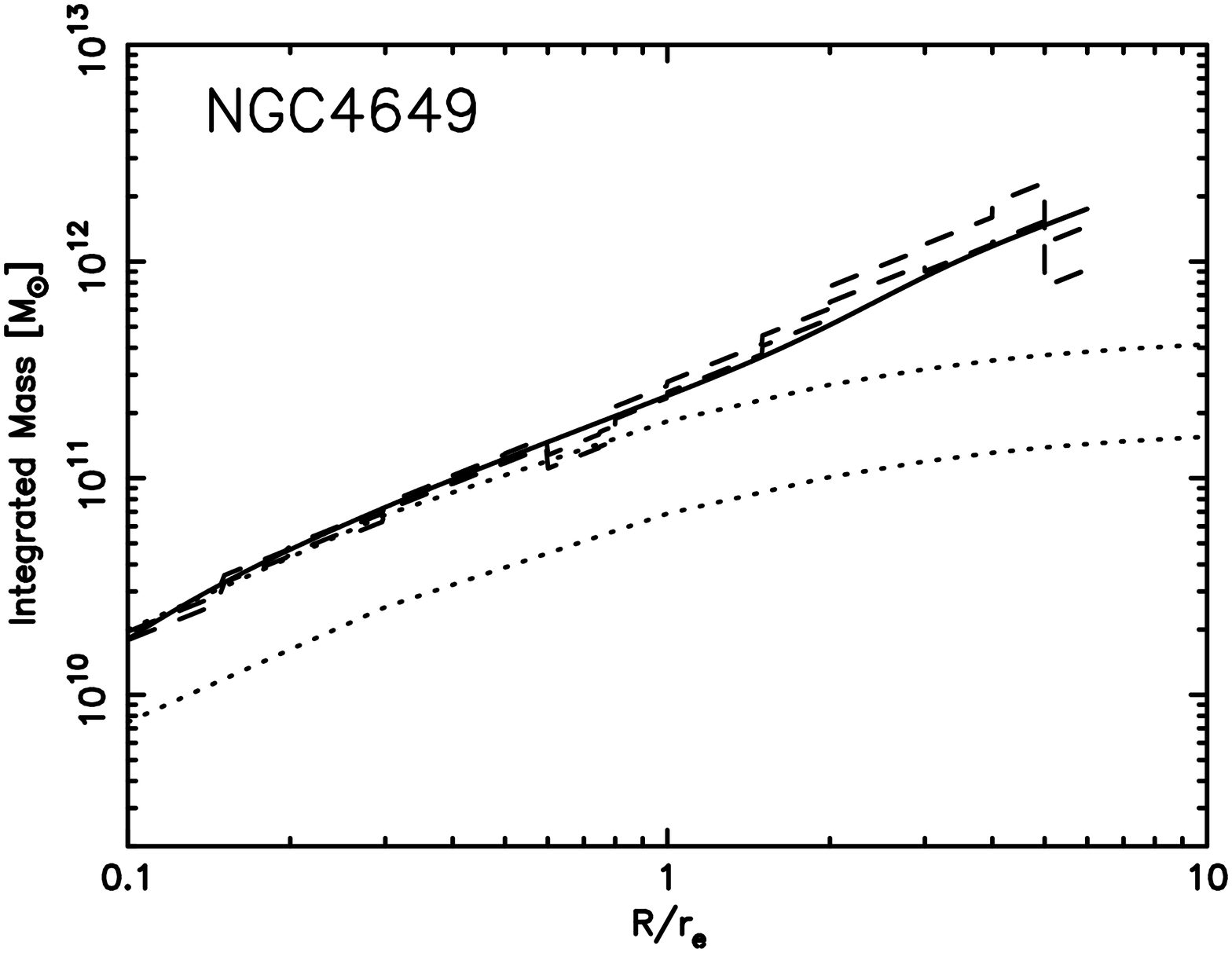}
    \includegraphics[width=4.5cm]{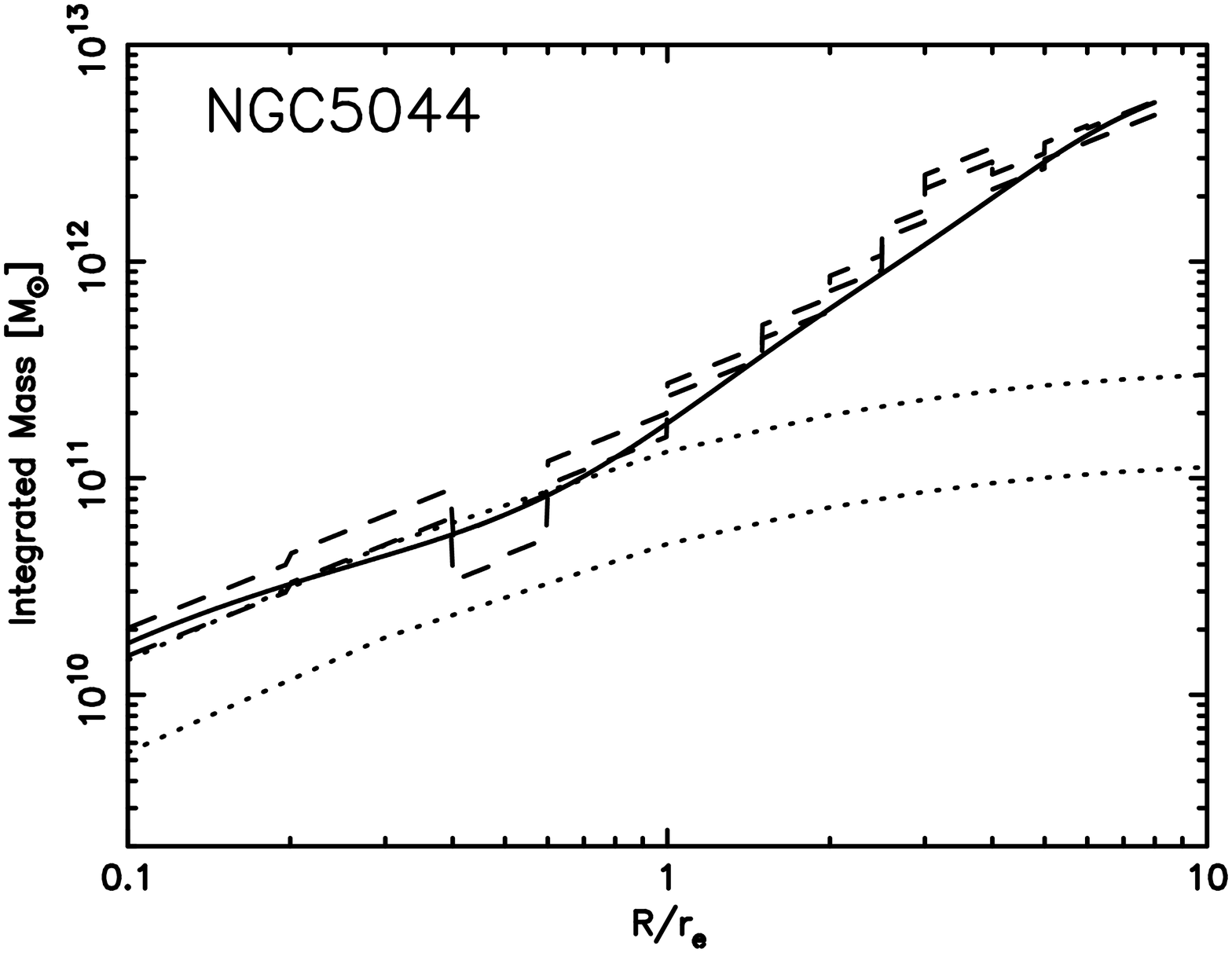}
    \includegraphics[width=4.5cm]{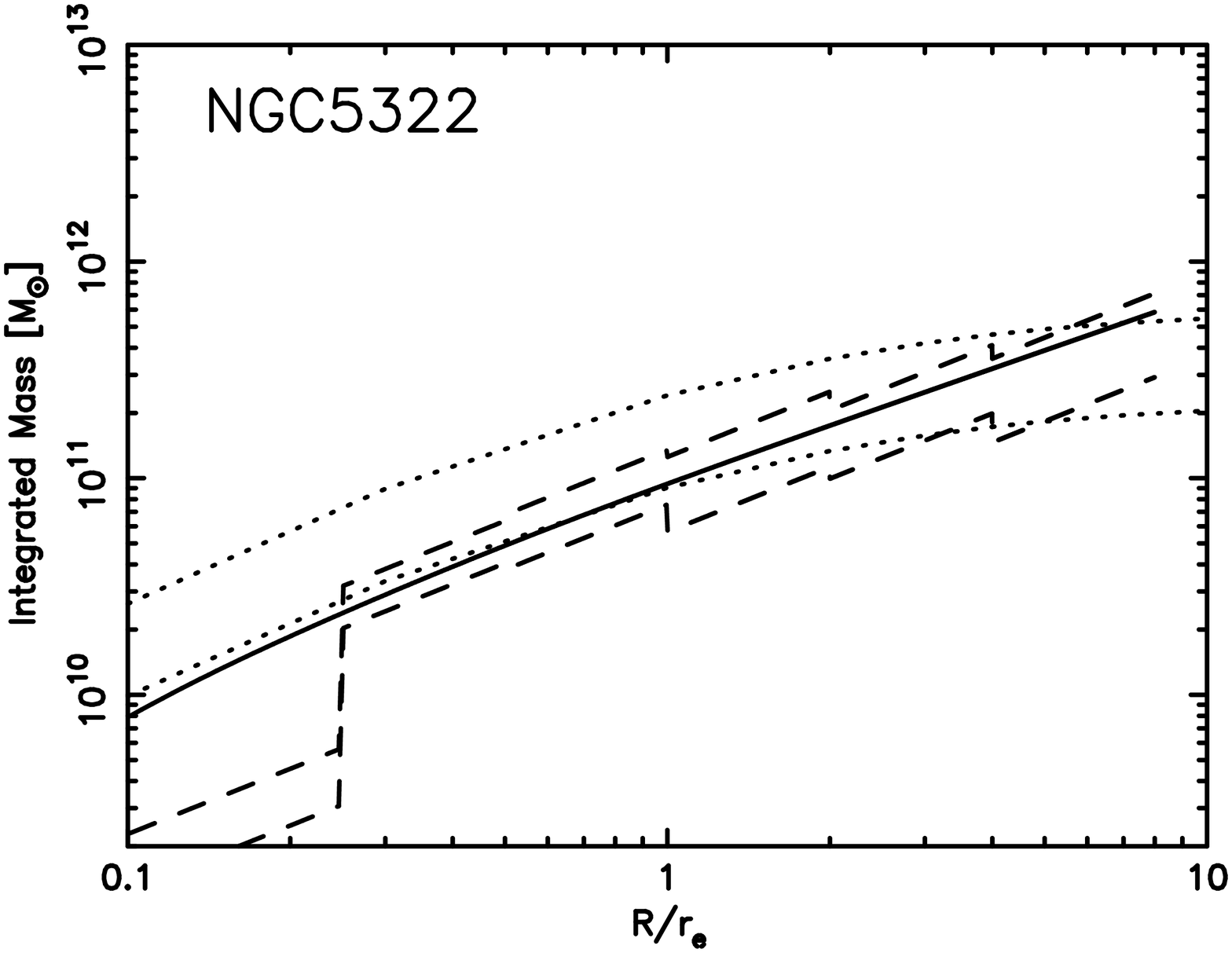}
    \includegraphics[width=4.5cm]{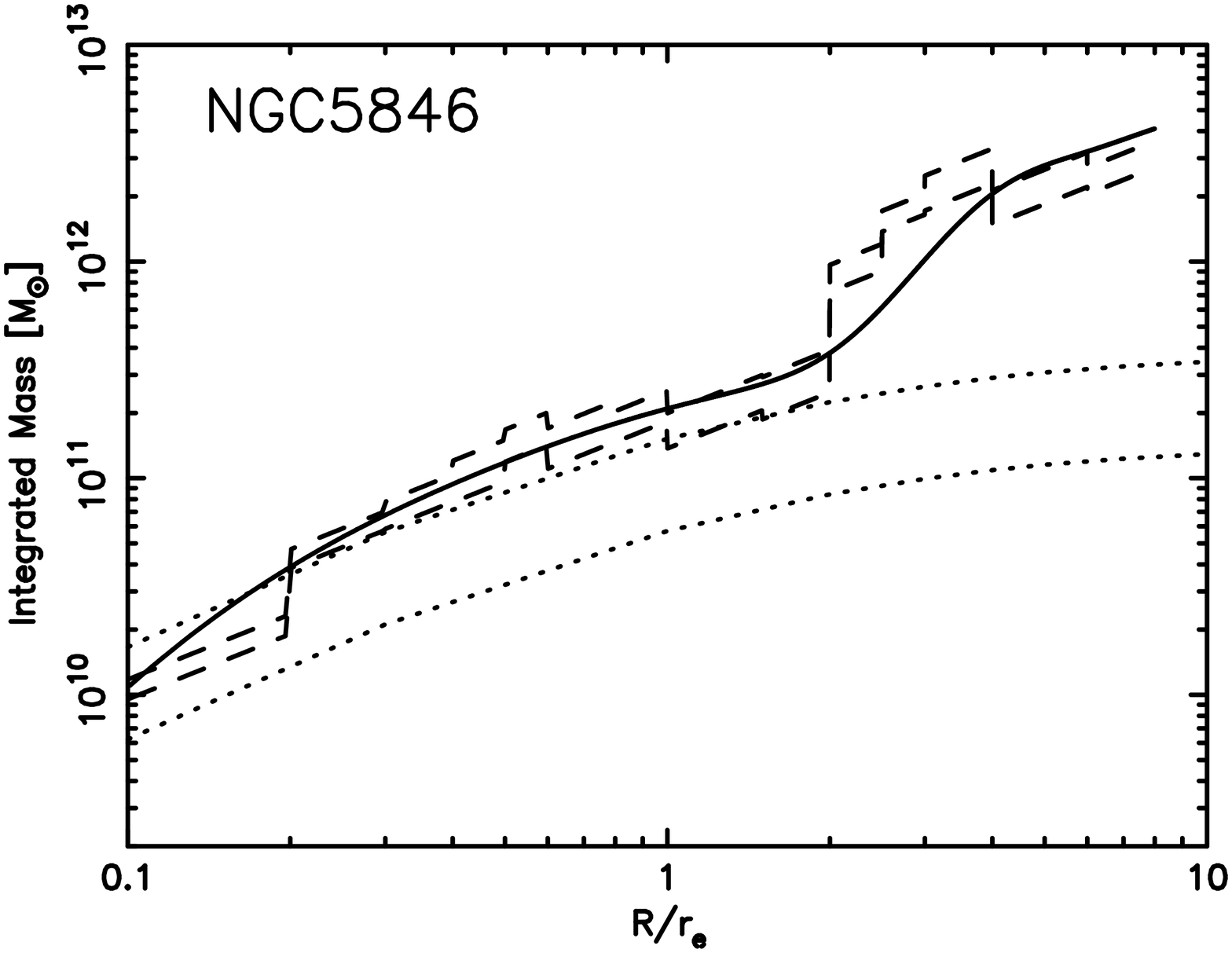}
    \caption{(continued)}
  \end{figure*}

\section{Discussion}

The observed temperature profiles and X-ray luminosities of the ISM
lead to a division of early-type galaxies into two categories: $X_E$
galaxies and $X_C$ galaxies. The $X_E$ galaxies have increasing temperature
profiles and $L_{{\rm ISM}}/L_{\sigma}>1$, whereas the $X_C$ galaxies have flat or negative
temperature gradients and $L_{{\rm ISM}}/L_{\sigma}\lesssim 1$. Here, $L_{\sigma}$
represents the expected energy input from stellar mass loss
\citep{Matsu2001}. In Section \ref{kTvssigma},
the derived ISM temperatures are compared with the stellar velocity
dispersions. In Sections \ref{mlb} and \ref{mlk}, we derive
gravitational-mass-to-light ratios in the B and K band, respectively,
and constrain contributions from stellar mass and differences in dark
mass between the $X_C$ and $X_E$ galaxies. Finally, in Section \ref{dm}, we
discuss dark matter distribution in early-type galaxies themselves and
their luminosity. 

\subsection{ISM temperature vs. stellar velocity dispersion}
\label{kTvssigma}

\begin{figure}[htbp]
    \centering
    \includegraphics[width=7.0cm]{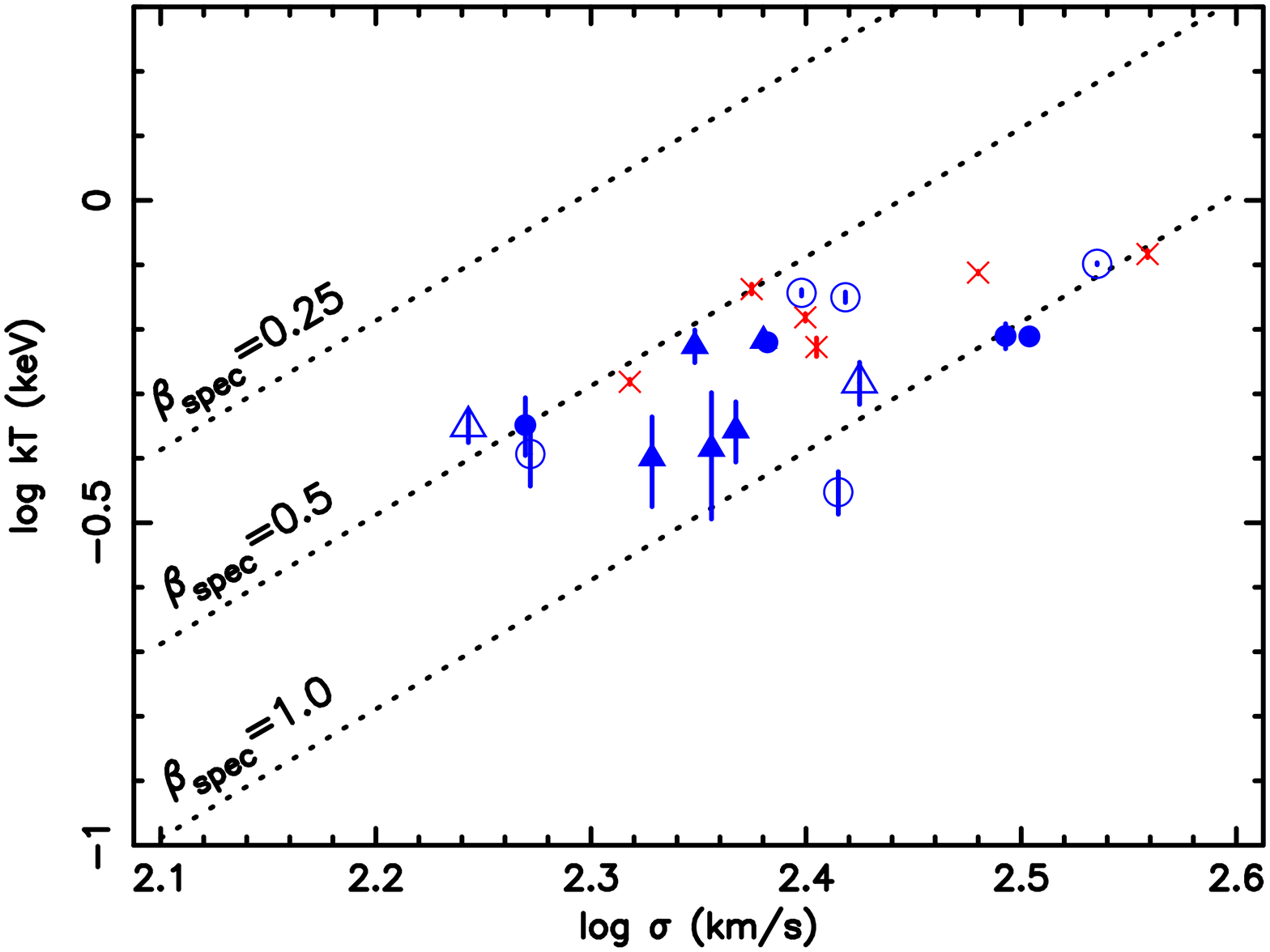}
  \caption{Central ISM temperatures plotted against central stellar
velocity dispersion, $\sigma$. Colors indicate galaxy categories for $X_E$
galaxies (red crosses), field $X_C$ galaxies (filled blue symbols), and
$X_C$ galaxies in clusters (open blue symbols). For $X_C$ galaxies, the
temperatures within 0.5$r_e$ (circles) and 1$r_e$ (triangles) are plotted
for the data with high and poor statistics, respectively. The dotted
lines correspond to $\beta_{spec}$ of 0.25, 0.5, and 1.0 from the top to
bottom.}
\label{fig:kts}
   \end{figure}

Figure \ref{fig:kts} shows the correlation between central ISM temperature and
central stellar velocity dispersion $\sigma^2$ (Table
\ref{tab:tar_list}). The temperature
roughly correlates with the stellar velocity dispersion. The parameter
$\beta_{spec}$ denotes the ratio of stellar velocity dispersion to ISM
temperature with $\beta_{spec}=\mu m_p \sigma^2/kT$, with $\mu$ indicating the mean
molecular weight in terms of proton mass $m_p$. For $X_C$
galaxies, $\beta_{spec}$ is about $\beta_{spec}=$0.5-1.0, which indicates that
the ISM temperatures are consistent with heating due to stellar motion
\citep{Matsu2001}. The galaxies with low $\sigma$ tend to have low $\beta$
values. It may be due to a selection effect, since brighter galaxies
with higher ISM temperatures were the first proposed for observation. 

\begin{figure}
    \centering
    \includegraphics[width=7.0cm]{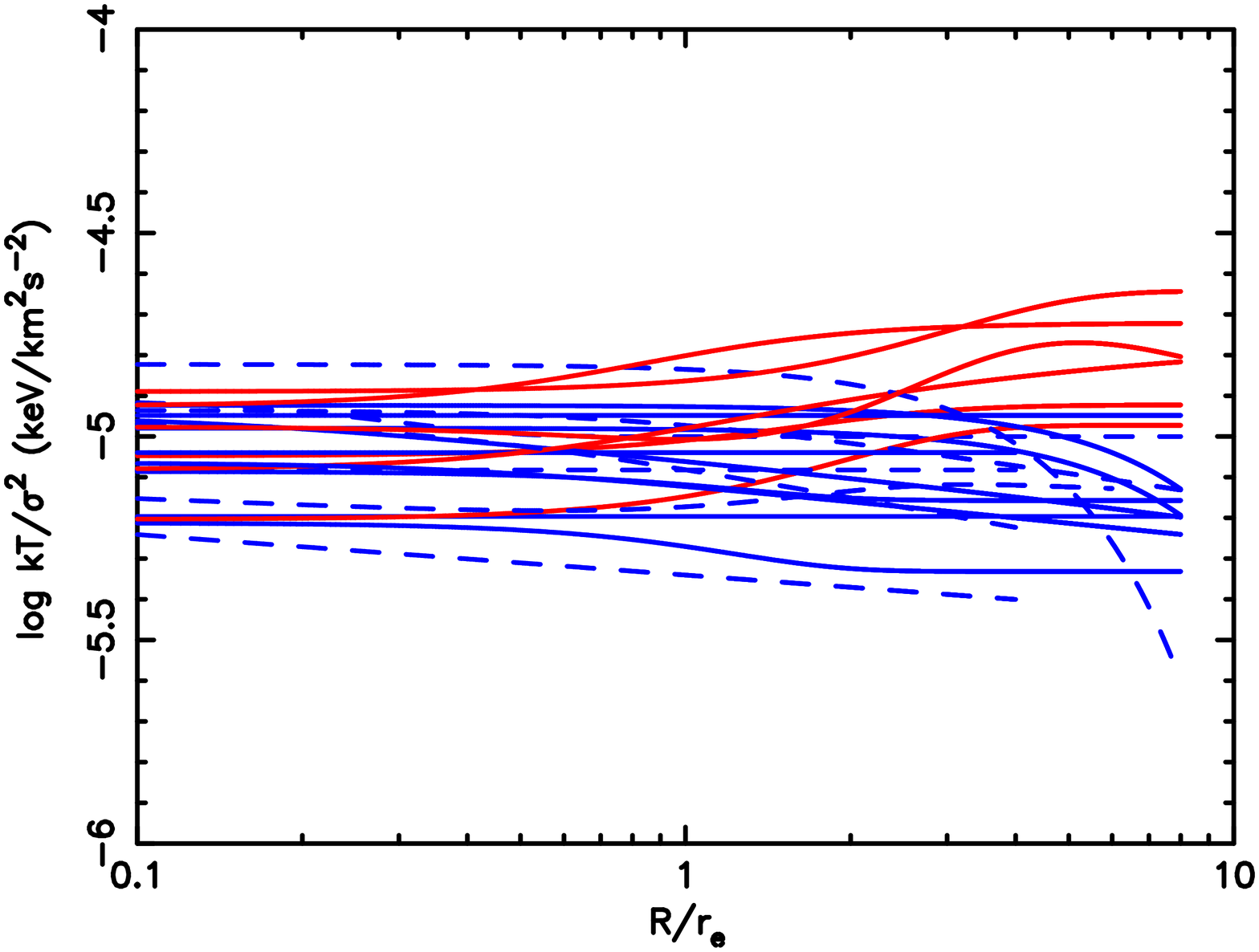}
  \caption{Temperature profiles of the ISM scaled with central stellar
velocity dispersion. The solid red, solid blue, and dashed blue lines
represent the $X_E$ galaxies, field $X_C$ galaxies, and $X_C$ galaxies in
clusters, respectively.}
\label{fig:kts2_all}
   \end{figure}

In Figure \ref{fig:kts2_all}, we plotted the temperature profiles of the sample
galaxies scaled with central stellar velocity dispersion. There is no
significant difference in the ISM temperature between the $X_C$ and $X_E$
galaxies at the central regions. Therefore, the ISM temperatures of
the $X_C$ and $X_E$ galaxies would reflect the same potential. However, in
the outer regions the $X_E$ galaxies have higher ISM temperatures than
the $X_C$ galaxies for the same stellar velocity dispersion. This is due
to the difference in potential due to the hot intra-group medium
around the $X_E$ galaxies. 

\subsection{Mass-to-light ratio in B-band}
\label{mlb}

\begin{figure*}
   \centering
    \includegraphics[width=6.5cm]{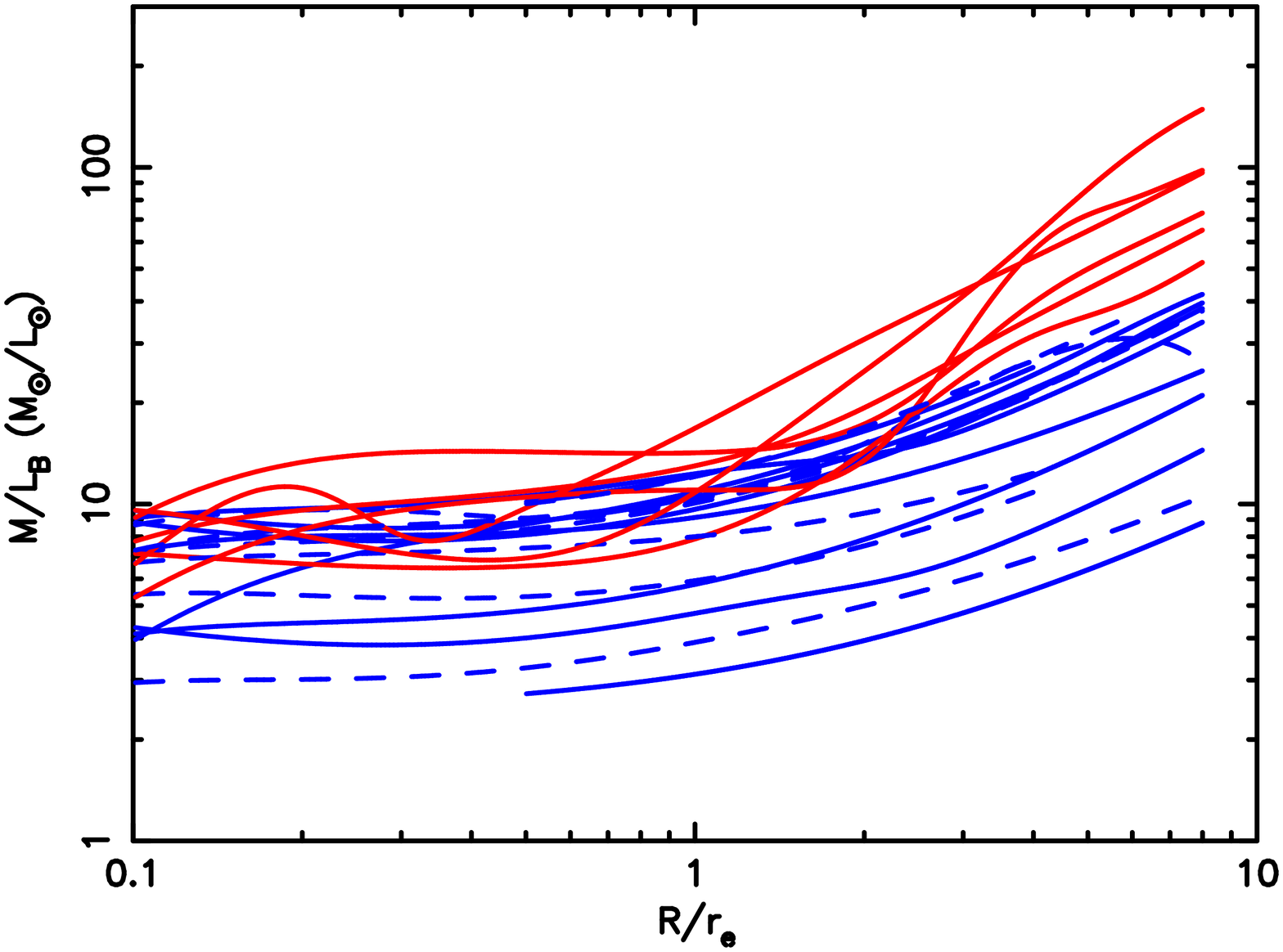}
    \includegraphics[width=6.5cm]{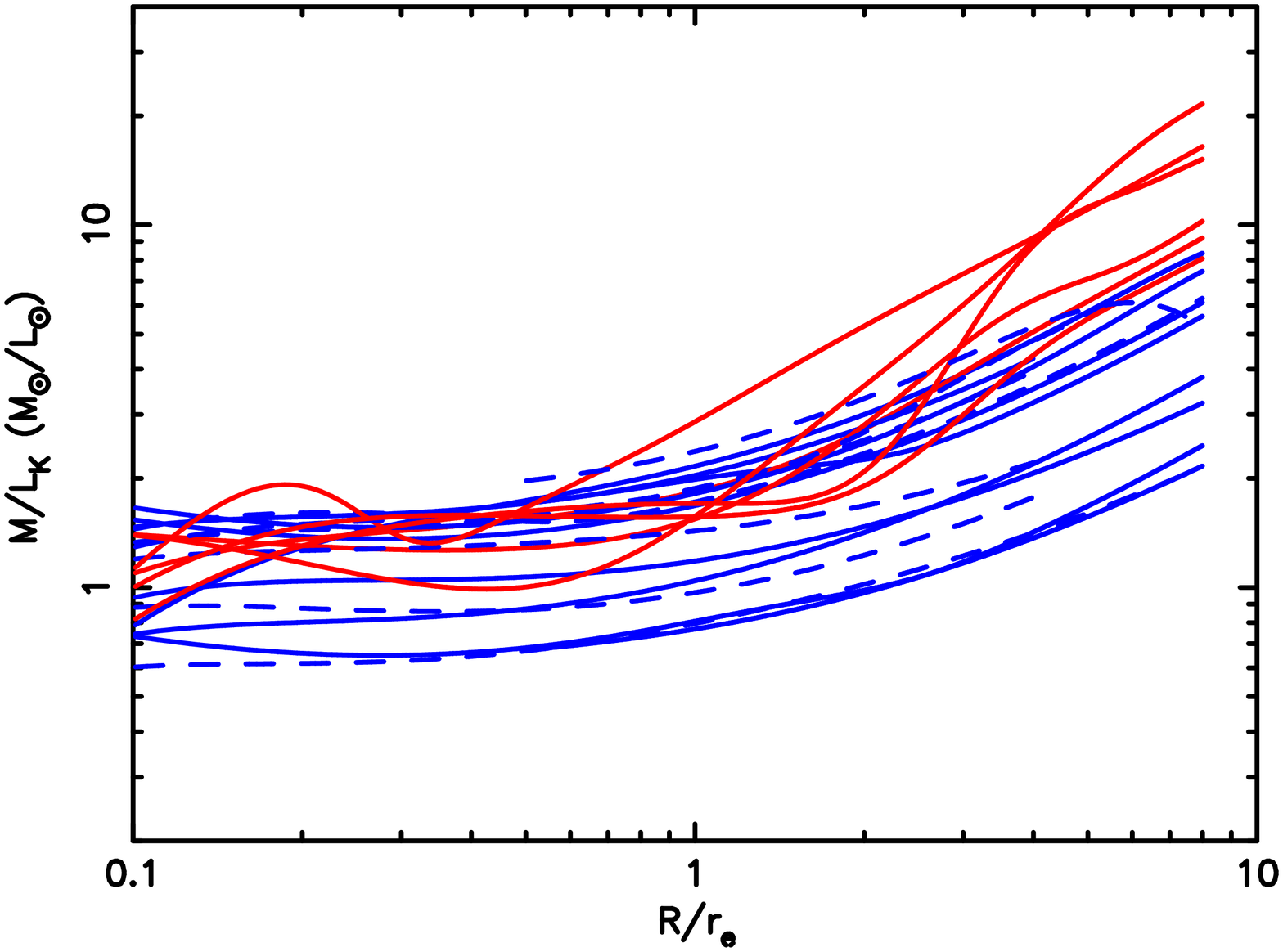}
  \caption{Profiles of integrated $M/L_B$ (left) and $M/L_K$
    (right). The meanings of colors and lines are the same as those in
    Figure \ref{fig:kts2_all}.}
\label{fig:mlsim_all}
    \end{figure*}

In order to compare the difference in dark matter profiles between the
$X_C$ and $X_E$ galaxies, we used the mass-to-light ratio, $M/L_B$. We assumed
the de Vaucouleurs law for the stellar distribution. Figure \ref{fig:mlsim_all}
summarizes the integrated $M/L_B$ profiles. For NGC 1549, NGC 4477, and
NGC 5322, profiles within 0.5$r_e$ were not plotted in the figure, since
we used only XMM-Newton data for those galaxies. From 0.2$r_e$ to 1$r_e$,
$M/L_B$ of each $X_C$ galaxy is nearly constant at 3--10, which is
consistent with typical values of stellar $M/L_B$. Even in the $X_E$
galaxies, $M/L_B$ within 0.5--1$r_e$ are also flat. These results
suggest that stellar mass dominates within $r<0.5$-$1r_e$. In NGC
4636, a wavy $M/L_B$ profile is seen at $r<0.3r_e$, which is artificially
caused by the complicated X-ray structures in the central regions
discovered by Chandra (\citealt{Jones2002}). Hereafter, we denote
integrated $M/L_B$ within $s r_e$ as $M/L_B(r< s r_e)$, where $s$ is a constant. 

\begin{figure*}
   \centering
    \includegraphics[width=6.5cm]{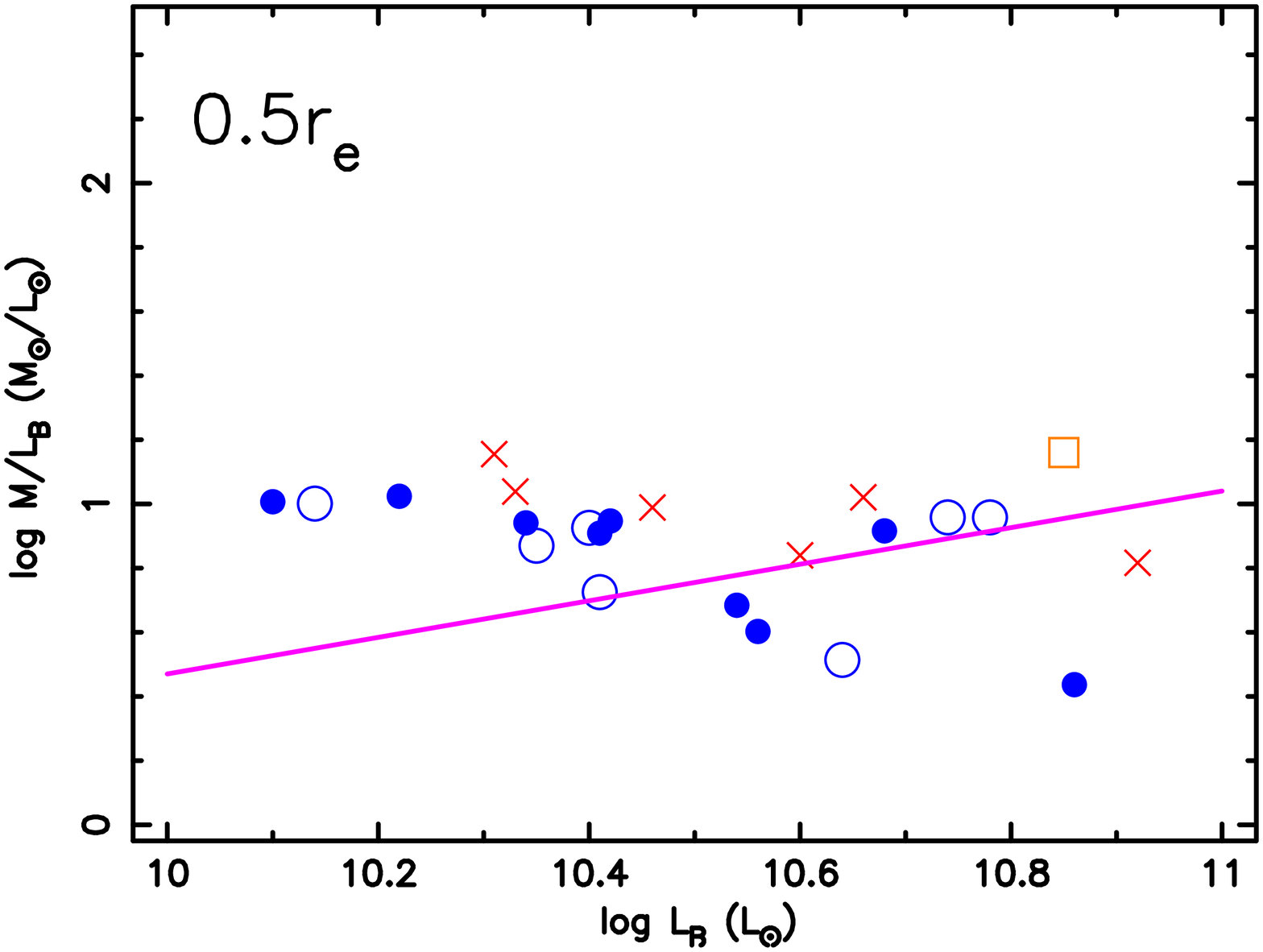}
    \includegraphics[width=6.5cm]{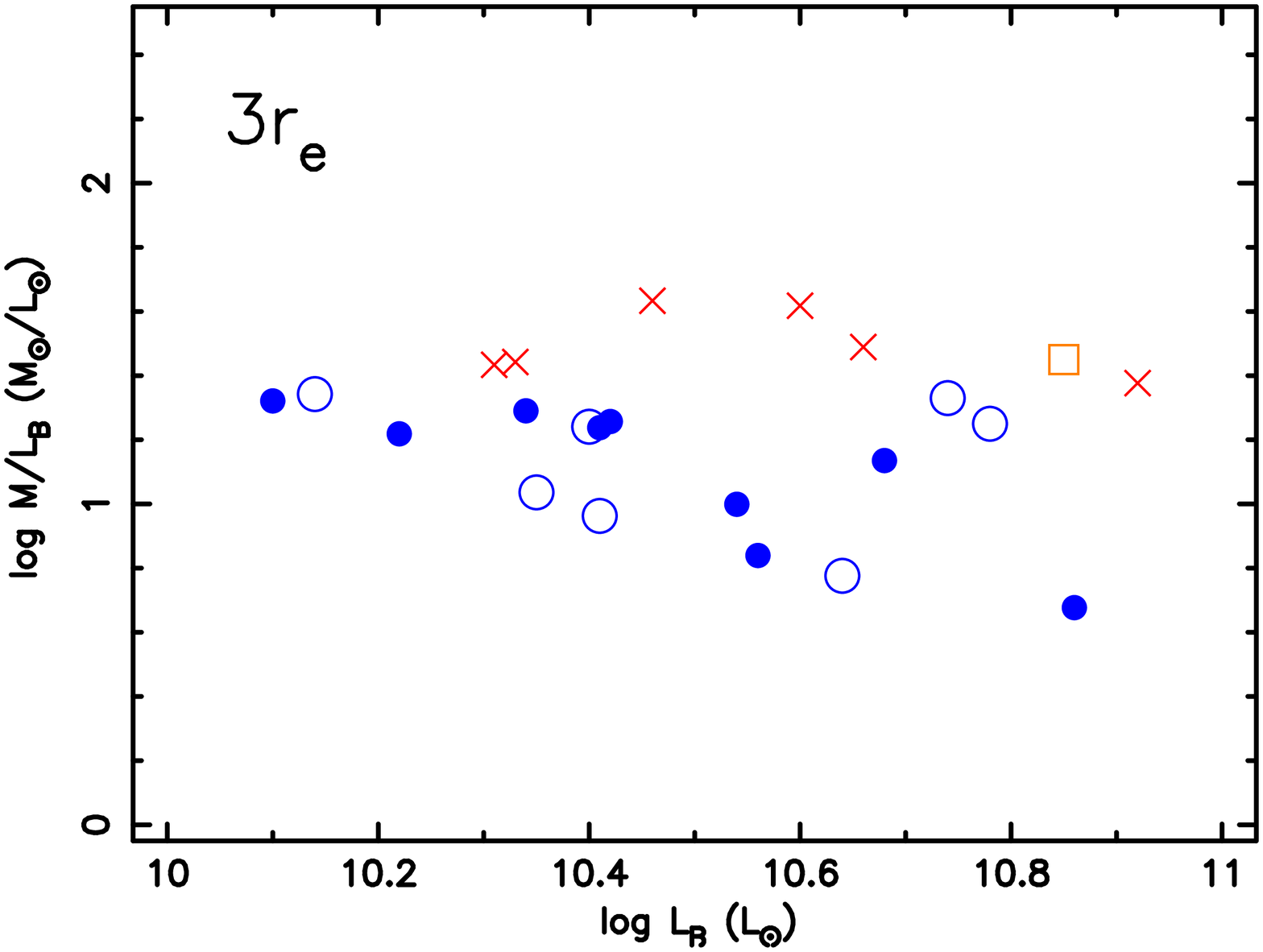}
    \includegraphics[width=6.5cm]{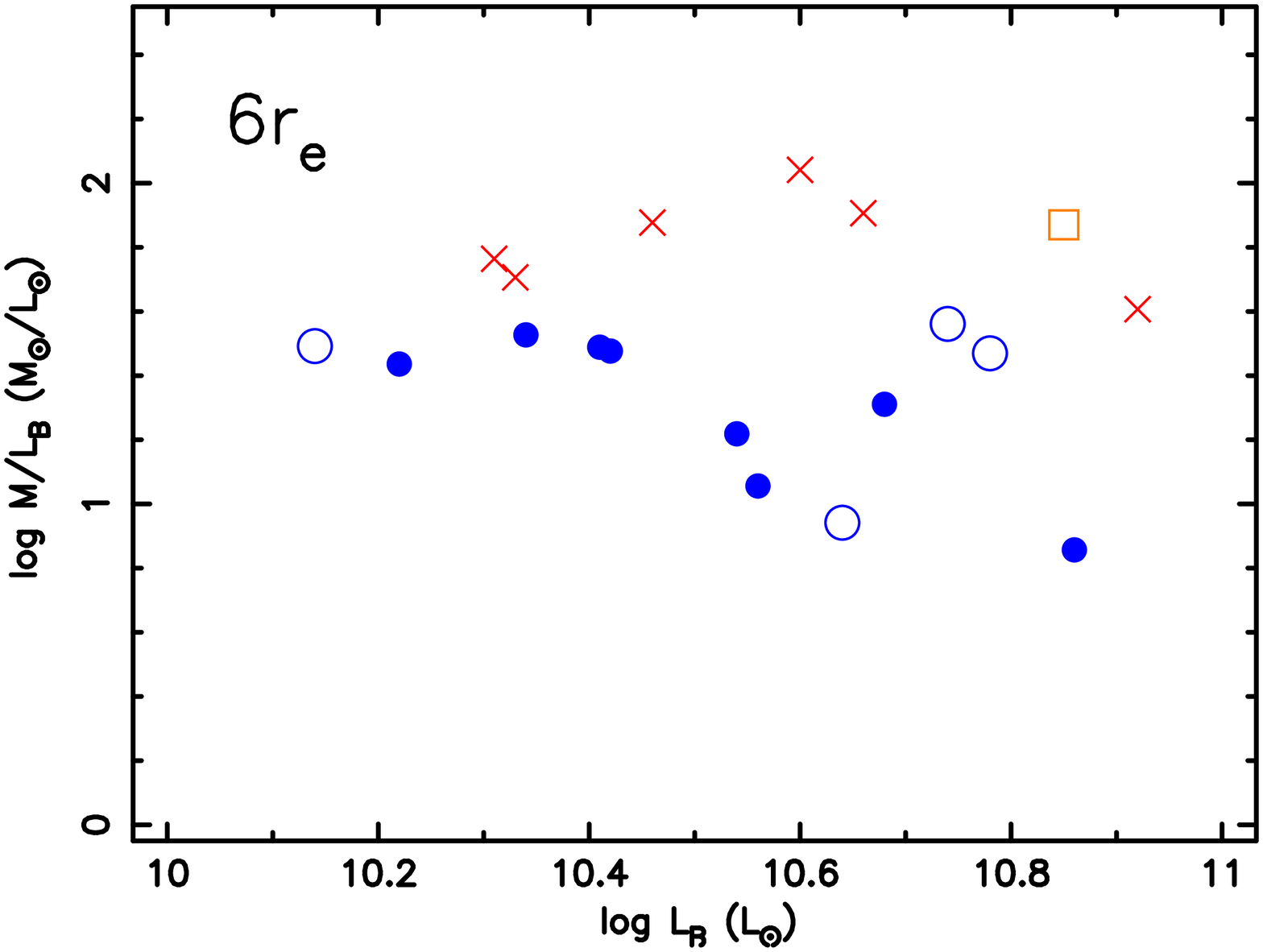}
  \caption{$M/L_B(r<0.5r_e)$ (top left), $M/L_B(r<3r_e)$ (top right),
    and $M/L_B(r<6r_e)$ (bottom) against $L_B$. The quadrangle represents
    the value of M87 by \cite{Matsu2002}. Meanings of other symbols are the
same as those in Figure \ref{fig:lxlbs2}. 
The solid line represents the correlation derived from the
stellar velocity dispersion by \citet{Gerh2001}.}
\label{fig:mllb_sim}
   \end{figure*}

The top left panel of Figure \ref{fig:mllb_sim} shows the relationship of
$M/L_B(r<0.5 r_e)$ to total B-band luminosity, $L_B$. 
In this plot, we add $M/L_B$ of  the cD  galaxy of the Virgo
cluster, M 87 obtained from XMM-Newton observation \citep{Matsu2002}.
$M/L_B(r<0.5 r_e)$ of the $X_E$ and $X_C$
galaxies is about 10, with significant scatter. A correlation between
$M/L_B$ and $L_B$ elliptical galaxies was found by optical measurements
\citep{Gerh2001},
where gravitational mass was derived from stellar velocity dispersion
at the central region of elliptical galaxies. Our $M/L_B$ and $L_B$ relation
scatters around the relation found by \citet{Gerh2001}.

On the other hand, at $r>1r_e$, the derived $M/L_B$ starts to increase
(Figure \ref{fig:mlsim_all}). The $X_C$ galaxies have similarly shaped $M/L_B$
profiles. $M/L_B(r<3 r_e)$ and $M/L_B(r<6 r_e)$ of the $X_C$ galaxies are about
6$\sim25 M_{\odot}/L_{\odot}$. The $X_E$ galaxies have systematically
larger $M/L_B$ values than the $X_C$ galaxies at $>3 r_e$. (Figure
\ref{fig:mllb_sim}). The $X_E$ galaxies have $M/L_B (r<3 r_e)$ and
$M/L_B(r<6 r_e)$ of $25\sim 40 M_{\odot}/L_{\odot}$ and $40\sim 100
M_{\odot}/L_{\odot}$, respectively. These results indicate that dark
matter is common in early-type galaxies. In addition, the $X_E$ galaxies
have more dark matter than the $X_C$ galaxies.

\begin{table*}
\caption{Integrated $M/L_B$ and $M/L_K$ at 0.5, 3 and 6$r_e$}
\label{tab:ml}
\centering
\begin{tabular}{lcccccc}
\hline \hline
Galaxy & $M/L_B$($<0.5r_e$) & $M/L_B$($<3r_e$) & $M/L_B$($<6r_e$) & $M/L_K$($<0.5r_e$) & $M/L_K$($<3r_e$) & $M/L_K$($<6r_e$) \\
       & $(M_{\odot}/L_{\odot})$ & $(M_{\odot}/L_{\odot})$ & $(M_{\odot}/L_{\odot})$ & $(M_{\odot}/L_{\odot})$ &$(M_{\odot}/L_{\odot})$ & $(M_{\odot}/L_{\odot})$\\\hline
IC1459 &   8.8 &  18.1 &  29.9 &   1.4 &   2.9 &   4.8 \\
NGC720 &   8.7 &  19.5 &  33.6 &   1.7 &   3.9 &   6.7 \\
NGC1316 &   9.1 &  17.8 &  29.5 &   1.5 &   3.0 &   4.9 \\
NGC1332 &  10.6 &  16.5 &  27.3 &   1.7 &   2.7 &   4.4 \\
NGC1395 &  10.9 &  27.7 &  50.9 &   1.5 &   3.9 &   7.2 \\
NGC1399 &  14.3 &  27.1 &  58.1 &   1.6 &   3.0 &   6.4 \\
NGC1549 &  10.2 &  20.9 &   --- &   1.7 &   3.5 &   --- \\
NGC3585 &   4.0 &   6.9 &  11.4 &   0.7 &   1.2 &   1.9 \\
NGC3607 &   8.1 &  17.3 &  30.8 &   1.5 &   3.2 &   5.8 \\
NGC3665 &   4.8 &  10.0 &  16.6 &   0.9 &   1.8 &   3.0 \\
NGC3923 &   8.2 &  13.7 &  20.4 &   1.1 &   1.8 &   2.6 \\
NGC4365 &   8.4 &  17.4 &   --- &   1.6 &   3.2 &   --- \\
NGC4382 &   3.3 &   6.0 &   8.7 &   0.7 &   1.2 &   1.8 \\
NGC4472 &   6.5 &  23.8 &  40.4 &   1.3 &   4.7 &   7.9 \\
NGC4477 &  10.0 &  22.0 &  31.0 &   2.0 &   4.3 &   6.1 \\
NGC4526 &   5.3 &   9.2 &   --- &   0.9 &   1.5 &   --- \\
NGC4552 &   7.4 &  10.9 &   --- &   1.3 &   1.9 &   --- \\
NGC4636 &   9.8 &  43.0 &  75.3 &   1.7 &   7.3 &  12.8 \\
NGC4649 &   9.1 &  21.4 &  36.4 &   1.6 &   3.8 &   6.5 \\
NGC5044 &   6.9 &  41.5 & 109.9 &   1.0 &   6.0 &  15.9 \\
NGC5322 &   2.7 &   4.8 &   7.2 &   0.7 &   1.2 &   1.8 \\
NGC5846 &  10.5 &  30.8 &  80.6 &   1.6 &   4.8 &  12.5 \\
\hline
\end{tabular}
\end{table*}

\subsection{Mass-to-light ratios in K band}
\label{mlk}

\begin{figure}
   \centering
    \includegraphics[width=7.0cm]{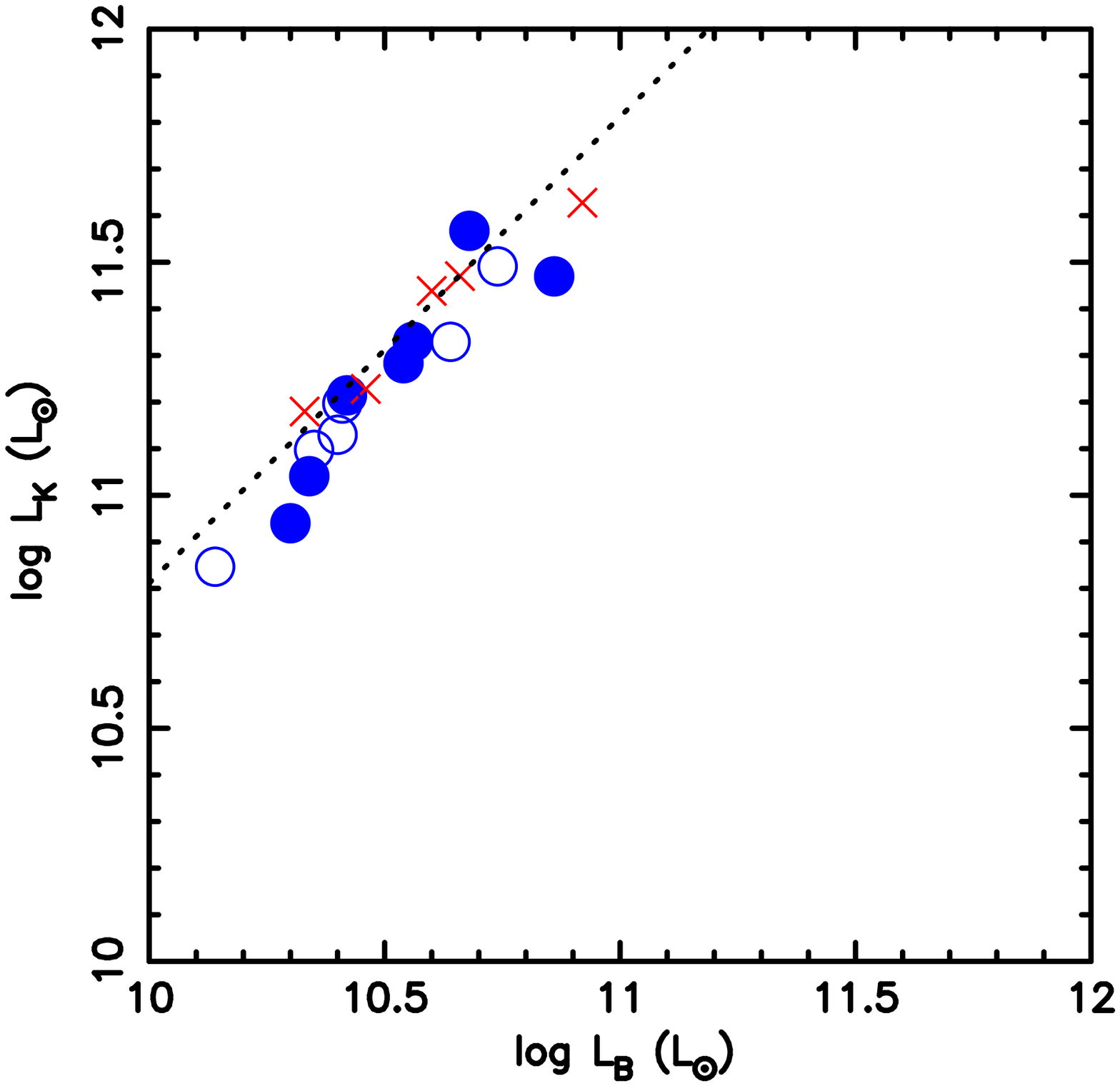}
  \caption{Relationship between B-band and K-band luminosity of the
sample galaxies. Meanings of the symbols are the same as those in
Figure \ref{fig:lxlbs2}. The dotted line corresponds to the appropriate
color $B-K=4.2$ for early-type galaxies \citep{Lin2004}.}
\label{fig:lblk}
    \end{figure}

Historically, the B-band mass-to-light ratio has been used in such
studies. However, K-band luminosity well describes the stellar
mass. Thus, we also derived the K-band mass-to-light ratio to study
the dark matter profiles. We calculated the K-band luminosity $L_K$ from
the Two Micron All Sky Survey (2MASS). The effect of Galactic
extinction was corrected using the NASA/IPAC Extragalactic Database
(NED). The relationship between $L_B$ and $L_K$ is close to that of
$B-K=4.2$, which is the appropriate color of stars in early-type galaxies
\citep{Lin2004}, with some scatter (Figure \ref{fig:lblk}). 

The right panel of Figure \ref{fig:mlsim_all} shows the integrated profiles of the
K-band mass-to-light ratio, $M/L_K$, of the sample galaxies. At the
central region, the $X_E$ and $X_C$ galaxies have similar $M/L_K$ at $\sim 1$. 

\begin{figure*}
   \centering
    \includegraphics[width=6.5cm]{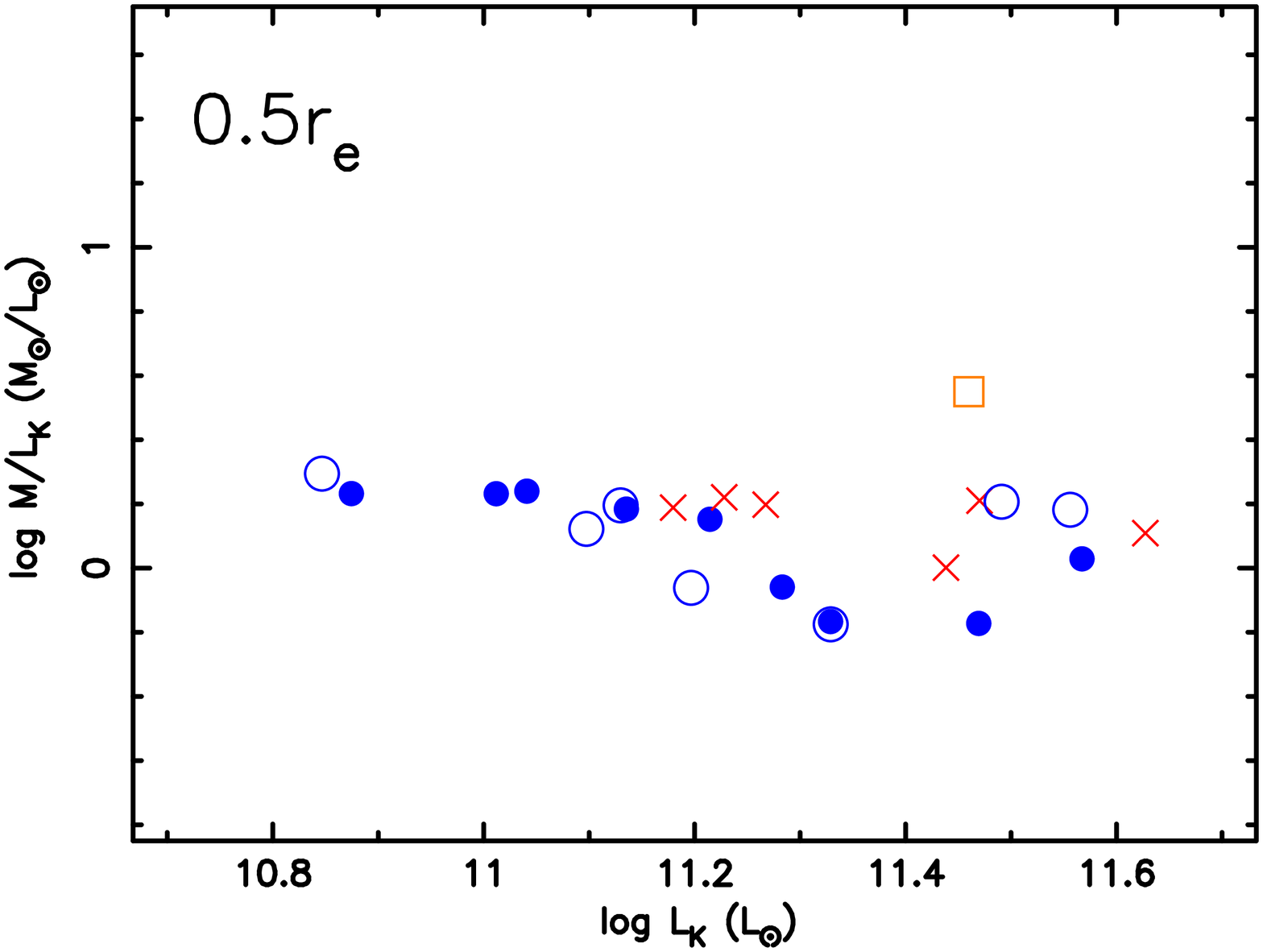}
    \includegraphics[width=6.5cm]{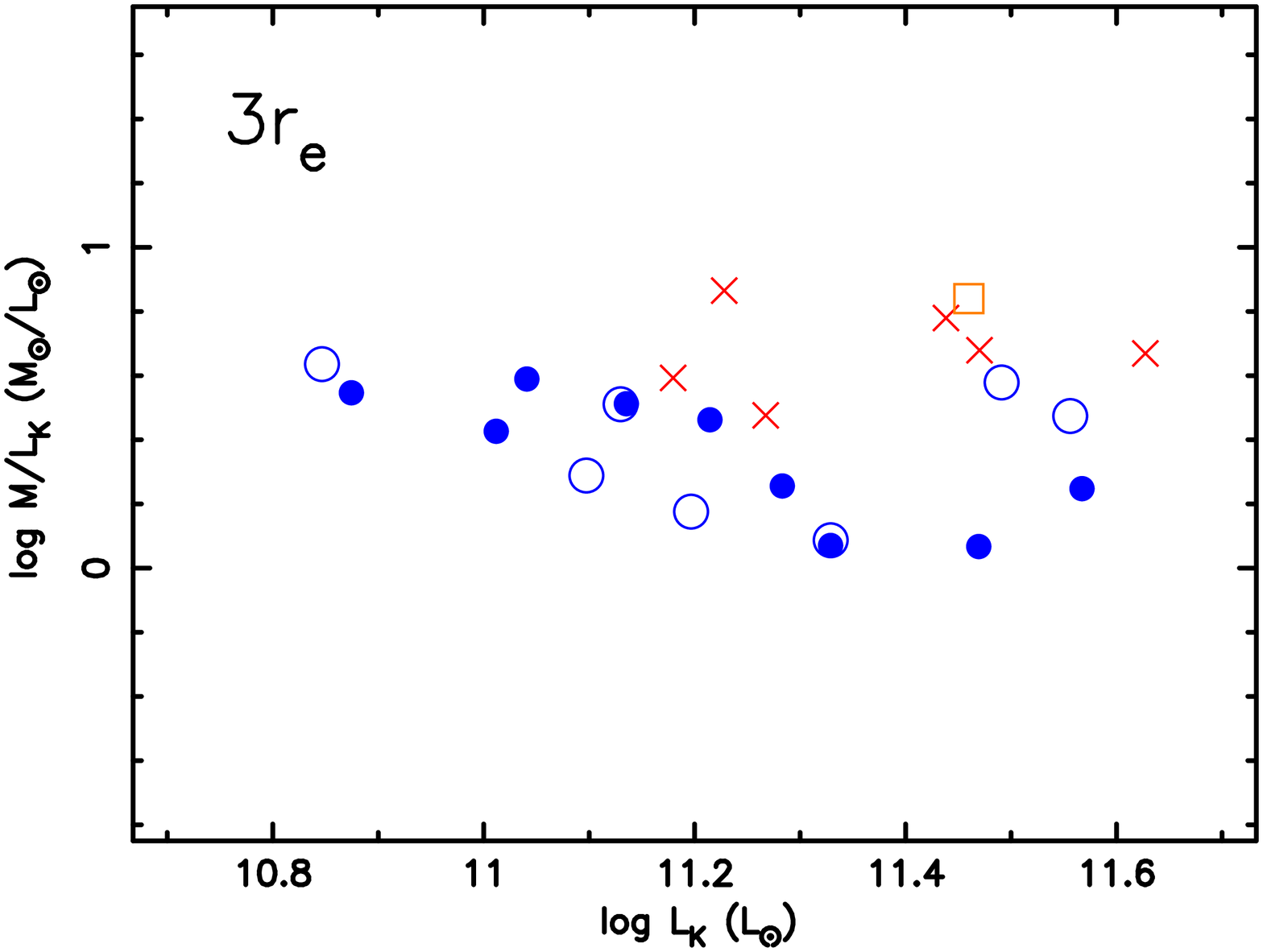}
    \includegraphics[width=6.5cm]{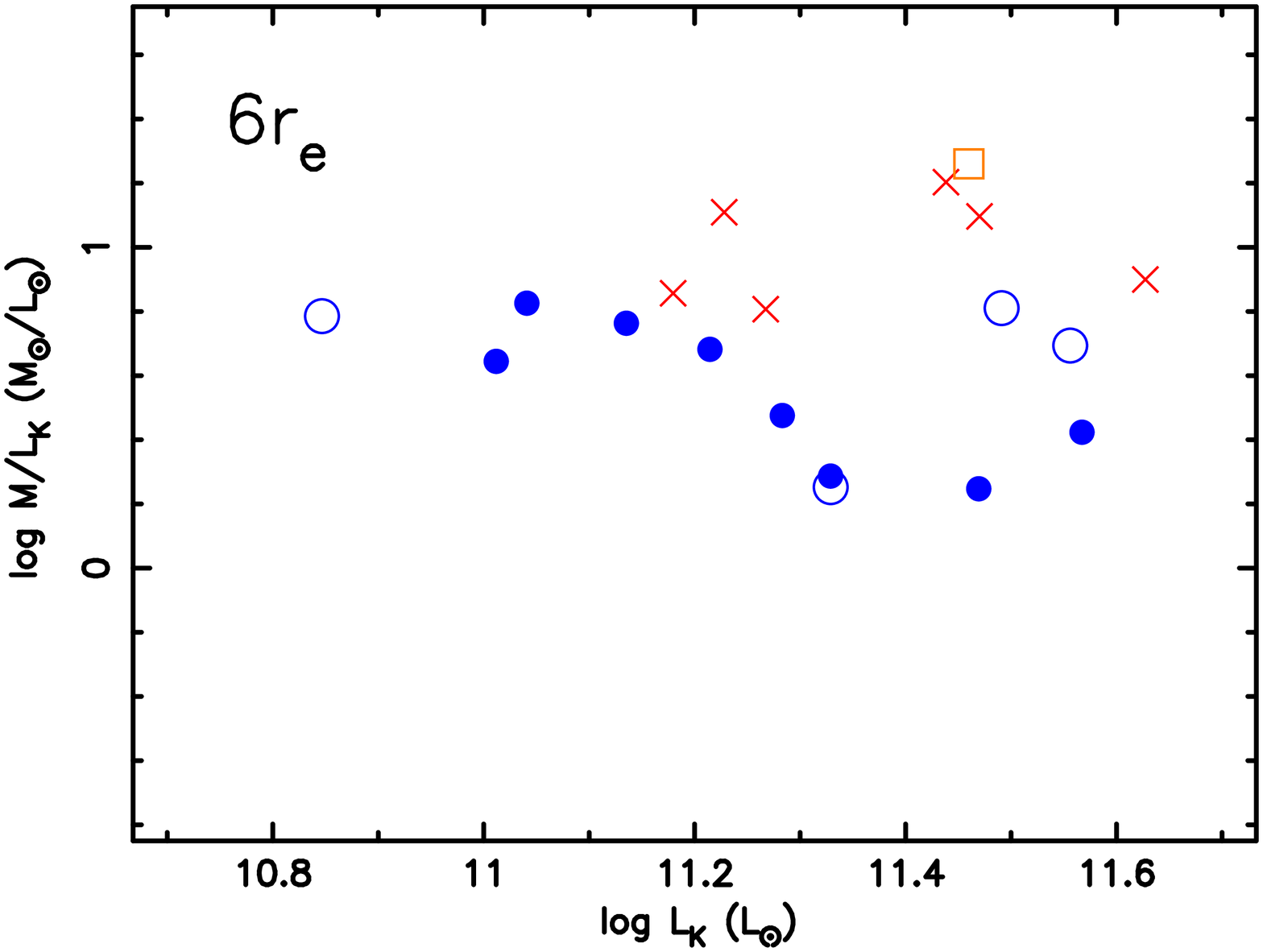}
  \caption{Integrated $M/L_K (r<0.5r_e)$ (top left), $M/L_K (r<3r_e)$ (top
right), and $M/L_K(r<6r_e)$ (bottom) against $L_K$. Meanings of the symbols
are the same as those in Figure \ref{fig:mllb_sim}.}
  \label{fig:mllk_sim}
    \end{figure*}

In Figure \ref{fig:mllk_sim}, we compared the relationship of
$M/L_K(r<0.5r_e)$, $M/L_K(r<3r_e)$, and $M/L_K(r<6r_e)$ with
$L_K$ except M 87. For early-type galaxies, the stellar K-band mass-to-light ratio,
$M/L_K$, is about $1M_{\odot}/L_{\odot}$. The scatter
of $M/L_K(r<0.5r_e)$ values becomes smaller than that of $M/L_B(r<0.5r_e)$
values (Figure \ref{fig:mllk_sim}). Both the $X_E$ and $X_C$ galaxies
have $M/L_K$ values of $\sim 1M_{\odot}/L_{\odot}$, and no correlation
with $L_K$. This result indicates that stars
dominate the mass within 0.5$r_e$, and we observed the stellar $M/L_K$ in
these region, except the cD galaxy M 87.
M 87  may contain  similar amount of dark matter with stellar mass 
within 0.5$r_e$, since $M/L_K$ is a factor of 2 larger than those of
 $X_E$ and $X_C$ galaxies.
 On the other hand, the $M/L_K(r<3r_e)$ values of the $X_E$
and $X_C$ galaxies are $4\sim 10$ and $1\sim 4 M_{\odot}/L_{\odot}$,
respectively (Figure \ref{fig:mllk_sim}). The $M/L_K (r<6r_e)$ values of
the $X_E$ and $X_C$ galaxies are $\sim 10$ and $2\sim 6
M_{\odot}/L_{\odot}$, respectively (Figure \ref{fig:mllk_sim}). 

\subsection{Dark matter in early-type galaxies}
\label{dm}

\begin{figure}
   \centering
    \includegraphics[width=7.0cm]{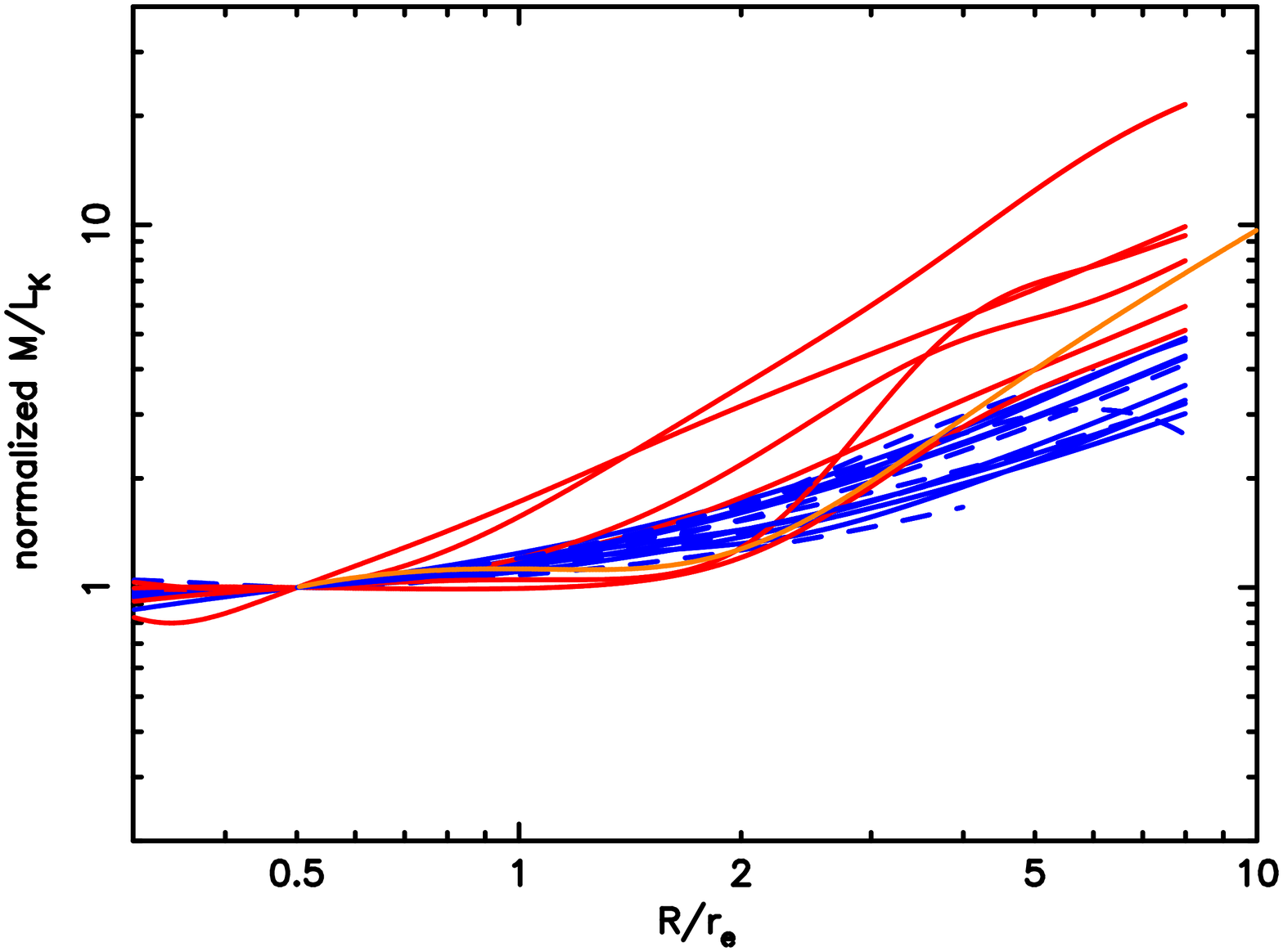}
  \caption{Profiles of $M/L_K$ normalized by $M/L_K(r<0.5 r_e)$. 
    We also plotted the $M/L_K$ profile of M87 by \cite{Matsu2002} 
    (orange solid line). The meanings of other colors and lines are the
    same as those in Figure \ref{fig:kts2_all}}\label{fig:ml_norm}
   \end{figure}

\begin{figure*}
   \centering
    \includegraphics[width=6.5cm]{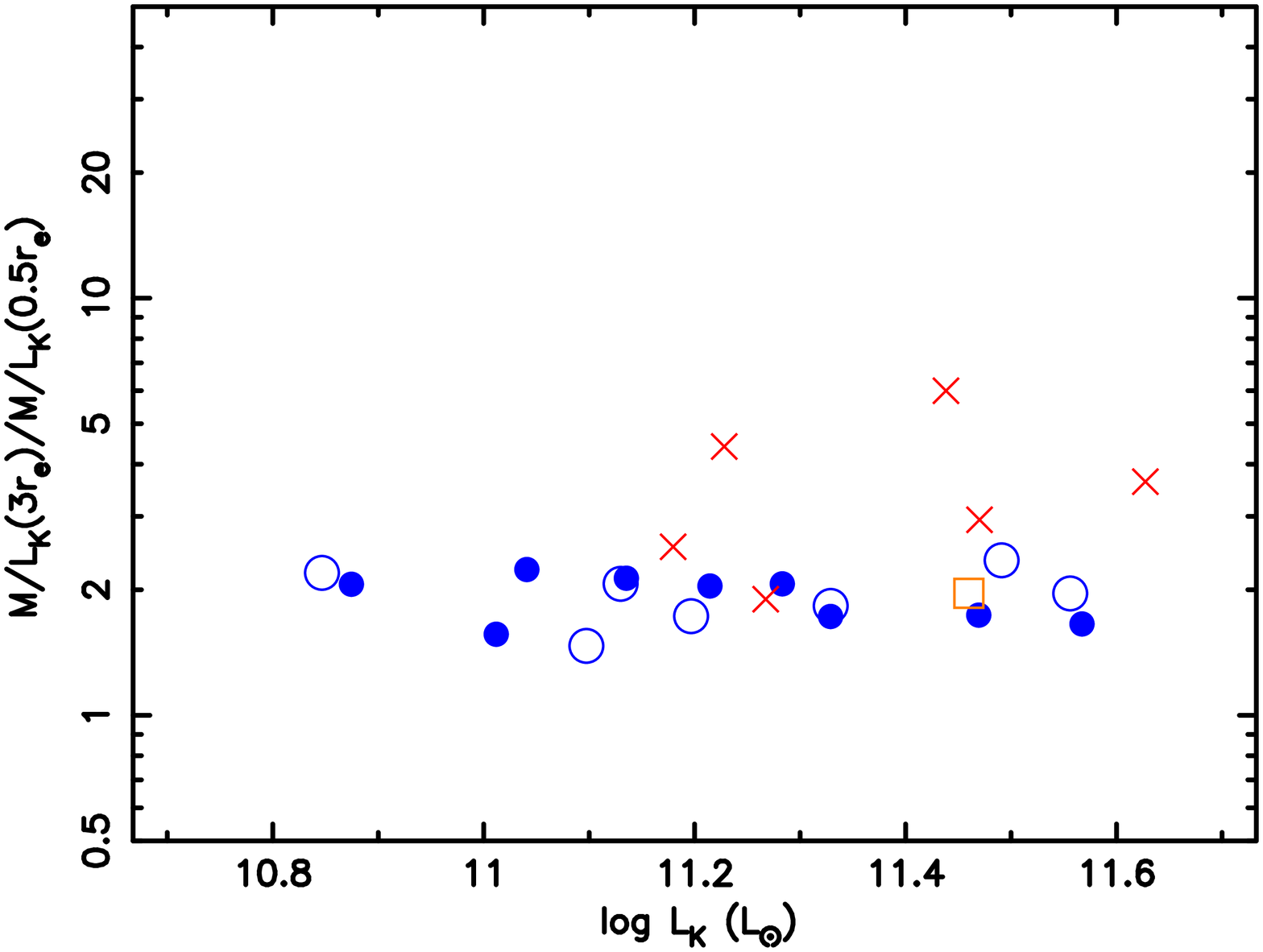}
    \includegraphics[width=6.5cm]{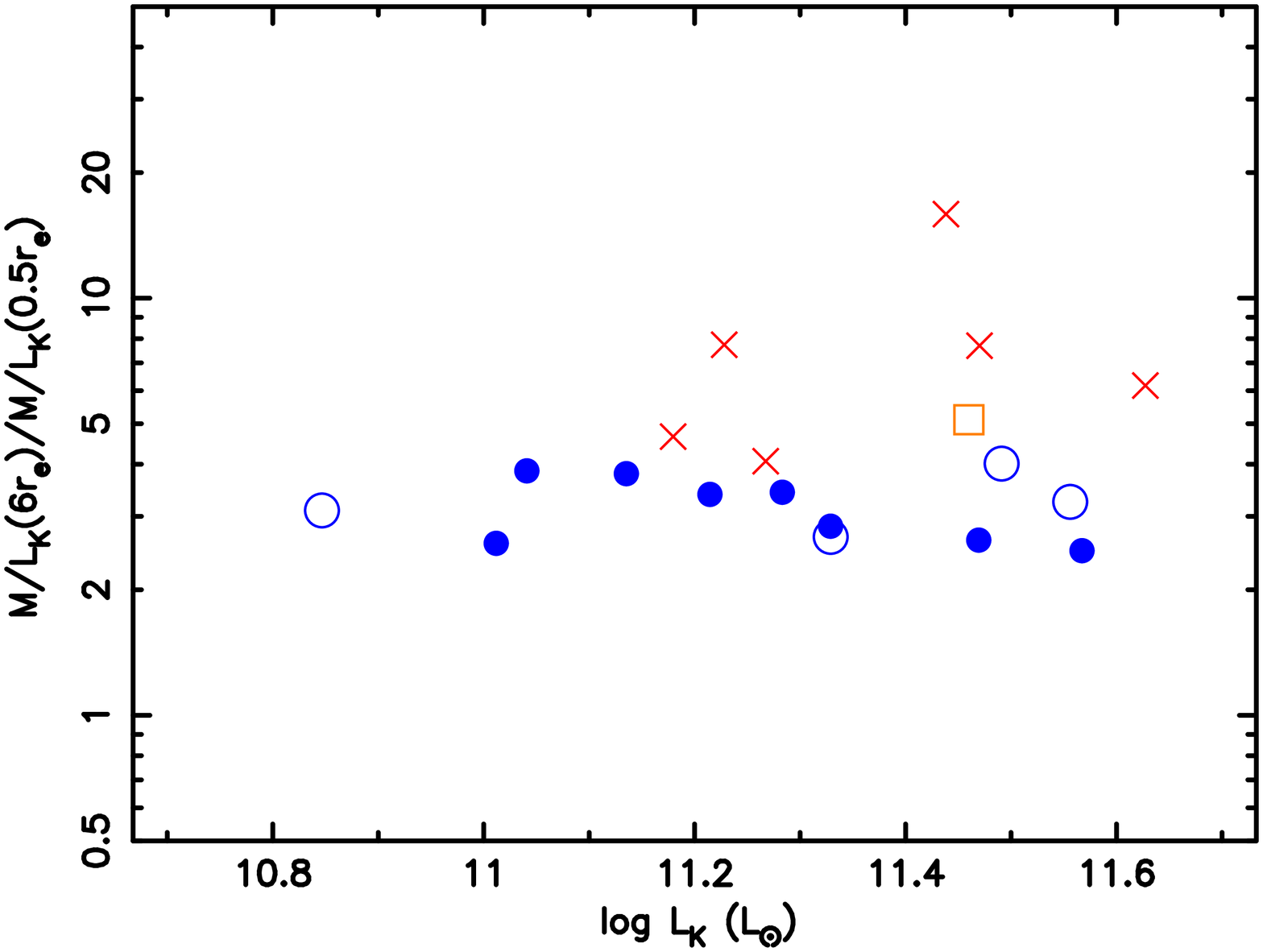}
  \caption{Ratio of the integrated $M/L_K (r<3r_e)$  and $M/L_K(r<6r_e)$
 to $M/L_K(r<0.5r_e)$
 plotted against $L_K$.  Meanings of the symbols are the same as
 those in Figure \ref{fig:mllb_sim}.}\label{fig:ml5_05}
    \end{figure*}

We normalized the  $M/L_K$ profile with $M/L_K(r<0.5r_e)$ for each
galaxy (Figure \ref{fig:ml_norm}). 
Then, the profiles of mass-to-light ratios of $X_C$ galaxies have 
similarities.
We also derived the ratio of $M/L_K (r<3r_e)$ and $M/L_K(r<6r_e)$
to $M/L_K(r<0.5r_e)$ for each galaxy. The ratios of $M/L_K (r<3r_e)$ and
$M/L_K(r<6r_e)$ to $M/L_K(r<0.5r_e)$ are similar among the $X_C$ galaxies at 
2 and $3\sim 4$, respectively (Figure \ref{fig:ml5_05}). Considering that stellar
mass dominates within 1$r_e$, and assuming that the stellar $M/L_K$ is
nearly constant within a galaxy, the $X_C$ galaxies contain similar
amounts of dark matter. Within $3r_e$, the mass of dark matter is similar
to the stellar mass, while within $6r_e$, the former is 2--3 times
larger than the latter. These ratios should reflect the potential of
early-type galaxies themselves. Thus, the dark mass distribution in
early-type galaxies is slightly more extended than that of stars. It
is thought that early-type galaxies have 10 times more dark mass than
stellar mass as in spiral galaxies (e.g., \citealt{Ciot1991}). However, 
the galaxies themselves may not contain such large
amounts of mass, at least within several times $r_e$. This result should
be important for the study of the origin of the dark matter content in
early-type galaxies and for the study of the formation and evolution
of these galaxies. 

\begin{figure*}
   \centering
    \includegraphics[width=6.5cm]{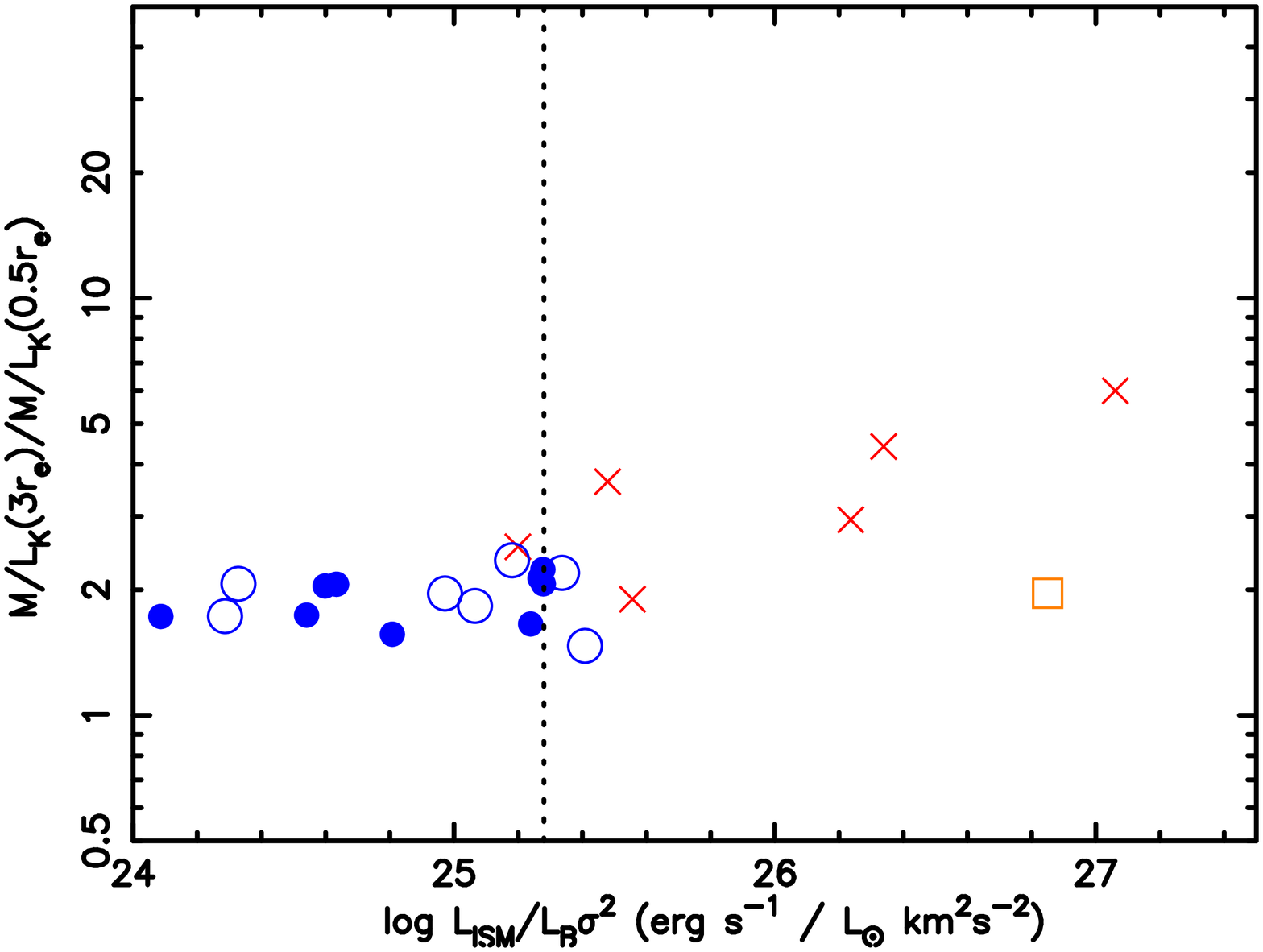}
    \includegraphics[width=6.5cm]{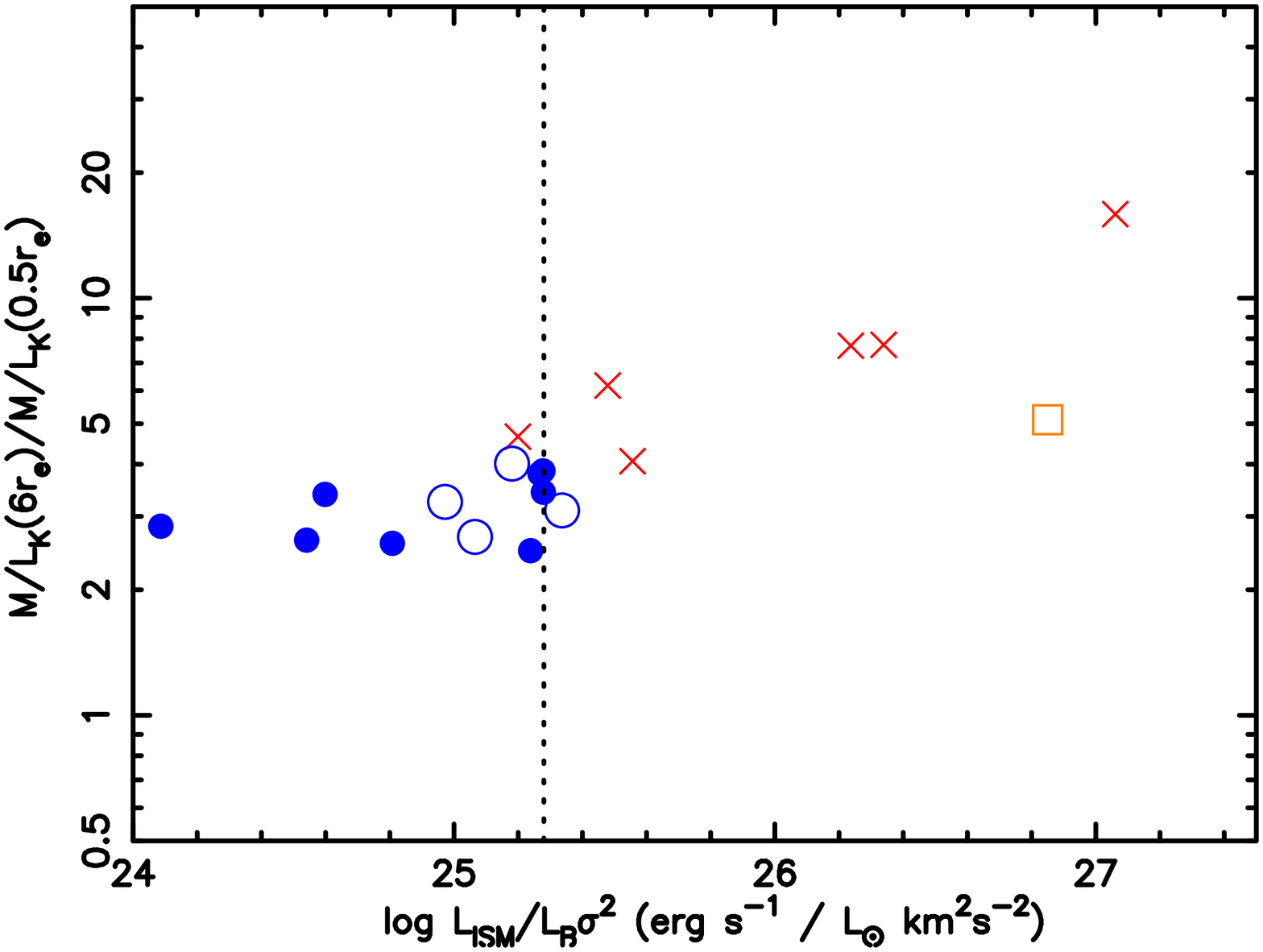}
  \caption{Ratio of the integrated $M/L_K (r<3r_e)$  and $M/L_K(r<6r_e)$
 to $M/L_K(r<0.5r_e)$ plotted against $L_{{\rm ISM}}/L_B\sigma^2$.
 Meanings of the symbols are the same as
 those in Figure \ref{fig:mllb_sim}. The dotted line represents the
    kinetic heating rate by stellar mass loss.}\label{fig:ml_lx}
    \end{figure*}

By contrast, the $X_E$ galaxies have systematically larger ratios; i.e., they
have more dark matter in their outer regions. Figure \ref{fig:ml_lx} shows that the
galaxies with a larger $L_{\rm X}/L_{\sigma}$ have more dark matter at
$r>1 r_e$. 
The ratio of $M 87$ may also in the relation between $L_{\rm
X}/L_{\sigma}$ and dark to stellar mass ratio, considering that
M 87 may contain similar amount of dark mass to stellar mass within
$0.5r_e$. 
In other words, the X-ray luminosity of the $X_E$ galaxies may
be determined in relation to their potential structure, as indicated by
\citet{Matsu2001} and \citet{Matsu2002}. The difference in temperature
profiles between the $X_E$ and $X_C$ galaxies would be due to differences in
the surrounding gravitational potential. These results suggest that
the X-ray luminous early-type galaxies commonly sit in the center of a
large-scale (a few hundred kpc) potential well, which leads to their
high luminosities. Other galaxies may lack such a large-scale
potential well and contain their own dark matter. 

\section{Conclusion}

We analyzed 22 early-type galaxies using XMM-Newton and Chandra
data. To derive the gravitational mass profiles, we obtained the
temperature and ISM density profiles through spectral fitting and
spatial analysis, respectively. We classified the galaxies into two
categories, $X_E$ and $X_C$ galaxies, on the basis of whether the
temperature gradient is positive or negative toward the outer
radius. The ISM luminosity of the $X_C$ galaxies is consistent with the
energy input from stellar mass loss. By contrast, the $X_E$ galaxies have
larger ISM luminosity. 

At the central regions, $r<0.5-1 r_e$, the derived integrated $M/L_K$
of both of the $X_C$ and $X_E$ galaxies are about 1 and have smaller scatter
than $M/L_B$. The values and profiles of $M/L_K$ indicate that stellar mass
dominates the total mass in these regions. In the outer regions, $M/L_B$
and $M/L_K$ of the $X_E$ galaxies are higher than those of the $X_C$ galaxies. 

On the basis of these results, we can conclude the following. The
normal early-type galaxies, $X_C$ galaxies, contain their own dark
matter at amounts that are 1.5$\sim$2 times larger than the stellar
mass within 5$r_e$. The $X_E$ galaxies are located as the central galaxy in
a larger scale potential structure, such as a galaxy group. This fact
directly affects the gravitational potential profile of the galaxy
itself, and causes it to contain significantly higher amounts of dark
matter than that in the $X_C$ galaxies. This difference in the
gravitational potential leads to the difference in the temperature
profile and X-ray ISM luminosity between the $X_E$ and $X_C$ galaxies.

\newpage

\longtab{3}{

{\scriptsize
\begin{longtable}{lcccccccc}
\caption{\label{tab:fit_pj}Spectral fitting results of the projected annular
 spectra.}\\
\hline \hline
Galaxy & ring$^a$ & $kT^b$ & O & Si & Fe & $\log L_{{\rm ISM}}$$^b$ & $\log L_{hard}$$^b$ & $\chi^2$/d.o.f. \\
       & ($r_e$) & (keV) & (solar) & (solar) & (solar) & (erg/s)    & (erg/s)             &                \\ \hline
\endfirsthead
\hline \hline
Galaxy & ring & $kT$ & O & Si & Fe & $\log L_{{\rm ISM}}$ & $\log L_{hard}$ & $\chi^2$/d.o.f. \\
       & ($r_e$) & (keV) & (solar) & (solar) & (solar) & (erg/s)    & (erg/s)             &                \\ \hline
\endhead
\endfoot
  \hline
\multicolumn{9}{l}{$^a$The inner and outer radii to integrated the spectrum.}\\
\multicolumn{9}{l}{$^b$$kT$ is ISM temperature. $L_{{\rm
 ISM}}$ and $L_{hard}$ are the X-ray luminosities of the thermal
 emission and the non}\\
\multicolumn{9}{l}{~~ thermal emission in the range of 0.3-2.0 keV, respectively.}\\
\multicolumn{9}{l}{$^{\ast}$fixed to the value.}\\
\endlastfoot
IC1459 & 0-0.5 & $ 0.606 _{- 0.022 }^{+ 0.022 } $ & $  0.56 _{-  0.14 }^{+  0.18 } $ & $  0.54 _{-  0.38 }^{+  0.42 } $ & $  0.36 _{-  0.06 }^{+  0.03 } $ &  39.60 & 40.07 &   242 /   201 \\
 & 0.5-1 & $ 0.622 _{- 0.026 }^{+ 0.026 } $ & --- & --- & --- &  39.32 & 39.54 & --- \\
 & 1-2 & $ 0.574 _{- 0.022 }^{+ 0.021 } $ & --- & --- & --- &  39.47 & 39.42 & --- \\
 & 2-4 & $ 0.589 _{- 0.023 }^{+ 0.022 } $ & --- & --- & --- &  39.48 & 39.12 & --- \\
 & 4-8 & $ 0.638 _{- 0.029 }^{+ 0.028 } $ & --- & --- & --- &  39.68 & 38.99 & --- \\
\hline
NGC720 & 0-1 & $ 0.601 _{- 0.010 }^{+ 0.010 } $ & $  0.49 _{-  0.08 }^{+  0.07 } $ & $  0.31 _{-  0.26 }^{+  0.24 } $ & $  0.50 _{-  0.03 }^{+  0.03 } $ &  40.01 & 39.46 &   207 /   157 \\
 & 1-2 & $ 0.564 _{- 0.012 }^{+ 0.013 } $ & --- & --- & --- &  39.93 & 39.23 & --- \\
 & 2-4 & $ 0.554 _{- 0.012 }^{+ 0.013 } $ & --- & --- & --- &  40.01 & 39.56 & --- \\
 & 4-8 & $ 0.480 _{- 0.034 }^{+ 0.038 } $ & --- & --- & --- &  39.94 & 39.54 & --- \\
\hline
NGC1316 & 0-0.25 & $ 0.728 _{- 0.005 }^{+ 0.006 } $ & $  0.48 _{-  0.03 }^{+  0.03 } $ & $  0.24 _{-  0.05 }^{+  0.04 } $ & $  0.36 _{-  0.01 }^{+  0.01 } $ &  40.09 & 39.45 &  1602 /   939 \\
 & 0.25-0.5 & $ 0.632 _{- 0.007 }^{+ 0.007 } $ & --- & --- & --- &  39.82 & 39.31 & --- \\
 & 0.5-1 & $ 0.599 _{- 0.006 }^{+ 0.007 } $ & --- & --- & --- &  40.00 & 39.61 & --- \\
 & 1-2 & $ 0.664 _{- 0.009 }^{+ 0.009 } $ & --- & --- & --- &  39.87 & 39.35 & --- \\
 & 2-4 & $ 0.373 _{- 0.013 }^{+ 0.013 } $ & --- & --- & --- &  39.81 & 39.68 & --- \\
 & 4-8 & $ 0.616 _{- 0.046 }^{+ 0.046 } $ & --- & --- & --- &  39.90 & 39.77 & --- \\
\hline
NGC1332 & 0-0.5 & $ 0.617 _{- 0.007 }^{+ 0.003 } $ & $  0.72 _{-  0.08 }^{+  0.09 } $ & $  0.89 _{-  0.18 }^{+  0.10 } $ & $  0.56 _{-  0.05 }^{+  0.05 } $ &  39.72 & 39.14 &   337 /   227 \\
 & 0.5-1 & $ 0.580 _{- 0.011 }^{+ 0.011 } $ & --- & --- & --- &  39.36 &  39.06 & --- \\
 & 1-2 & $ 0.486 _{- 0.014 }^{+ 0.015 } $ & --- & --- & --- &  39.40 & 39.15 & --- \\
 & 2-4 & $ 0.495 _{- 0.015 }^{+ 0.008 } $ & --- & --- & --- &  39.49 & 39.18 & --- \\
 & 4-8 & $ 0.542 _{- 0.020 }^{+ 0.015 } $ & --- & --- & --- &  39.34 & 38.89 & --- \\
\hline
NGC1395 & 0-0.5 & $ 0.597 _{- 0.010 }^{+ 0.009 } $ & $  0.79 _{-  0.10 }^{+  0.23 } $ & $  1.15 _{-  0.23 }^{+  0.27 } $ & $  0.80 _{-  0.02 }^{+  0.02 } $ &  39.63 & 39.34 &   381 /   342 \\
 & 0.5-1 & $ 0.607 _{- 0.009 }^{+ 0.009 } $ & --- & --- & --- &  39.67 & 39.37 & --- \\
 & 1-2 & $ 0.710 _{- 0.010 }^{+ 0.010 } $ & --- & --- & --- &  39.89 & 39.69 & --- \\
 & 2-4 & $ 0.748 _{- 0.010 }^{+ 0.010 } $ & --- & --- & --- &  40.02 & 39.71 & --- \\
 & 4-8 & $ 0.762 _{- 0.021 }^{+ 0.020 } $ & --- & --- & --- &  40.02 & 39.94 & --- \\
\hline
NGC1399 & 0-0.5 & $ 0.846 _{- 0.007 }^{+ 0.007 } $ & $  0.72 _{-  0.13 }^{+  0.14 } $ & $  1.24 _{-  0.14 }^{+  0.15 } $ & $  1.07 _{-  0.08 }^{+  0.09 } $ &  40.52 & 39.56 &   374 /   250 \\
 & 0.5-1 & $ 0.987 _{- 0.013 }^{+ 0.013 } $ & --- & --- & --- &  40.36 & 39.44 & --- \\
 & 1-2 & $ 1.169 _{- 0.022 }^{+ 0.022 } $ & --- & --- & --- &  40.41 & 39.93 & --- \\
 & 2-4 & $ 1.342 _{- 0.011 }^{+ 0.011 } $ & --- & --- & --- &  40.73 & 39.77 & --- \\
 & 4-6 & $ 1.440 _{- 0.043 }^{+ 0.036 } $ & --- & --- & --- &  40.62 & 39.85 & --- \\
 & 6-8 & $ 1.535 _{- 0.043 }^{+ 0.052 } $ & --- & --- & --- &  40.43 & 39.87 & --- \\
\hline
NGC1549 & 0-1 & $ 0.398 _{- 0.051 }^{+ 0.075 } $ & $  0.25 _{-  0.09 }^{+  0.13 } $ & $ 1.00^{\ast} $ & $  0.27 _{-  0.09 }^{+  0.06 } $ &  39.14 & 39.07 &    93 /   111 \\
 & 1-4 & $ 0.441 _{- 0.076 }^{+ 0.088 } $ & --- & --- & --- &  39.03 & 37.73 & --- \\
\hline
NGC3585 & 0-1 & $ 0.412 _{- 0.076 }^{+ 0.107 } $ & $  0.78 _{-  0.49 }^{+  0.60 } $ & $ 1.00^{\ast} $ & $  1.03 _{-  0.74 }^{+ 33.81 } $ &  38.69 & 39.12 &   241 /   159 \\
 & 1-2 & $ 0.382 _{- 0.089 }^{+ 0.131 } $ & --- & --- & --- &  38.70 & 39.00 & --- \\
 & 2-4 & $ 0.338 _{- 0.039 }^{+ 0.057 } $ & --- & --- & --- &  39.11 & 38.14 & --- \\
 & 4-8 & $ 0.399 _{- 0.067 }^{+ 0.068 } $ & --- & --- & --- &  39.41 & 39.06 & --- \\
\hline
NGC3607 & 0-1 & $ 0.595 _{- 0.034 }^{+ 0.034 } $ & $  0.09 _{-  0.06 }^{+  0.07 } $ & $  0.18 _{-  0.18 }^{+  0.45 } $ & $  0.24 _{-  0.04 }^{+  0.03 } $ &  39.68 & 39.51 &   177 /   193 \\
 & 1-2 & $ 0.550 _{- 0.050 }^{+ 0.032 } $ & --- & --- & --- &  39.73 & 39.26 & --- \\
 & 2-4 & $ 0.566 _{- 0.022 }^{+ 0.022 } $ & --- & --- & --- &  40.13 & 39.34 & --- \\
 & 4-8 & $ 0.449 _{- 0.021 }^{+ 0.035 } $ & --- & --- & --- &  40.20 & 40.18 & --- \\
\hline
NGC3665 & 0-0.5 & $ 0.448 _{- 0.044 }^{+ 0.048 } $ & $  0.15 _{-  0.03 }^{+  0.05 } $ & $  0.01 _{-  0.01 }^{+  0.41 } $ & $  0.23 _{-  0.03 }^{+  0.04 } $ &  39.47 & 39.41 &   275 /   223 \\
 & 0.5-1 & $ 0.382 _{- 0.019 }^{+ 0.025 } $ & --- & --- & --- &  39.65 & 39.29 & --- \\
 & 1-2 & $ 0.378 _{- 0.017 }^{+ 0.022 } $ & --- & --- & --- &  39.82 &  39.38 & --- \\
 & 2-4 & $ 0.377 _{- 0.017 }^{+ 0.022 } $ & --- & --- & --- &  39.94 &  39.34 & --- \\
 & 4-8 & $ 0.435 _{- 0.023 }^{+ 0.034 } $ & --- & --- & --- &  39.82 &   0.00 & --- \\
\hline
NGC3923 & 0-0.25 & $ 0.613 _{- 0.008 }^{+ 0.008 } $ & $  0.37 _{-  0.04 }^{+  0.05 } $ & $  0.24 _{-  0.10 }^{+  0.14 } $ & $  0.37 _{-  0.02 }^{+  0.02 } $ &  40.16 & 39.12 &   319 /   220 \\
 & 0.25-0.5 & $ 0.559 _{- 0.009 }^{+ 0.009 } $ & --- & --- & --- &  40.03 & 39.27 & --- \\
 & 0.5-1 & $ 0.479 _{- 0.013 }^{+ 0.012 } $ & --- & --- & --- &  40.04 & 39.52 & --- \\
 & 1-2 & $ 0.491 _{- 0.016 }^{+ 0.013 } $ & --- & --- & --- &  40.00 & 39.65 & --- \\
 & 2-4 & $ 0.562 _{- 0.018 }^{+ 0.017 } $ & --- & --- & --- &  39.91 & 39.88 & --- \\
 & 4-8 & $ 0.471 _{- 0.103 }^{+ 0.154 } $ & --- & --- & --- &  39.77 & 39.83 & --- \\
\hline
NGC4365 & 0-1 & $ 0.522 _{- 0.036 }^{+ 0.042 } $ & $  1.00 _{-  0.32 }^{+  0.80 } $ & $  0.96 _{-  0.80 }^{+  1.24 } $ & $  0.47 _{-  0.10 }^{+  0.11 } $ &  39.15 & 39.47 &   177 /   190 \\
 & 1-4 & $ 0.624 _{- 0.030 }^{+ 0.030 } $ & --- & --- & --- &  39.37 & 39.64 & --- \\
\hline
NGC4382 & 0-0.5 & $ 0.404 _{- 0.023 }^{+ 0.065 } $ & $  0.17 _{-  0.04 }^{+  0.07 } $ & $  0.16 _{-  0.16 }^{+  0.44 } $ & $  0.33 _{-  0.07 }^{+  0.06 } $ &  39.36 & 39.20  &   244 /   211 \\
 & 0.5-1 & $ 0.384 _{- 0.019 }^{+ 0.025 } $ & --- & --- & --- &  39.53 & 39.09 & --- \\
 & 1-2 & $ 0.349 _{- 0.011 }^{+ 0.017 } $ & --- & --- & --- &  39.76 & 39.07 & --- \\
 & 2-4 & $ 0.315 _{- 0.014 }^{+ 0.008 } $ & --- & --- & --- &  39.80 & 39.45 & --- \\
 & 4-8 & $ 0.266 _{- 0.025 }^{+ 0.031 } $ & --- & --- & --- &  39.70 & 39.43 & --- \\
\hline
NGC4472 & 0-0.25 & $ 0.782~~~~~~~~ $ & $  0.77~~~~~~~ $ & $  1.15~~~~~~~ $ & $  0.92~~~~~~~ $ &  40.61 &  39.53&   782 /   272 \\
 & 0.25-0.5 & $ 0.903~~~~~~~~ $ & $  0.75~~~~~~~ $ & $  0.97~~~~~~~ $ & $  0.77~~~~~~~ $ &  40.47 &  39.49&   964 /   216 \\
 & 0.5-0.75 & $ 0.995~~~~~~~~ $ & $  1.01~~~~~~~ $ & $  1.53~~~~~~~ $ & $  1.11~~~~~~~ $ &  40.40 & 39.58 &   667 /   229 \\
 & 0.75-1 & $ 1.031~~~~~~~~ $ & $  0.86~~~~~~~ $ & $  1.68~~~~~~~ $ & $  1.10~~~~~~~ $ &  40.36 &   39.52 &   541 /   195 \\
 & 1-1.5 & $ 1.044~~~~~~~~ $ & $  0.94~~~~~~~ $ & $  1.76~~~~~~~ $ & $  1.22~~~~~~~ $ &  40.59 &  39.76 &   808 /   194 \\
 & 1.5-2 & $ 1.160~~~~~~~~ $ & $  0.58~~~~~~~ $ & $  1.38~~~~~~~ $ & $  1.06~~~~~~~ $ &  40.50 &  39.65 &   625 /   186 \\
 & 2-3 & $ 1.227~~~~~~~~ $ & $  0.58~~~~~~~ $ & $  1.35~~~~~~~ $ & $  1.00~~~~~~~ $ &  40.61 &  39.78 &   427 /   193 \\
 & 3-4 & $ 1.270 _{- 0.013 }^{+ 0.012 } $ & $  0.39 _{-  0.14 }^{+  0.15 } $ & $  1.27 _{-  0.16 }^{+  0.20 } $ & $  0.67 _{-  0.05 }^{+  0.05 } $ &  40.38 & 39.42 &   201 /   115 \\
\hline
NGC4477 & 0-1 & $ 0.446 _{- 0.027 }^{+ 0.023 } $ & $  0.25 _{-  0.05 }^{+  0.04 } $ & $ 1.00^{\ast} $ & $  0.17 _{-  0.02 }^{+  0.02 } $ &  39.71 & 39.01 &    97 /    79 \\
 & 1-2 & $ 0.452 _{- 0.029 }^{+ 0.034 } $ & --- & --- & --- &  39.56 & 38.60 & --- \\
 & 2-4 & $ 0.273 _{- 0.193 }^{+ 0.271 } $ & --- & --- & --- &  38.67 & 39.32 & --- \\
 & 4-8 & $ 0.122 _{- 0.041 }^{+ 0.241 } $ & --- & --- & --- &  38.99 & 39.28 & --- \\
\hline
NGC4526 & 0-0.5 & $ 0.353 _{- 0.024 }^{+ 0.029 } $ & $  0.29 _{-  0.06 }^{+  0.08 } $ & $ 1.00^{\ast} $ & $  0.40 _{-  0.06 }^{+  0.06 } $ &  39.21 & 39.29 &    77 /   117 \\
 & 0.5-1 & $ 0.322 _{- 0.024 }^{+ 0.035 } $ & --- & --- & --- &  38.92 & 38.92 & --- \\
 & 1-4 & $ 0.280 _{- 0.031 }^{+ 0.030 } $ & --- & --- & --- &  38.97 & 39.20& --- \\
\hline
NGC4552 & 0-0.5 & $ 0.679 _{- 0.011 }^{+ 0.020 } $ & $  0.52 _{-  0.06 }^{+  0.05 } $ & $  0.56 _{-  0.14 }^{+  0.13 } $ & $  0.41 _{-  0.02 }^{+  0.02 } $ &  40.21 & 39.81 &   275 /   182 \\
 & 0.5-1 & $ 0.582 _{- 0.009 }^{+ 0.010 } $ & --- & --- & --- &  40.08 & 39.54 & --- \\
 & 1-2 & $ 0.500 _{- 0.014 }^{+ 0.015 } $ & --- & --- & --- &  39.97 & 39.63 & --- \\
 & 2-4 & $ 0.528 _{- 0.034 }^{+ 0.025 } $ & --- & --- & --- &  39.66 & 39.63 & --- \\
\hline
NGC4636 & 0-0.25 & $ 0.561~~~~~~~~ $ & $  0.45~~~~~~~~ $ & $  0.59~~~~~~~~ $ & $  0.46~~~~~~~~ $ &  40.63 &  38.95&   881 /   335 \\
 & 0.25-0.5 & $ 0.599 _{- 0.002 }^{+ 0.002 } $ & $  0.61 _{-  0.02 }^{+  0.03 } $ & $  0.88 _{-  0.05 }^{+  0.06 } $ & $  0.68 _{-  0.02 }^{+  0.02 } $ &  40.66 & 38.94 &   656 /   350 \\
 & 0.5-0.75 & $ 0.653 _{- 0.002 }^{+ 0.002 } $ & $  0.93 _{-  0.05 }^{+  0.05 } $ & $  1.36 _{-  0.09 }^{+  0.08 } $ & $  0.96 _{-  0.03 }^{+  0.03 } $ &  40.65 & 39.15 &   549 /   293 \\
 & 0.75-1 & $ 0.706 _{- 0.002 }^{+ 0.002 } $ & $  0.91 _{-  0.06 }^{+  0.07 } $ & $  1.51 _{-  0.06 }^{+  0.10 } $ & $  0.99 _{-  0.04 }^{+  0.04 } $ &  40.52 & 39.17 &   417 /   252 \\
 & 1-1.5 & $ 0.742 _{- 0.002 }^{+ 0.002 } $ & $  0.82 _{-  0.06 }^{+  0.04 } $ & $  1.45 _{-  0.09 }^{+  0.11 } $ & $  1.05 _{-  0.03 }^{+  0.04 } $ &  40.61 & 39.47 &   535 /   297 \\
 & 1.5-2 & $ 0.777 _{- 0.003 }^{+ 0.003 } $ & $  0.79 _{-  0.09 }^{+  0.09 } $ & $  1.37 _{-  0.07 }^{+  0.06 } $ & $  1.07 _{-  0.04 }^{+  0.05 } $ &  40.46 & 39.47 &   306 /   254 \\
 & 2-3 & $ 0.806 _{- 0.003 }^{+ 0.003 } $ & $  0.87 _{-  0.11 }^{+  0.13 } $ & $  1.32 _{-  0.12 }^{+  0.17 } $ & $  1.12 _{-  0.05 }^{+  0.05 } $ &  40.58 & 39.65 &   390 /   288 \\
 & 3-4 & $ 0.818 _{- 0.005 }^{+ 0.005 } $ & $  0.98 _{-  0.22 }^{+  0.25 } $ & $  1.01 _{-  0.22 }^{+  0.21 } $ & $  0.99 _{-  0.08 }^{+  0.09 } $ &  40.27 & 39.53 &   263 /   227 \\
 & 4-6 & $ 0.845 _{- 0.011 }^{+ 0.011 } $ & $  0.83 _{-  0.22 }^{+  0.36 } $ & $  0.28 _{-  0.28 }^{+  0.32 } $ & $  0.43 _{-  0.07 }^{+  0.08 } $ &  40.21 & 37.80 &   119 /   111 \\
\hline
NGC4649 & 0-0.25 & $ 0.810~~~~~~~~ $ & $  1.11~~~~~~~~ $ & $  1.81~~~~~~~~ $ & $  1.50~~~~~~~~ $ &  40.42 & 39.47 &   757 /   343 \\
 & 0.25-0.5 & $ 0.782 _{- 0.002 }^{+ 0.002 } $ & $  1.24 _{-  0.11 }^{+  0.13 } $ & $  2.29 _{-  0.18 }^{+  0.20 } $ & $  1.55 _{-  0.08 }^{+  0.09 } $ &  40.34 & 39.43 &   410 /   222 \\
 & 0.5-0.75 & $ 0.791 _{- 0.003 }^{+ 0.003 } $ & $  1.23 _{-  0.17 }^{+  0.20 } $ & $  2.18 _{-  0.24 }^{+  0.29 } $ & $  1.60 _{-  0.09 }^{+  0.10 } $ &  40.10 & 39.33 &   259 /   158 \\
 & 0.75-1 & $ 0.805 _{- 0.004 }^{+ 0.004 } $ & $  0.82 _{-  0.13 }^{+  0.16 } $ & $  1.65 _{-  0.21 }^{+  0.25 } $ & $  1.15 _{-  0.06 }^{+  0.06 } $ &  39.95 & 39.29 &   187 /   119 \\
 & 1-1.5 & $ 0.821 _{- 0.004 }^{+ 0.004 } $ & $  1.13 _{-  0.17 }^{+  0.21 } $ & $  2.12 _{-  0.25 }^{+  0.31 } $ & $  1.34 _{-  0.05 }^{+  0.05 } $ &  40.10 & 39.70 &   258 /   188 \\
 & 1.5-2 & $ 0.828 _{- 0.005 }^{+ 0.005 } $ & $  1.09 _{-  0.22 }^{+  0.32 } $ & $  2.34 _{-  0.37 }^{+  0.52 } $ & $  1.28 _{-  0.04 }^{+  0.04 } $ &  39.88 & 39.53 &   163 /   131 \\
 & 2-4 & $ 0.836 _{- 0.008 }^{+ 0.008 } $ & $  0.36 _{-  0.10 }^{+  0.12 } $ & $  0.40 _{-  0.15 }^{+  0.16 } $ & $  0.34 _{-  0.04 }^{+  0.05 } $ &  40.05 & 39.88 &   250 /   193 \\
 & 4-6 & $ 0.863 _{- 0.054 }^{+ 0.119 } $ & $  0.50 _{-  0.36 }^{+  1.02 } $ & $  0.00 _{-  0.00 }^{+  0.20 } $ & $  0.13 _{-  0.05 }^{+  0.15 } $ &  39.37 & 39.46 &    72 /    53 \\
\hline
NGC5044 & 0-0.5 & $ 0.771 _{- 0.004 }^{+ 0.004 } $ & $  0.61 _{-  0.06 }^{+  0.07 } $ & $  0.82 _{-  0.08 }^{+  0.09 } $ & $  0.65 _{-  0.03 }^{+  0.03 } $ &  41.48 & 40.14 &   214 /   126 \\
 & 0.5-1 & $ 0.798 _{- 0.003 }^{+ 0.002 } $ & $  0.69 _{-  0.05 }^{+  0.06 } $ & $  0.99 _{-  0.07 }^{+  0.07 } $ & $  0.78 _{-  0.03 }^{+  0.03 } $ &  41.76 & 40.27 &   609 /   356 \\
 & 1-1.5 & $ 0.833 _{- 0.003 }^{+ 0.002 } $ & $  0.74 _{-  0.06 }^{+  0.06 } $ & $  1.05 _{-  0.07 }^{+  0.08 } $ & $  0.75 _{-  0.03 }^{+  0.03 } $ &  41.74 & 40.24 &   978 /   520 \\
 & 1.5-2 & $ 0.929 _{- 0.005 }^{+ 0.004 } $ & $  0.54 _{-  0.05 }^{+  0.05 } $ & $  0.80 _{-  0.06 }^{+  0.06 } $ & $  0.67 _{-  0.02 }^{+  0.02 } $ &  41.68 & 40.22 &   676 /   349 \\
 & 2-3 & $ 1.004 _{- 0.003 }^{+ 0.003 } $ & $  0.50 _{-  0.04 }^{+  0.05 } $ & $  0.80 _{-  0.05 }^{+  0.05 } $ & $  0.68 _{-  0.02 }^{+  0.02 } $ &  41.86 & 40.43  &   787 /   491 \\
 & 3-4 & $ 1.196 _{- 0.007 }^{+ 0.007 } $ & $  0.41 _{-  0.06 }^{+  0.06 } $ & $  0.64 _{-  0.05 }^{+  0.05 } $ & $  0.65 _{-  0.03 }^{+  0.03 } $ &  41.65 & 40.00  &   334 /   194 \\
 & 4-6 & $ 1.247 _{- 0.006 }^{+ 0.006 } $ & $  0.33 _{-  0.06 }^{+  0.06 } $ & $  0.68 _{-  0.05 }^{+  0.05 } $ & $  0.62 _{-  0.02 }^{+  0.03 } $ &  41.79 & 40.50  &   316 /   249 \\
 & 6-8 & $ 1.231 _{- 0.013 }^{+ 0.012 } $ & $  0.54 _{-  0.11 }^{+  0.12 } $ & $  0.57 _{-  0.09 }^{+  0.09 } $ & $  0.47 _{-  0.03 }^{+  0.03 } $ &  41.53 & 40.48  &   192 /   176 \\
\hline
NGC5322 & 0-1 & $ 0.440 _{- 0.056 }^{+ 0.038 } $ & $  0.13 _{-  0.04 }^{+  0.06 } $ & $ 1.00^{\ast} $ & $  0.15 _{-  0.03 }^{+  0.02 } $ &  39.93 & 39.33 &    67 /    71 \\
 & 1-2 & $ 0.373 _{- 0.054 }^{+ 0.059 } $ & --- & --- & --- &  39.37 & 39.29 & --- \\
 & 2-4 & $ 0.391 _{- 0.088 }^{+ 0.036 } $ & --- & --- & --- &  39.46 & 39.39 & --- \\
 & 4-8 & $ 0.306 _{- 0.060 }^{+ 0.064 } $ & --- & --- & --- &  39.73 & 39.60 & --- \\
\hline
NGC5846 & 0-0.5 & $ 0.652 _{- 0.005 }^{+ 0.005 } $ & $  0.72 _{-  0.05 }^{+  0.06 } $ & $  0.77 _{-  0.08 }^{+  0.09 } $ & $  0.68 _{-  0.02 }^{+  0.02 } $ &  41.05 & 39.87 &   803 /   450 \\
 & 0.5-1 & $ 0.644 _{- 0.006 }^{+ 0.006 } $ & --- & --- & --- &  41.06 & 39.76 & --- \\
 & 1-1.5 & $ 0.699 _{- 0.006 }^{+ 0.006 } $ & --- & --- & --- &  40.97 & 39.75 & --- \\
 & 1.5-2 & $ 0.749 _{- 0.007 }^{+ 0.007 } $ & --- & --- & --- &  40.91 & 39.79 & --- \\
 & 2-3 & $ 0.838 _{- 0.007 }^{+ 0.007 } $ & --- & --- & --- &  40.91 & 39.94 & --- \\
 & 3-4 & $ 1.160 _{- 0.035 }^{+ 0.032 } $ & --- & --- & --- &  40.56 & 39.38 & --- \\
 & 4-8 & $ 1.189 _{- 0.070 }^{+ 0.054 } $ & --- & --- & --- &  40.70 & 40.30 & --- \\
\hline
\end{longtable}
}

}

\longtab{4}{

{\scriptsize
\begin{longtable}{lcccccccc}
\caption{\label{tab:fit_dpj}Spectral fitting results of the deprojected annular spectra.}\\
\hline \hline
Galaxy & ring$^a$ & $kT^b$ & O & Si & Fe & $\log L_{{\rm ISM}}$$^b$ & $\log L_{hard}$$^b$ & $\chi^2$/d.o.f. \\
       & ($r_e$) & (keV) & (solar) & (solar) & (solar) & (erg/s)    & (erg/s)             &                \\ \hline
\endfirsthead
\hline \hline
Galaxy & ring & $kT$ & O & Si & Fe & $\log L_{{\rm ISM}}$ & $\log L_{hard}$ & $\chi^2$/d.o.f. \\
       & ($r_e$) & (keV) & (solar) & (solar) & (solar) & (erg/s)    & (erg/s)             &                \\ \hline
\endhead
\endfoot
  \hline
\multicolumn{9}{l}{$^a$The inner and outer radii to integrated the spectrum.}\\
\multicolumn{9}{l}{$^b$$kT$ is ISM temperature. $L_{{\rm
 ISM}}$ and $L_{hard}$ are the X-ray luminosities of the thermal
 emission and the non}\\
\multicolumn{9}{l}{~~thermal emission in the range of 0.3-2.0 keV, respectively.}\\
\multicolumn{9}{l}{$^c$We fitted the spectra with 2T model.}\\
\multicolumn{9}{l}{$^{\ast}$fixed to the value.}\\
\endlastfoot
IC1459 & 0-0.5 & $ 0.616 _{- 0.027 }^{+ 0.028 } $ & $  0.35 _{-  0.13 }^{+  0.27 } $ & $  0.09 _{-  0.09 }^{+  0.51 } $ & $  0.28 _{-  0.03 }^{+  0.04 } $ &  39.54 & 40.02 &   174 /   197 \\
 & 0.5-1 & $ 0.642 _{- 0.065 }^{+ 0.058 } $ & --- & --- & --- &  39.20 & 39.55 & --- \\
 & 1-2 & $ 0.557 _{- 0.059 }^{+ 0.047 } $ & --- & --- & --- &  39.41 & 39.52 & --- \\
 & 2-4 & $ 0.523 _{- 0.068 }^{+ 0.085 } $ & --- & --- & --- &  39.38 & 39.18 & --- \\
 & 4-8 & $ 0.644 _{- 0.032 }^{+ 0.031 } $ & --- & --- & --- &  39.72 & 39.28 & --- \\
\hline
NGC720 & 0-1 & $ 0.606 _{- 0.015 }^{+ 0.015 } $ & $  0.67 _{-  0.19 }^{+  0.27 } $ & $ 1.00^{\ast} $ & $  0.70 _{-  0.06 }^{+  0.07 } $ &  39.88 & 39.37 &   108 /    97 \\
 & 1-2 & $ 0.570 _{- 0.034 }^{+ 0.027 } $ & --- & --- & --- &  39.84 & 39.32 & --- \\
 & 2-4 & $ 0.566 _{- 0.022 }^{+ 0.022 } $ & --- & --- & --- &  39.97 & 39.62 & --- \\
 & 4-8 & $ 0.454 _{- 0.032 }^{+ 0.031 } $ & --- & --- & --- &  40.00 & 39.67 & --- \\
\hline
NGC1316 & 0-0.25 & $ 0.748 _{- 0.007 }^{+ 0.007 } $ & $  0.54 _{-  0.06 }^{+  0.08 } $ & $  0.22 _{-  0.09 }^{+  0.11 } $ & $  0.44 _{-  0.01 }^{+  0.02 } $ &  39.99 & 39.39 &   586 /   615 \\
 & 0.25-0.5 & $ 0.657 _{- 0.019 }^{+ 0.019 } $ & --- & --- & --- &  39.68 & 39.30& --- \\
 & 0.5-1 & $ 0.586 _{- 0.013 }^{+ 0.014 } $ & --- & --- & --- &  39.96 & 39.73 & --- \\
 & 1-2 & $ 0.702 _{- 0.041 }^{+ 0.025 } $ & --- & --- & --- &  39.83 & 39.52 & --- \\
 & 2-4 & $ 0.639 _{- 0.048 }^{+ 0.115 } $ & --- & --- & --- &  39.68 & 39.63 & --- \\
 & 4-8 & $ 0.740 _{- 0.031 }^{+ 0.028 } $ & --- & --- & --- &  40.17 & 40.10 & --- \\
\hline
NGC1332 & 0-0.5 & $ 0.615 _{- 0.008 }^{+ 0.008 } $ & $  0.74 _{-  0.12 }^{+  0.16 } $ & $  0.82 _{-  0.27 }^{+  0.29 } $ & $  0.65 _{-  0.02 }^{+  0.05 } $ &  39.62 & 39.06 &   266 /   260 \\
 & 0.5-1 & $ 0.598 _{- 0.021 }^{+ 0.021 } $ & --- & --- & --- &  39.31 & 39.06 & --- \\
 & 1-2 & $ 0.468 _{- 0.027 }^{+ 0.029 } $ & --- & --- & --- &  39.30 & 39.12 & --- \\
 & 2-4 & $ 0.465 _{- 0.024 }^{+ 0.025 } $ & --- & --- & --- &  39.42 & 39.27 & --- \\
 & 4-8 & $ 0.511 _{- 0.032 }^{+ 0.033 } $ & --- & --- & --- &  39.26 & 39.15 & --- \\
\hline
NGC1395 & 0-0.5 & $ 0.592 _{- 0.019 }^{+ 0.019 } $ & $  0.80 _{-  0.12 }^{+  0.51 } $ & $ 1.00^{\ast} $ & $  0.87 _{-  0.05 }^{+  0.06 } $ &  39.41 & 39.19 &   132 /   173 \\
 & 0.5-1 & $ 0.520 _{- 0.035 }^{+ 0.050 } $ & --- & --- & --- &  39.54 & 39.23 & --- \\
 & 1-2 & $ 0.698 _{- 0.028 }^{+ 0.029 } $ & --- & --- & --- &  39.78 & 39.71 & --- \\
 & 2-4 & $ 0.743 _{- 0.019 }^{+ 0.019 } $ & --- & --- & --- &  39.99 & 39.19 & --- \\
 & 4-8 & $ 0.771 _{- 0.017 }^{+ 0.018 } $ & --- & --- & --- &  40.13 & 40.02 & --- \\
\hline
NGC1399 & 0-0.5 & $ 0.826 _{- 0.011 }^{+ 0.010 } $ & $  0.58 _{-  0.25 }^{+  0.36 } $ & $  0.89 _{-  0.38 }^{+  0.37 } $ & $  1.18 _{-  0.09 }^{+  0.10 } $ &  40.36 & 39.47 &   158 /   163 \\
 & 0.5-1 & $ 0.934 _{- 0.028 }^{+ 0.034 } $ & --- & --- & --- &  40.25 & 39.62 & --- \\
 & 1-2 & $ 0.995 _{- 0.034 }^{+ 0.039 } $ & --- & --- & --- &  40.15 & 39.86 & --- \\
 & 2-4 & $ 1.332 _{- 0.026 }^{+ 0.023 } $ & --- & --- & --- &  40.62 & 39.84 & --- \\
 & 4-6 & $ 1.352 _{- 0.030 }^{+ 0.062 } $ & --- & --- & --- &  40.59 & 39.85 & --- \\
 & 6-8 & $ 1.461 _{- 0.109 }^{+ 0.087 } $ & --- & --- & --- &  40.68 & 40.35 & --- \\
\hline
NGC3923 & 0-0.25 & $ 0.627 _{- 0.011 }^{+ 0.012 } $ & $  0.46 _{-  0.10 }^{+  0.11 } $ & $ 1.00^{\ast} $ & $  0.47 _{-  0.02 }^{+  0.02 } $ &  40.04 & 38.80 &   194 /   212 \\
 & 0.25-0.5 & $ 0.575 _{- 0.017 }^{+ 0.017 } $ & --- & --- & --- &  39.99 & 39.14 & --- \\
 & 0.5-1 & $ 0.482 _{- 0.025 }^{+ 0.028 } $ & --- & --- & --- &  40.03 & 39.42 & --- \\
 & 1-2 & $ 0.514 _{- 0.029 }^{+ 0.042 } $ & --- & --- & --- &  39.97 & 39.35 & --- \\
 & 2-4 & $ 0.490 _{- 0.037 }^{+ 0.043 } $ & --- & --- & --- &  39.87 & 39.95 & --- \\
 & 4-8 & $ 0.367 _{- 0.023 }^{+ 0.028 } $ & --- & --- & --- &  40.06 & 39.82 & --- \\
\hline
NGC4472 & 0-0.25 & $ 0.738 _{- 0.013 }^{+ 0.013 } $ & $  0.70 _{-  0.12 }^{+  0.12 } $ & $  1.19 _{-  0.16 }^{+  0.12 } $ & $  1.13 _{-  0.08 }^{+  0.09 } $ &  40.49 & 39.39 &   237 /   159 \\
(2T model)$^c$ & & $ 1.263 _{- 0.104 }^{+ 0.090 } $ & --- & --- & --- & --- & --- & --- \\
\hline
NGC4472 & 0-0.25 & $ 0.764 _{- 0.003 }^{+ 0.003 } $ & $  0.72 _{-  0.09 }^{+  0.08 } $ & $  1.15 _{-  0.13 }^{+  0.12 } $ & $  1.01 _{-  0.05 }^{+  0.05 } $ &  40.48 & 39.47 &   281 /   161 \\
 & 0.25-0.5 & $ 0.784 _{- 0.007 }^{+ 0.007 } $ & $  0.76 _{-  0.18 }^{+  0.24 } $ & $  1.04 _{-  0.26 }^{+  0.29 } $ & $  0.87 _{-  0.09 }^{+  0.11 } $ &  40.33 & 39.41 &    95 /    88 \\
 & 0.5-0.75 & $ 0.920 _{- 0.019 }^{+ 0.025 } $ & $  1.42 _{-  0.88 }^{+  0.65 } $ & $  1.51 _{-  0.63 }^{+  0.56 } $ & $  1.35 _{-  0.28 }^{+  0.26 } $ &  40.24 & 39.62 &    44 /    61 \\
 & 0.75-1 & $ 1.013 _{- 0.021 }^{+ 0.019 } $ & $  0.60 _{-  0.34 }^{+  0.62 } $ & $  1.27 _{-  0.49 }^{+  0.76 } $ & $  0.96 _{-  0.15 }^{+  0.21 } $ &  40.26 & 39.65 &    28 /    66 \\
 & 1-1.5 & $ 1.020 _{- 0.010 }^{+ 0.011 } $ & $  0.88 _{-  0.26 }^{+ 0.47 } $ & $  1.51 _{-  0.38 }^{+  0.51 } $ & $  1.21 _{-  0.14 }^{+ 0.18 } $ &  40.52 & 39.77 &    53 /    68 \\
 & 1.5-2 & $ 1.045 _{- 0.013 }^{+ 0.012 } $ & $  0.34 _{-  0.31 }^{+  0.38 } $ & $  1.17 _{-  0.40 }^{+  0.54 } $ & $  1.04 _{-  0.13 }^{+  0.26 } $ &  40.49 & 39.86 &    49 /    67 \\
 & 2-3 & $ 1.210 _{- 0.017 }^{+ 0.016 } $ & $  0.38 _{-  0.26 }^{+  0.36 } $ & $  1.05 _{-  0.29 }^{+  0.31 } $ & $  1.08 _{-  0.13 }^{+  0.11 } $ &  40.65 & 39.99 &    52 /    91 \\
 & 3-4 & $ 1.232 _{- 0.044 }^{+ 0.037 } $ & $  0.14 _{-  0.14 }^{+  0.43 } $ & $  0.89 _{-  0.36 }^{+  0.43 } $ & $  0.51 _{-  0.13 }^{+  0.11 } $ &  40.50 & 39.36 &    22 /    48 \\
 & 4-6 & $ 1.334 _{- 0.015 }^{+ 0.014 } $ & $  0.05 _{-  0.05 }^{+  0.16 } $ & $  0.99 _{-  0.16 }^{+  0.18 } $ & $  0.47 _{-  0.05 }^{+  0.07 } $ &  40.79 & 39.97 &    98 /   107 \\
 & 6-8 & $ 1.392 _{- 0.044 }^{+ 0.048 } $ & $  0.00 _{-  0.00 }^{+  0.09 } $ & $  0.39 _{-  0.10 }^{+  0.10 } $ & $  0.34 _{-  0.04 }^{+  0.03 } $ &  40.68 & 40.04 &   118 /   102 \\
\hline
NGC4552 & 0-0.5 & $ 0.708 _{- 0.013 }^{+ 0.013 } $ & $  0.53 _{-  0.10 }^{+  0.07 } $ & $  0.89 _{-  0.24 }^{+  0.27 } $ & $  0.46 _{-  0.03 }^{+  0.05 } $ &  40.08 &39.72 &   162 /   159 \\
 & 0.5-1 & $ 0.610 _{- 0.016 }^{+ 0.016 } $ & --- & --- & --- &  40.07 & 39.43 & --- \\
 & 1-2 & $ 0.500 _{- 0.023 }^{+ 0.021 } $ & --- & --- & --- &  40.01 & 39.65 & --- \\
 & 2-4 & $ 0.469 _{- 0.037 }^{+ 0.039 } $ & --- & --- & --- &  39.73 & 39.73 & --- \\
\hline
NGC4636 & 0-0.25 & $ 0.438 _{- 0.014 }^{+ 0.045 } $ & $  0.40 _{-  0.03 }^{+  0.03 } $ & $  0.68 _{-  0.09 }^{+  0.10 } $ & $  0.49 _{-  0.03 }^{+  0.04 } $ &  40.45 & 38.89 &   416 /   338 \\
(2T model) & & $ 0.657 _{- 0.042 }^{+ 0.042 } $ & --- & --- & --- & --- & --- & --- \\
\hline
NGC4636 & 0-0.25 & $ 0.526 _{- 0.005 }^{+ 0.005 } $ & $  0.42 _{-  0.03 }^{+  0.03 } $ & $  0.57 _{-  0.08 }^{+  0.09 } $ & $  0.41 _{-  0.02 }^{+  0.02 } $ &  40.45 & 38.81 &   448 /   340 \\
 & 0.25-0.5 & $ 0.520 _{- 0.008 }^{+ 0.009 } $ & $  0.47 _{-  0.06 }^{+  0.06 } $ & $  0.91 _{-  0.24 }^{+  0.25 } $ & $  0.56 _{-  0.05 }^{+  0.06 } $ &  40.48 & 0.00 &   189 /   222 \\
 & 0.5-0.75 & $ 0.618 _{- 0.006 }^{+ 0.006 } $ & $  0.76 _{-  0.08 }^{+  0.25 } $ & $  0.97 _{-  0.23 }^{+  0.43 } $ & $  0.89 _{-  0.06 }^{+  0.16 } $ &  40.61 & 39.21 &   125 /   192 \\
 & 0.75-1 & $ 0.671 _{- 0.007 }^{+ 0.007 } $ & $  1.02 _{-  0.26 }^{+  0.35 } $ & $  1.60 _{-  0.39 }^{+  0.67 } $ & $  1.11 _{-  0.15 }^{+  0.20 } $ &  40.54 & 39.34 &    95 /   180 \\
 & 1-1.5 & $ 0.711 _{- 0.007 }^{+ 0.007 } $ & $  0.77 _{-  0.17 }^{+  0.37 } $ & $  1.33 _{-  0.35 }^{+  0.41 } $ & $  1.04 _{-  0.10 }^{+  0.17 } $ &  40.58 & 39.54  &    74 /   156 \\
 & 1.5-2 & $ 0.749 _{- 0.012 }^{+ 0.013 } $ & $  0.94 _{-  0.54 }^{+  0.97 } $ & $  1.28 _{-  0.78 }^{+  1.01 } $ & $  1.13 _{-  0.26 }^{+  0.23 } $ &  40.46 & 39.57  &    33 /   122 \\
 & 2-3 & $ 0.799 _{- 0.007 }^{+ 0.007 } $ & $  0.56 _{-  0.16 }^{+  0.85 } $ & $  0.95 _{-  0.34 }^{+  0.65 } $ & $  0.84 _{-  0.08 }^{+  0.21 } $ &  40.68 & 39.59 &    38 /   106 \\
 & 3-4 & $ 0.808 _{- 0.015 }^{+ 0.015 } $ & $  1.10 _{-  0.76 }^{+ 27.28 } $ & $  0.77 _{-  0.77 }^{+  0.93 } $ & $  1.78 _{-  0.67 }^{+  1.58 } $ &  40.38 & 39.91 &    22 /    77 \\
 & 4-6 & $ 0.839 _{- 0.019 }^{+ 0.018 } $ & $  0.64 _{-  0.29 }^{+  0.55 } $ & $  0.00 _{-  0.00 }^{+  0.58 } $ & $  0.39 _{-  0.08 }^{+  0.23 } $ &  40.47 &  0.00 &    21 /    48 \\
 & 6-8 & $ 0.783 _{- 0.022 }^{+ 0.022 } $ & $  0.19 _{-  0.08 }^{+  0.08 } $ & $  0.04 _{-  0.04 }^{+  0.16 } $ & $  0.12 _{-  0.02 }^{+  0.01 } $ &  40.68 & 39.42 &   101 /    83 \\
\hline
NGC4649 & 0-0.25 & $ 0.810 _{- 0.004 }^{+ 0.003 } $ & $  0.96 _{-  0.09 }^{+  0.19 } $ & $  1.38 _{-  0.16 }^{+  0.26 } $ & $  1.73 _{-  0.10 }^{+  0.21 } $ &  40.32 & 0.00 &   265 /   197 \\
(2T model) & & $ 2.766 _{- 0.199 }^{+ 0.198 } $ & --- & --- & --- & --- & --- & --- \\
\hline
NGC4649 & 0-0.25 & $ 0.819 _{- 0.003 }^{+ 0.003 } $ & $  1.11 _{-  0.14 }^{+  0.17 } $ & $  1.63 _{-  0.18 }^{+  0.21 } $ & $  1.54 _{-  0.05 }^{+  0.12 } $ &  40.27 & 39.38 &   375 /   199 \\
 & 0.25-0.5 & $ 0.776 _{- 0.005 }^{+ 0.004 } $ & $  1.21 _{-  0.21 }^{+  0.28 } $ & $  2.29 _{-  0.31 }^{+  0.43 } $ & $  1.61 _{-  0.16 }^{+  0.22 } $ &  40.31 & 39.40 &   189 /   206 \\
 & 0.5-0.75 & $ 0.778 _{- 0.009 }^{+ 0.009 } $ & $  1.75 _{-  0.61 }^{+  1.69 } $ & $  3.09 _{-  0.90 }^{+  0.96 } $ & $  2.29 _{-  0.49 }^{+  0.53 } $ &  40.08 & 39.23 &    83 /   147 \\
 & 0.75-1 & $ 0.792 _{- 0.015 }^{+ 0.014 } $ & $  0.55 _{-  0.32 }^{+  0.58 } $ & $  1.04 _{-  0.68 }^{+  0.74 } $ & $  0.98 _{-  0.15 }^{+  0.24 } $ &  39.92 & 39.25 &    68 /   108 \\
 & 1-1.5 & $ 0.816 _{- 0.009 }^{+ 0.009 } $ & $  1.04 _{-  0.28 }^{+  0.90 } $ & $  1.77 _{-  0.53 }^{+  1.01 } $ & $  1.22 _{-  0.11 }^{+  0.24 } $ &  40.14 & 39.68 &    96 /   186 \\
 & 1.5-2 & $ 0.830 _{- 0.013 }^{+ 0.012 } $ & $  0.49 _{-  0.32 }^{+  0.82 } $ & $  1.00 _{-  1.00 }^{+  1.00 } $ & $  1.02 _{-  0.09 }^{+  0.13 } $ &  39.97 & 39.70 &    77 /   166 \\
 & 2-4 & $ 0.913 _{- 0.023 }^{+ 0.021 } $ & $  0.18 _{-  0.12 }^{+  0.13 } $ & $  0.42 _{-  0.17 }^{+  0.16 } $ & $  0.26 _{-  0.03 }^{+  0.04 } $ &  40.20 & 39.82 &   167 /   206 \\
 & 4-6 & $ 0.864 _{- 0.041 }^{+ 0.063 } $ & $  0.00 _{-  0.00 }^{+  0.11 } $ & $  0.00 _{-  0.00 }^{+  0.14 } $ & $  0.07 _{-  0.02 }^{+  0.04 } $ &  39.81 & 39.79 &   149 /   157 \\
\hline
NGC5044 & 0-0.5 & $ 0.729 _{- 0.012 }^{+ 0.013 } $ & $  0.57 _{-  0.17 }^{+  0.16 } $ & $  0.44 _{-  0.28 }^{+  0.29 } $ & $  0.53 _{-  0.07 }^{+  0.07 } $ &  41.09 & 39.96 &    64 /    68 \\
 & 0.5-1 & $ 0.762 _{- 0.007 }^{+ 0.007 } $ & $  0.62 _{-  0.15 }^{+  0.18 } $ & $  0.76 _{-  0.21 }^{+  0.25 } $ & $  0.83 _{-  0.08 }^{+  0.10 } $ &  41.57 & 40.26 &    97 /   148 \\
 & 1-1.5 & $ 0.794 _{- 0.008 }^{+ 0.008 } $ & $  0.81 _{-  0.20 }^{+  0.34 } $ & $  1.29 _{-  0.29 }^{+  0.46 } $ & $  0.92 _{-  0.11 }^{+  0.16 } $ &  41.65 & 40.35 &   113 /   169 \\
 & 1.5-2 & $ 0.833 _{- 0.009 }^{+ 0.009 } $ & $  0.67 _{-  0.23 }^{+  0.34 } $ & $  0.95 _{-  0.29 }^{+  0.22 } $ & $  0.84 _{-  0.11 }^{+  0.14 } $ &  41.62 & 40.57 &   110 /   162 \\
 & 2-3 & $ 0.944 _{- 0.004 }^{+ 0.008 } $ & $  0.54 _{-  0.12 }^{+  0.13 } $ & $  0.82 _{-  0.12 }^{+  0.14 } $ & $  0.76 _{-  0.06 }^{+  0.04 } $ &  41.88 & 40.61 &   150 /   161 \\
 & 3-4 & $ 1.138 _{- 0.024 }^{+ 0.026 } $ & $  0.31 _{-  0.14 }^{+  0.18 } $ & $  0.42 _{-  0.14 }^{+  0.15 } $ & $  0.58 _{-  0.07 }^{+  0.06 } $ &  41.67 & 40.22 &    86 /   111 \\
 & 4-6 & $ 1.248 _{- 0.010 }^{+ 0.011 } $ & $  0.23 _{-  0.10 }^{+  0.11 } $ & $  0.72 _{-  0.09 }^{+  0.09 } $ & $  0.70 _{-  0.05 }^{+  0.05 } $ &  41.84 & 40.49 &   242 /   246 \\
 & 6-8 & $ 1.224 _{- 0.015 }^{+ 0.014 } $ & $  0.63 _{-  0.13 }^{+  0.14 } $ & $  0.56 _{-  0.09 }^{+  0.10 } $ & $  0.48 _{-  0.04 }^{+  0.04 } $ &  41.78 & 40.67 &   285 /   261 \\
\hline
NGC5846 & 0-0.5 & $ 0.659 _{- 0.008 }^{+ 0.009 } $ & $  0.48 _{-  0.08 }^{+  0.08 } $ & $  0.45 _{-  0.17 }^{+  0.11 } $ & $  0.52 _{-  0.03 }^{+  0.02 } $ &  40.89 & 39.83 &   351 /   359 \\
 & 0.5-1 & $ 0.621 _{- 0.012 }^{+ 0.012 } $ & --- & --- & --- &  40.96 & 39.66 & --- \\
 & 1-1.5 & $ 0.636 _{- 0.018 }^{+ 0.018 } $ & --- & --- & --- &  40.89 & 39.75 & --- \\
 & 1.5-2 & $ 0.682 _{- 0.019 }^{+ 0.022 } $ & --- & --- & --- &  40.97 & 39.79 & --- \\
 & 2-3 & $ 0.811 _{- 0.013 }^{+ 0.013 } $ & --- & --- & --- &  41.05 & 39.81 & --- \\
 & 3-4 & $ 1.060 _{- 0.051 }^{+ 0.120 } $ & --- & --- & --- &  40.65 & 39.89 & --- \\
 & 4-8 & $ 1.057 _{- 0.037 }^{+ 0.020 } $ & --- & --- & --- &  40.83 & 40.66 & --- \\
\end{longtable}
}

}

\appendix

\begin{figure*}
  \centering
    \includegraphics[width=4.5cm]{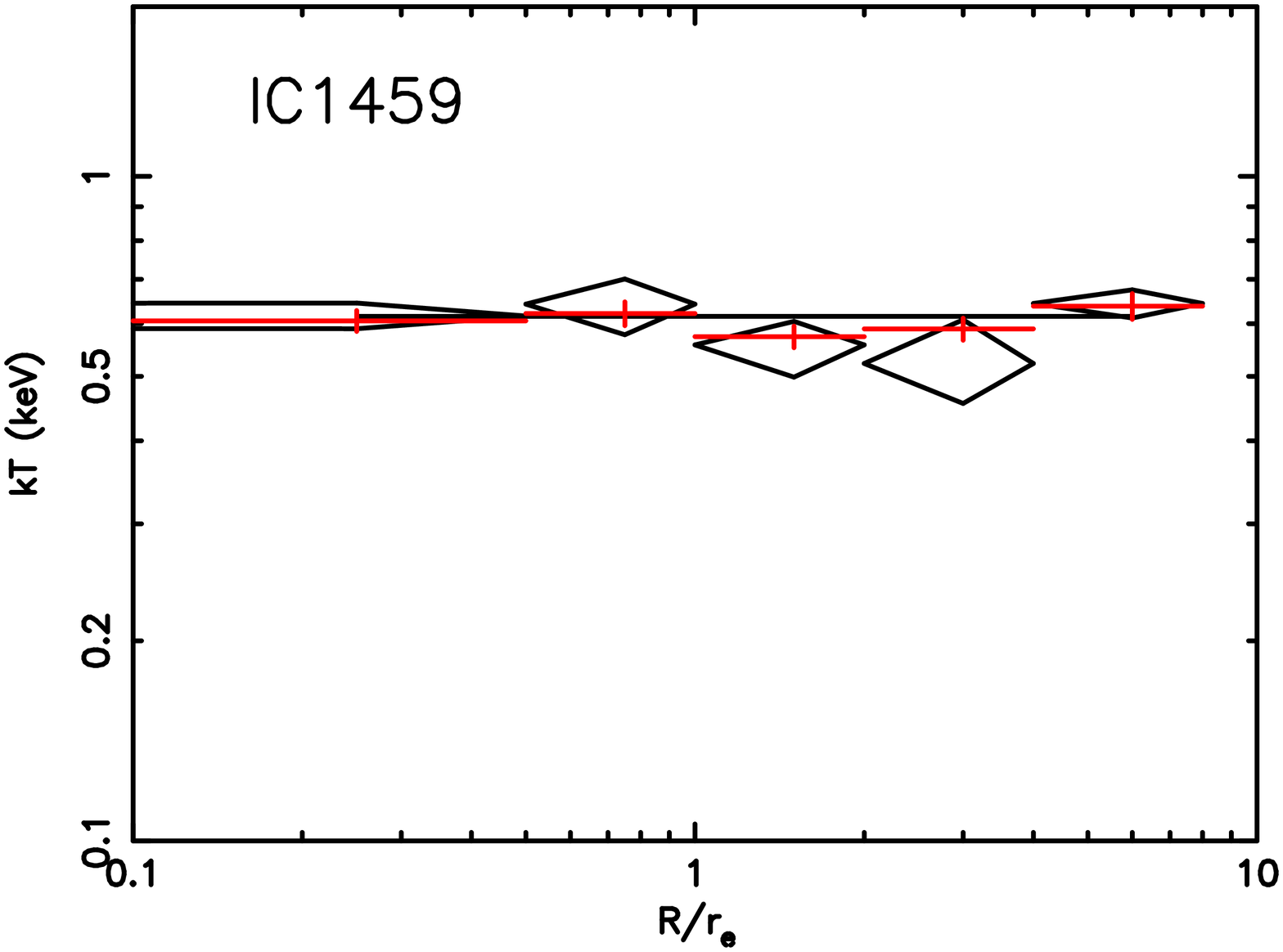}
    \includegraphics[width=4.5cm]{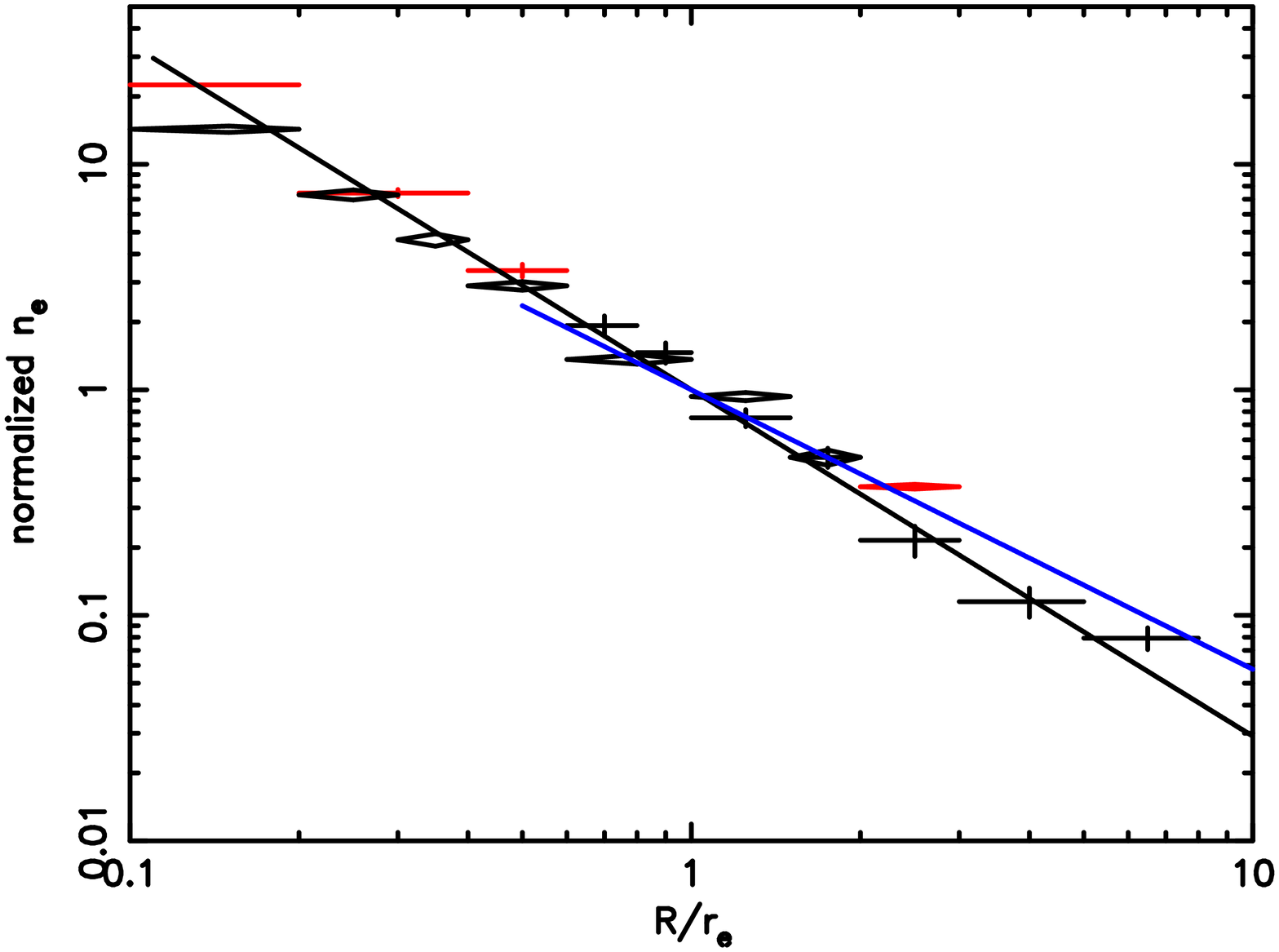}
    \includegraphics[width=4.5cm]{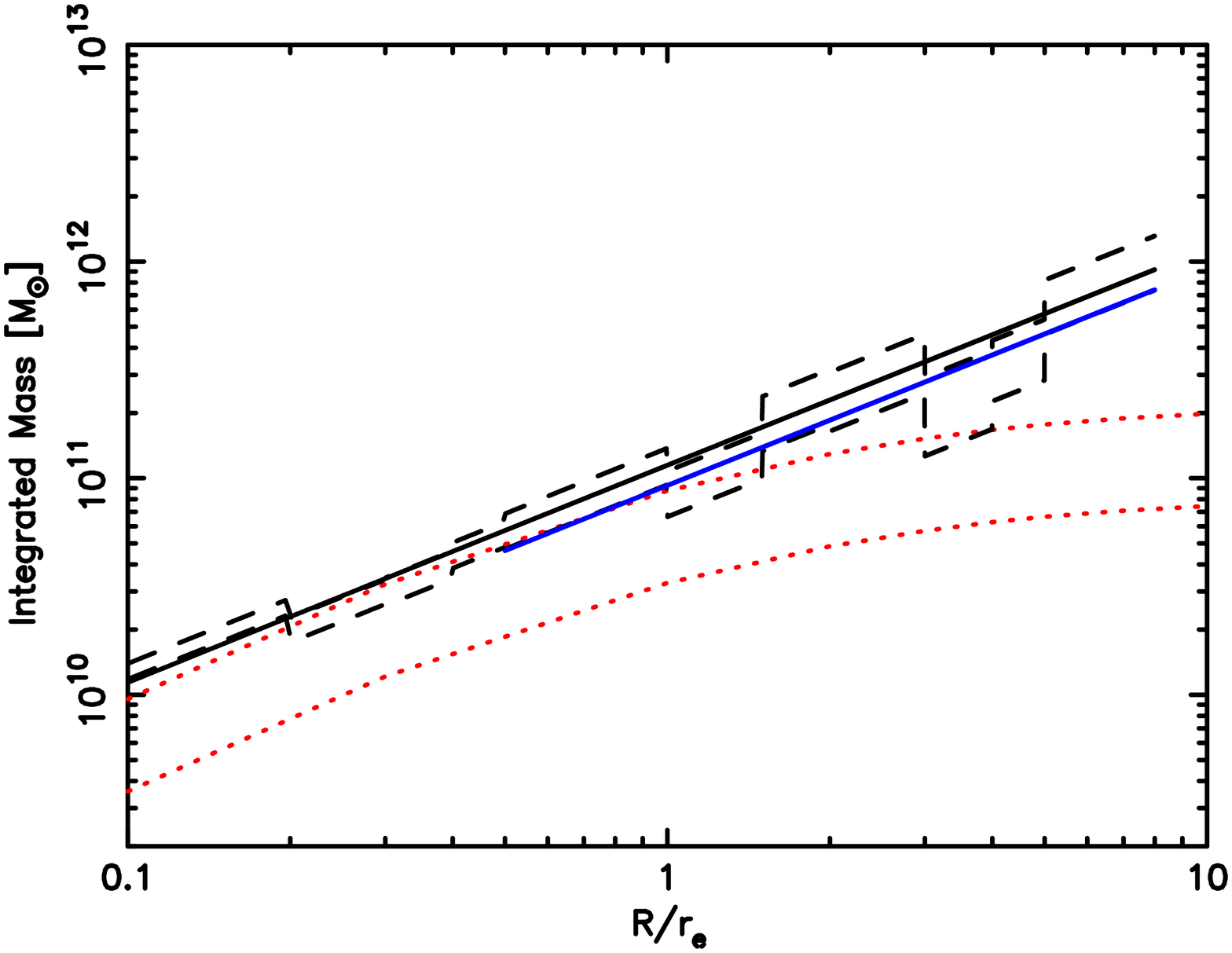}\\
    \includegraphics[width=4.5cm]{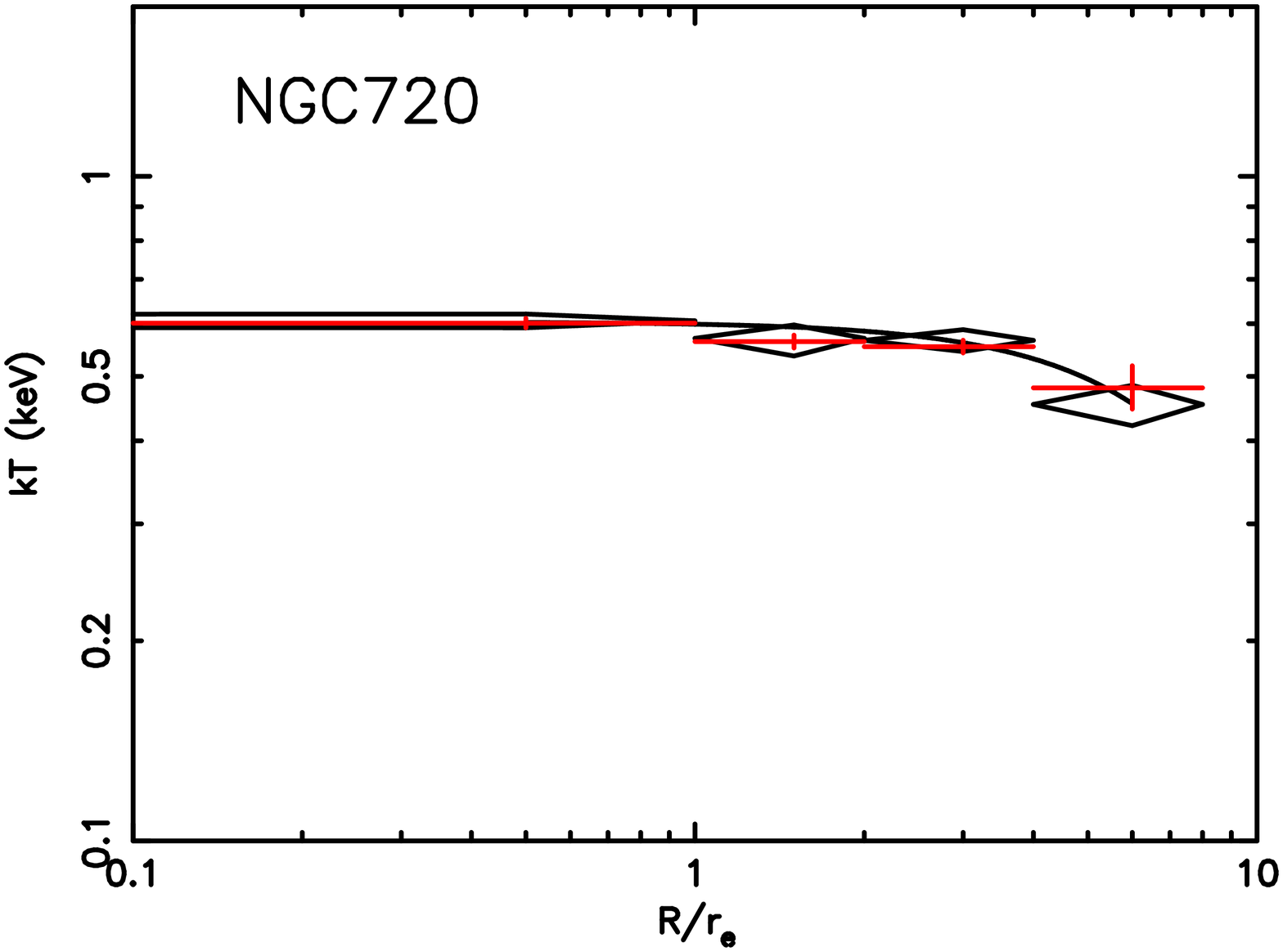}
    \includegraphics[width=4.5cm]{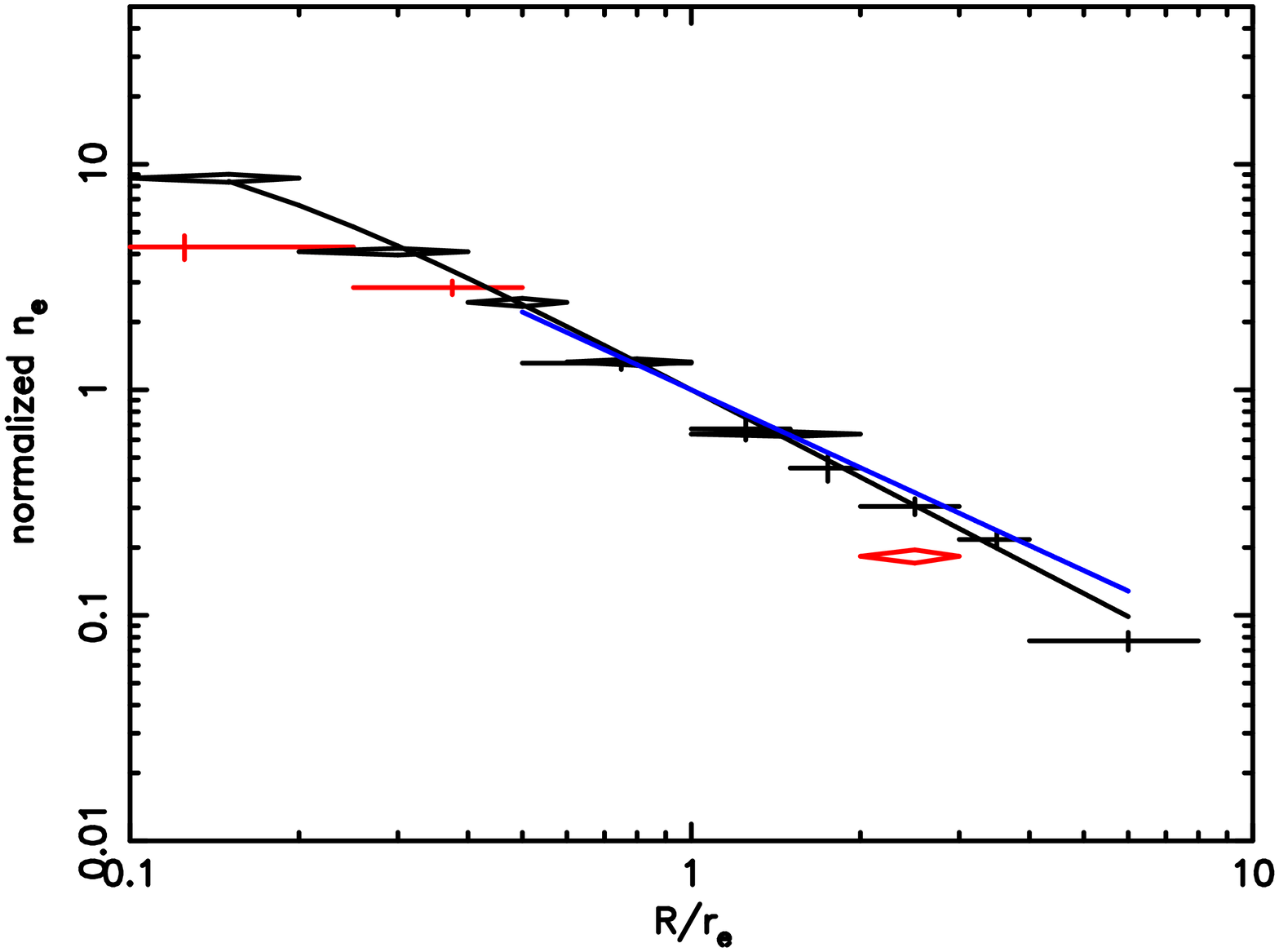}
    \includegraphics[width=4.5cm]{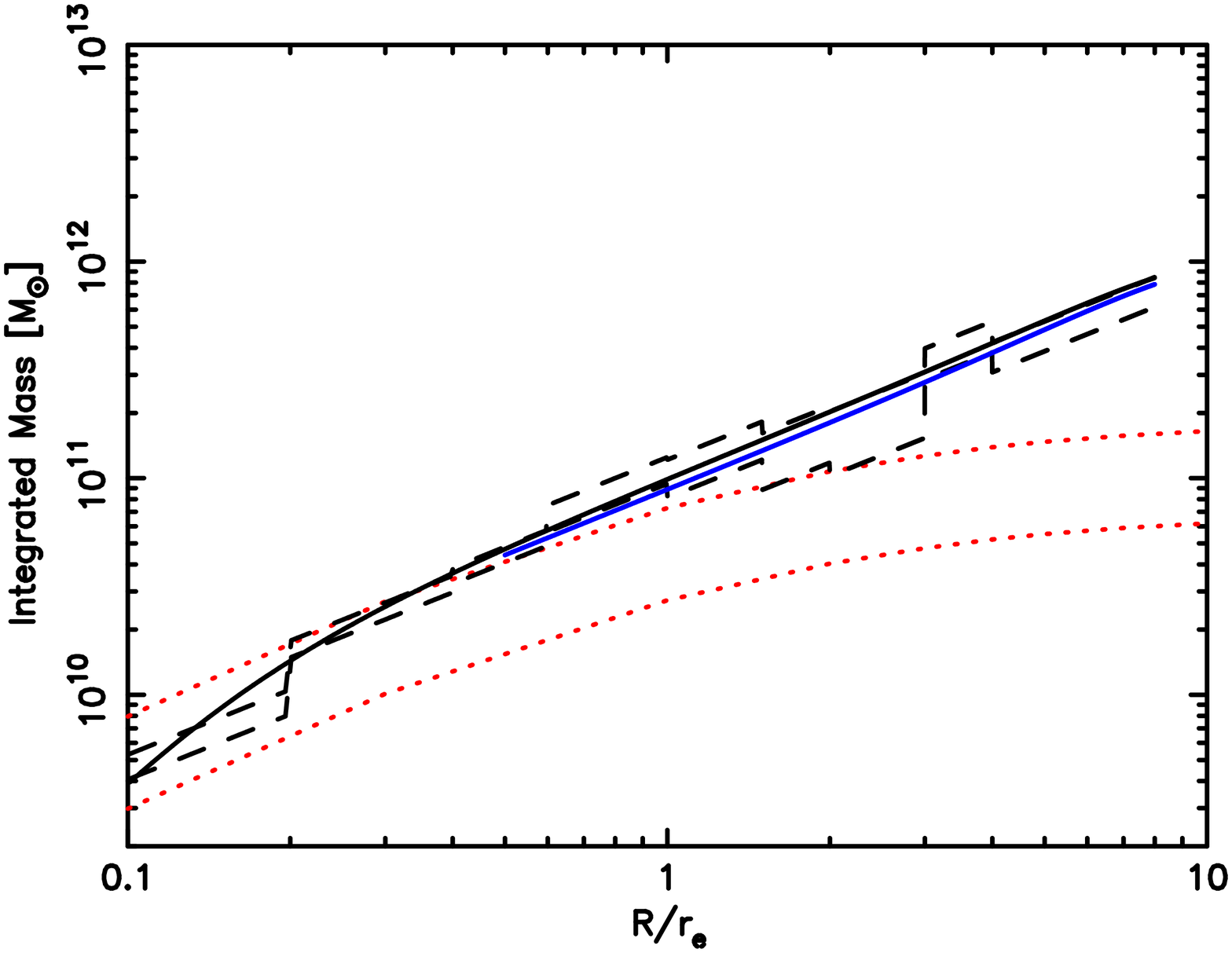}\\
    \includegraphics[width=4.5cm]{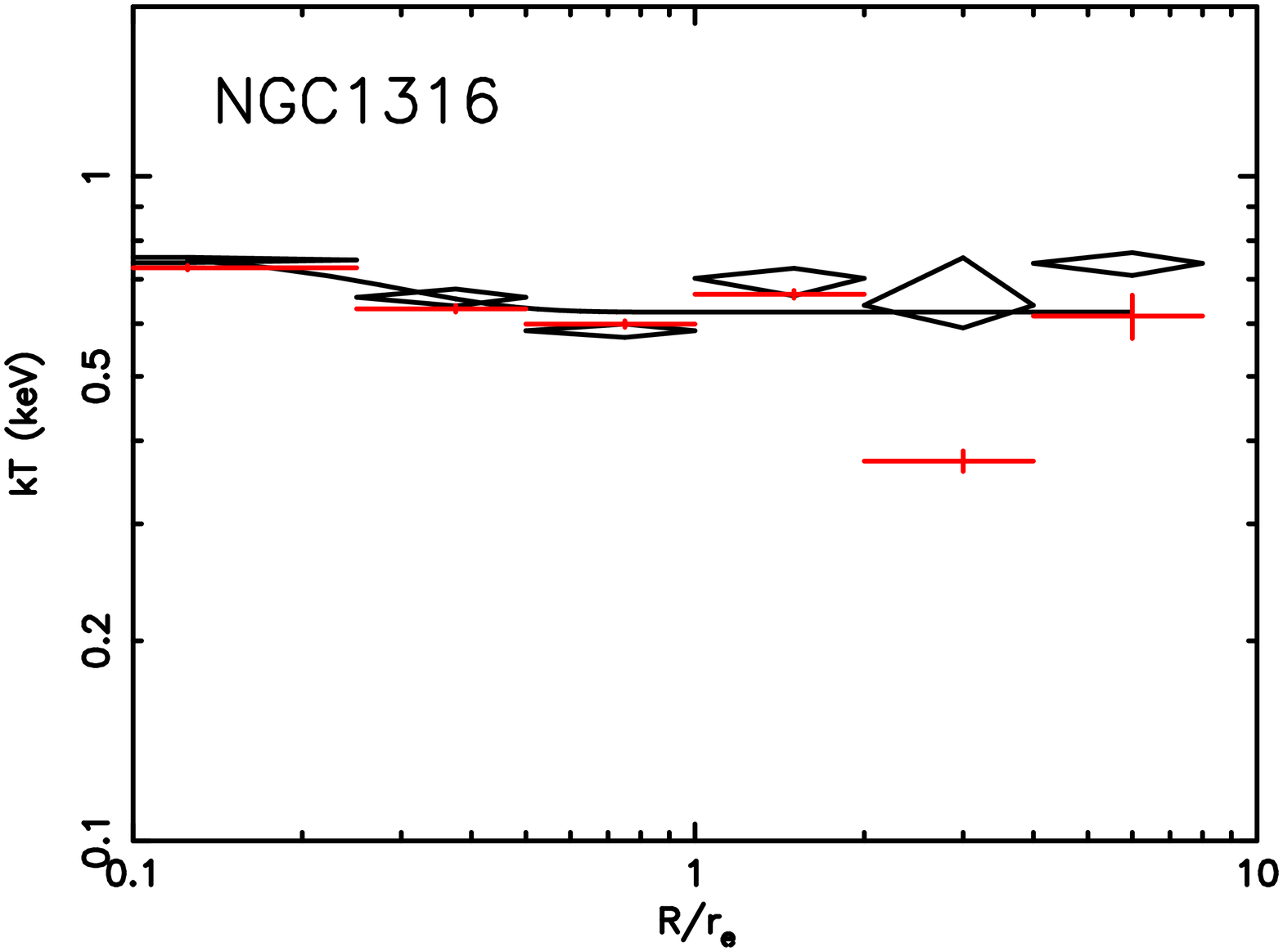}
    \includegraphics[width=4.5cm]{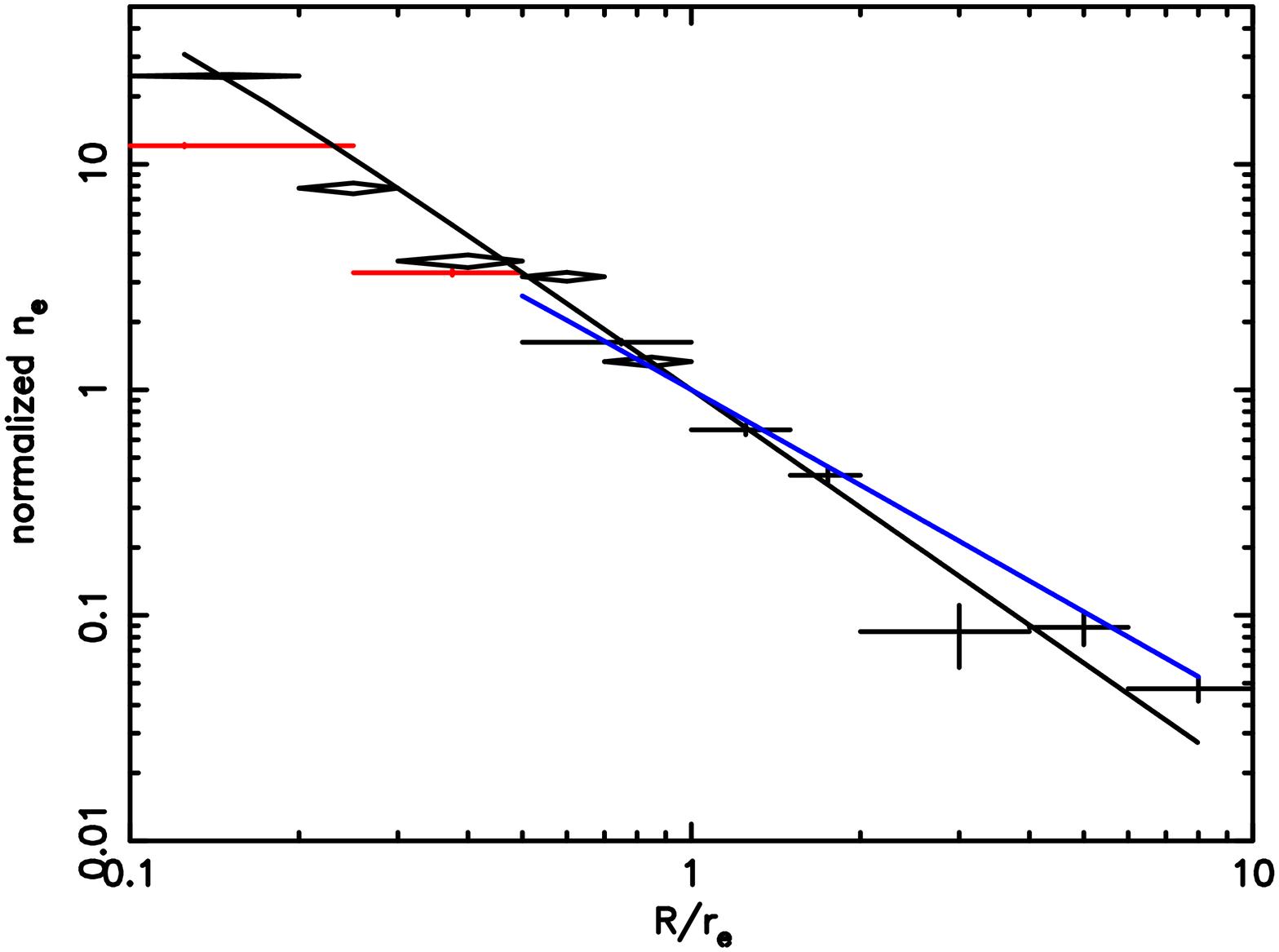}
    \includegraphics[width=4.5cm]{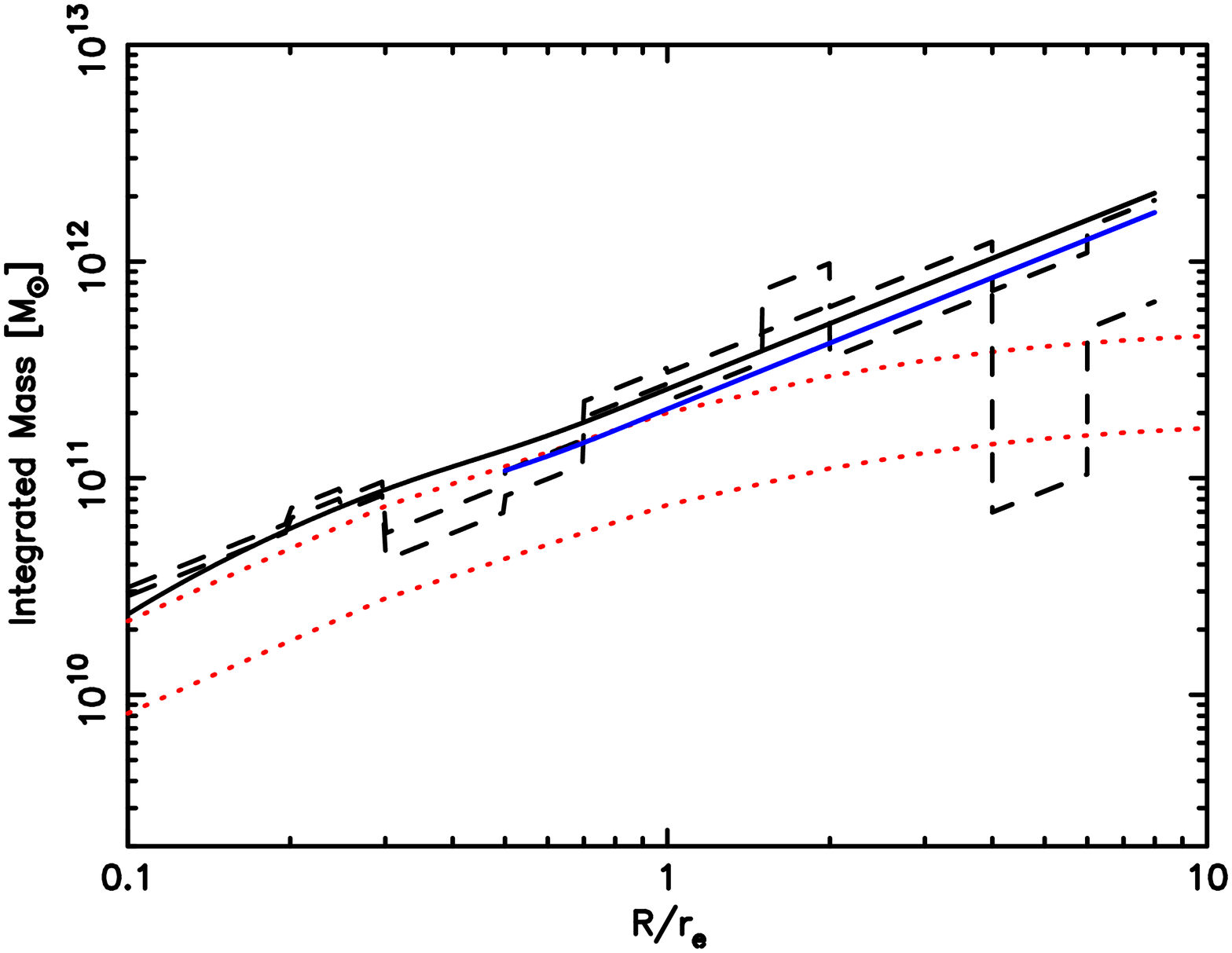}\\
    \includegraphics[width=4.5cm]{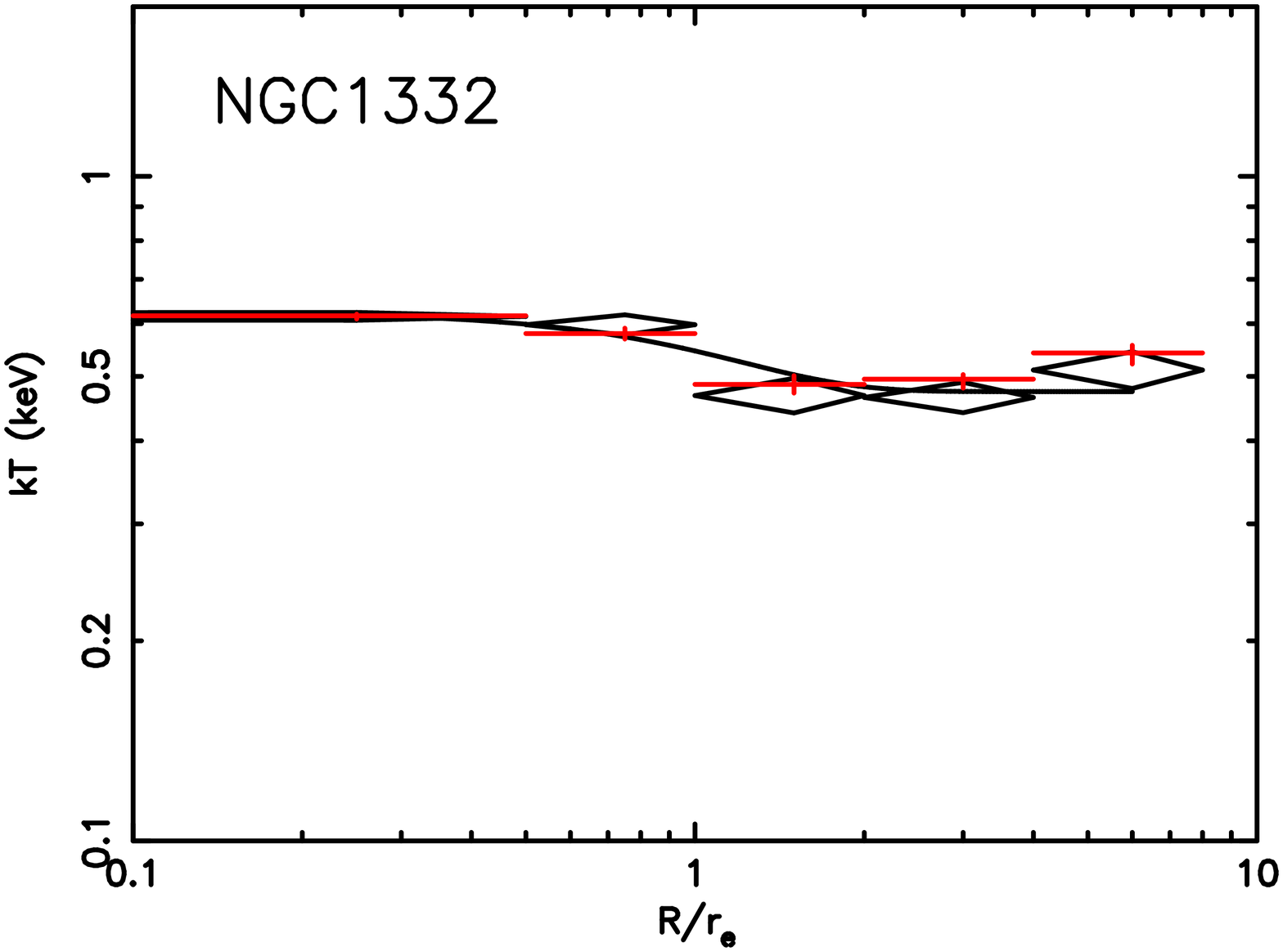}
    \includegraphics[width=4.5cm]{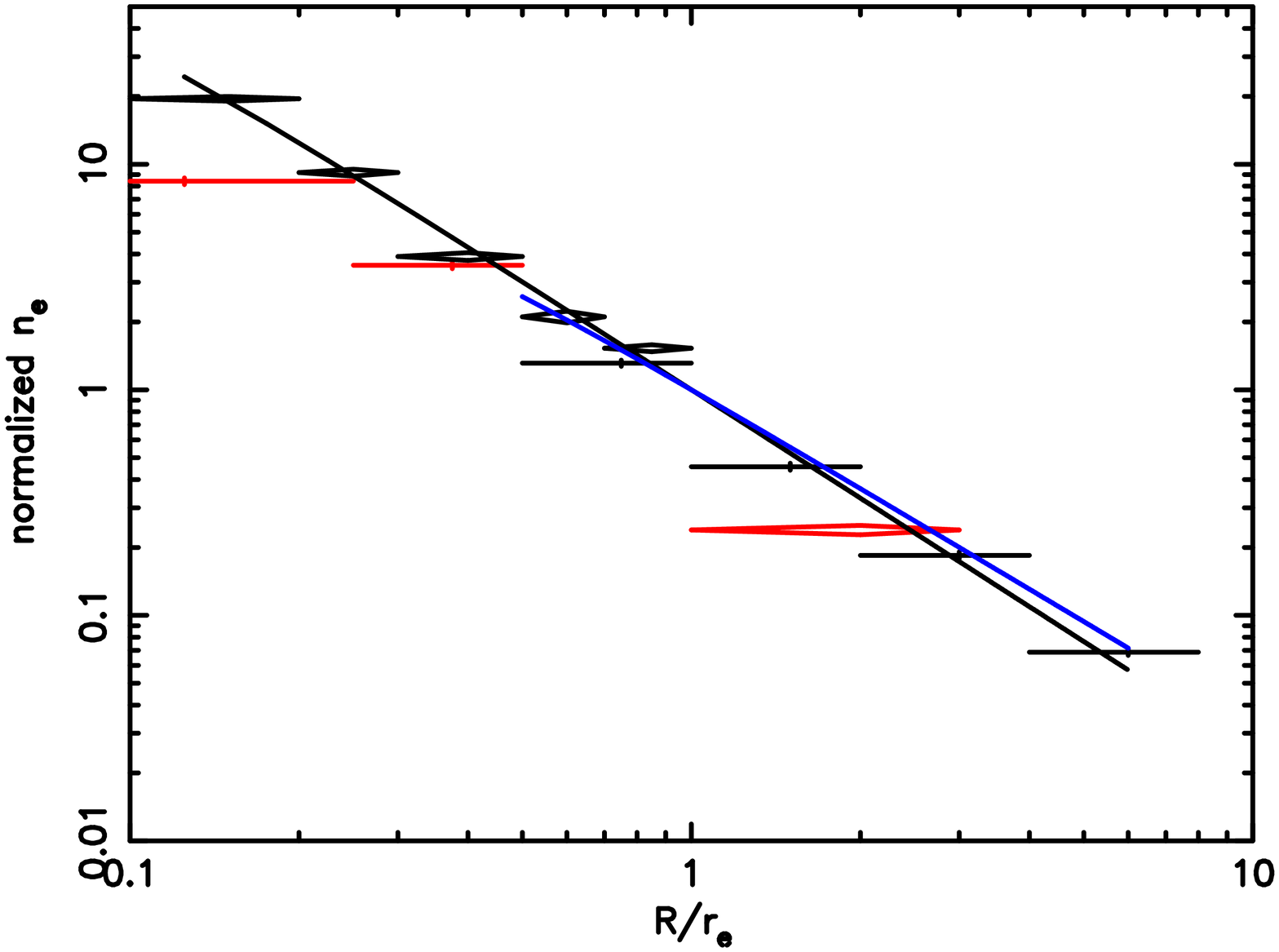}
    \includegraphics[width=4.5cm]{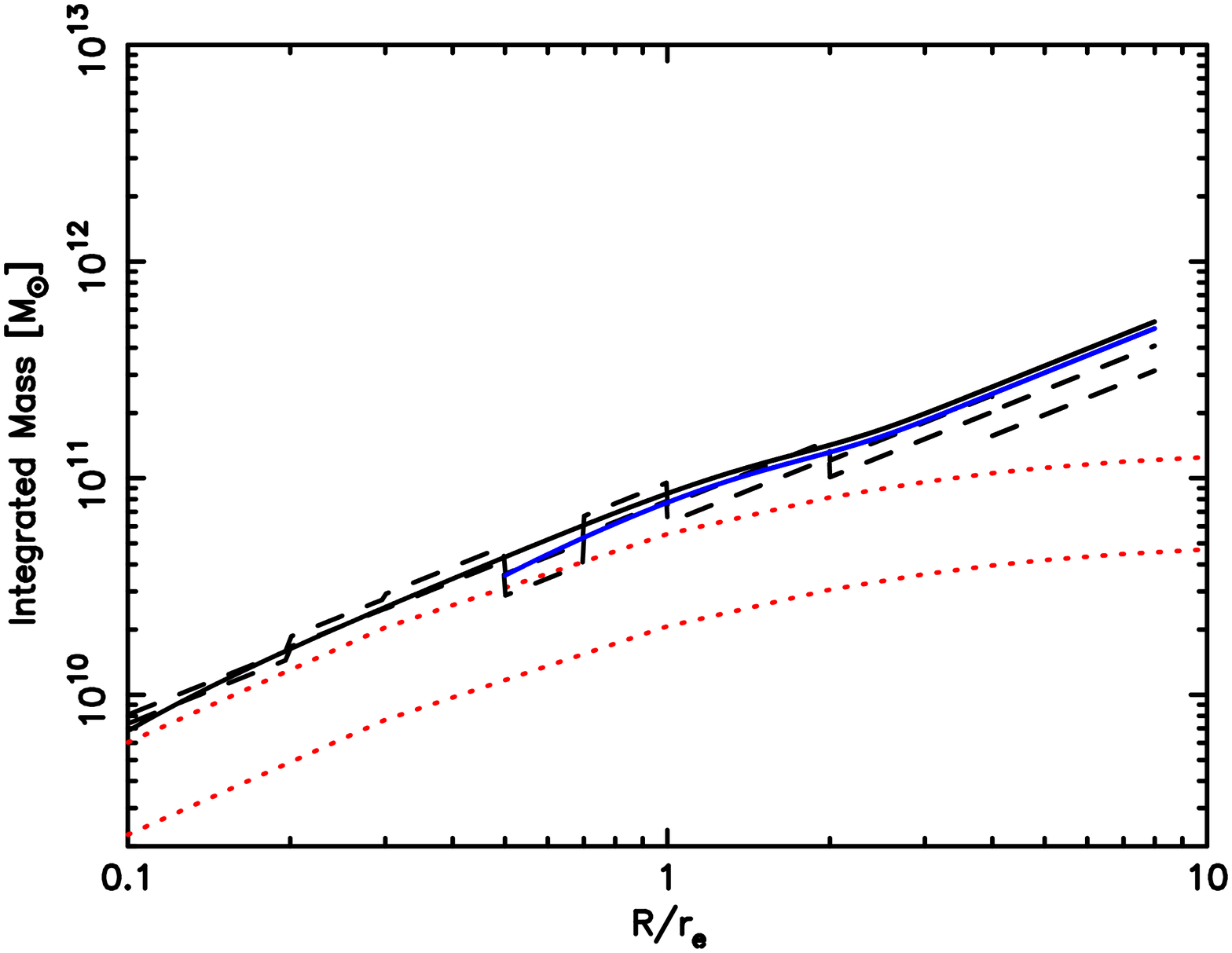}\\
    \includegraphics[width=4.5cm]{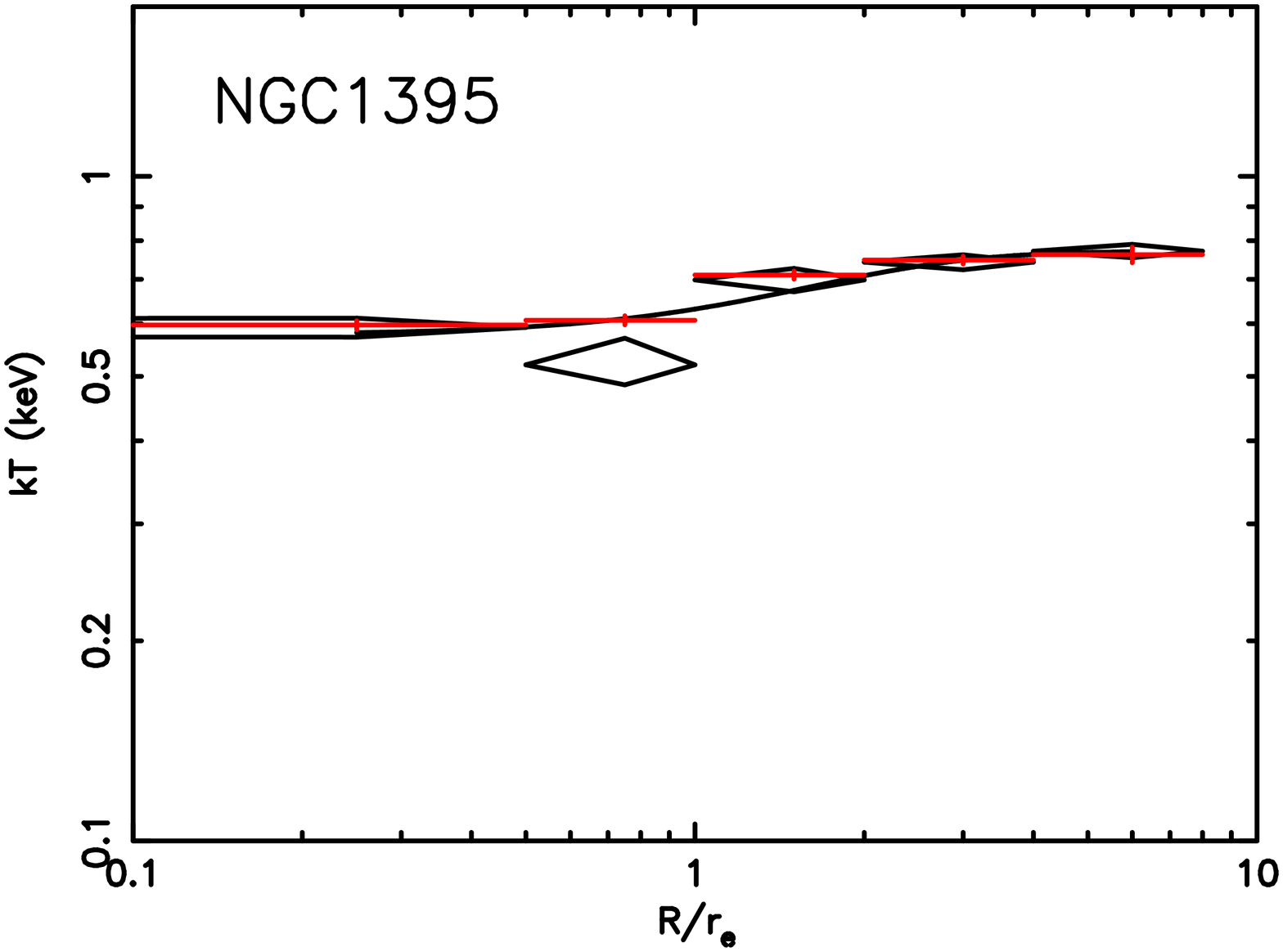}
    \includegraphics[width=4.5cm]{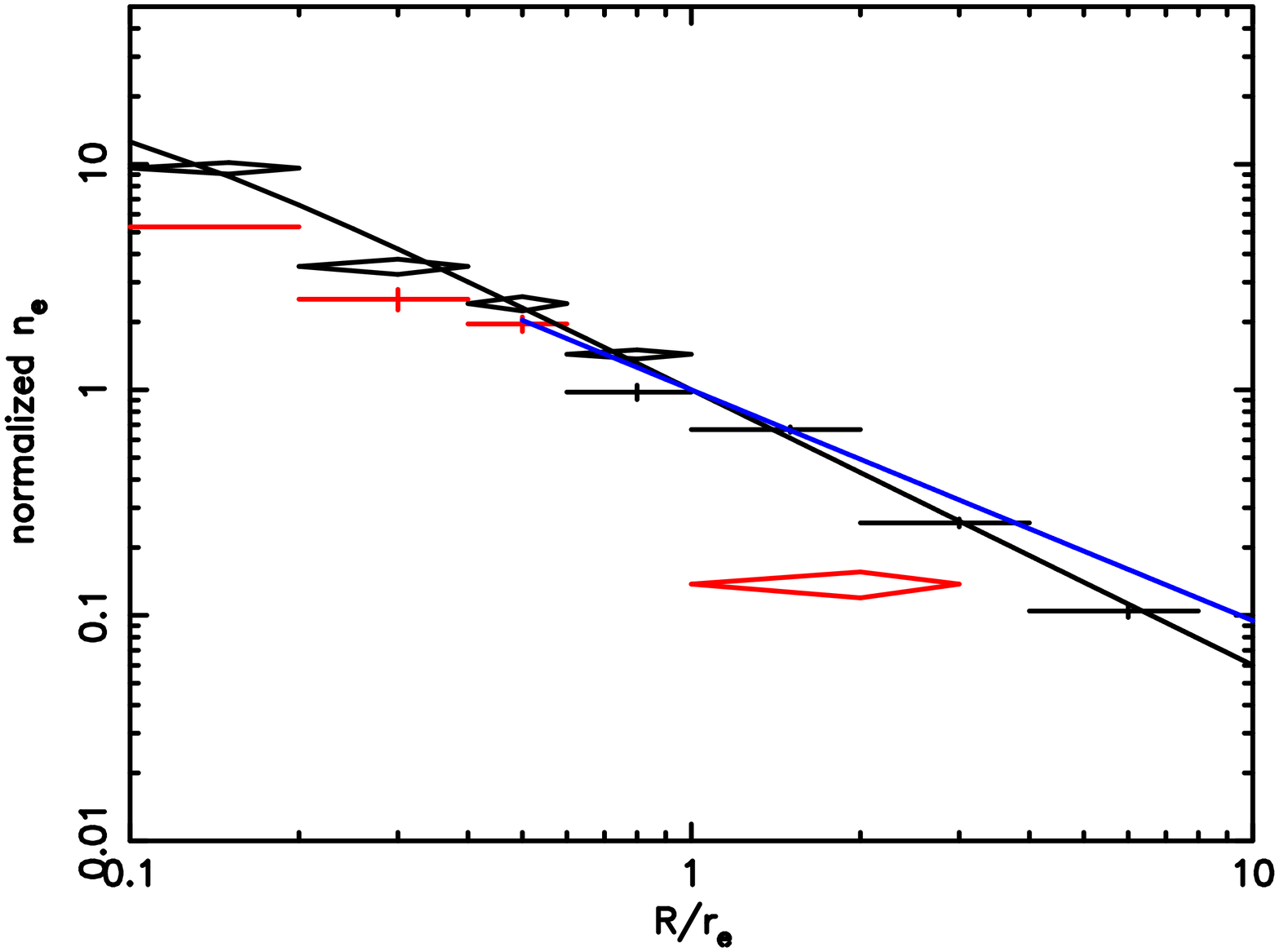}
    \includegraphics[width=4.5cm]{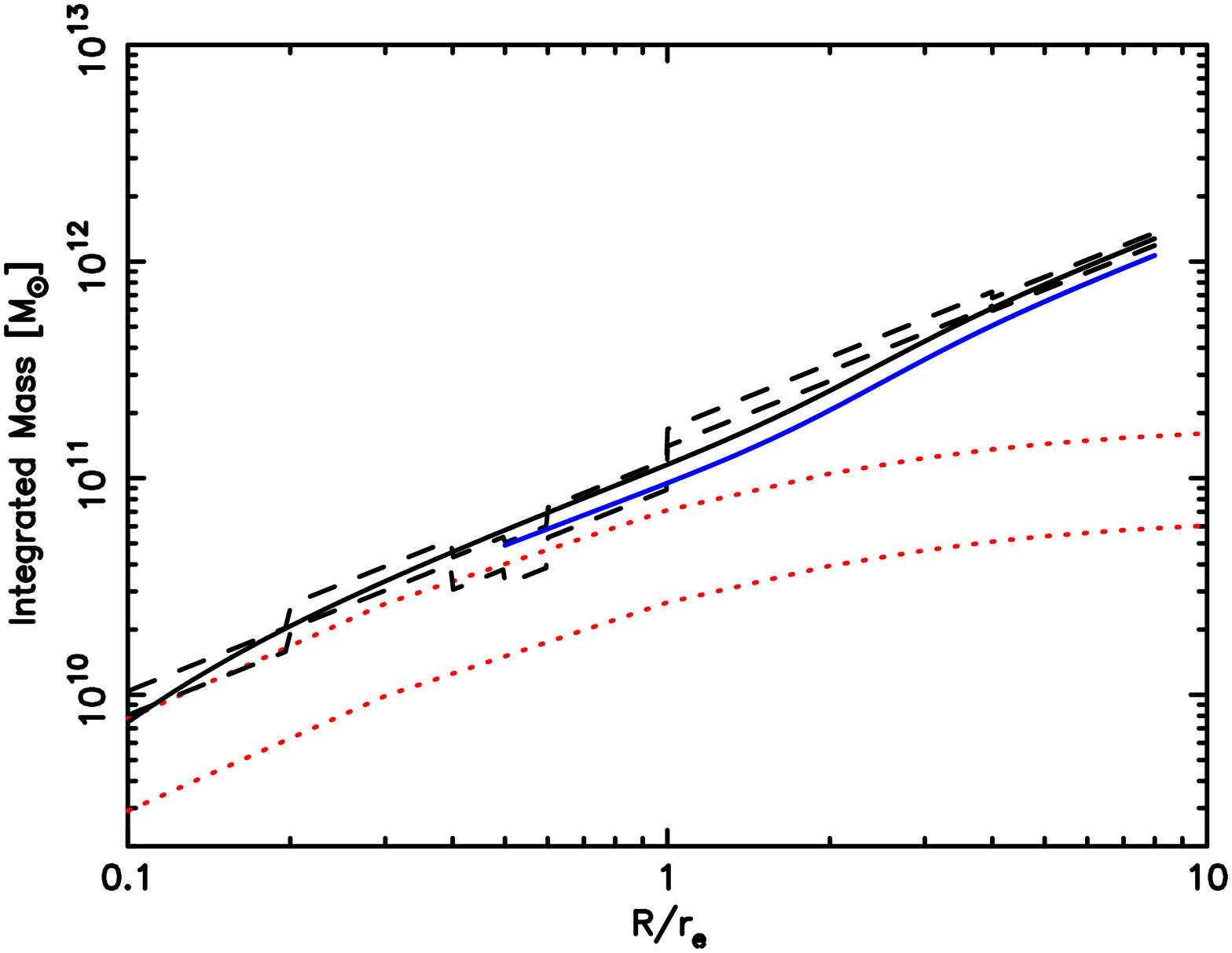}\\
 \flushleft
 {\bf Fig.~A.1.~~}(Left) The ISM temperature profiles of sample galaxies. The
 cross data points represent temperature profile derived from projected spectra, 
 and the diamond data points are from deprojected spectra. The solid lines
 represent each best-fit function. The black data points are used to fit
 the function, but the red data points are not. (Center) The ISM
 density profiles. The cross data points represent XMM-Newton
 data, and the diamond data points are Chandra data obtained from
 spatial analysis. The black solid lines represent each best-fit
 function. Meanings of the red and black colors are the same as left
 panel. The blue solid lines show best-fit function of ISM density
 profile obtained from spectral fit. These ISM density profiles are
 normalized by the value at $1r_e$. (Right) The integrated mass
 profiles. The solid lines are total mass
 obtained from the best-fit function of temperature and ISM density
 profiles from the surface brightness (black) and spectral fit (blue). The
 dashed lines are upper and lower limit. We also plotted
 the stellar mass profiles assuming stellar $M/L_B$ to be 3 and 8 (red
 dotted lines).
\end{figure*}

\begin{figure*}
  \centering
    \includegraphics[width=4.5cm]{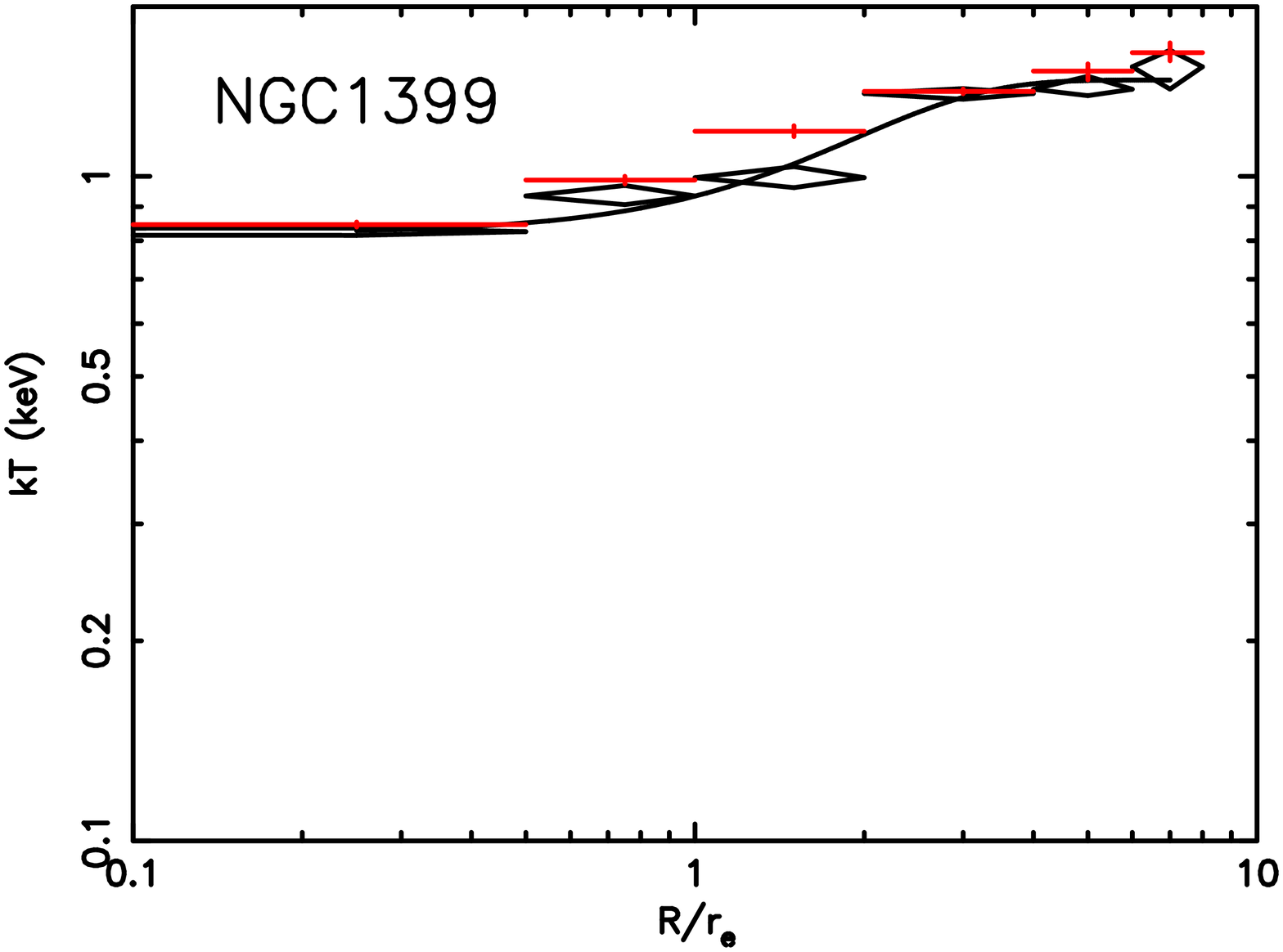}
    \includegraphics[width=4.5cm]{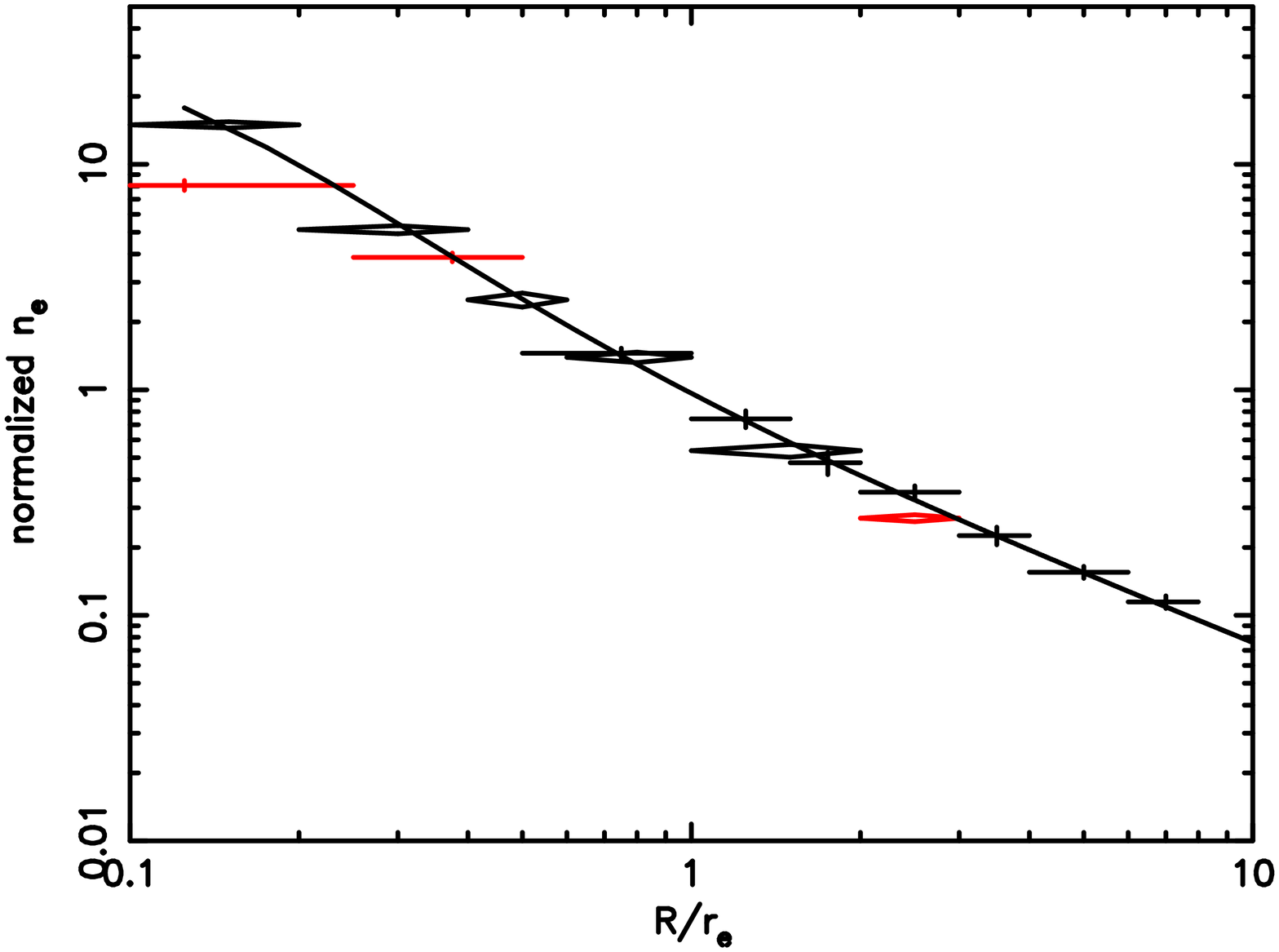}
    \includegraphics[width=4.5cm]{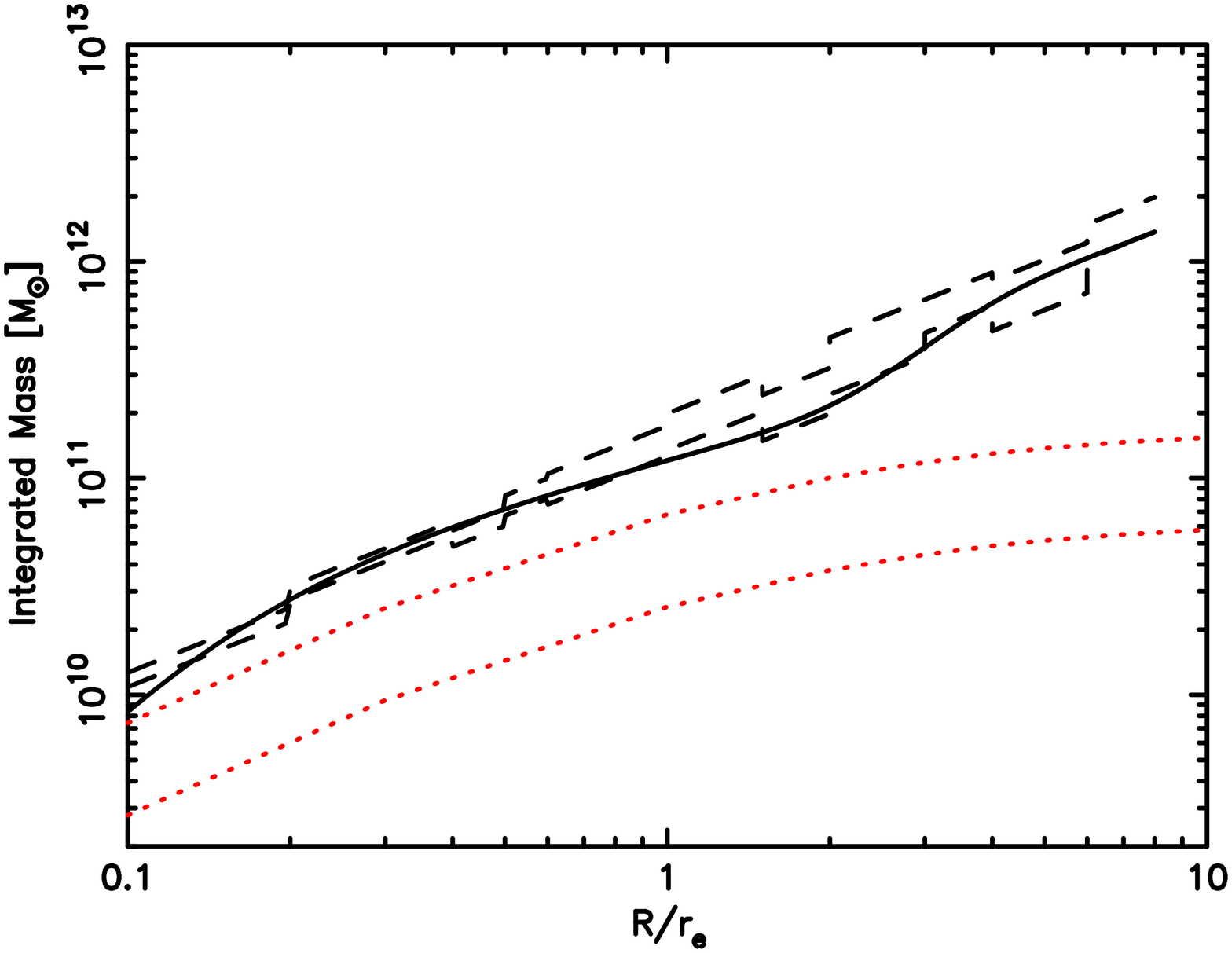}\\
    \includegraphics[width=4.5cm]{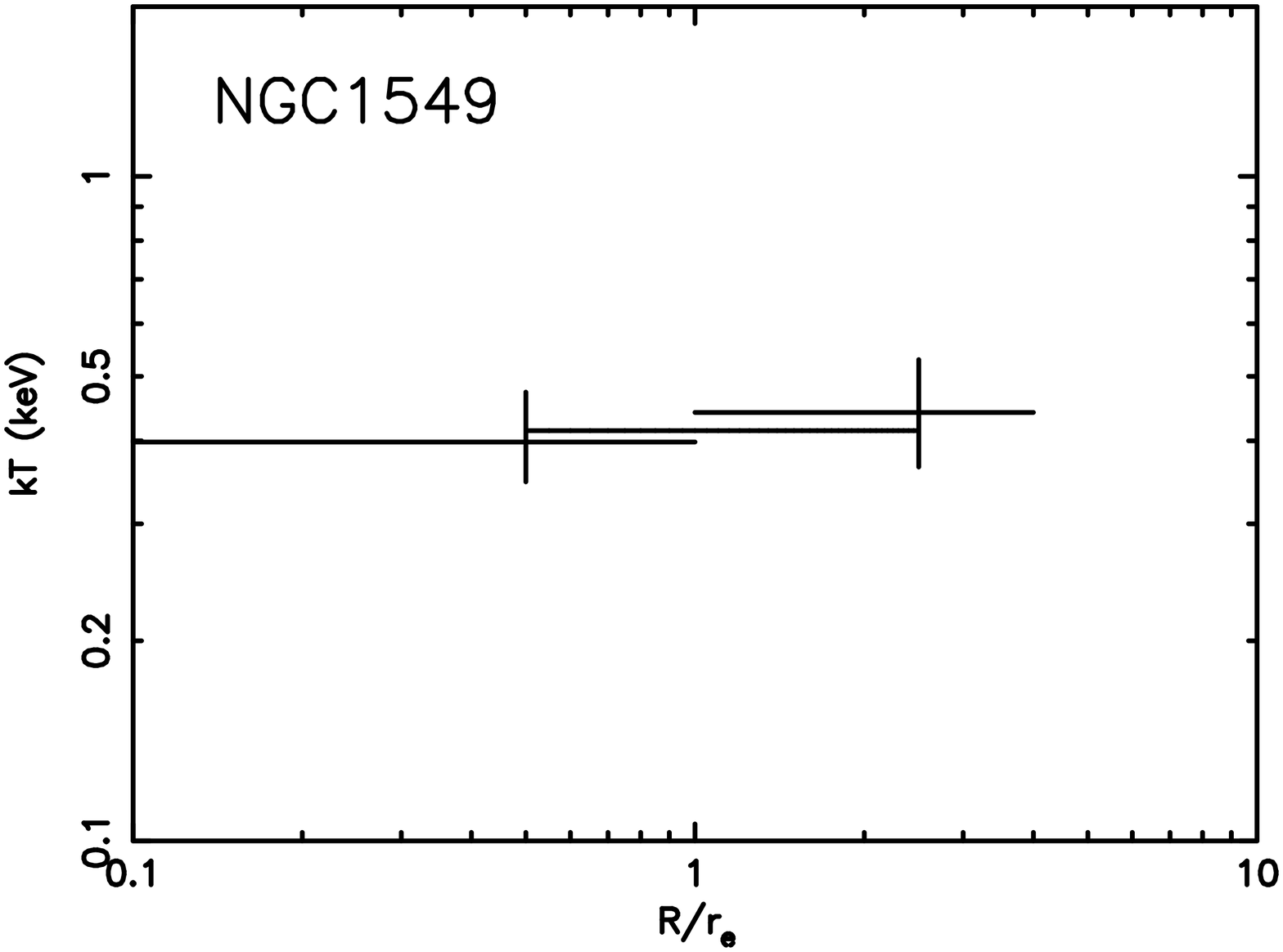}
    \includegraphics[width=4.5cm]{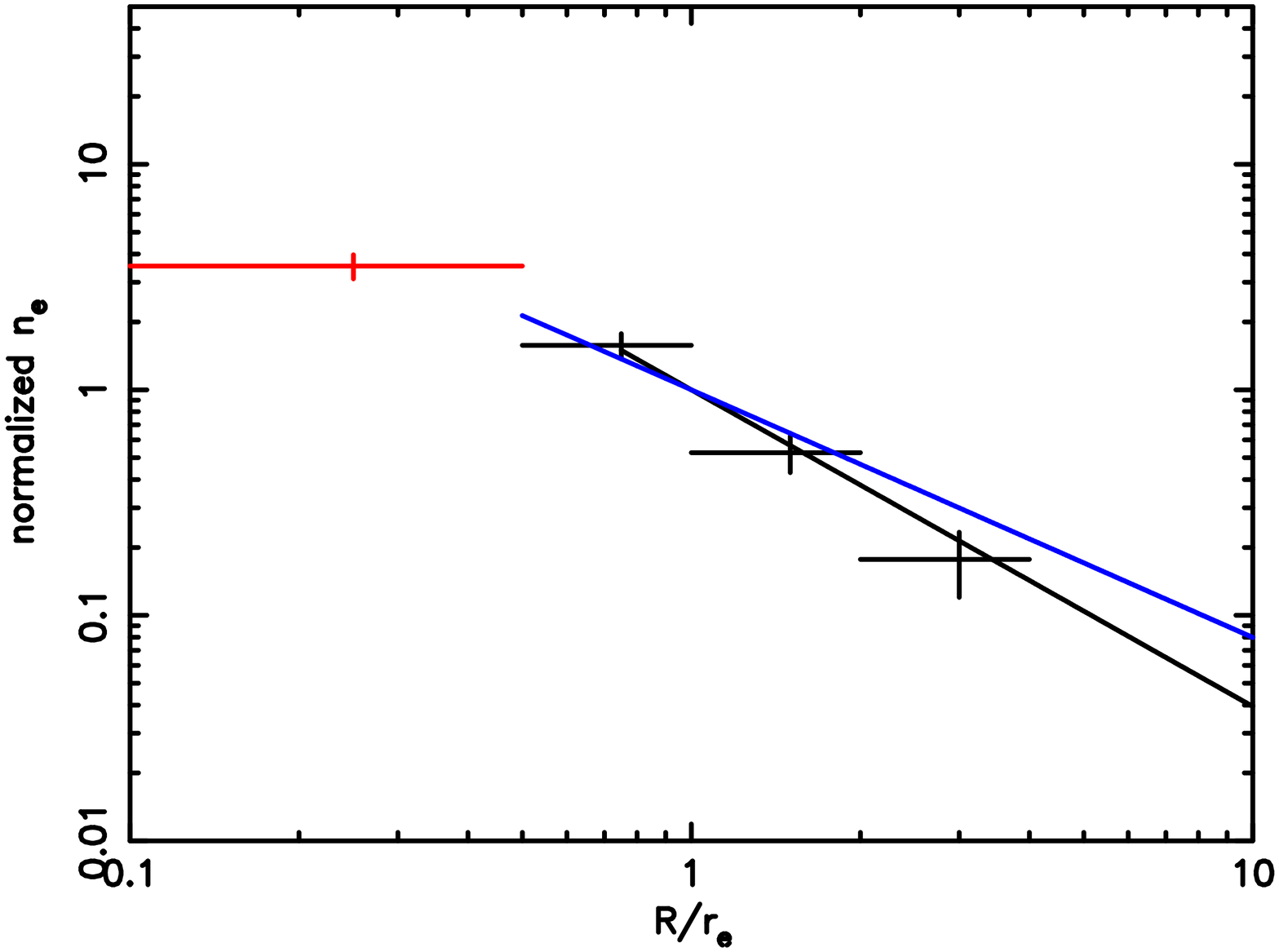}
    \includegraphics[width=4.5cm]{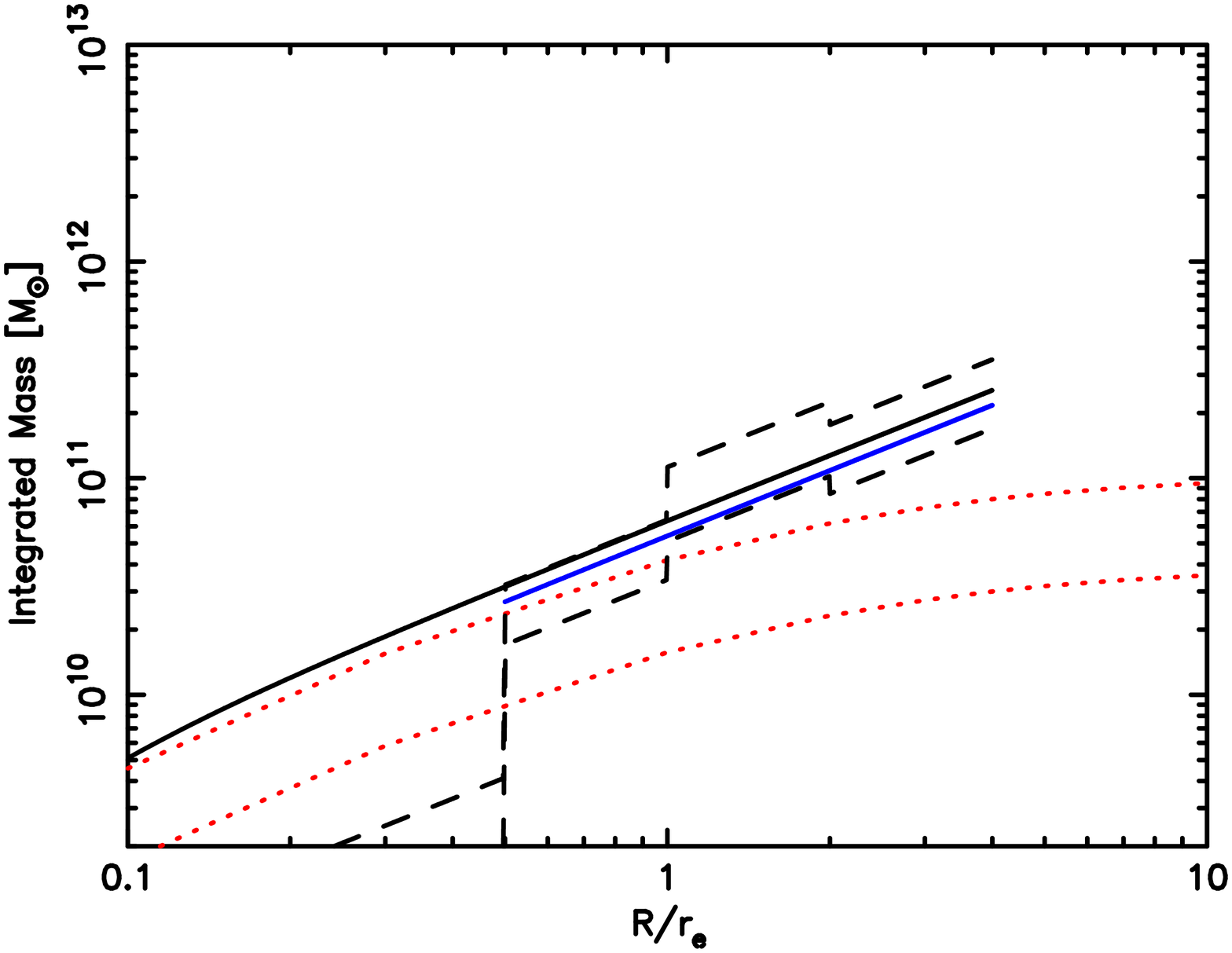}\\
    \includegraphics[width=4.5cm]{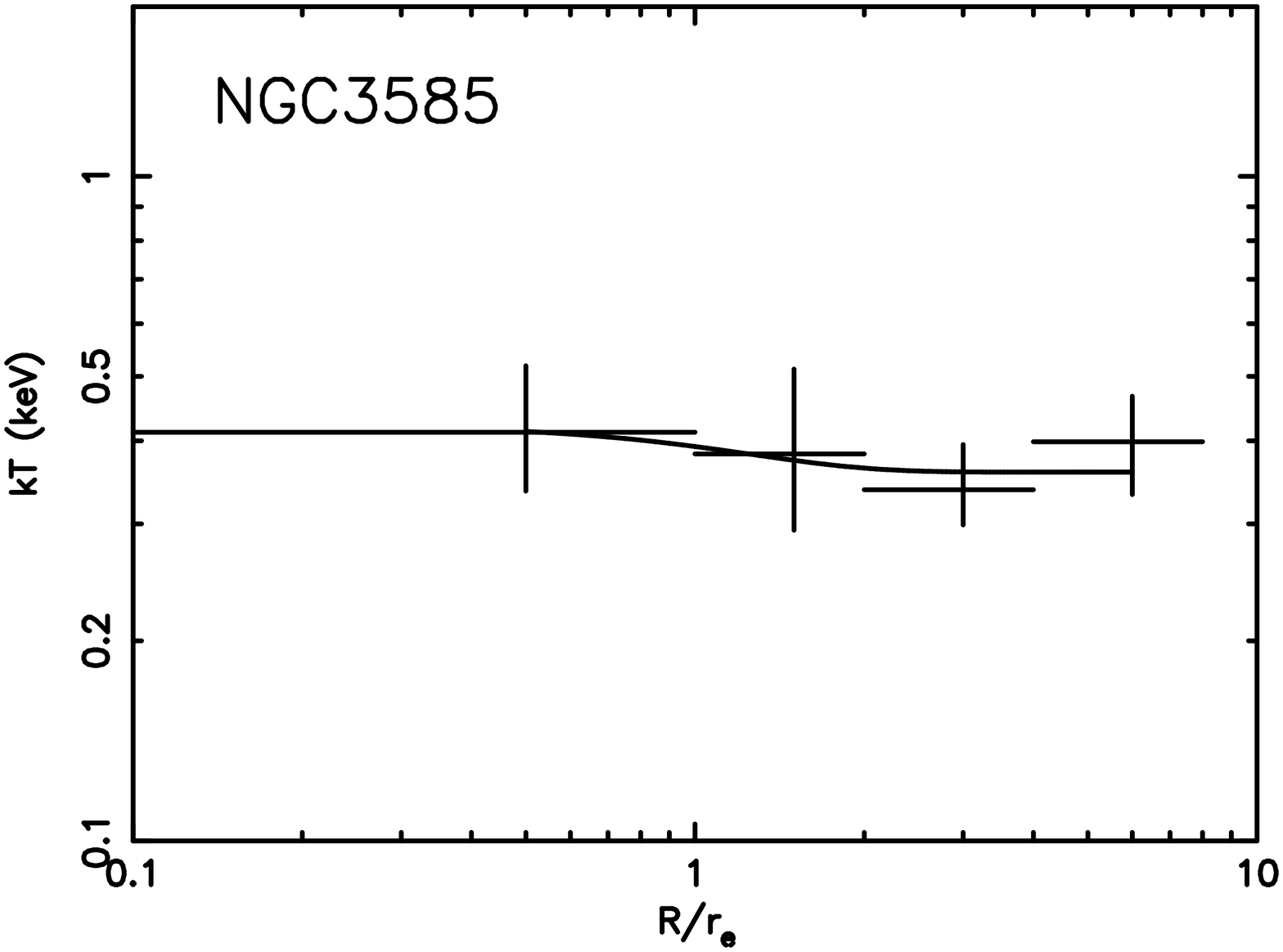}
    \includegraphics[width=4.5cm]{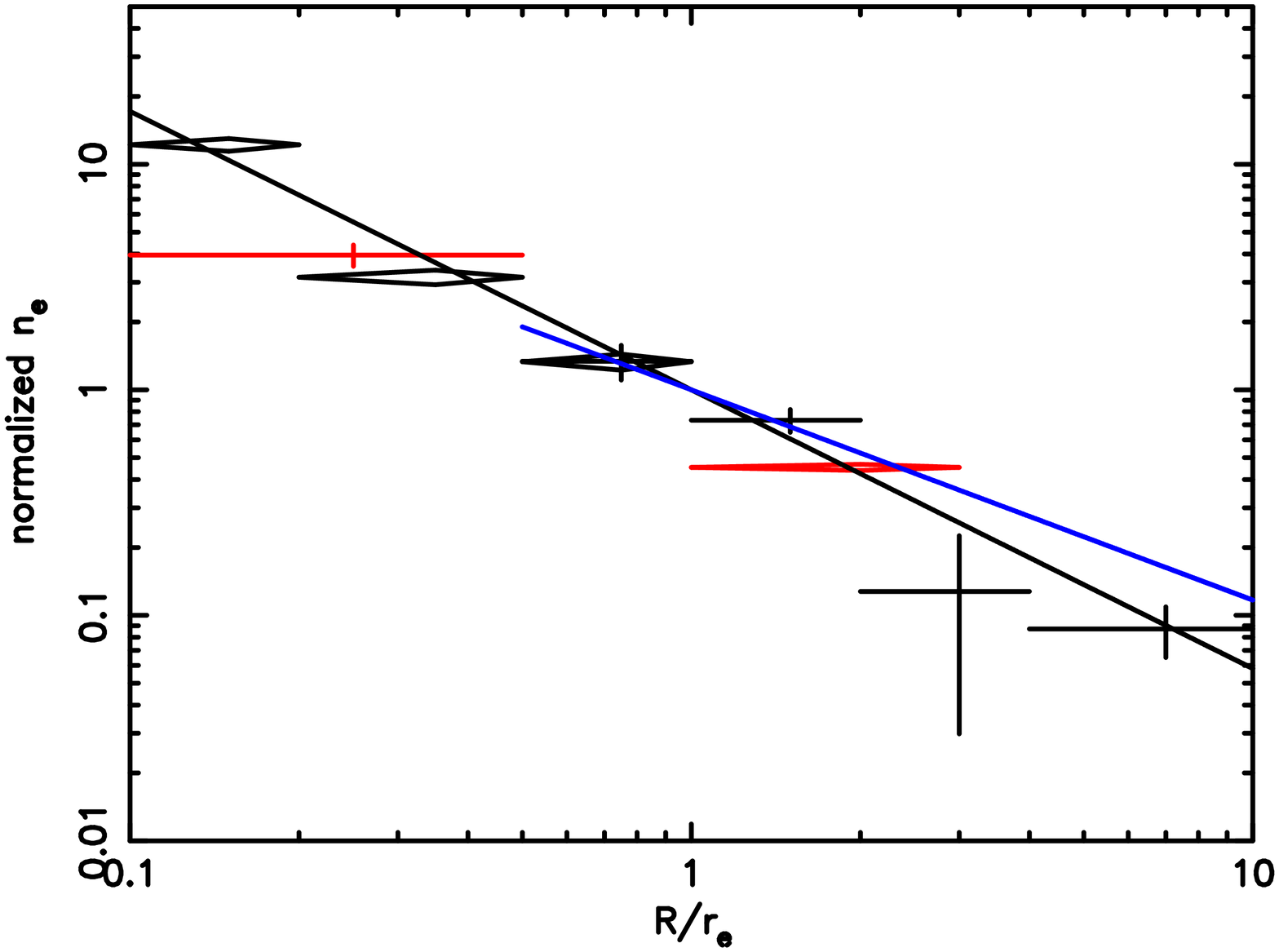}
    \includegraphics[width=4.5cm]{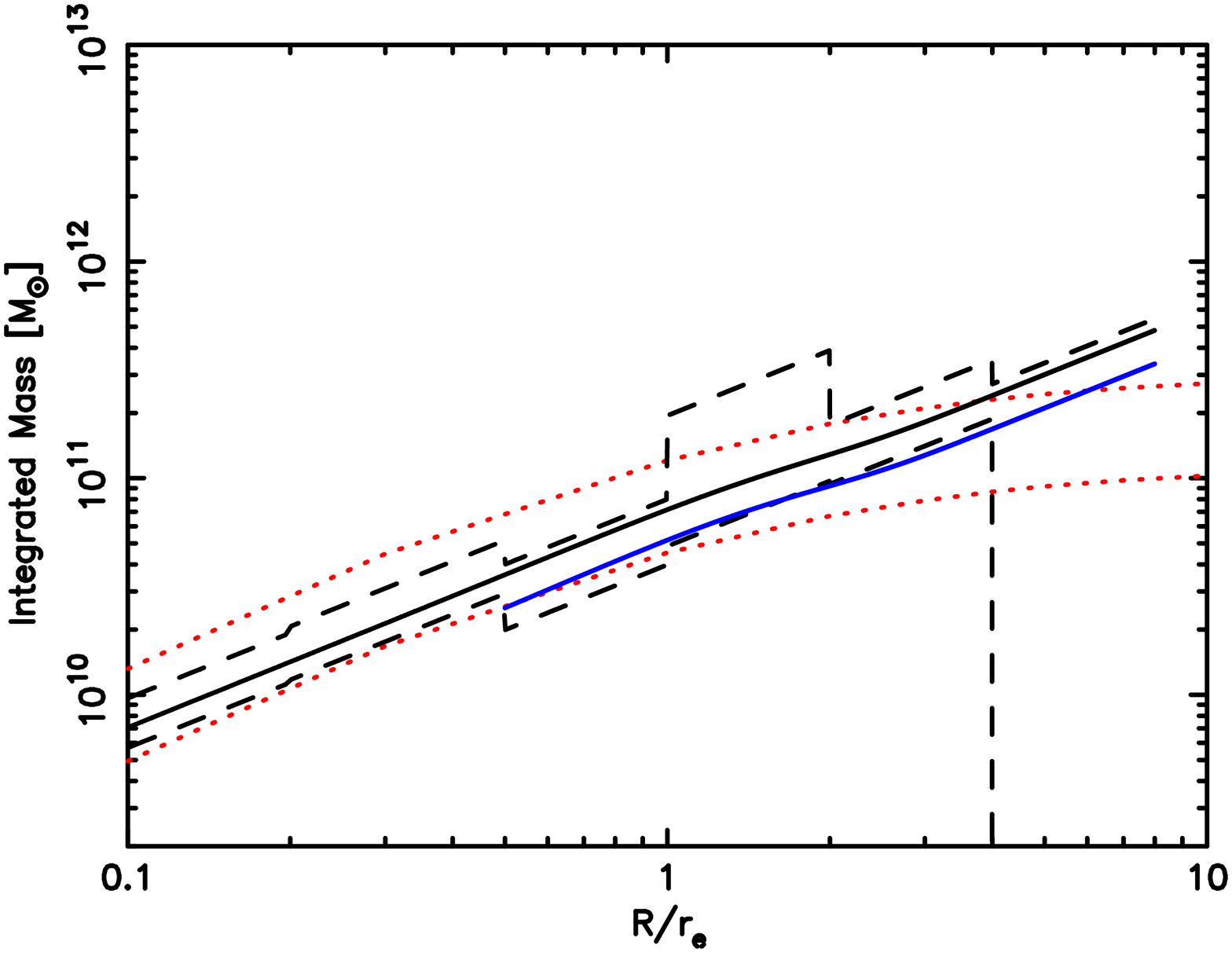}\\
    \includegraphics[width=4.5cm]{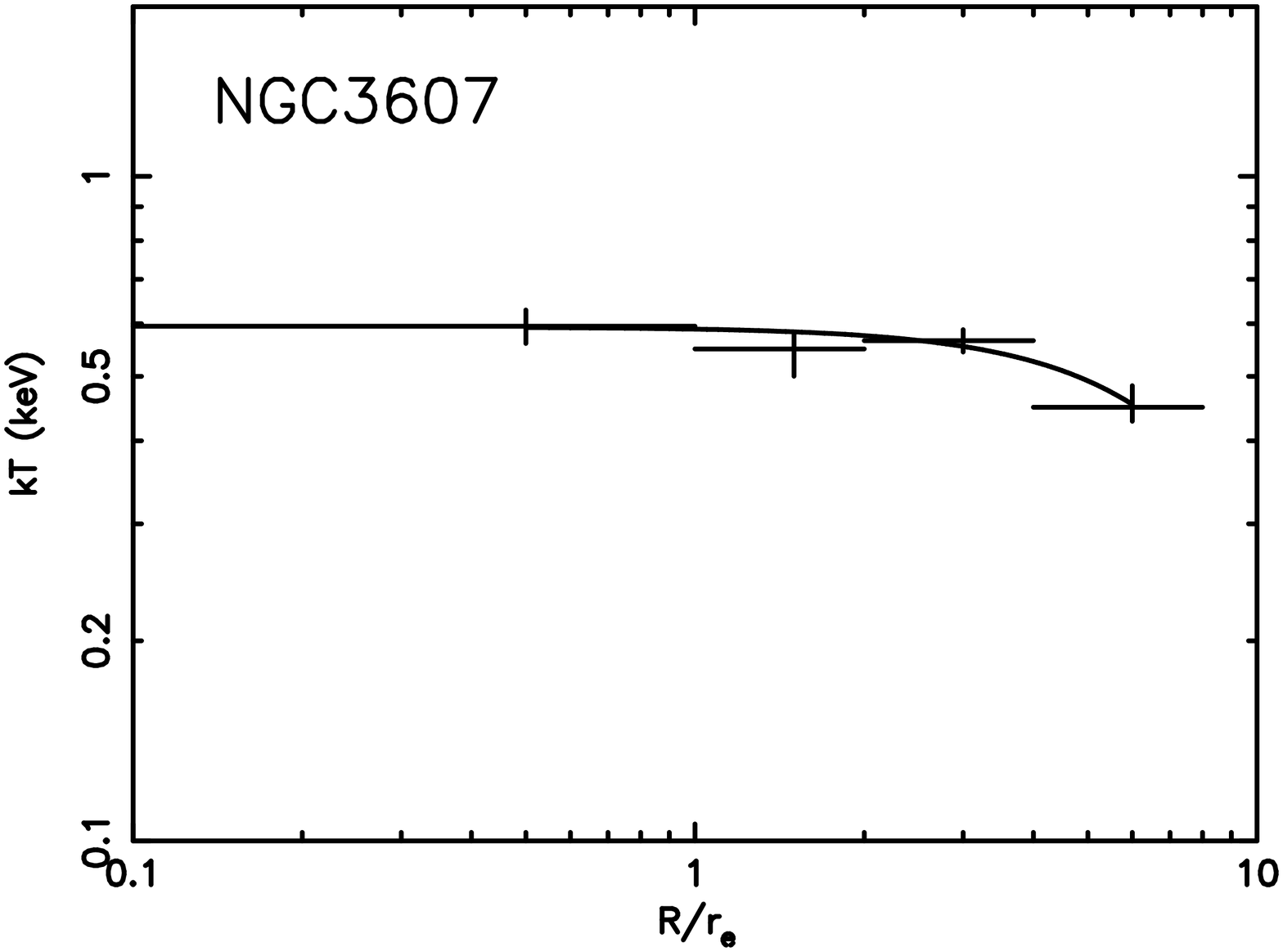}
    \includegraphics[width=4.5cm]{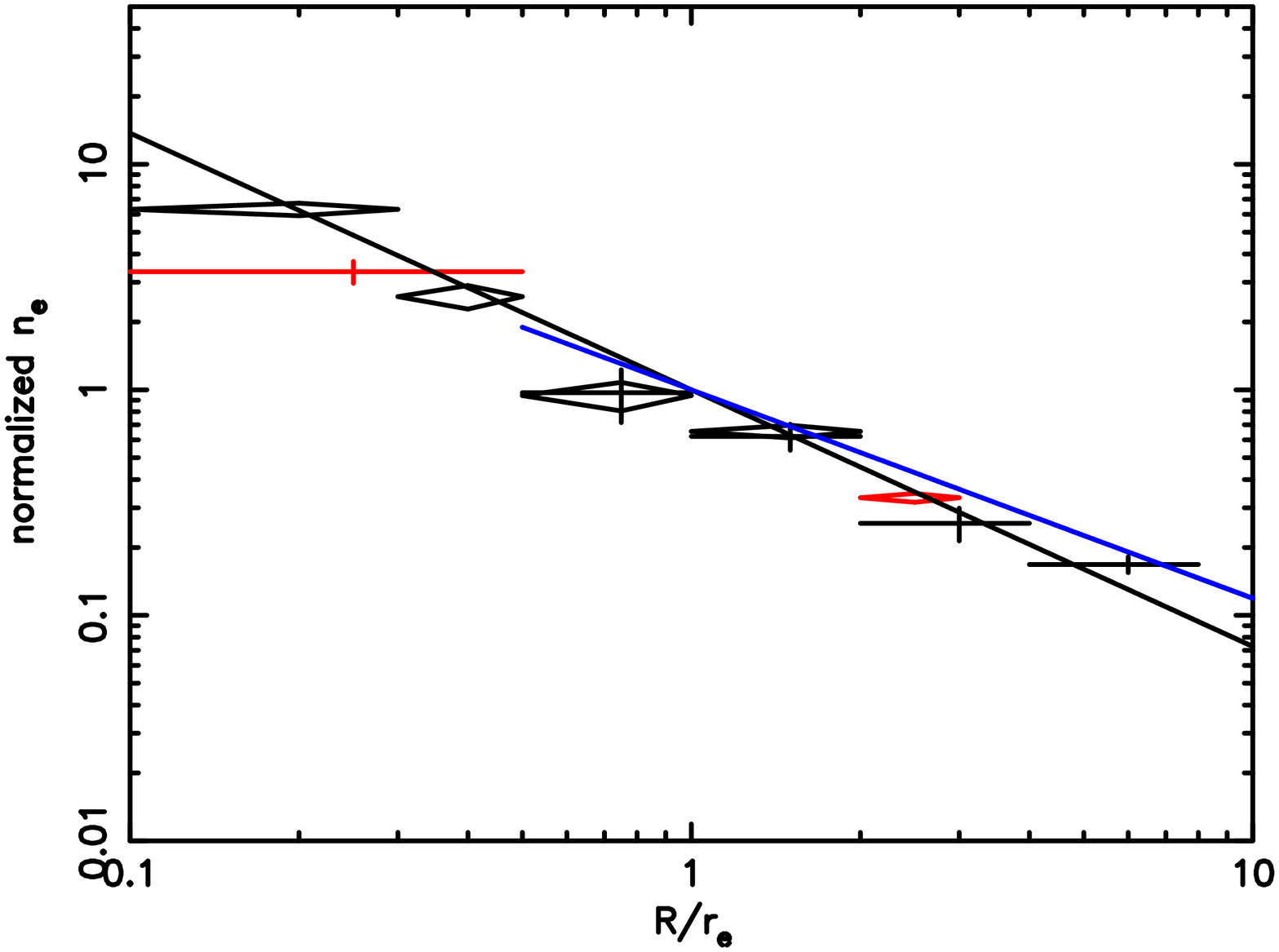}
    \includegraphics[width=4.5cm]{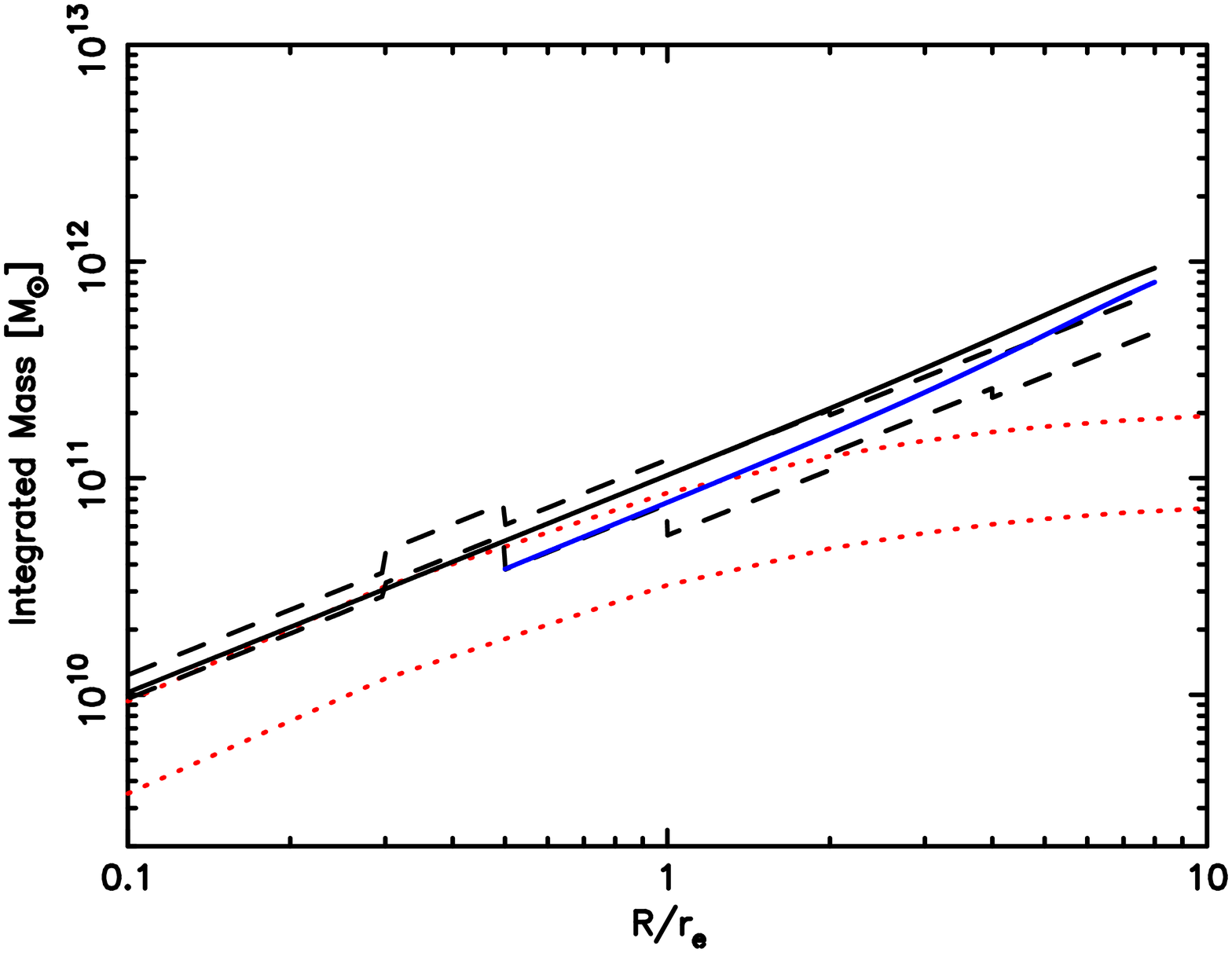}\\
    \includegraphics[width=4.5cm]{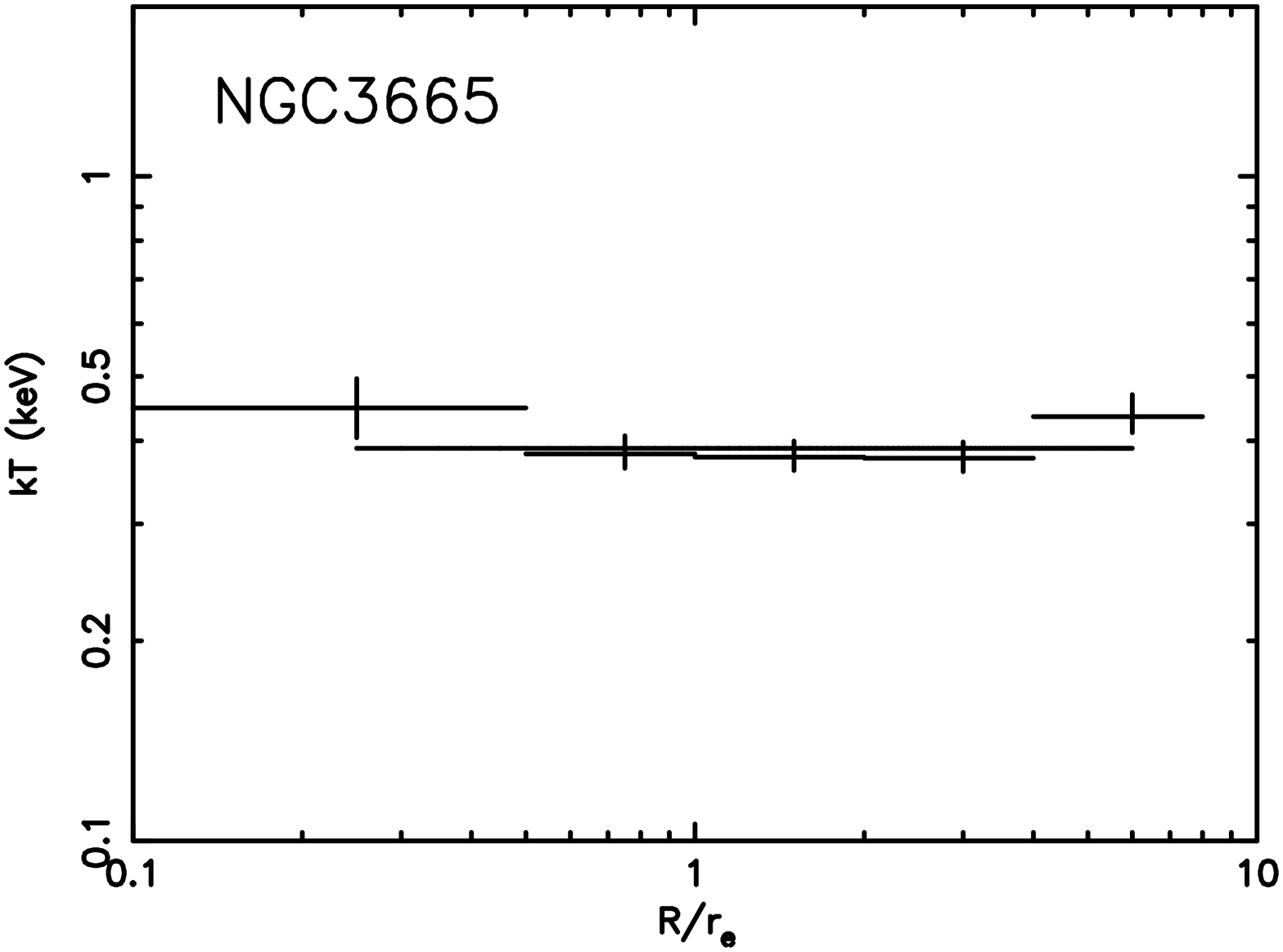}
    \includegraphics[width=4.5cm]{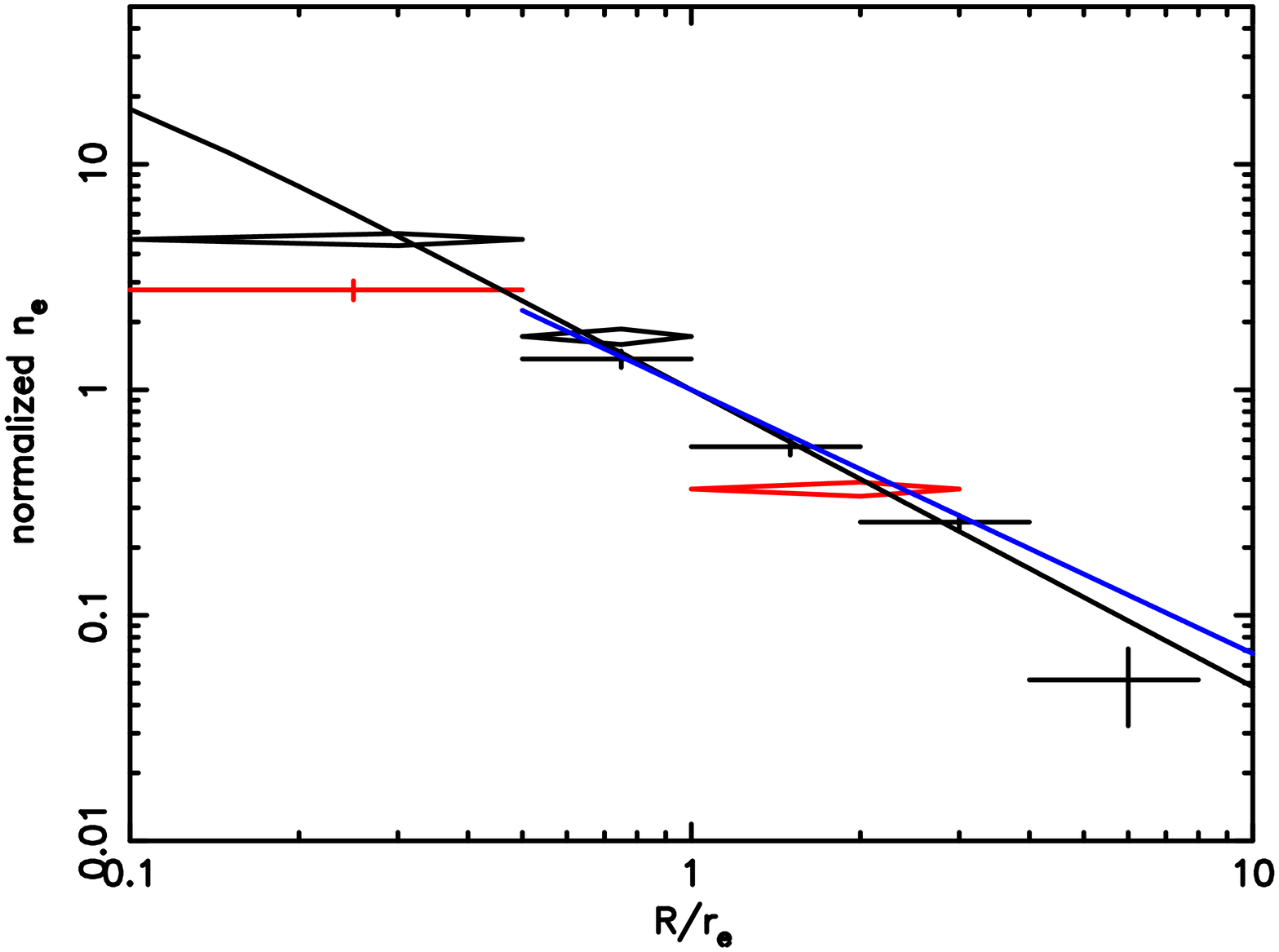}
    \includegraphics[width=4.5cm]{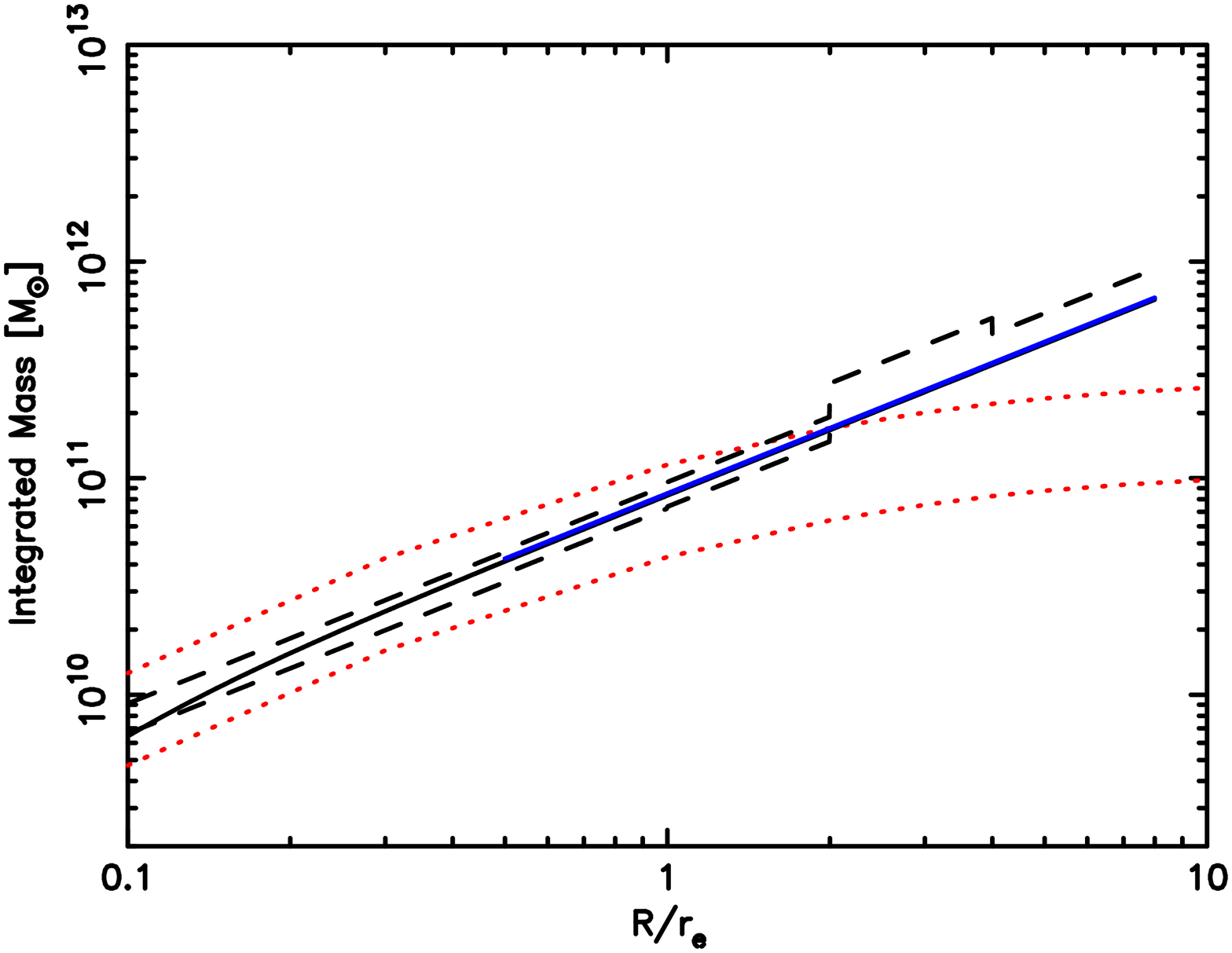}\\
    \includegraphics[width=4.5cm]{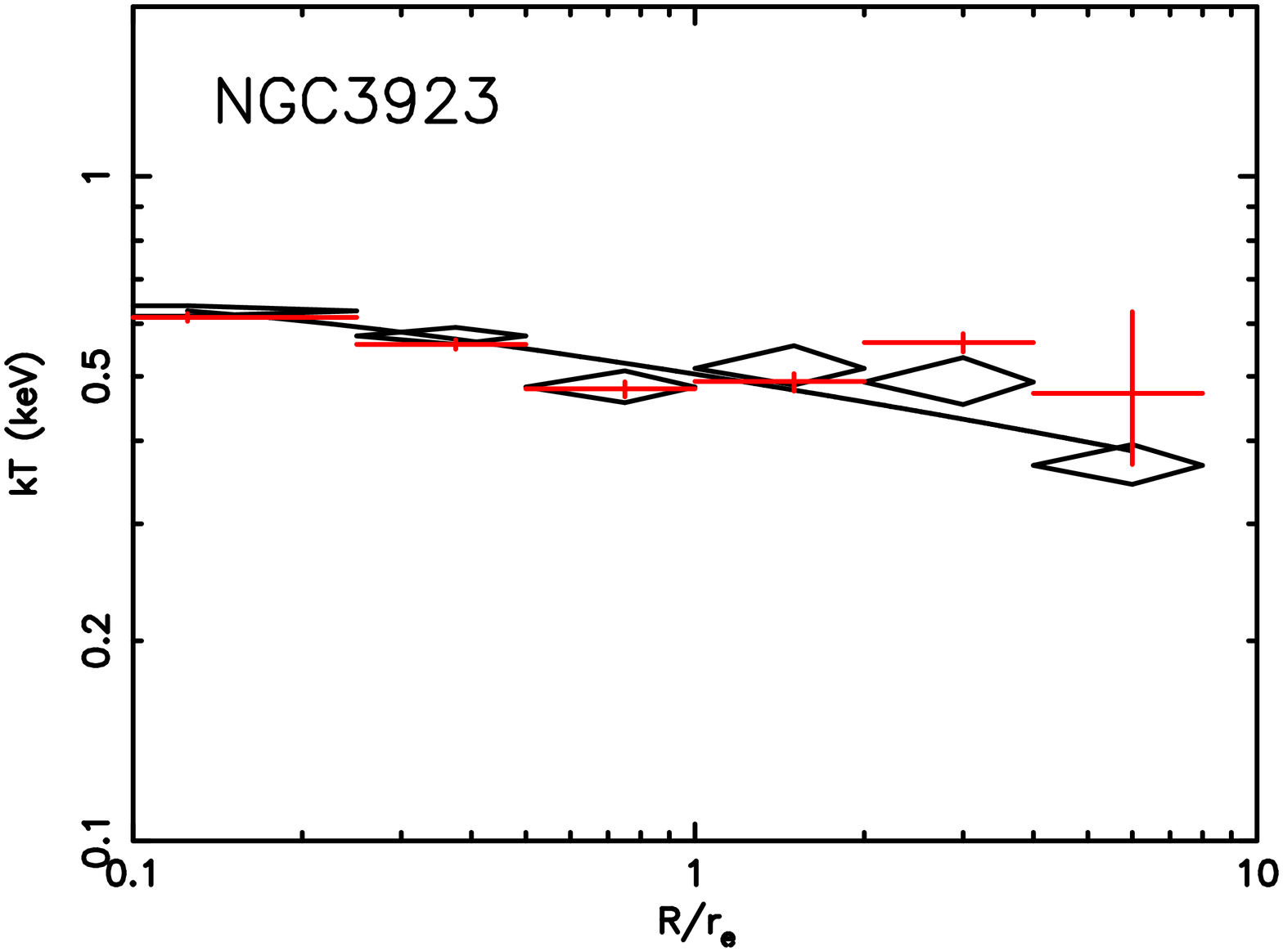}
    \includegraphics[width=4.5cm]{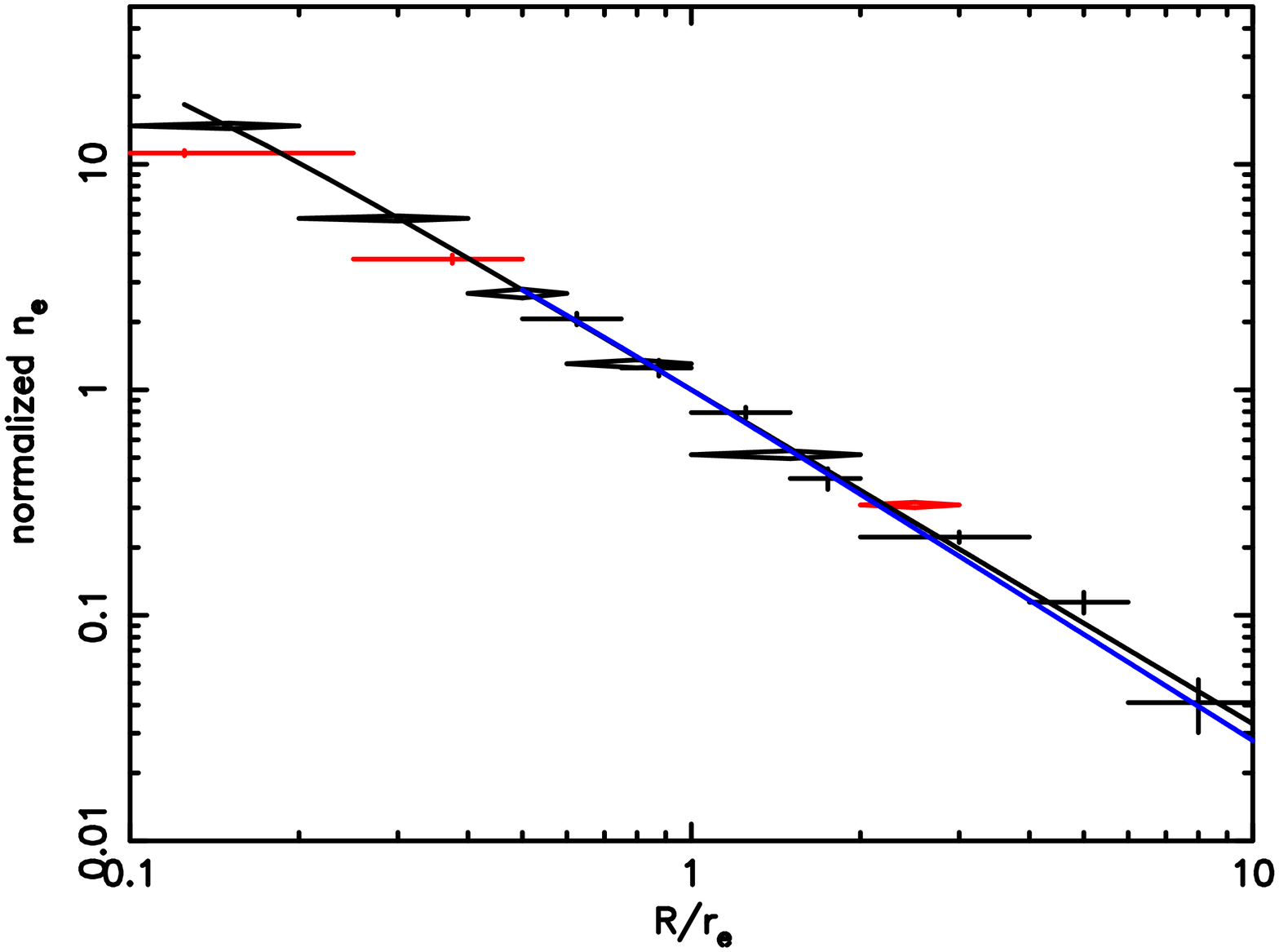}
    \includegraphics[width=4.5cm]{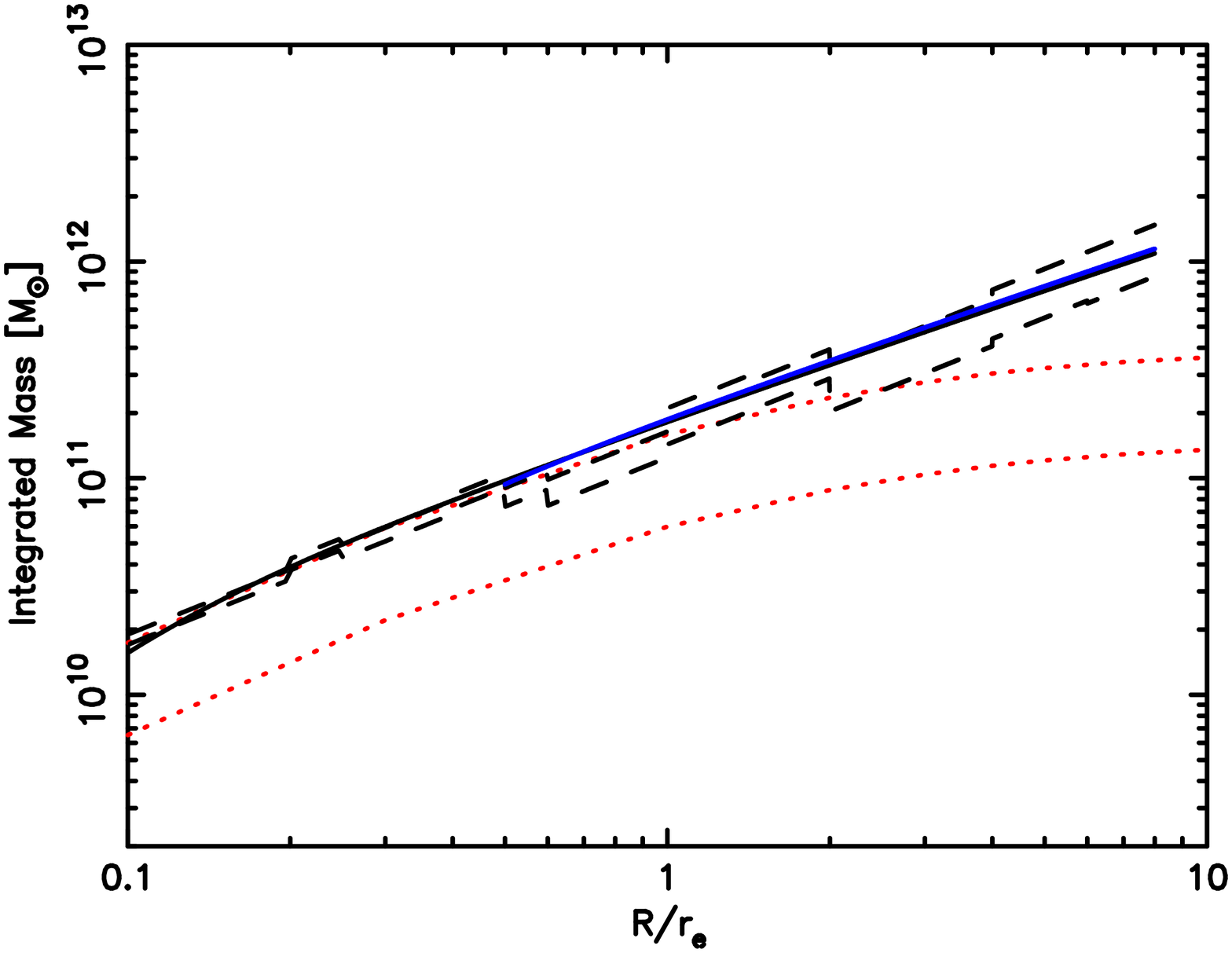}\\
 \flushleft
 {\bf Fig.~A.1.~~}(continued)
\end{figure*}

\begin{figure*}
  \centering
    \includegraphics[width=4.5cm]{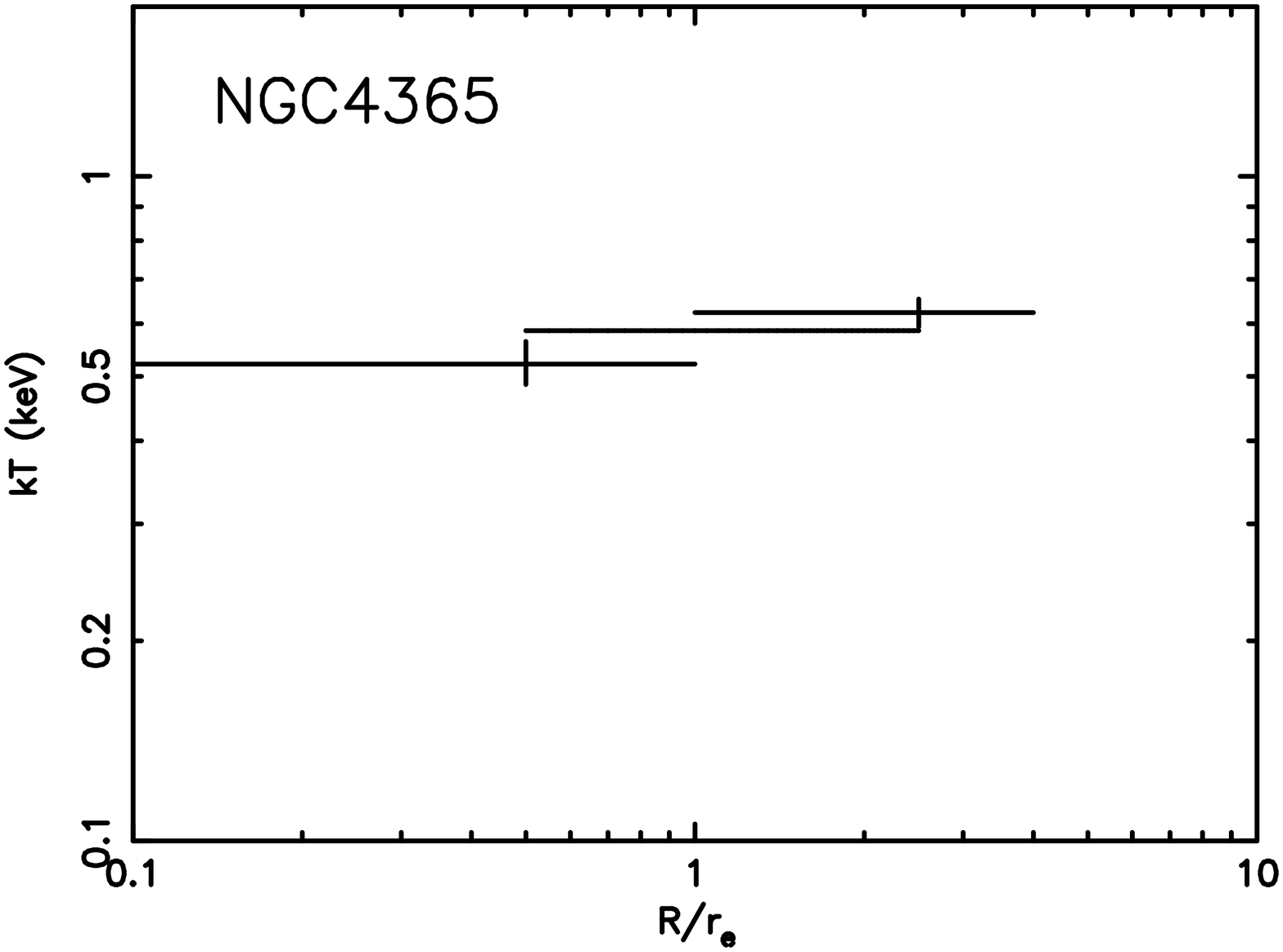}
    \includegraphics[width=4.5cm]{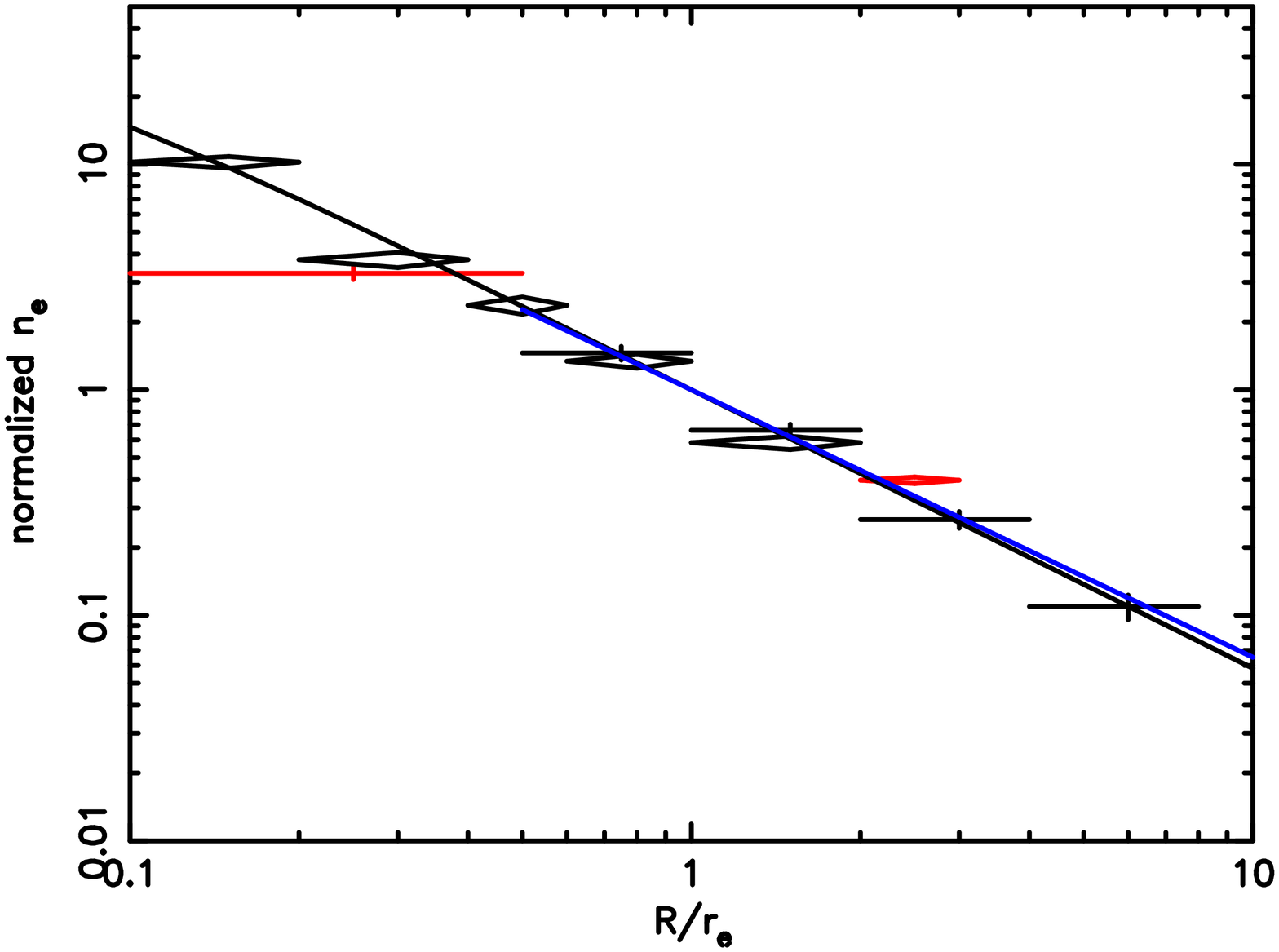}
    \includegraphics[width=4.5cm]{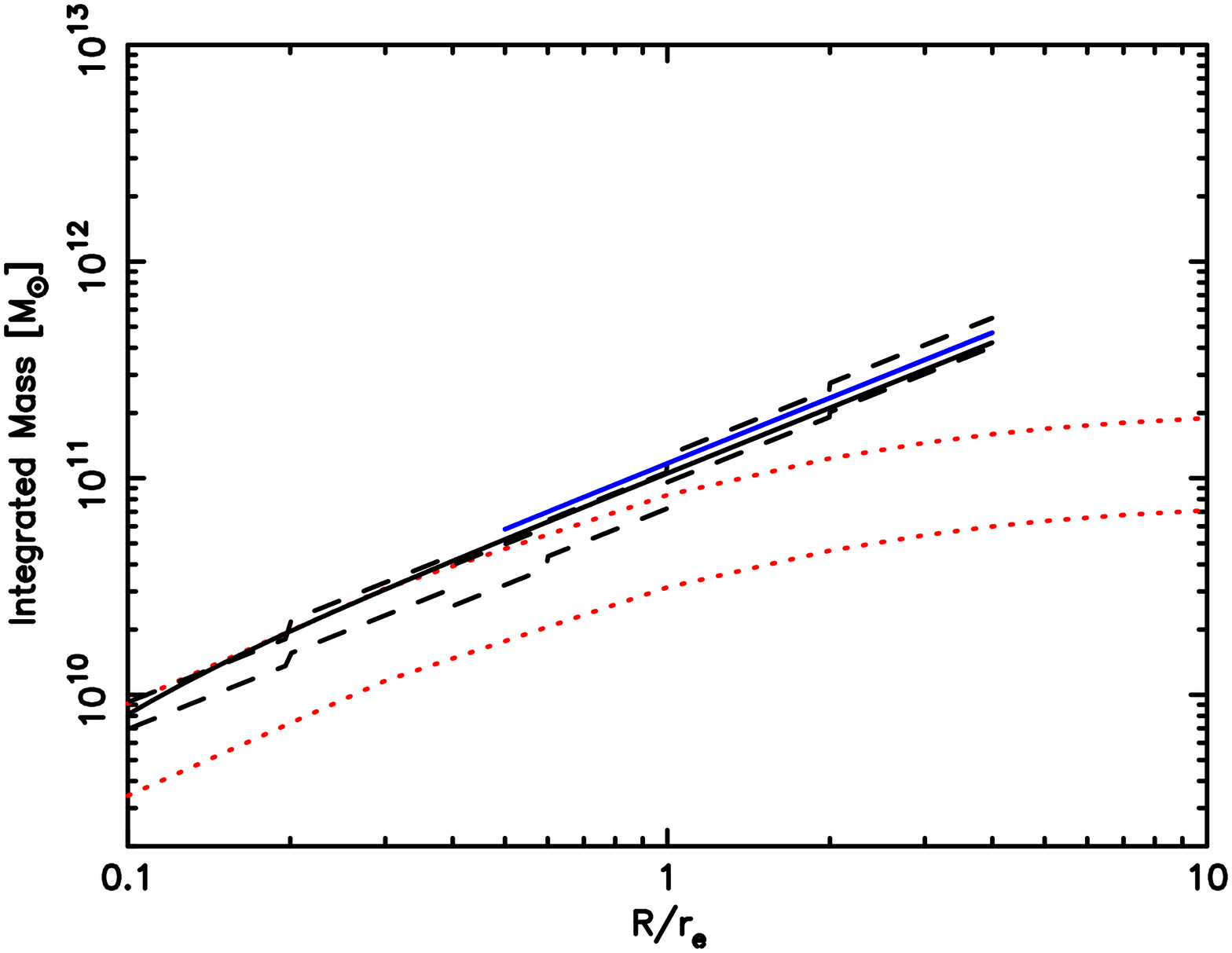}\\
    \includegraphics[width=4.5cm]{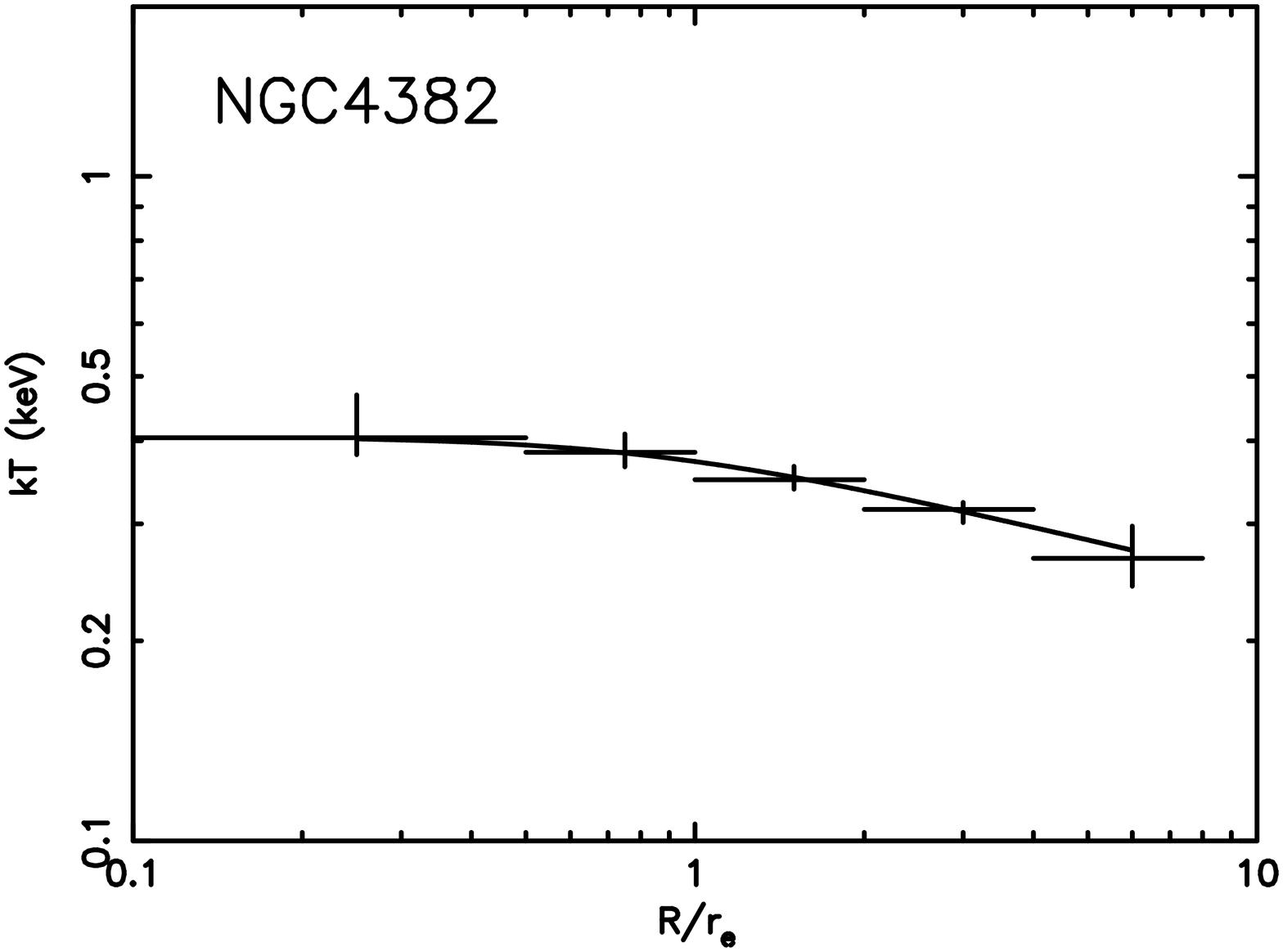}
    \includegraphics[width=4.5cm]{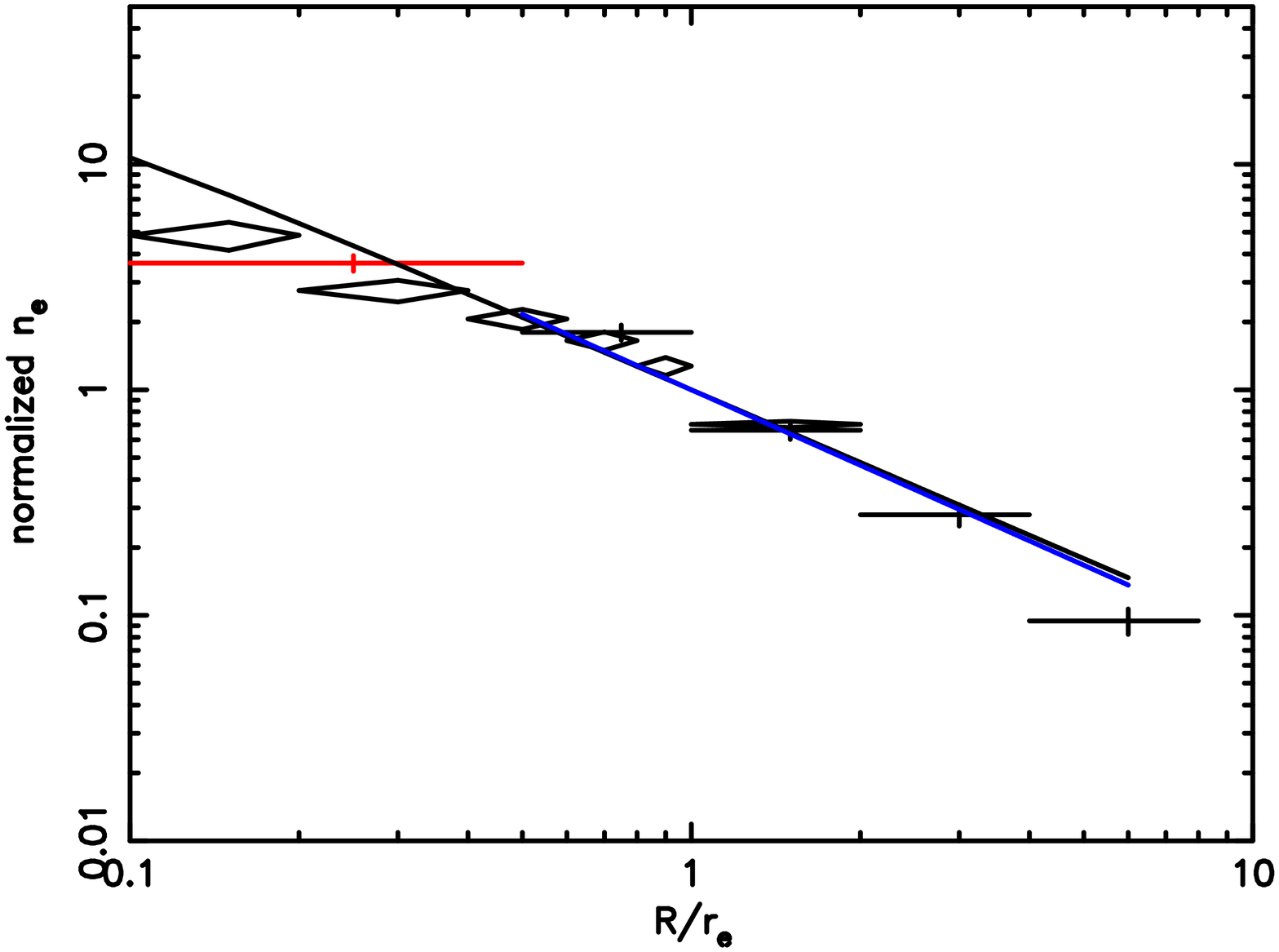}
    \includegraphics[width=4.5cm]{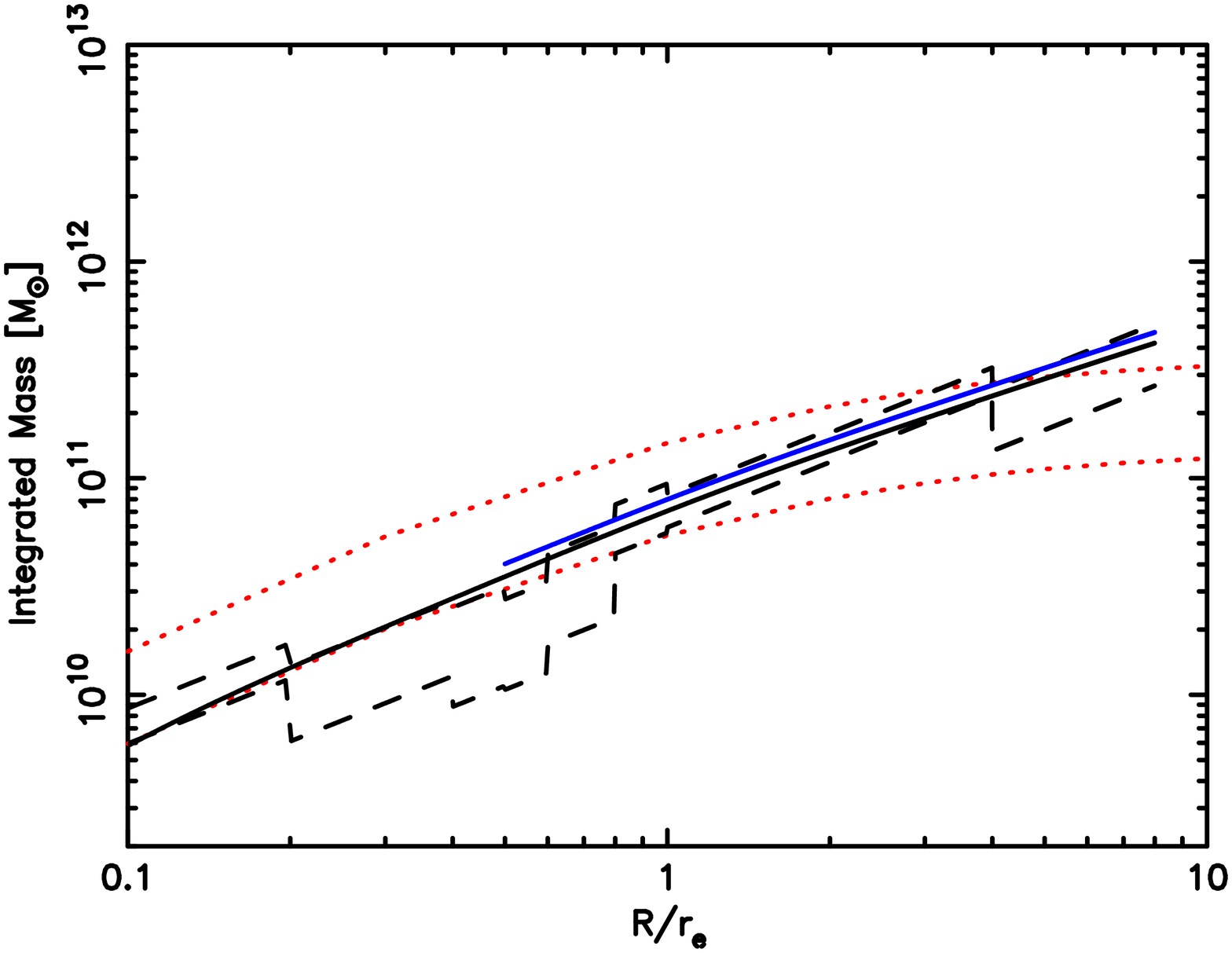}\\
    \includegraphics[width=4.5cm]{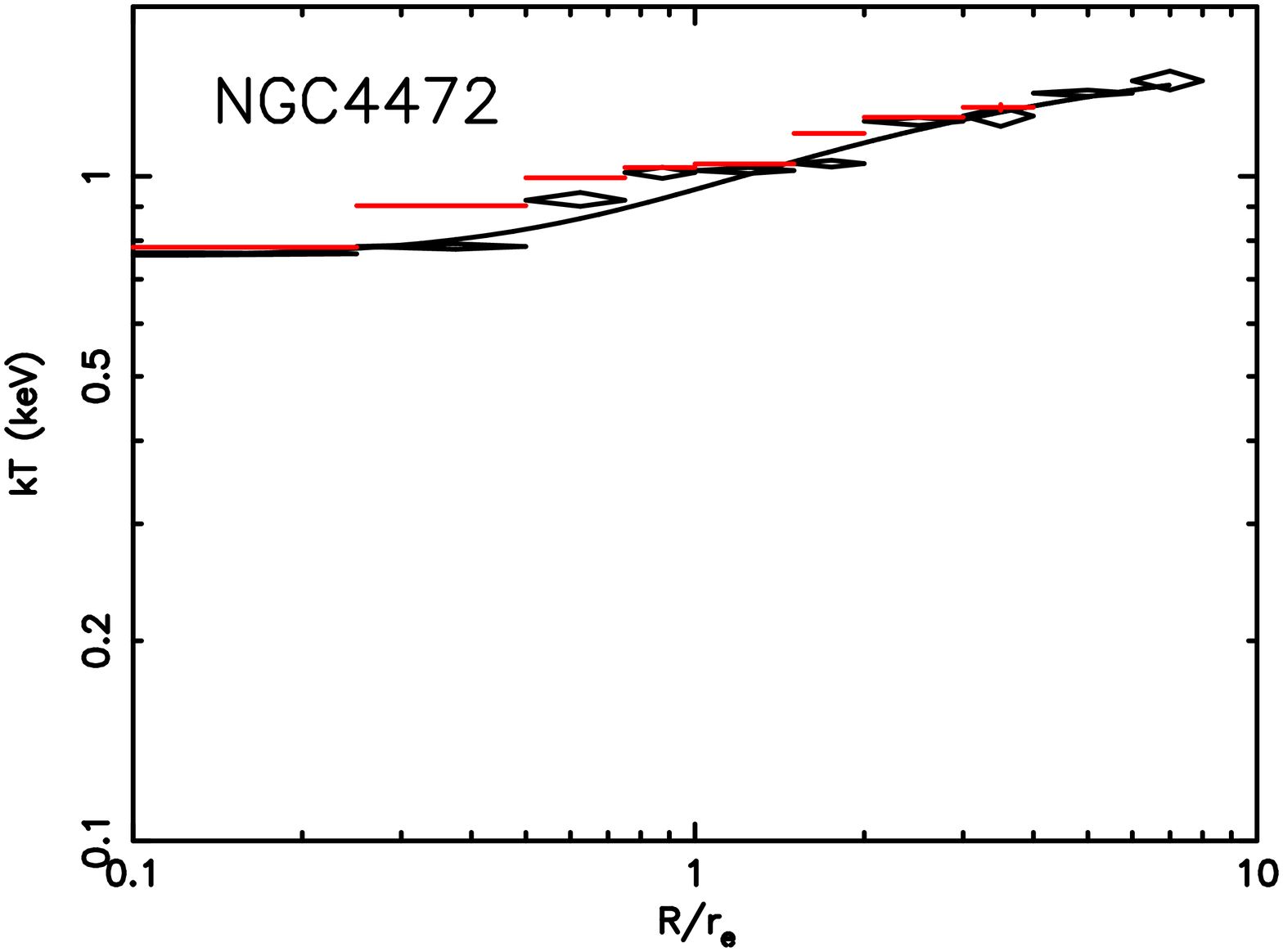}
    \includegraphics[width=4.5cm]{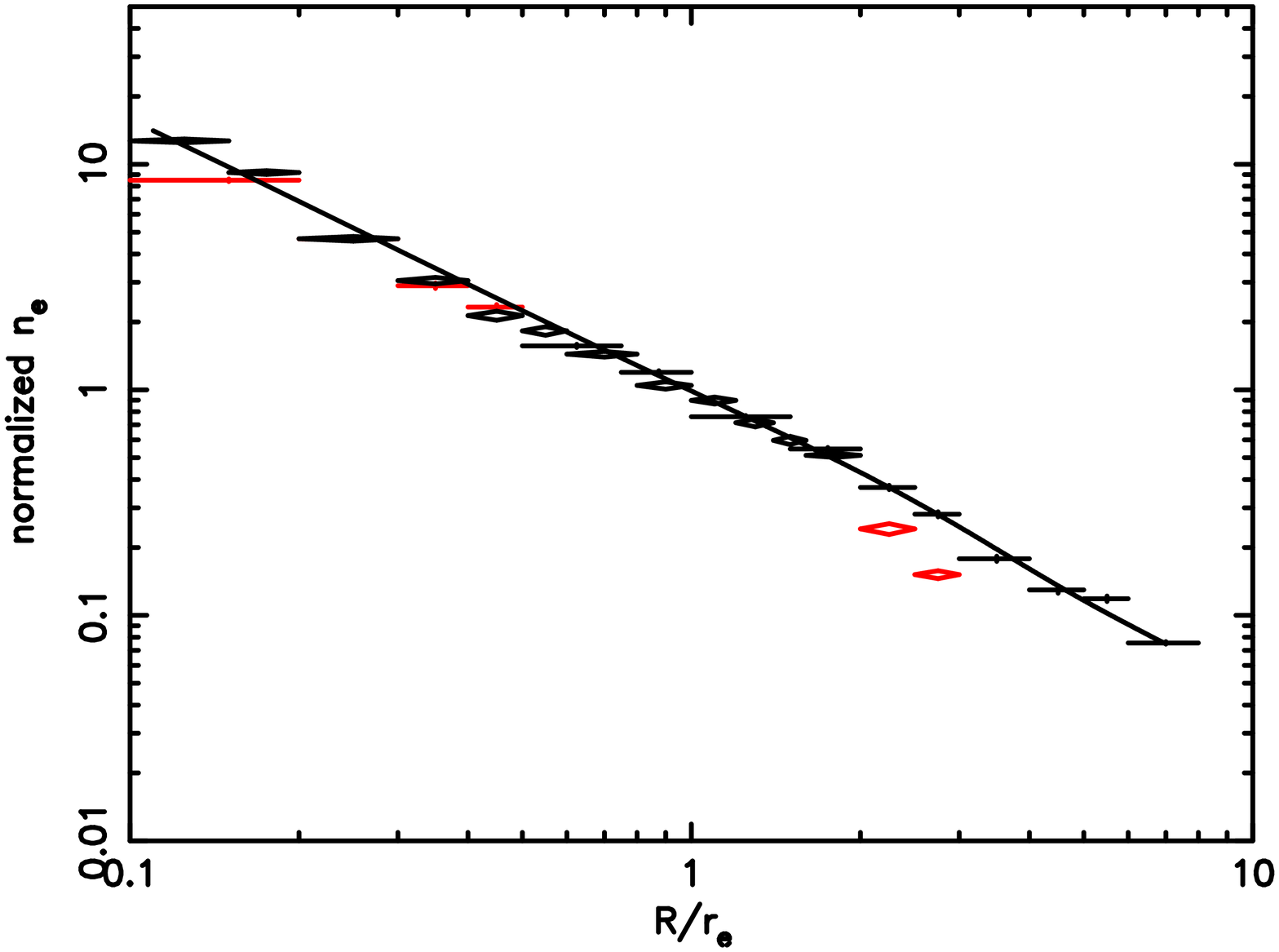}
    \includegraphics[width=4.5cm]{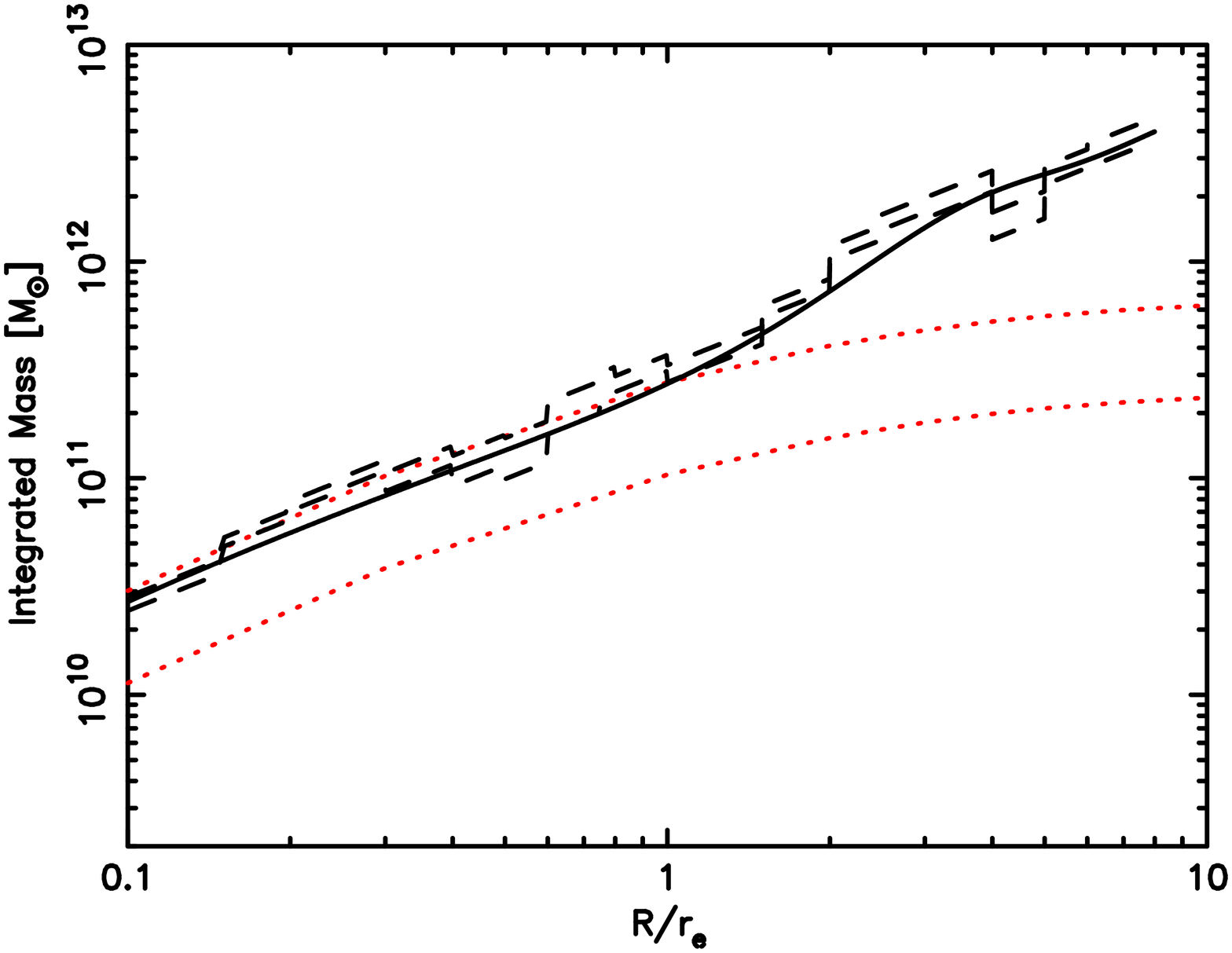}\\
    \includegraphics[width=4.5cm]{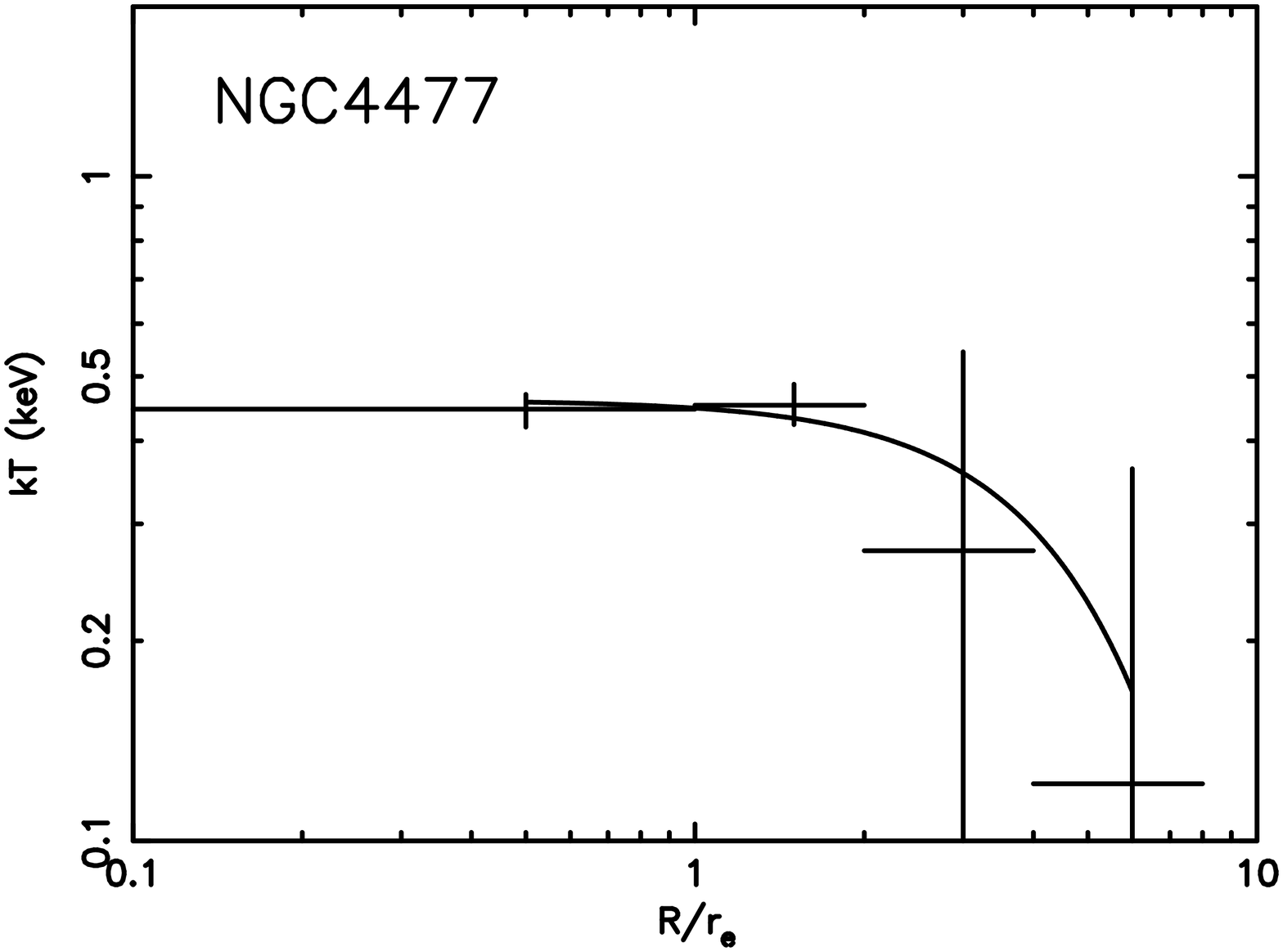}
    \includegraphics[width=4.5cm]{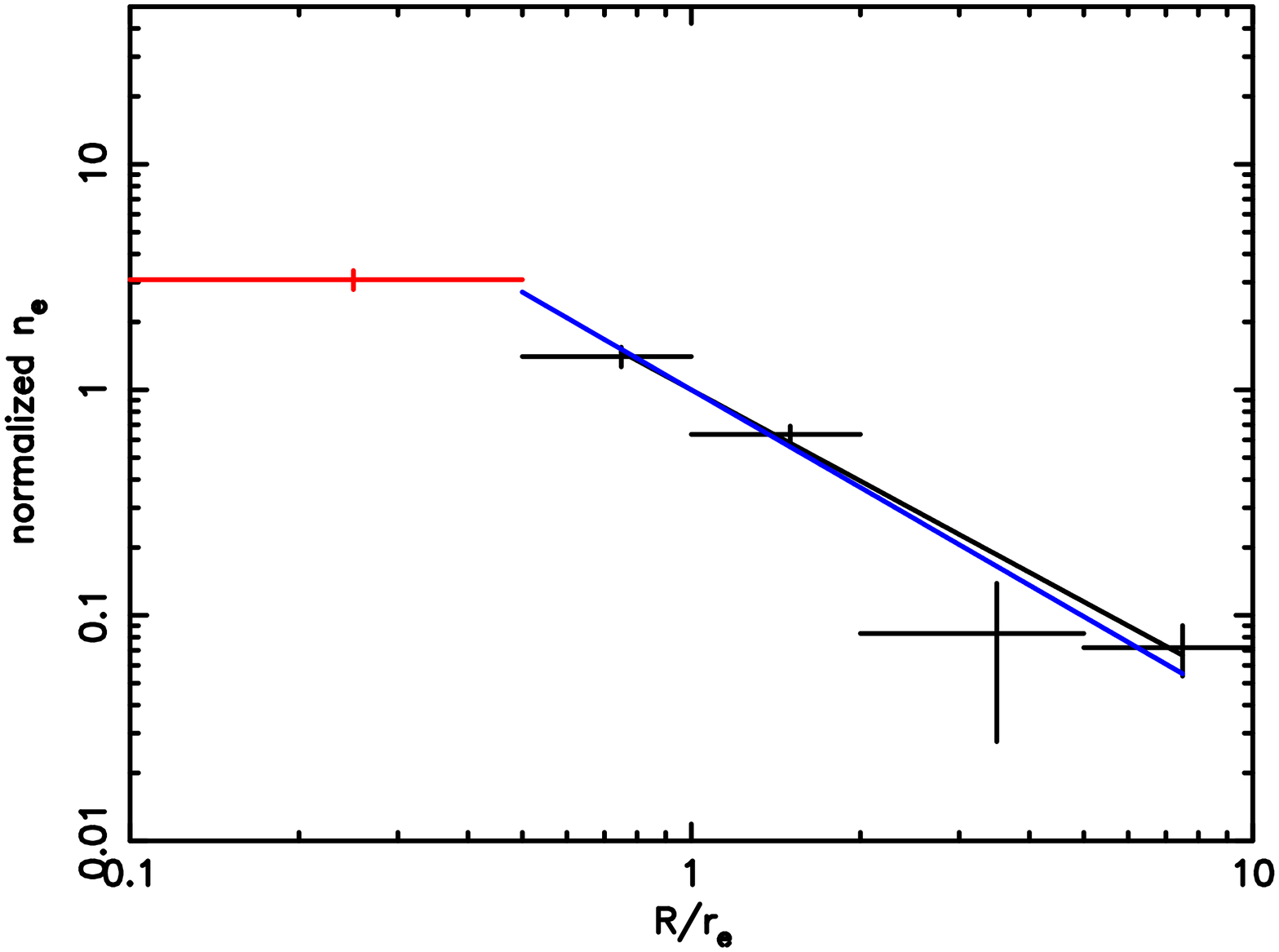}
    \includegraphics[width=4.5cm]{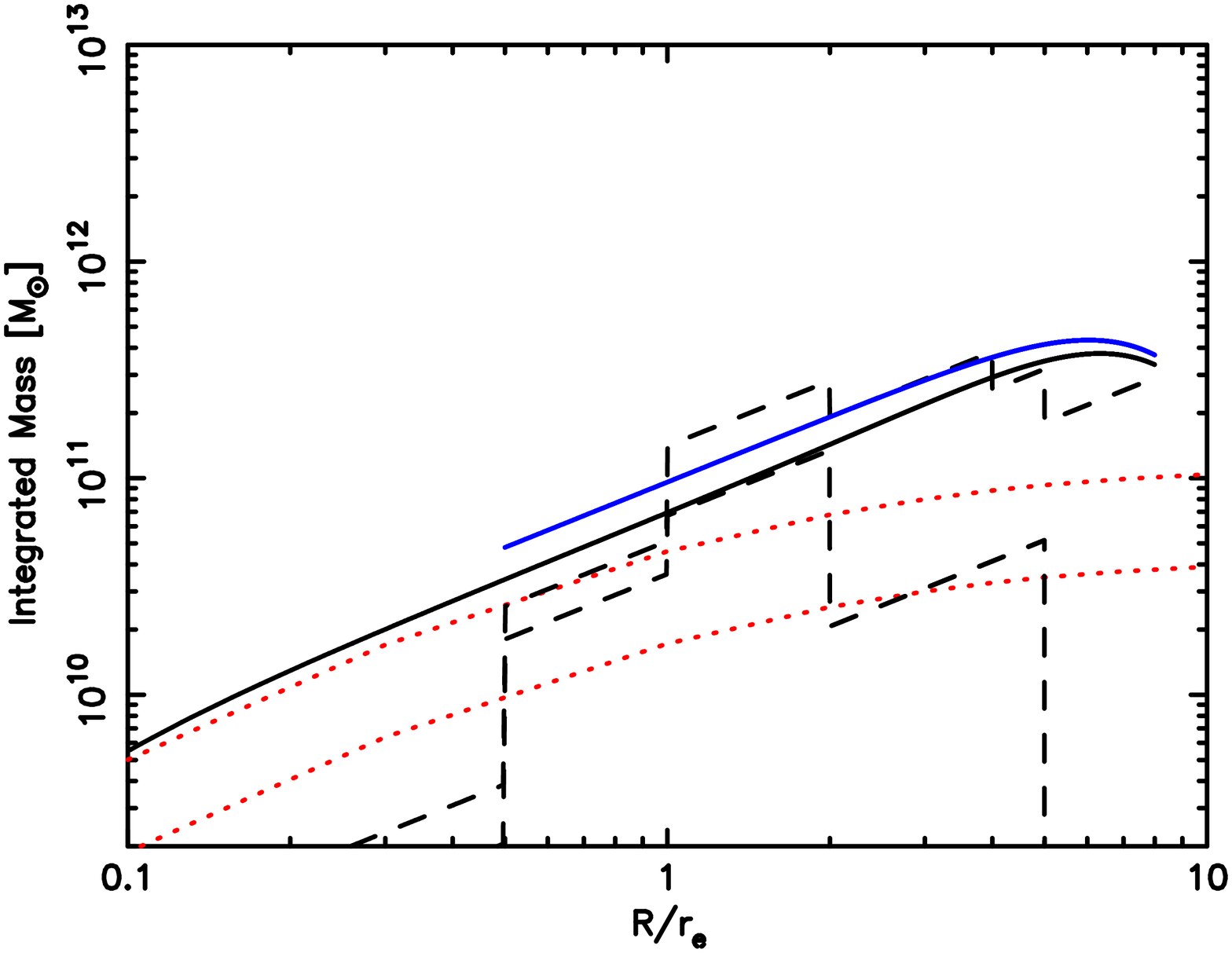}\\
    \includegraphics[width=4.5cm]{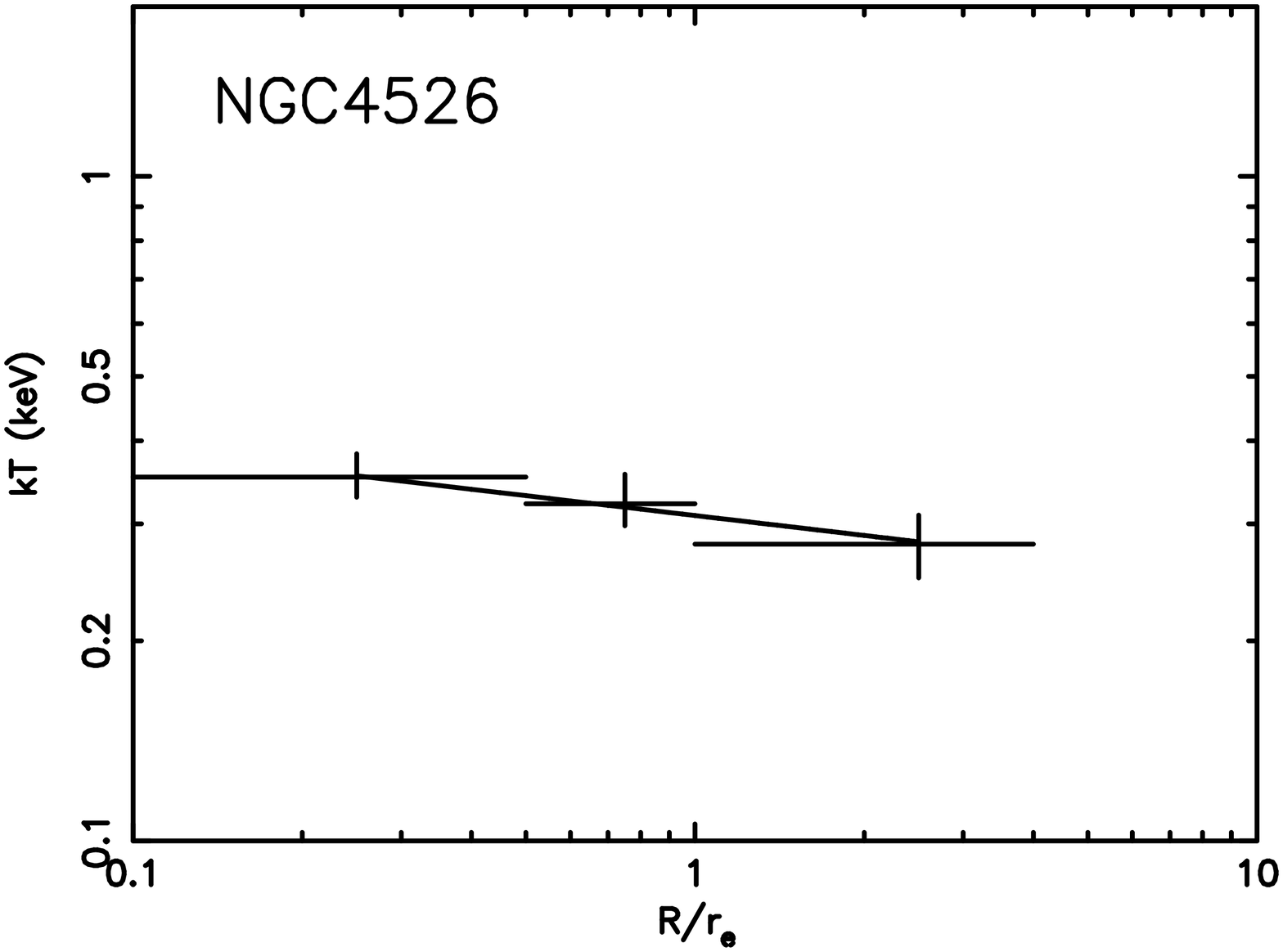}
    \includegraphics[width=4.5cm]{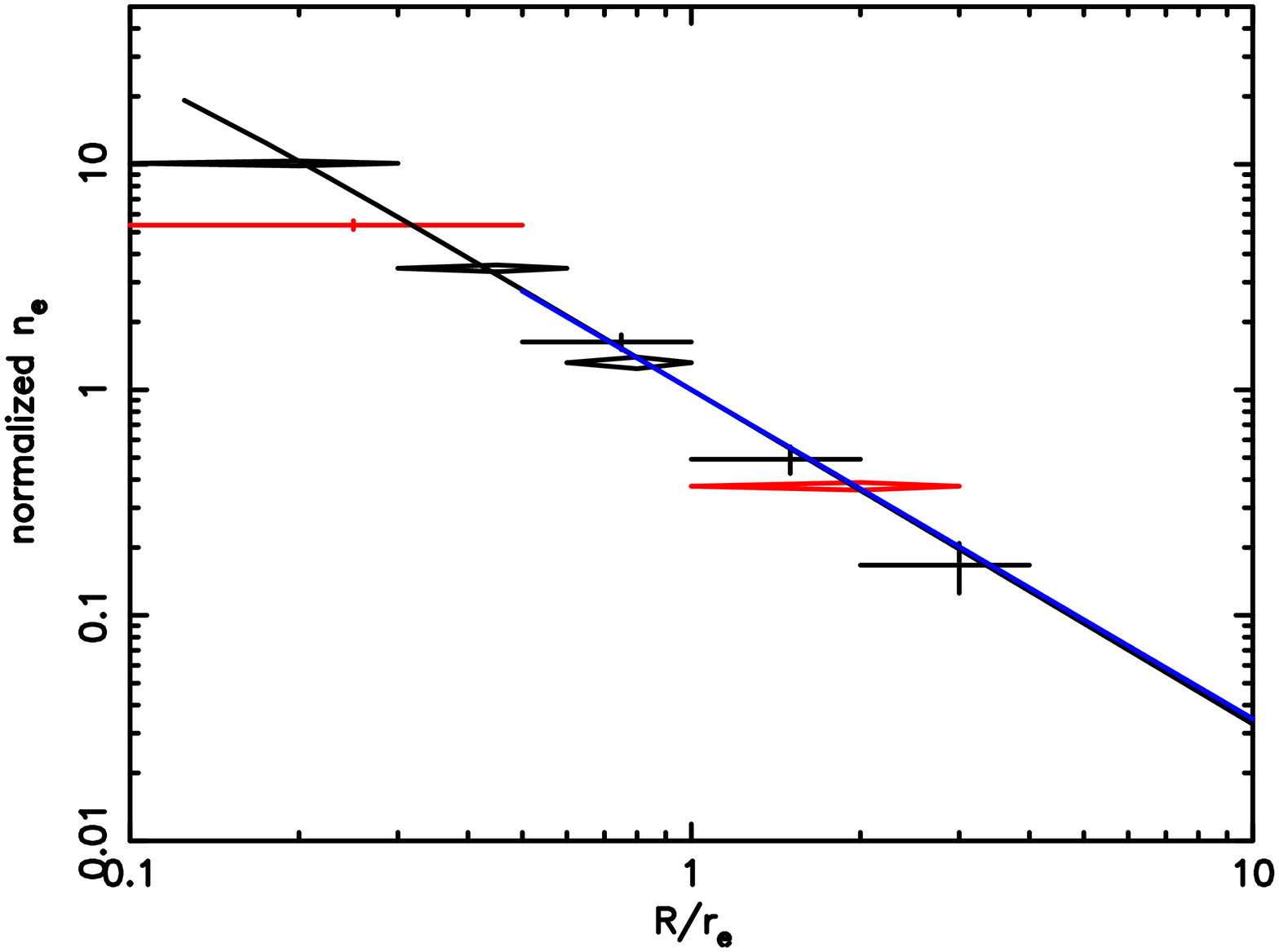}
    \includegraphics[width=4.5cm]{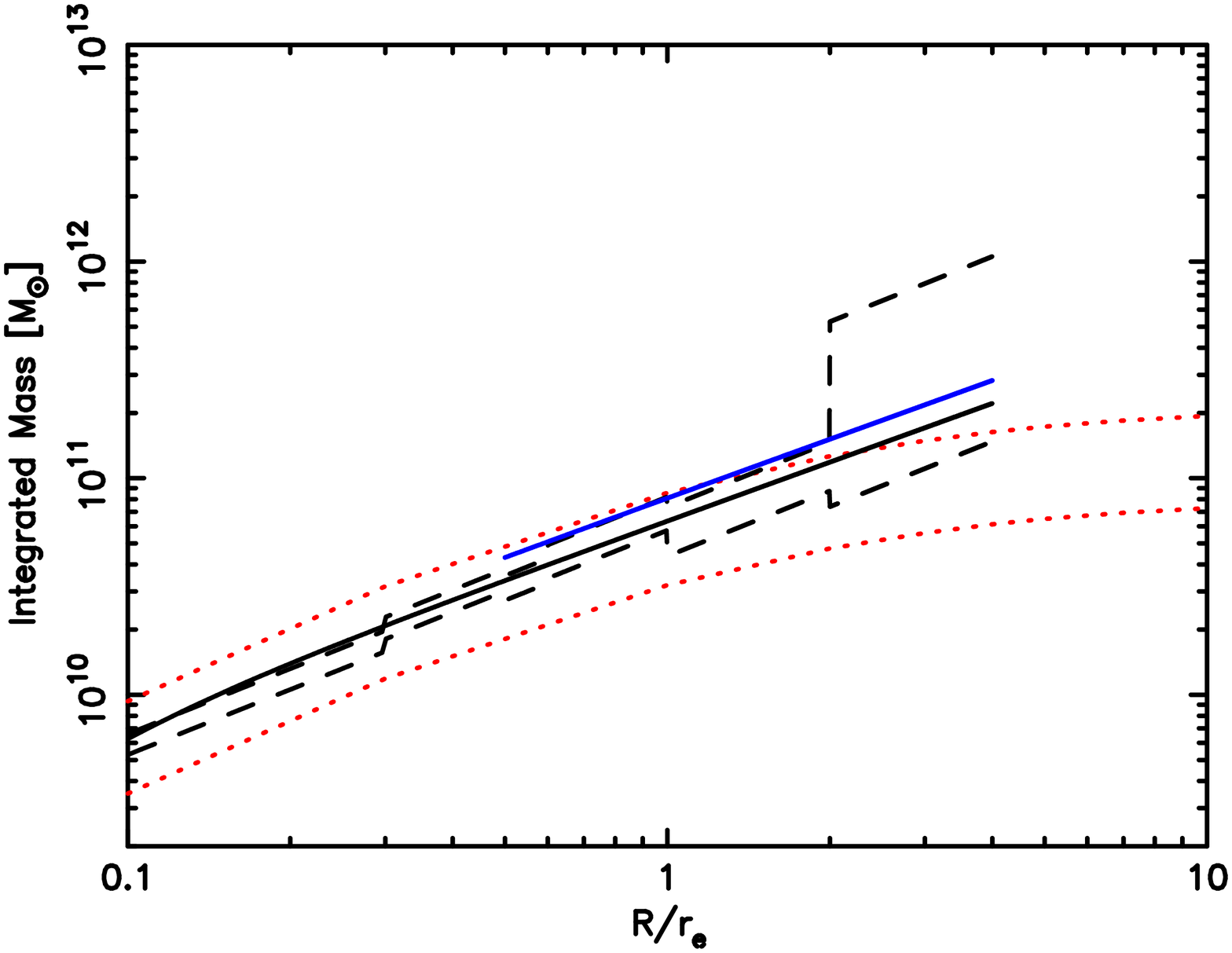}\\
    \includegraphics[width=4.5cm]{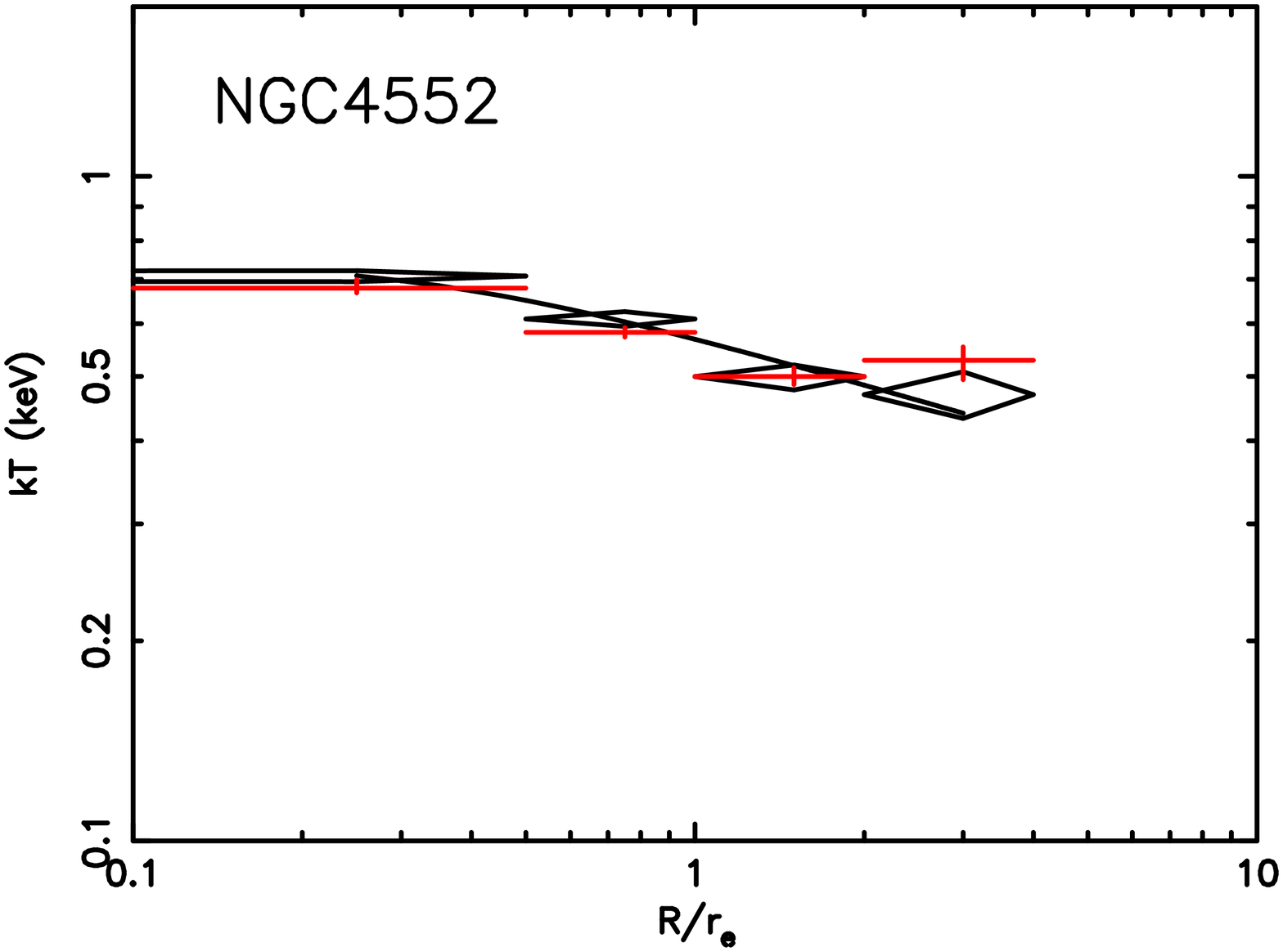}
    \includegraphics[width=4.5cm]{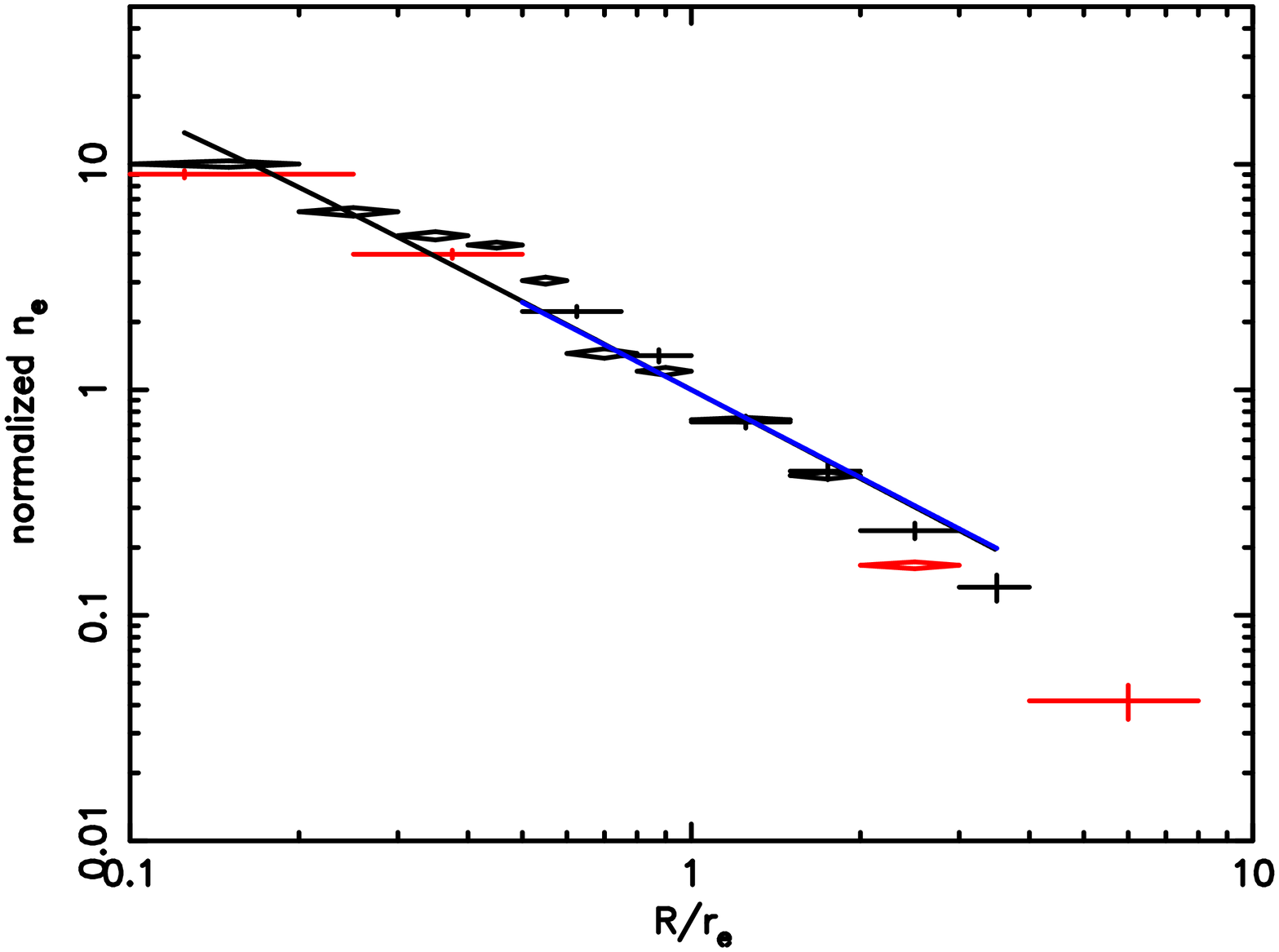}
    \includegraphics[width=4.5cm]{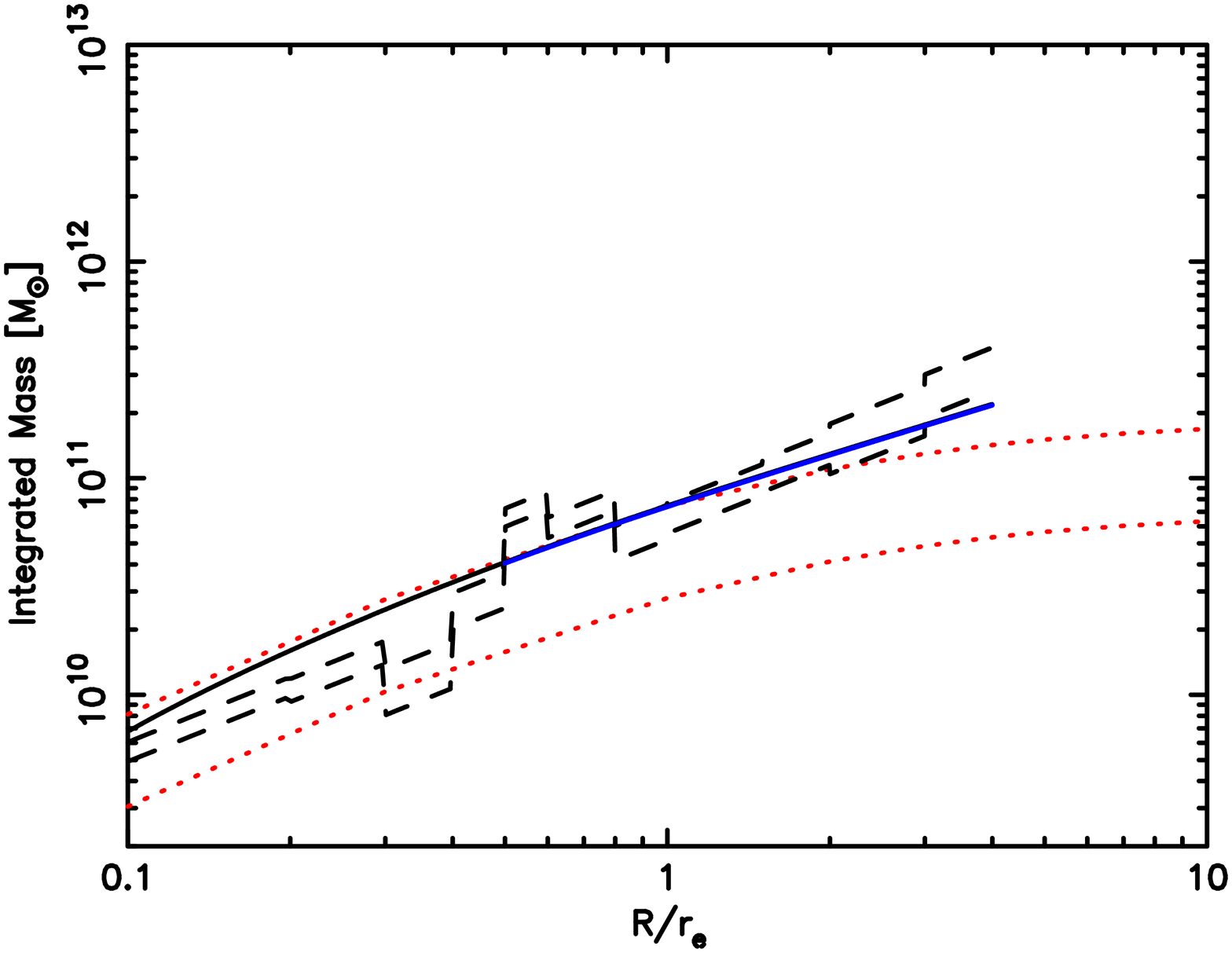}\\
 \flushleft
 {\bf Fig.~A.1.~~}(continued)
\end{figure*}

\begin{figure*}
  \centering
    \includegraphics[width=4.5cm]{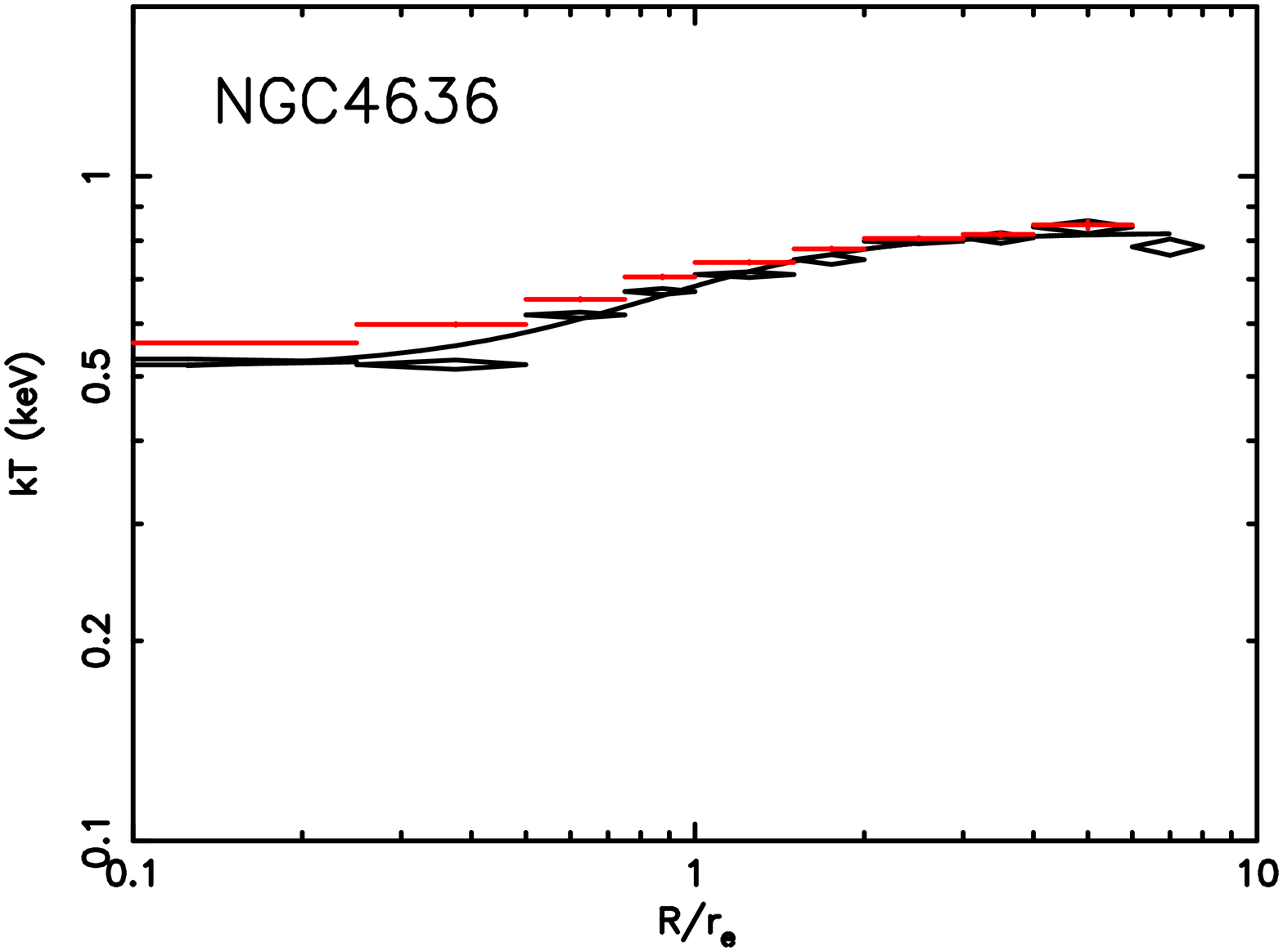}
    \includegraphics[width=4.5cm]{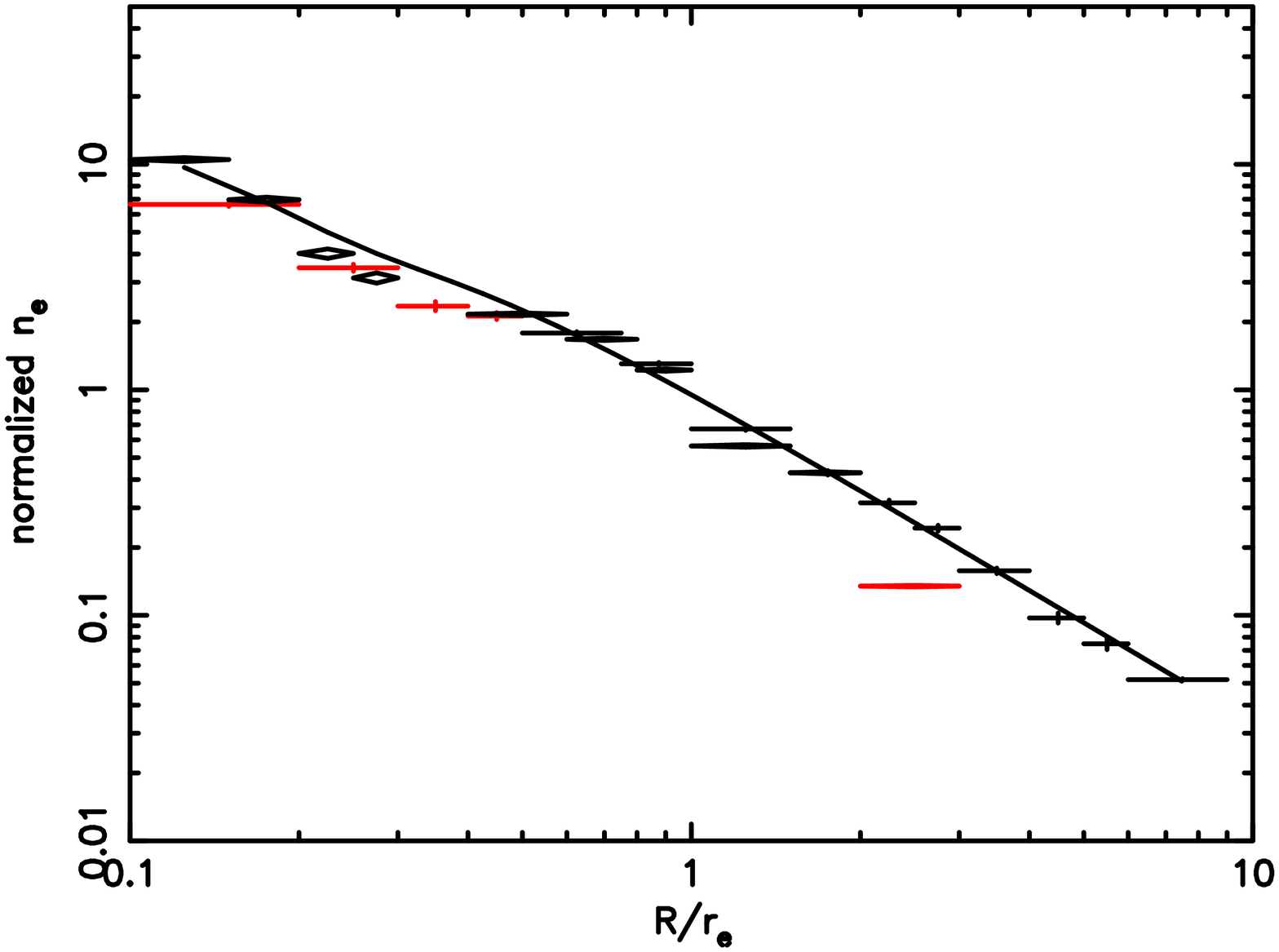}
    \includegraphics[width=4.5cm]{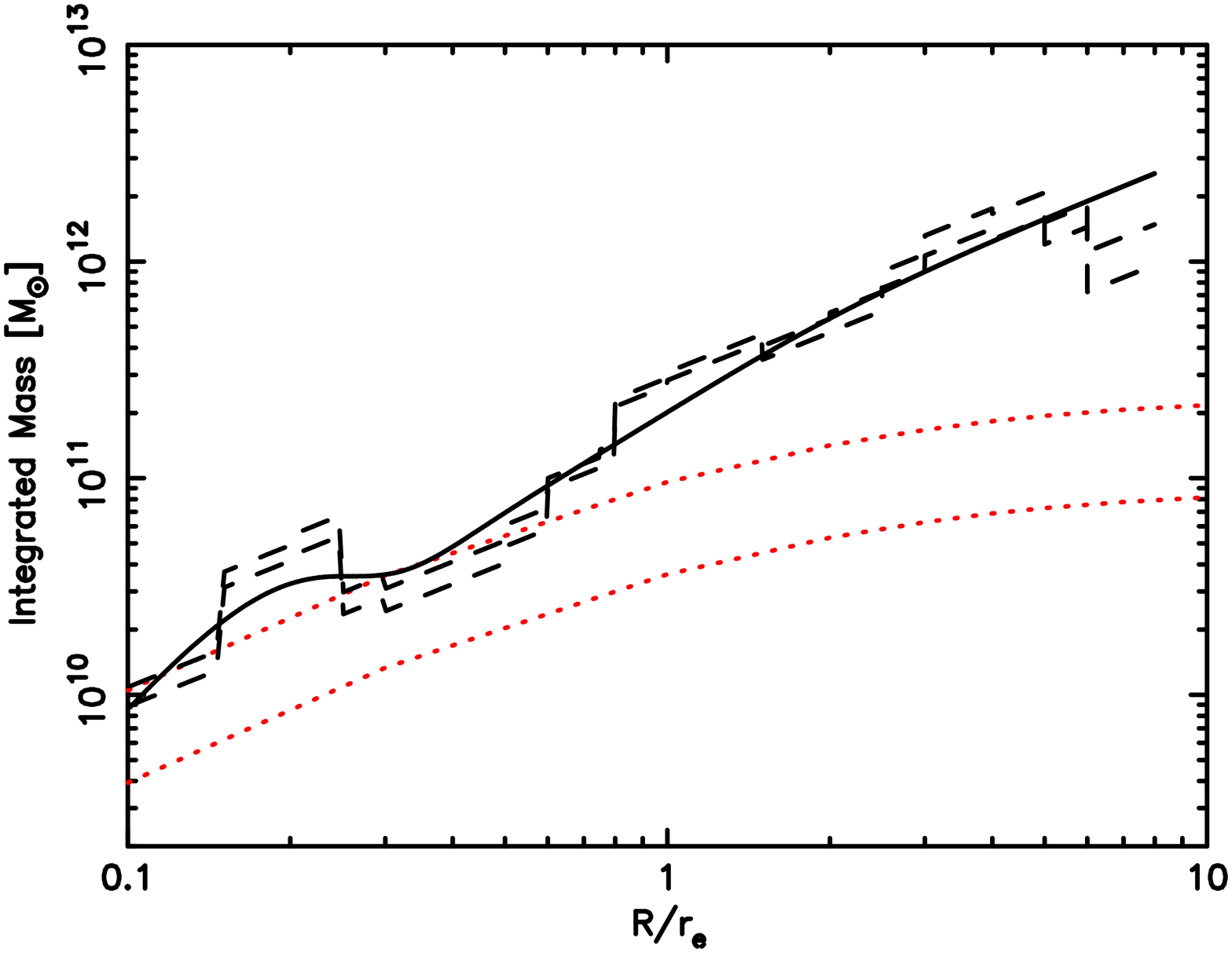}\\
    \includegraphics[width=4.5cm]{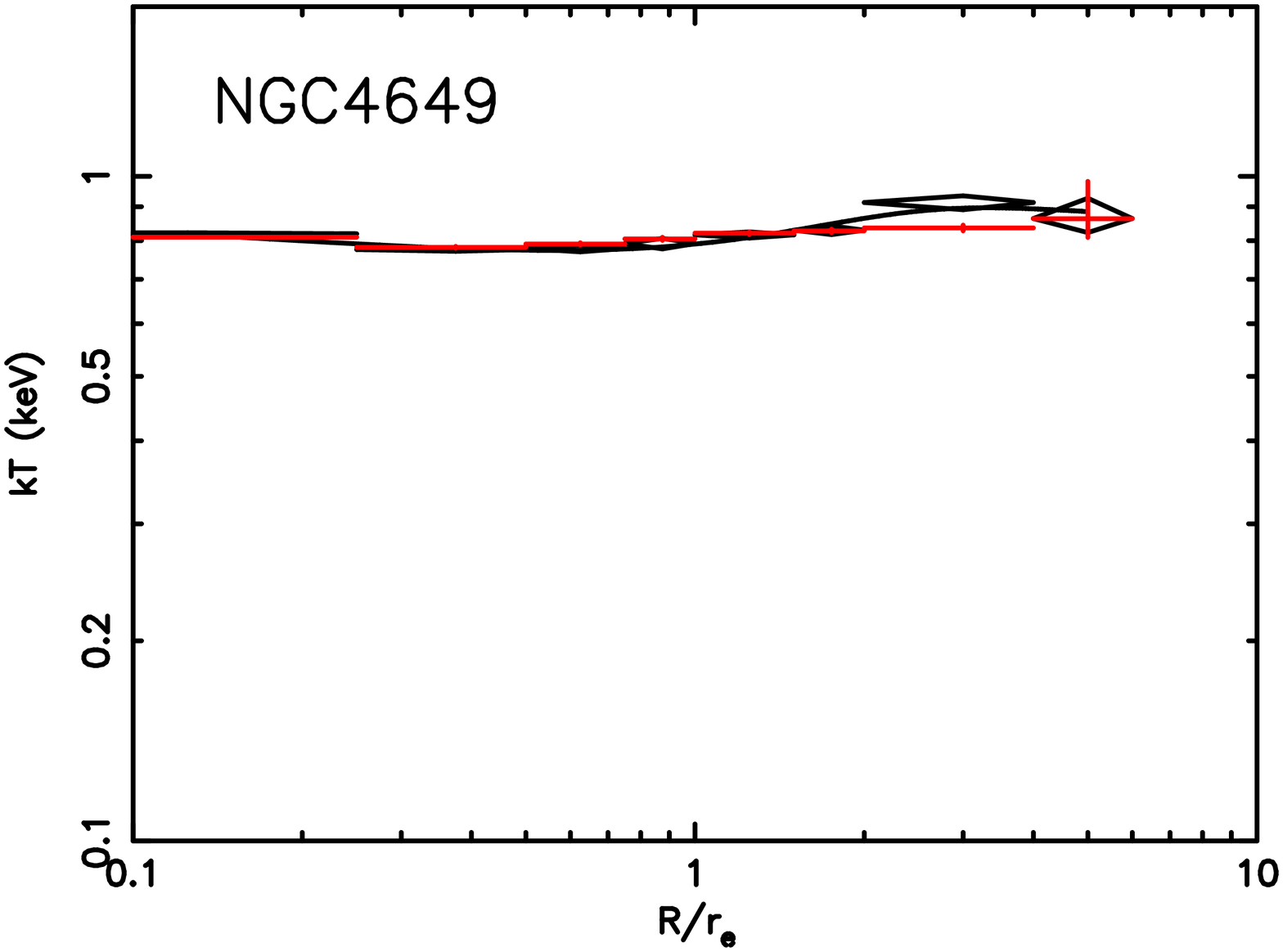}
    \includegraphics[width=4.5cm]{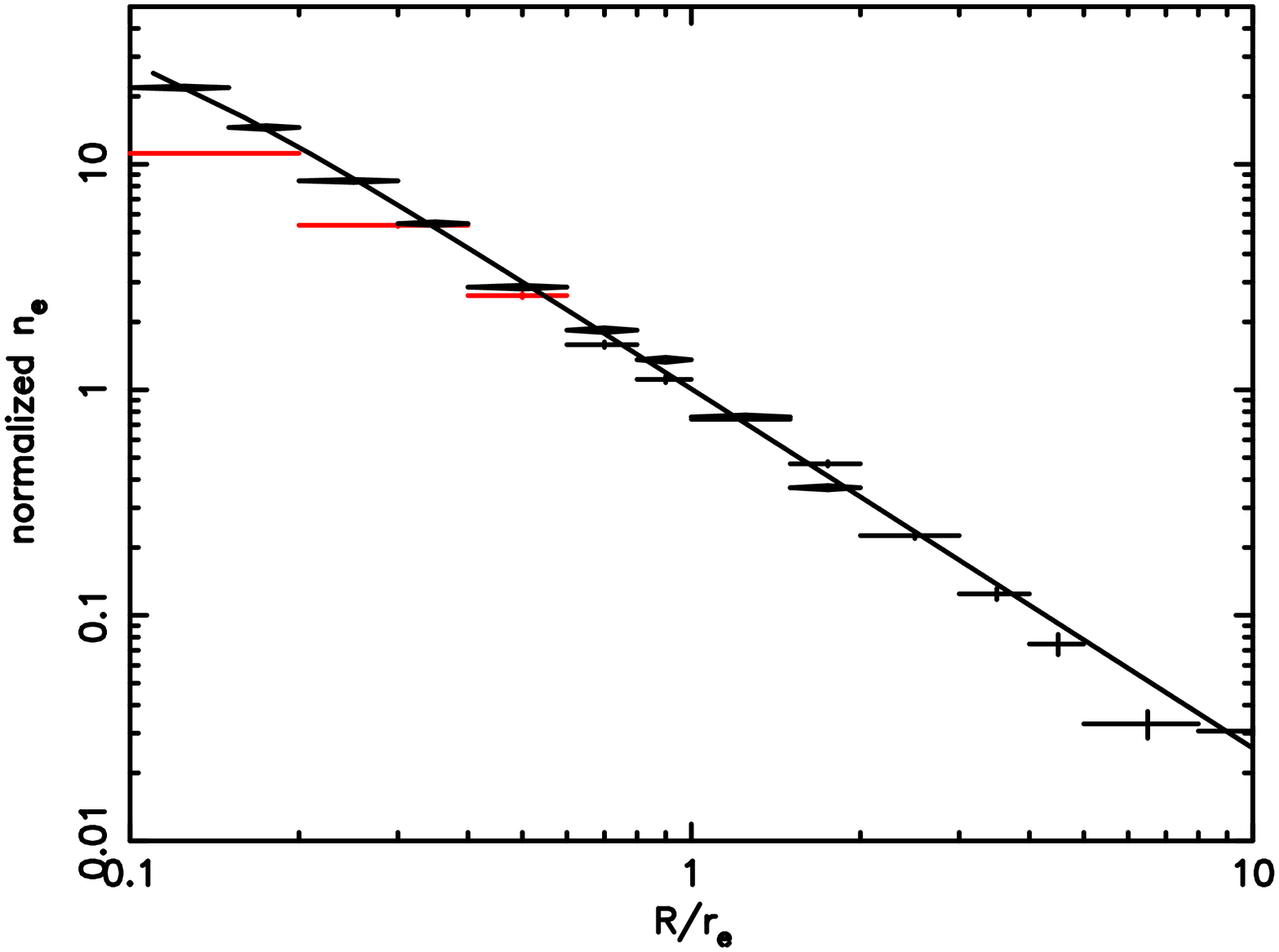}
    \includegraphics[width=4.5cm]{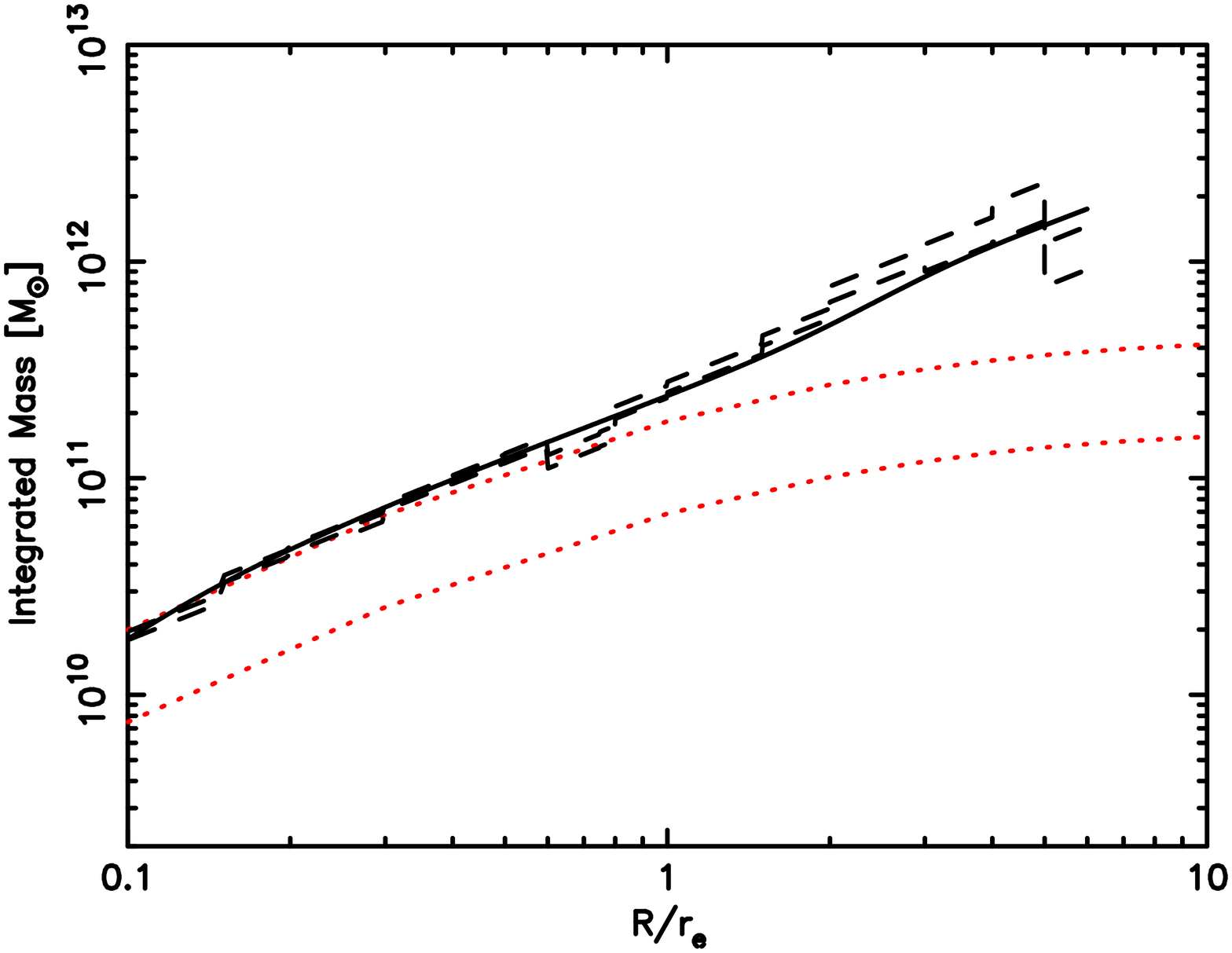}\\
    \includegraphics[width=4.5cm]{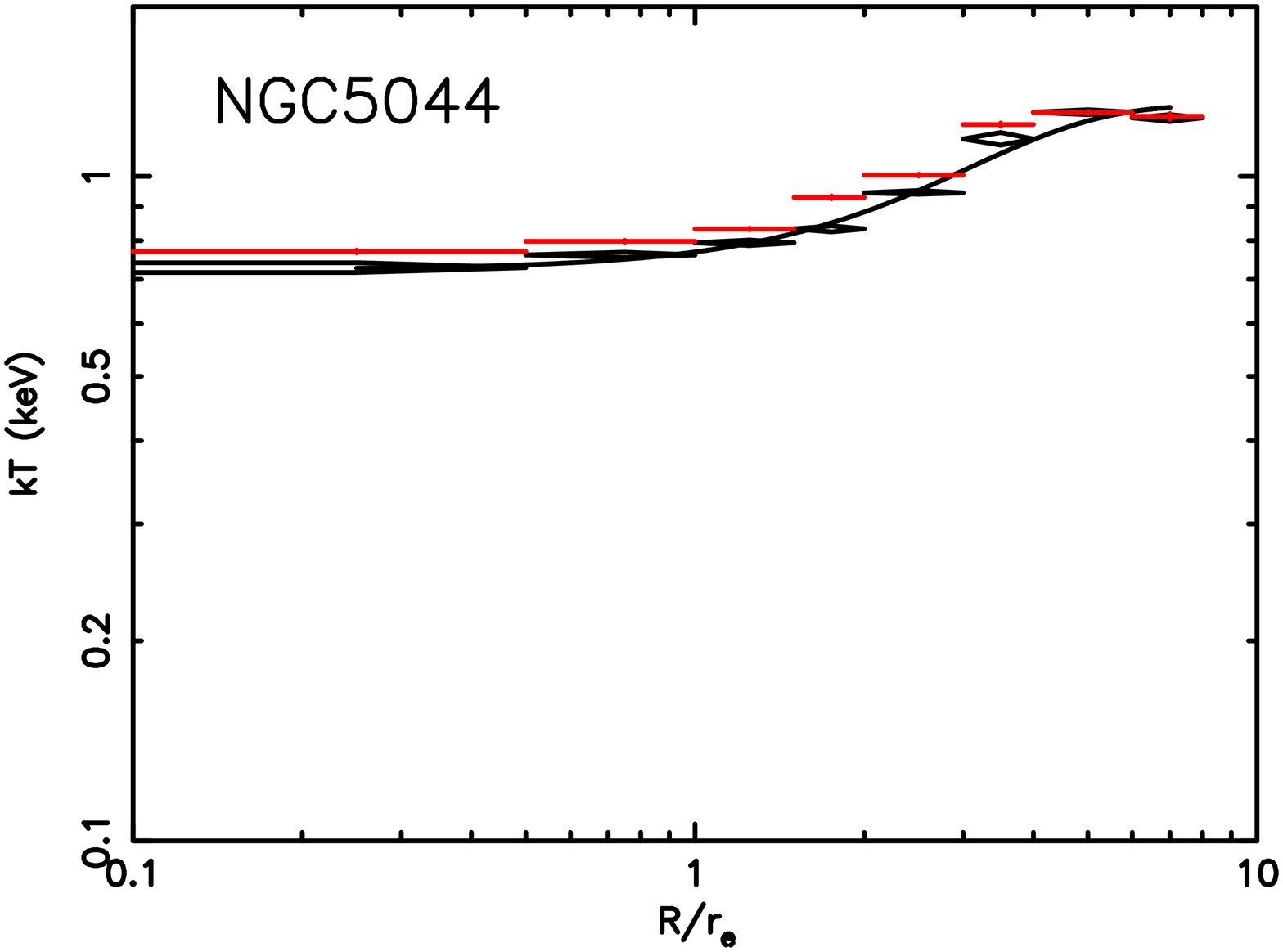}
    \includegraphics[width=4.5cm]{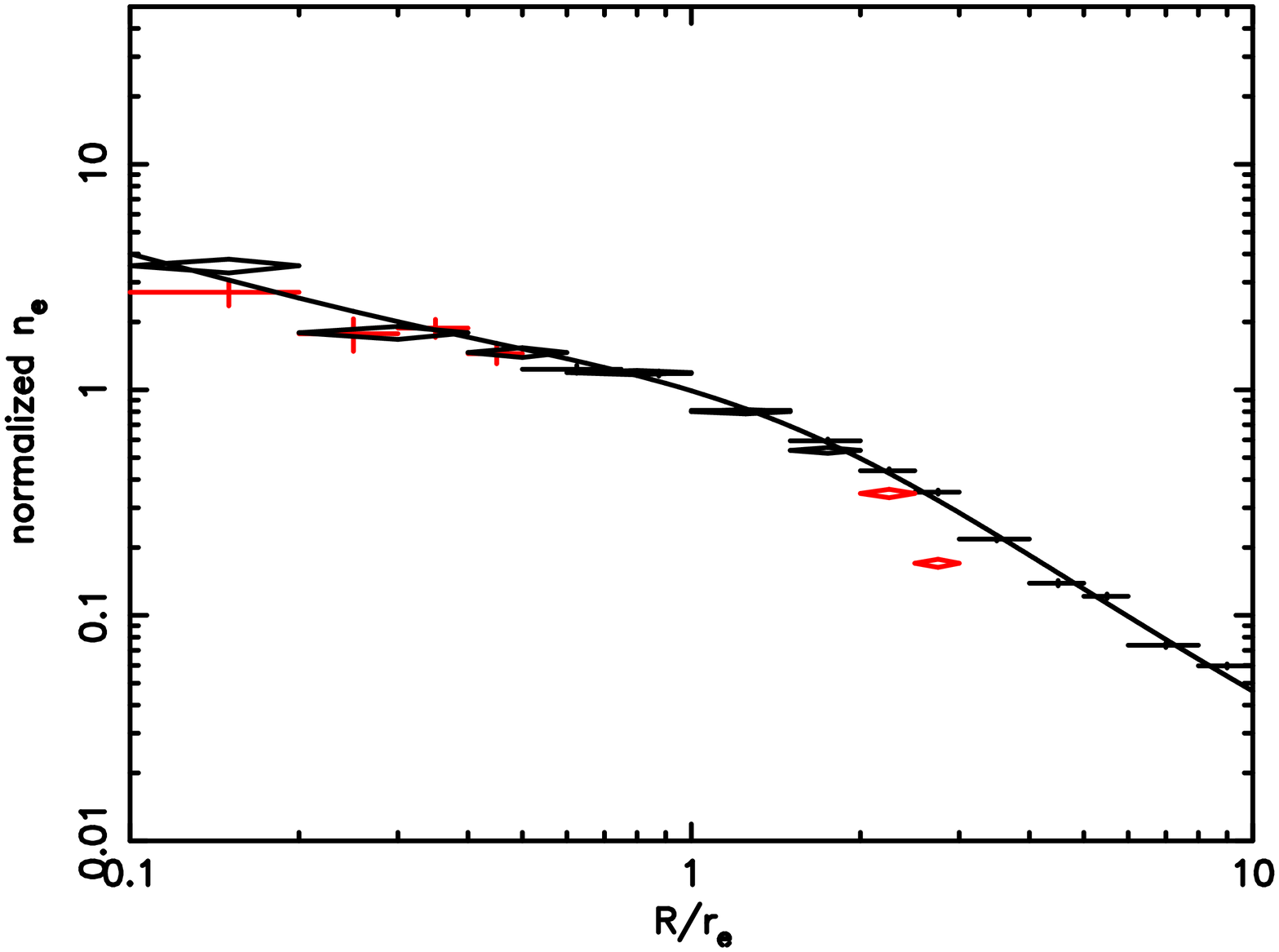}
    \includegraphics[width=4.5cm]{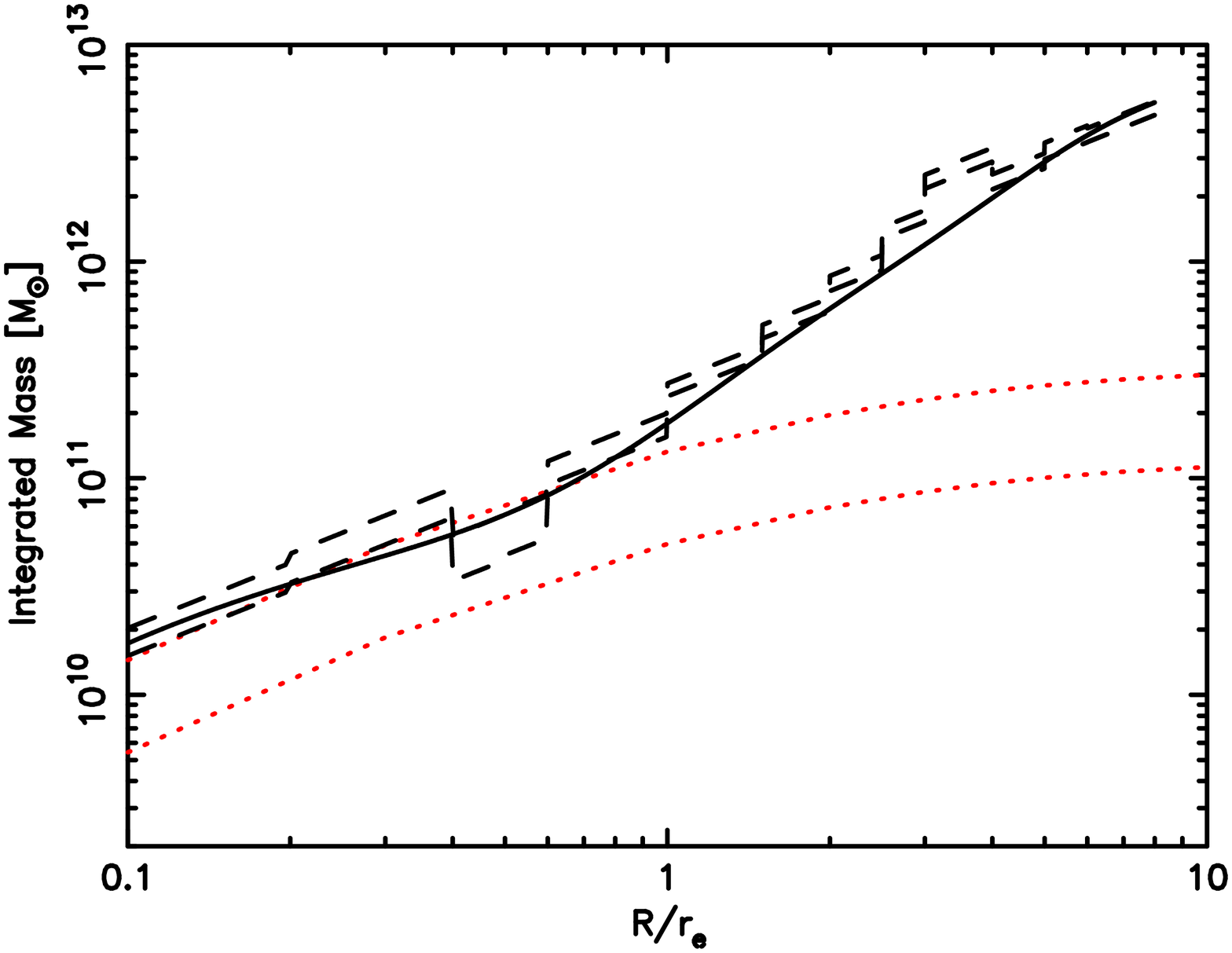}\\
    \includegraphics[width=4.5cm]{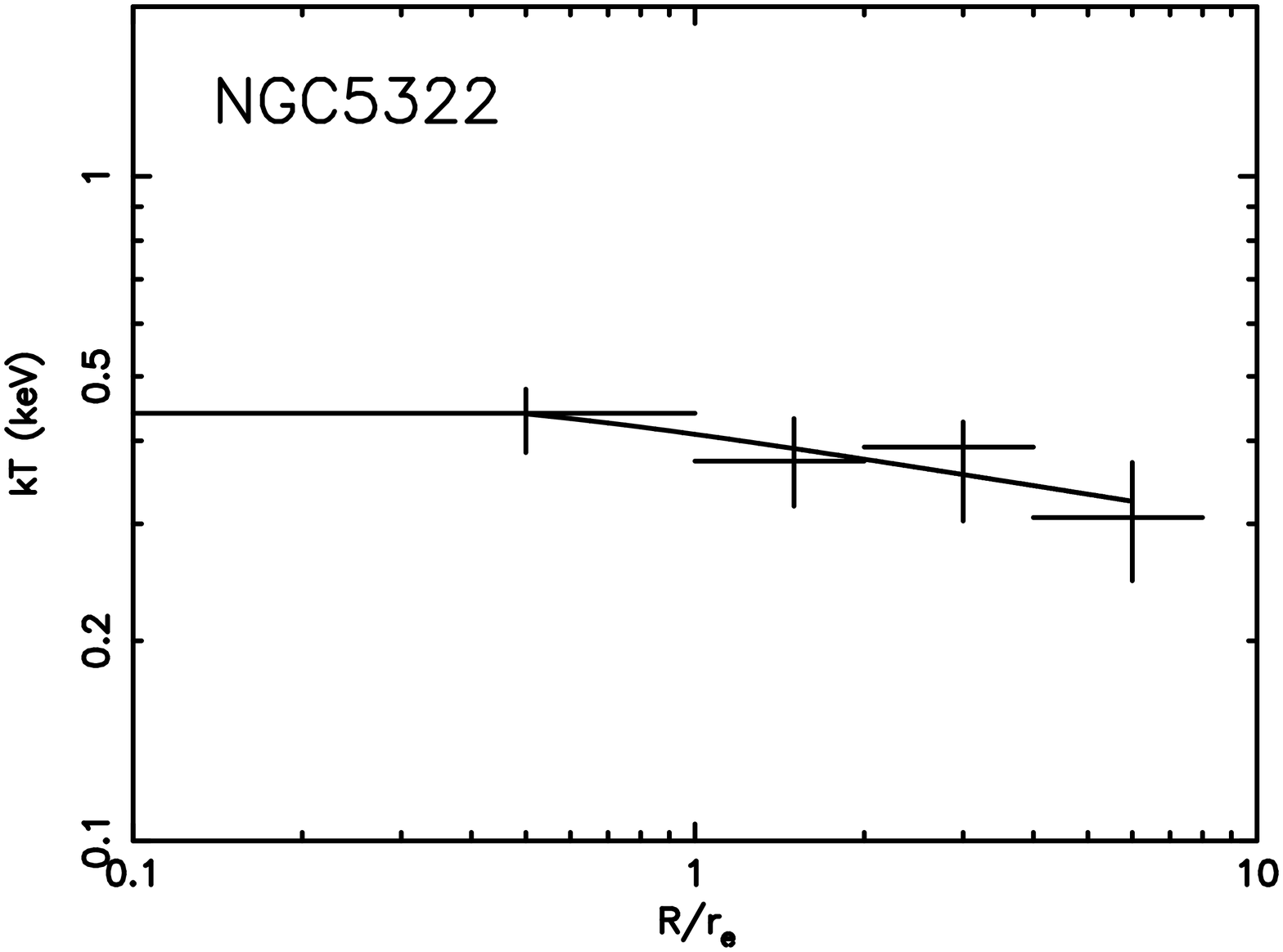}
    \includegraphics[width=4.5cm]{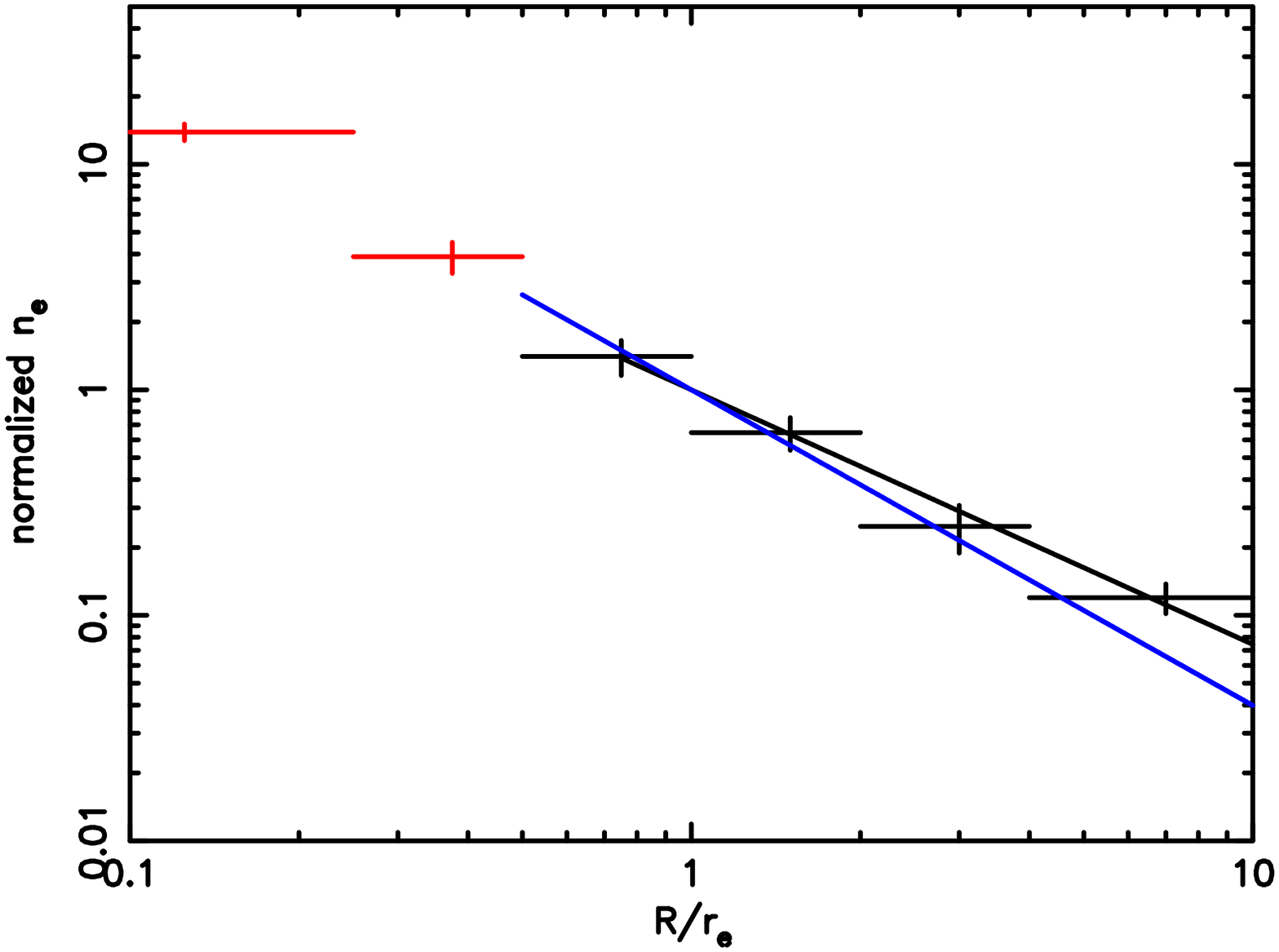}
    \includegraphics[width=4.5cm]{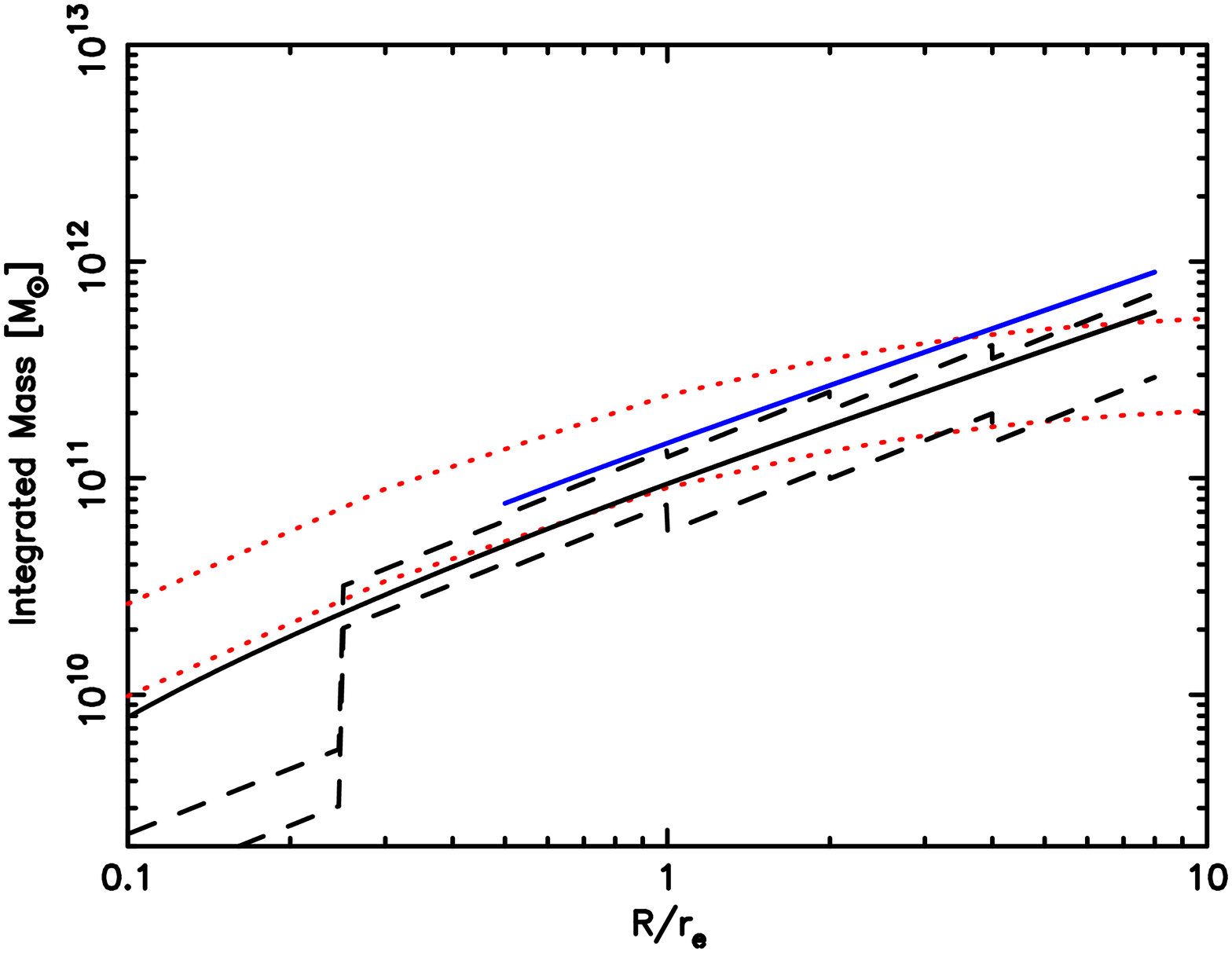}\\
    \includegraphics[width=4.5cm]{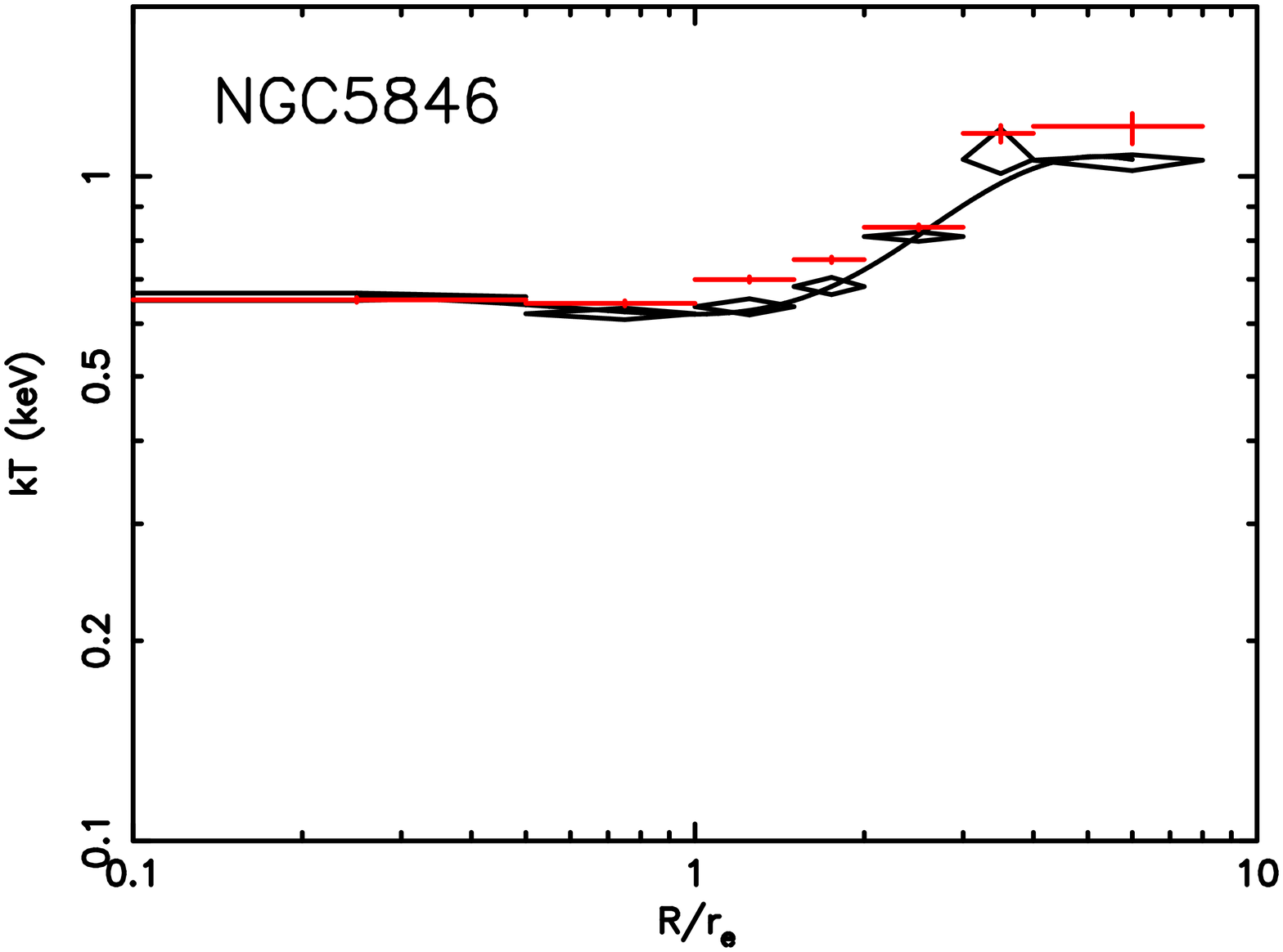}
    \includegraphics[width=4.5cm]{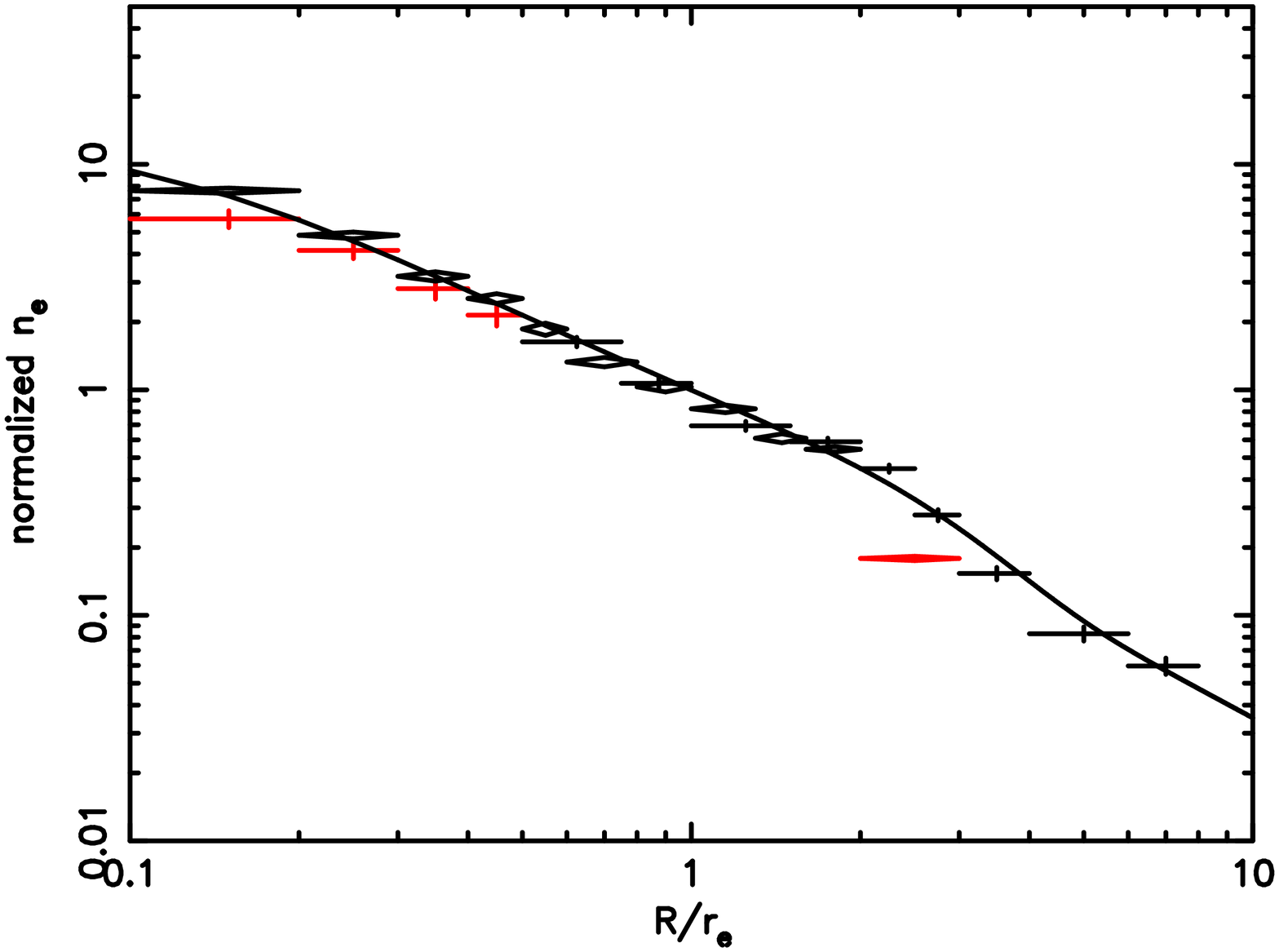}
    \includegraphics[width=4.5cm]{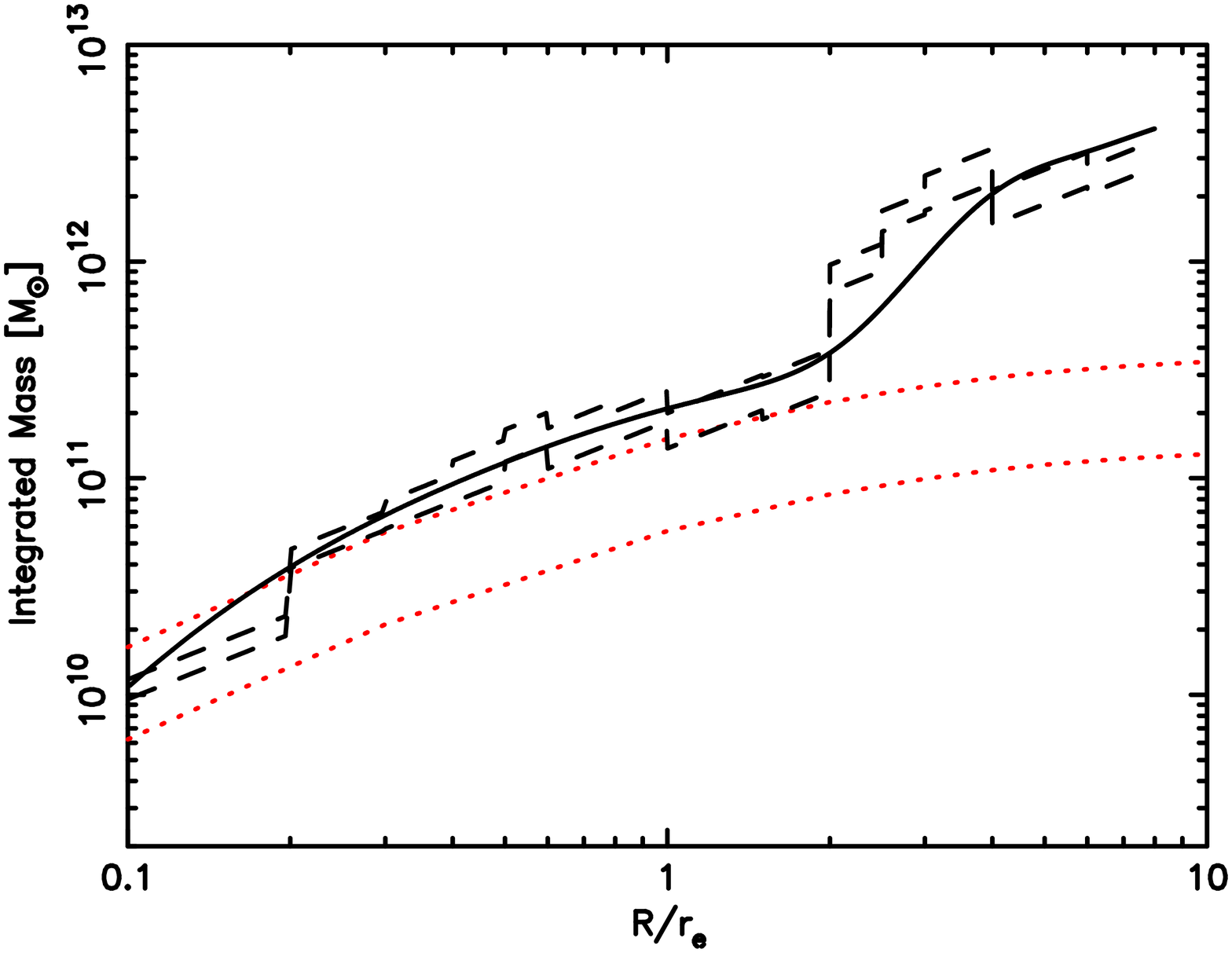}\\
 \flushleft
 {\bf Fig.~A.1.~~}(continued)
\end{figure*}

\end{document}